\documentclass{dune}
\pdfoutput=1

\input{common/preamble}
\newcommand \eos{\textsc{Eos} }
\newcommand \theia{\textsc{Theia} }

\begin{document}

\pagestyle{titlepage}


\pagestyle{titlepage}

\date{}

\title{\scshape\Large Snowmass Neutrino Frontier: \\
NF10 Topical Group Report\\
\normalsize Neutrino Detectors
\vskip -10pt
\snowmasstitle
}


\renewcommand\Authfont{\scshape\small}
\renewcommand\Affilfont{\itshape\footnotesize}

\newcommand*\samethanks[1][\value{footnote}]{\footnotemark[#1]}

\author[1] {Josh~Klein\thanks{Editor}} 
\author[2]{Ana Machado\samethanks} 
\author[3]{David Schmitz\samethanks} 
\author[4]{Raimund Strauss\samethanks\vspace{0.3cm}}

\affil[1]{University of Pennsylvania, Philadelphia, PA, USA}
\affil[2]{University of Campinas, Brazil}
\affil[3]{University of Chicago, Chicago, IL, USA}
\affil[4]{Technical University Munich, Munich, Germany}
\affil[5]{Brookhaven National Laboratory, Upton, NY, USA}
\affil[6]{Pacific Northwest National Laboratory, Richland, WA, USA}
\affil[7]{Laborat\'{o}rio de Instrumenta\c{c}\~{a}o e F\'{\i}sica Experimental de Part\'{\i}culas - LIP, 1649-003 Lisboa, Portugal    and    Faculdade de Ci\^{e}ncias - FCUL, Universidade de Lisboa, Portuga}
\affil[8]{University of Liverpool, Liverpool, UK}
\affil[9]{University College Londond, London, UK}
\affil[10]{Fermi National Accelerator Laboratory, Batavia IL, USA}
\affil[11]{Stanford Linear Accelerator Center, Stanford, CA, USA}

\author[5]{\authorcr Milind Diwan\thanks{Contributor}}
\author[6]{Christopher Jackson\samethanks}
\author[7]{Jose~Maneira\samethanks}
\author[8]{Kostantinos~Mavrokoridis\samethanks}
\author[9]{Nicola~McConkey\samethanks}
\author[10]{Tanaz Mohai\samethanks}
\author[11]{Gianluca Petrillo\samethanks}
\author[10]{Joseph Zennamo\samethanks}


\maketitle

\renewcommand{\familydefault}{\sfdefault}
\renewcommand{\thepage}{\roman{page}}
\setcounter{page}{0}

\pagestyle{plain} 
\clearpage
\textsf{\tableofcontents}





\renewcommand{\thepage}{\arabic{page}}
\setcounter{page}{1}

\pagestyle{fancy}

\fancyhead{}
\fancyhead[RO]{\textsf{\footnotesize \thepage}}
\fancyhead[LO]{\textsf{\footnotesize \rightmark}}

\fancyfoot{}
\fancyfoot[RO]{\textsf{\footnotesize Snowmass 2021}}
\fancyfoot[LO]{\textsf{\footnotesize NF10 Topical Group Report}}
\fancypagestyle{plain}{}

\renewcommand{\headrule}{\vspace{-4mm}\color[gray]{0.5}{\rule{\headwidth}{0.5pt}}}



\clearpage

\section*{Executive Summary} 

Neutrino physics spans an enormous range of energies and scales: from detection of low-energy cosmic neutrinos; to keV-scale recoils in coherent neutrino scattering; to MeV-scale solar, reactor, and neutrinoless double beta decay events; to GeV and TeV-scale detection of neutrinos from accelerators and the atmosphere; to cosmic sources in the PeV--ZeV range.  While any particular experiment tends to focus on just one or two detection approaches, the great breadth of neutrino physics means that there is an equally broad spectrum of neutrino detection technologies and methodologies.

At any one time, there are a dozen or more medium- to large-scale neutrino detectors operating worldwide, several more in the design or construction phases, and many future detectors planned. Beyond this are a diverse set of smaller scale prototypes distributed across universities and labs. 

The focus in this report is on new technologies and approaches that will enable future neutrino detectors, and thus experiments that are already built and running, are under construction, or for which technical designs exist are not discussed in great detail.

\paragraph{Recommendations for Next-Generation Neutrino Detectors}

While there are many exciting detectors and enabling technologies described in this report, there are a few ideas that have had a particularly large community interest, and we formulate these into a set of recommendations below:
\begin{itemize}
    \item {\bf \ul{Broaden the Noble Liquid and Gas Physics Program:}} Many ideas are being pursued for improving LArTPCs, including new charge readout technologies, the use of underground argon for low-background physics, the addition of various dopants (xenon, photo-ionizing) to increase photon or charge yields, or light traps like the ARAPUCAs for improved light detection.  Many of these ideas may enable a broader physics program than can be done with existing detector designs.
    \item {\bf \ul{Pursue hybrid Scintillation/Cherenkov Detectors:}} Many different technologies are being developed for these, including water-based liquid scintillator, slow fluors, fast timing with LAPPDs and other devices, and spectral photon sorting with dichroicons.  At very large scales like the proposed Theia detector, these could have very broad physics programs.
     \item {\bf \ul{Optimize low-threshold neutrino detectors:}} To expand the ever-growing CE$\nu$NS  program and to fully exploit the physics reach of CE$\nu$NS in the next decade requires not just lowering energy thresholds but improving background rejection techniques, understanding detector responses at the eV-scale, and moving toward larger detector masses. The enabling technologies have many synergies with direct neutrino mass measurements and recoil-imaging directional dark matter detectors.
    \item {\bf \ul{Develop Technologies for Neutrino Detection at the TeV Scale and Beyond:}} Observations of high-energy neutrinos at large neutrino telescopes have provided a wealth of physics and multi-messenger astrophysics, and new opportunities for studying neutrino interactions at the LHC, including possibly tagging the production vertex, are particularly exciting.  Enabling technologies for neutrino telescopes include radar echo detection, Askaryan effect detection, and ever-larger scale optical detection, while at the LHC they include picosecond timing synchronizations, intelligent triggering, and high-resolution tracking.
    
    \item {\bf \ul{Co-develop neutrino and dark-matter detectors:}} Both the development of noble liquid and gas detectors and low-threshold detectors of various technologies are likely to lead to 
    new technologies that are useful for both the Neutrino Frontier and the Cosmic Frontier.
\end{itemize}

\paragraph{Detector and Technology Summary} 
Through the Snowmass LOI process, plus several community discussions, it has become clear that many ideas for new neutrino technologies and detectors fall into a few broad classes:
\begin{itemize}
    \item {\bf Noble gas and liquids}: New detector ideas here typically focus on improved reconstruction (e.g., by using pixellated TPCs or high-pressure gas), and broad physics programs accomplished by effectively moving to lower energy thresholds.
    \item {\bf``Photon-based'' detectors}: Enabling technologies here are various approaches to development of hybrid Cherenkov/scintillation detectors, new approaches to segmentation, and new isotopic loading techniques for particle ID, neutrino detection, and neutrinoless double beta decay.
    \item {\bf Low-threshold neutrino detectors}: These include a broad array of technologies many of which are aimed at detection of low-energy nuclear recoils created by coherent neutrino-nucleus scattering (CE$\nu$NS).
    \item {\bf High-energy neutrino detection}: New approaches here include large-scale detectors for neutrinos from cosmic sources including radar echo technology, and space-based geo-observations, and new techniques for detecting and vertexing neutrinos from the LHC.
    \item {\bf Novel Detector Ideas}: Some of the detectors fall into no particular technical category but are nevertheless novel in themselves, putting together in some cases existing technologies to create new neutrino physics capabilities.
\end{itemize}

\paragraph{Liquid Noble and Gas Detectors}

Since the last Snowmass, there has been enormous progress and diversity in detectors that use either noble liquids or gases as their target material. Such detectors have been used for dark matter searches by experiments like DarkSide~\cite{darkside}, DEAP~\cite{DEAP}, and LZ~\cite{LZ}, for neutrinoless double beta decay with experiments like EXO-200~\cite{EXO-200}, and as part of both the Short-Baseline Neutrino program at FNAL~\cite{MicroBooNE:2015bmn} with MicroBooNE, SBND, and ICARUS, and the DUNE experiment at SURF~\cite{DUNE:2020lwj}.  There are many new ideas for doing more with these detectors, from scaling upwards (e.g., nEXO~\cite{nEXO:2017nam}), to improving reconstruction capabilities by using pixelated charge readout in LArTPCs (e.g., LArPix, QPix) or 
moving to high-pressure (e.g. ND-GAr, NEXT) or atmospheric-pressure (e.g. CYGNUS) TPCs, to lowering thresholds by using underground sources of LAr.  There is also interest in new approaches to detecting scintillation light in these detectors, including pixelated SiPM arrays, photo-ionizing dopants, or by increasing light coverage with ARAPUCA light-traps deployed on TPC cathodes using power-over-fiber. A common theme in the LArTPC community is the pursuit of a broader physics program than currently planned, including low-energy solar neutrinos or neutrinoless double beta decay, much of which would be leveraged by underground or low-background argon.  

    Thus some of the highest technical priorities in this area are:
    \begin{itemize}
    \item Larger-scale production of underground argon sources, or ways to remove $^{39}$Ar and $^{42}$Ar
    \item Development of pixelated charge readout for TPCs
    \item Development of pixelated light readout for TPCs
    \item Investigations of new ways for photon detection in liquid noble detectors, including photo-ionizing dopants
    \item Advanced triggering schemes at low-energies, including machine learning techniques
    \end{itemize}

\paragraph{Photon-based Detectors} Neutrino detectors that use photons as their primary carrier of neutrino interaction information have an incredibly successful history in neutrino physics. They include Cherenkov detectors like IMB, Kamiokande, SNO, and Super-Kamiokande, and scintillation detectors like KamLAND, Double CHOOZ, Daya Bay, RENO, BOREXINO, NOVA, PROSPECT, and SNO+. Of particular interest in the past decade or so are {\it hybrid} Cherenkov/scintillation detectors, which can detect and discriminate between Cherenov and scintillation photons (``chertons'' and ``scintons'') in the same detector, thus allowing a very broad neutrino physics program.  Enabling technologies that would allow this have been and continue to be developed over the past decade, including new materials like water-based liquid scintillator and slow fluors, new and faster devices like LAPPDs, and spectral photon sorting with devices like dichroicons.  These new technologies, employed either separately or in concert, make large-scale hybrid detectors a real possibility.  The most developed of these ideas is the proposed Theia experiment, which could sit in the LBNF beam at SURF.  

In addition to hybrid Cherenkov/scintillation detectors, new ideas for segemented detectors have also been developed.  The most developed of these is LiquidO, which would use scintillator with short scattering lengths and an array of fiber optics to create a ``self-segmented'' detector that could allow precision track reconstruction at both high and low energies.  The SLIPS idea, which removes physical segmentation by floating a scintillator volume, could allow very low-background experiments by eliminating radioactive sources in the volume.

    Looking even further ahead, some of the highest technical priorities seen in this area, which could be explored between this Snowmass and the next, are:
    \begin{itemize}
        \item Lower-cost, large-area, high-quantum efficiency ($\sim$40\%) photon sensors
        \item Lower-cost, fast-timing ($\leq$100 ps) photon detectors
        \item Dichroic filters that can be deposited on non-flat surfaces and with sharper cut-on/cut-off curves, even at high incidence angles
        \item Narrow-band fluors for liquid scintillators
        \item High-yield scintillators with attenuation lengths $>$ 40~m
        \item High-yield ``slow'' ($\geq10$ ns risetime) fluors
        \item Low-background fiber optics
        \item New approaches to radiologic background reductions, beyond levels seen in Borexino
    \end{itemize}

\paragraph{Low-threshold Neutrino Detectors}

The development of low-threshold neutrino detectors with eV-scale resolution has become a priority in neutrino physics during the last decade since it has opened up new portals for the study of neutrino properties and the search for new physics. The COHERENT program has pioneered a detector technology which enabled the first observation of CE$\nu$NS and is driving since then a blooming new research field under US leadership. This technological breakthrough triggered R\&D activities worldwide of low-threshold neutrino detectors based on a wide range of technologies. A broad and complementary CE$\nu$NS program based on small-scale experimental projects will enable precision measurements and pave the way for applications. The community will profit form multiple technological synergies with direct neutrino mass measurements and a variety of proposed approaches. To achieve the technological goals the following challenges have to be addressed:  
\begin{itemize}
    \item Improve detector thresholds towards the eV scale. 
    \item  Develop advanced techniques for suppressing backgrounds, including the community-wide observed low-energy excess. 
    \item Establish multiplexing techniques to scale-up active detector mass. 
    \item Understand the detector response at the eV scale.
    \item Increase level of automatization for applications in science, industry, and for society. 
\end{itemize}
Exploiting the strong technological connections with direct dark matter (DM) searches is essential and of mutual interest for both communities. 
The CEvNS community will profit from the ongoing R\&D efforts on low-threshold directional recoil detectors for DM searches.  
Multi-ton DM detectors will play a crucial role for the measurements of solar and supernova neutrinos in the next decade.  

\paragraph{High-Energy and Ultra-High-Energy Neutrino Detectors}

Detection of neutrinos at the TeV scale and beyond has been pioneered by big neutrino telescopes like ICECUBE and KM3NET.  The amount of physics and astrophysics and multi-messenger possibilities from these detectors is remarkably broad, and they have been exceptionally successful.  Future plans for these detectors focus on moving toward even higher energies, into the EeV and ZeV regimes, which require scales going well beyond km$^3$ or new technologies, exploiting the Askarayan effect or radar echoes off ionization trails.  In many cases, the enabling ``technology'' for these telescopes is a piece of geography: polar or Greenland ice sheets, mountain ranges that can be used as targets, etc.  At the same time, there is a new opportunity at the LHC with the ``Forward Physics Facility'' (FPF) to detect neutrinos produced in collisions, perhaps with the possibility of tagging the neutrino production vertex in a collider detector.  High priorities over the next several years for these detectors are:
\begin{itemize}
    \item  Develop further the capability of radio detection of neutrinos interacting in ice or the atmosphere.
    \item  Demonstrate at larger scales the detection of neutrinos via radar echos off ionization cascades.
    \item  Create low-cost ways of scaling to ever-larger telescopes sizes.
    \item  Create intelligent triggers for background rejection at the FPF.
    \item  Create larger-scale high-resolution tracking options for FPF neutrino events.
\end{itemize}

\clearpage
\cleardoublepage

\section{Introduction}
\label{sec:intro}

Neutrino physics spans an enormous range of energies and scales: from detection of low-energy cosmic neutrinos; to keV-scale recoils in coherent neutrino scattering; to MeV-scale solar, reactor, and neutrinoless double beta decay events; to GeV and TeV-scale detection of neutrinos from accelerators and the atmosphere; to cosmic sources in the PeV--ZeV range.  While any particular neutrino experiment tends to focus on just one or two detection technologies---say, ionization and photon detection---the great breadth of neutrino physics means that there is an equally broad spectrum of detection technologies and methodologies. At any one time, there are a dozen medium- to large-scale neutrino detectors operating worldwide, several more in the planning or construction phases, and many future detectors planned.  Beyond this are a diverse set of smaller scale prototypes distributed across universities and labs.  
    
The focus of NF10, ``Neutrino Detectors,'' is on enabling technologies in the context of neutrino detection.  We have in this report focused on new approaches that will enable {\it future neutrino detectors} and thus, experiments that already exist, or are under construction or for which advanced technical designs already exist are not discussed in great detail.  Table \ref{tbl:wps} lists the contributed Snowmass White Papers relevant to the NF10 topical group.  In addition, during the summer of 2020 the neutrino physics community submitted nearly a hundred relevant LOIs that can be found linked from the \url{snowmass21.org} website.  

\vspace{7ex}

\begin{table}[hbt!]
\begin{centering}
\small
\begin{tabular}{l|p{13cm}|l}
\hline\hline
1 & Future Advances in Photon-Based Neutrino Detectors & \href{https://arxiv.org/abs/2203.07479}{arXiv:2203.07479} \\
2 & Coherent elastic neutrino-nucleus scattering: Terrestrial and astrophysical applications & \href{https://arxiv.org/abs/2203.07361}{arXiv:2203.07361} \\
3 & Recoil imaging for dark matter, neutrinos, and physics beyond the Standard Model & \href{https://arxiv.org/abs/2203.05914}{arXiv:2203.05914} \\
4 & Low-Energy Physics in Neutrino LArTPCs & \href{https://arxiv.org/abs/2203.00740}{arXiv:2203.00740} \\
5 & SoLAr: Solar Neutrinos in Liquid Argon & \href{https://arxiv.org/abs/2203.07501}{arXiv:2203.07501} \\
6 & Low Background kTon-Scale Liquid Argon Time Projection Chambers & \href{https://arxiv.org/abs/arXiv:2203.08821}{arXiv:2203.08821} \\
7 & Measuring the Neutrino Event Time in Liquid Argon by a Post-Reconstruction One-parameter Fit & \href{https://arxiv.org/abs/2004.00580}{arXiv:2004:00580} \\
8 & Adding Stroboscopic Muon Information For Reduction of Systematic Uncertainties in DUNE & \href{https://www.slac.stanford.edu/~mpeskin/Snowmass2021/Muon_monitoring_note.pdf}{link} \\
9 & DUNE Software and High Performance Computing & \href{https://arxiv.org/abs/2203.06104}{arXiv:2203.06104} \\
10 & High-pressure TPCs in pressurized caverns: opportunities in dark matter and neutrino physics & \href{https://arxiv.org/abs/2203.06262}{arXiv:2203.06262} \\
11 & Passive low energy nuclear recoil detection with color centers – PALEOCCENE & \href{https://arxiv.org/abs/2203.05525}{arXiv:2203.05525} \\
12 & Bubble Chamber Detectors with Light Nuclear Targets & \href{https://arxiv.org/abs/2203.11319}{arXiv:2203.11319} \\
13 & The COHERENT Experimental Program & \href{https://arxiv.org/abs/2204.04575}{arXiv:2204.04575} \\

\hline\hline
\end{tabular}
\caption{Table of submitted White Papers relevant to the NF10 topical group on Neutrino Detectors.\label{tbl:wps}}
\end{centering}
\end{table}
\cleardoublepage

\section{Noble Element Detectors}

Noble element detectors play a significant role in neutrino physics today.  Several new experimental programs have been built and operated during the past few years, each one advancing the technology further, and efforts are underway now to build detectors at massive scale and take advantage of this technology's capabilities for precision neutrino physics.  Of particular interest are liquid argon time projection chambers (LArTPCs), discussed next. 

\subsection{Advanced Technologies for Liquid Argon TPCs}



Liquid argon time project chamber (LArTPC) detectors offer the possibility of individual particle tracking and fine-grained calorimetry over very large volumes, making them an ideal technology for precision neutrino physics and searches for rare phenomena.  
LArTPCs have been used in frontier experiments in both accelerator neutrino physics and dark matter searches (e.g.\;ICARUS, ArgoNeuT, LArIAT, MicroBooNE, ProtoDUNE, SBND, LArIAT, WArP, ArDM, DarkSide). Over the years, many different types of charge and light readout have been developed or proposed: single phase TPC (only liquid argon), dual phase TPC (liquid and gaseous argon which profits from charge amplification in gaseous phase), wire or PCB-based charge readout, light detection with photon multipliers and silicon sensors, etc.

The international particle physics community is currently working to build a next generation neutrino observatory and long-baseline oscillation experiment, DUNE (the Deep Underground Neutrino Experiment), based on this technology.  The DUNE design includes four massive detector modules, each containing 10-kilotons of active argon mass, located in a deep underground cavern at the Sanford Underground Research Facility (SURF) in South Dakota. The 2014 P5 report strongly endorsed this program, and in addition recommended several efforts to pave the way to DUNE by executing a set of smaller-scale liquid argon experiments to develop the technology and build up the international community and expertise for the DUNE program. This recommendation has been realized in the SBN program at Fermilab, the ProtoDUNE program at CERN, and other experimental R\&D efforts.  The eight years since the 2014 P5 Report have seen tremendous progress in the design, construction, operation, and physics outputs of the liquid argon time projection chamber neutrino detector.  

To realize the full potential of the LArTPC technology, R\&D efforts continue. Large LArTPCs are well-matched to physics at accelerator beam neutrino energies (GeV-scale). A particular motivator for many current R\&D efforts is to push the reach of the detectors to lower and lower energies (MeV-scale) to optimize the performance for supernova neutrinos, solar neutrinos, and even the possibility of searches for neutrinoless double beta decay in a large-scale detector. DUNE is comprised of four 10-kiloton LArTPC modules (active mass).  Construction has started on the first modules, but the fourth ``module of opportunity'' presents a platform to further enhance the physics capabilities of the experiment in exciting ways through successful R\&D efforts, now underway. This section highlights some of those efforts.    

\subsubsection{New Charge Readout Technologies}

Charge readout is fundamental to the TPC and improved performance (signal-to-boise, granularity, etc.) can lead directly to improved particle reconstruction, energy resolution, and general physics performance.  Many ideas are being actively pursued within the neutrino research community.  

\begin{figure}[t!]
\centering 
\includegraphics[width=0.7\textwidth]{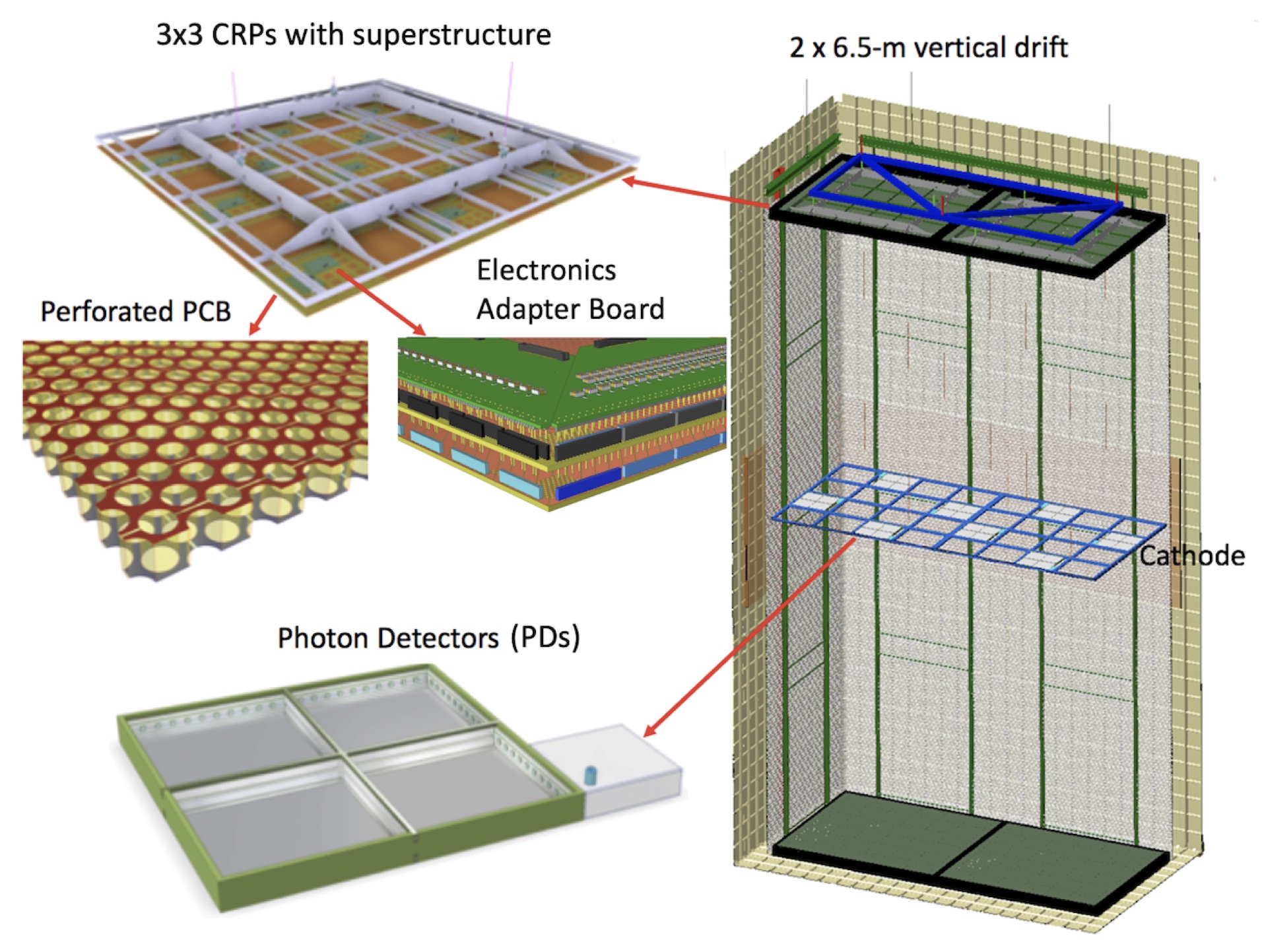}
\caption{DUNE vertical drift design with PCB-based charge readout.} 
\label{fig:vertical-drift}
\end{figure}

\paragraph{Charge Readout Planes (CRPs) and the Vertical Drift:} The first DUNE far detector module will use a single-phase TPC technology, horizontal drift field, and wire-based charge readout, as has been already successfully demonstrated in ICARUS, MicroBooNE, ProtoDUNE-SP~\cite{}, and other detectors. For the second DUNE far detector module, however, a new design, known as Vertical Drift, has been recently under development (see Snowmass LOI~\cite{verticaldrift:2020loi}). Figure~\ref{fig:vertical-drift} shows the concept.  A horizontal cathode plane is placed at mid-height in the active volume of the cryostat, dividing it into two vertically stacked equal volumes, each 6.5m in height. The anode planes are constructed of perforated printed circuit boards (PCBs) with etched electrodes forming a three-view charge readout. The top anode plane is placed close to the cryostat top, just below the surface of the LAr, and the other is located as close to the bottom of the cryostat as possible. Ionization electrons will drift vertically towards the anode plane at the end of the drift volume in which they are released. The vertical drift design offers a slightly larger instrumented volume compared to the first module design as well as a substantially simpler, more cost-effective construction and installation due to its geometry and structure.  

\begin{figure}[t!]
\centering 
\includegraphics[width=0.3\textwidth]{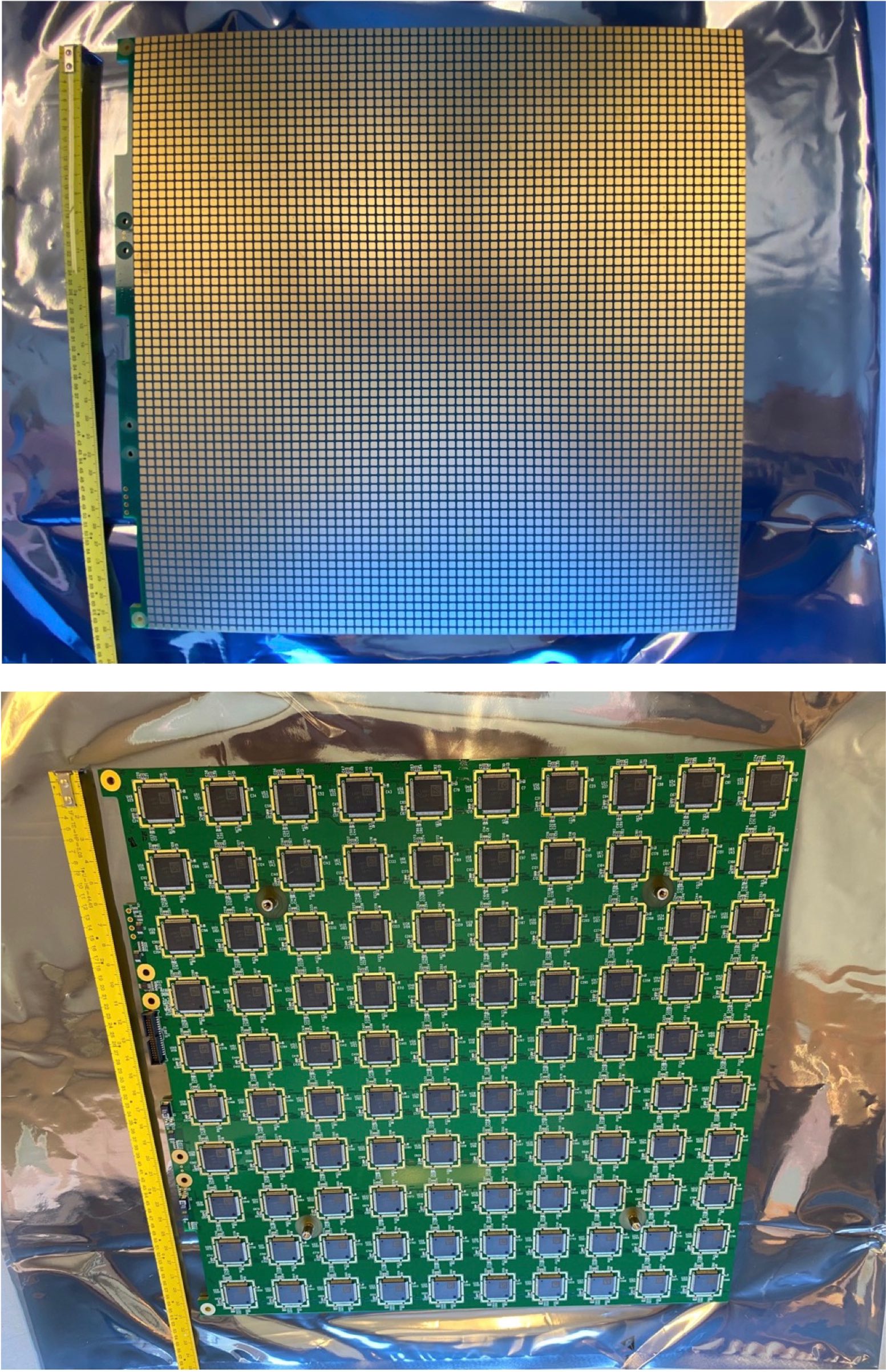}
\includegraphics[width=0.65\textwidth]{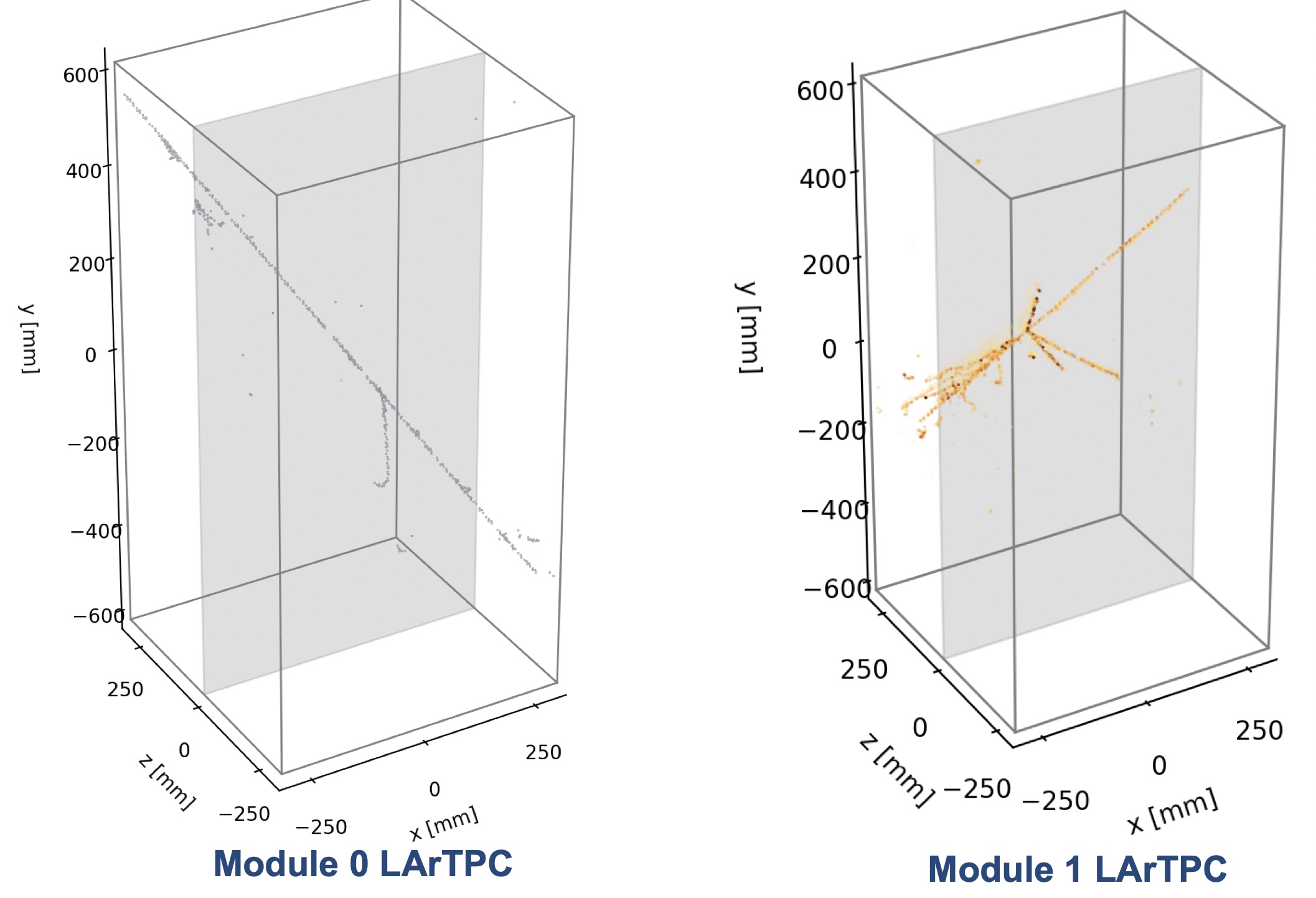}
\caption{Left: A single LArPix-v2 pixel anode panel. Right: Typical raw data from the LArPix charge readout system as deployed in the ArgonCube prototype detectors.} 
\label{fig:argoncube}
\end{figure}

\paragraph{LArPix:} Pixelated charge readout is a highly attractive alternative to traditional wire-based readouts as pixels can provide unambiguous 3D imaging in LArTPCs. To prove viable in a large LArTPC, such as the DUNE far detectors, pixelated readout systems must meet stringent requirements on noise, power, reliability in a cryogenic environment, and scalability to systems with order $10^9$ channels. The first-generation LArPix system provided a proof-of-concept for pixelated LArTPC readout by 2018~\cite{Dwyer_2018}, and since then the R\&D effort has focused on scalability and use of commercial production of components.  The LArPix system is now being tested at a scale relevant for the DUNE near detector in the ArgonCube prototype detectors. Figure~\ref{fig:argoncube} shows a LArPix anode panel and two examples of typical raw data from cosmic ray interactions in ArgonCube.  Further R\&D and prototyping will be needed to prepare a pixel-based readout system for use at DUNE far detector scale.   

\paragraph{Q-Pix:} Another approach to pixelated LArTPC readout is the Q-Pix design. The Q-Pix approach aims at low readout thresholds, vast reduction in data rates, and maximizing the protection against single point failures.  The design explicitly targets an underground environment, such as a DUNE far detector, where most of the time there is nothing of interest happening, but you need to be able to instantly capture any signals coming above threshold.  Technical details are available in~\cite{Nygren:2018rbl}, and impacts on the physics performance of pixel-based detectors vs. wire readout have been studied in~\cite{Adams_2020} and \cite{Q-Pix:2022zjm}.    

Consortia of institutions have formed around advancing the LArPix and Q-Pix designs for pixel charge readout, and in a joint Snowmmass submission these groups advocate for a coherent R\&D collaboration to achieve the scalable designs needed for future large-scale LArTPC detectors~\cite{pixels:2020loi}.

\begin{figure}[h!]
\centering 
\includegraphics[width=0.6\textwidth]{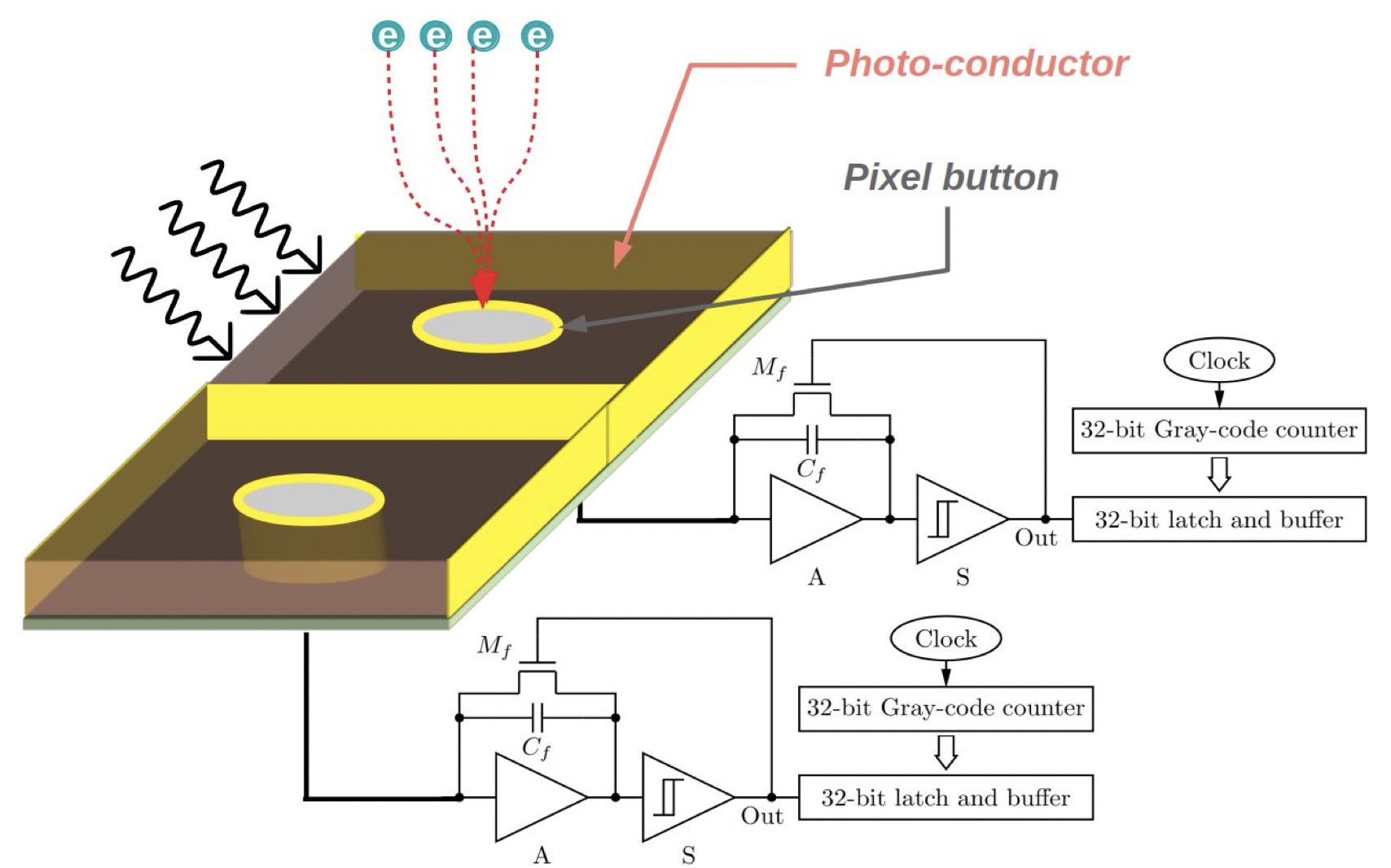}
\caption{Q-Pix pixel readout combined with a photoconducting material coating could lead to a pixel system that is simultaneously sensitive to ionization charge and VUV photons.} 
\label{fig:multi-modal}
\end{figure}

\paragraph{Multi-modal Pixels:} One challenge presented by pixel-based charge readouts is that the anode is then opaque, preventing the standard approach (in wire-based TPCs) of detecting scintillation light from behind the anode plane.  One rather elegant solution would be a pixel plane that is simultaneously sensitive to both VUV photons and charge, and R\&D is ongoing~\cite{multimodalpixels:2020loi}. Figure~\ref{fig:multi-modal} shows a basic concept starting from the Q-Pix pixel design.  A central pixel is sensitive to ionization charge, but there is an additional material coating on the pixel plane that converts VUV light also to charge, and then the same (or similar) electronics reads out the signal from the photoconverted charged.  Further R\&D is needed to produce a viable design, but such a dual mode pixel system would be an important advance for TPCs.   

\paragraph{Electron Multiplication:} Another potential enhancement to charge readout in large LArTPCs that is under development involves multiplication of the drifting electrons before the amplification stage of the readout electronics~\cite{multimodalpixels:2020loi}.  The goal of such multiplication is to improve the energy sensitivity of a large LArTPC to very low-energy physics signals down to the keV scale.  Multiplication has been used previously in gaseous TPCs, and an R\&D effort has been initiated to pursue controlled and stable electron proportional multiplication of drift electrons directly in liquid argon. The approach consists of using sub-micrometric anodic electrodes, a scaled down version of the geometries successfully adopted in gaseous TPCs, in order to generate a local electric field large enough (> 100 kV/cm) to trigger the proportional multiplication of charge carriers.  Different anode geometries are being explored in a controlled test-stand to quantify gain at different strengths of the electric field, and a simulation toolkit capable of studying the potential for amplification in a variety of anode geometries has also been developed, with promising results. Transitioning from proof-of-principle demonstrations with simplified anode geometries to scalable readout sensors, and appropriate readout electronics, will require continued R\&D going forward.  

\paragraph{Dual Readout (ion detection):} The dual-readout TPC concept refers to a high pressure gaseous TPC capable of collecting charge from both the ionization electrons at the anode and the positive ions at the cathode. Due to the lack of diffusion effects on the heavier ions, the use of positive ions collected at the cathode would push the intrinsic physical resolution of such a chamber in the 10-100 micron region. The challenge associated with this scheme is the development of a sensor that can reliably detect slow positive ions with the required granularity. Possible technology solutions are discussed in~\cite{dual-readout:2020loi}. Detection at micron-scale pitches in large detectors has the potential to push dark matter detection under the neutrino floor and could enable an exploration of $\nu_\tau$ interactions, which is essential to testing the unitarity of the neutrino mixing matrix.


\subsubsection{Dopants to Increase Charge Yields}

When energy is deposited into liquid argon (LAr), it is split between ionization and scintillation signals. In pure LAr at a fixed electric field, the ratio of ionization to scintillation depends on the amount of energy deposited. For low energy, stopping particles, or very heavy particles, understanding the fraction of energy that creates scintillation light is critical for reconstructing the particle's energy. The NEST Collaboration~\cite{nesteres} has shown that LArTPCs can improve the energy reconstruction of MeV-scale electrons by augmenting the charge measurements with a light collection efficiency near 50\%. Collecting 50\% of the scintillation light in a large LArTPC is challenging due to the large detector size, isotropic emission, and the requirement to wavelength shift the light. A different concept has been proposed in Refs~\cite{Mastbaum:2022rhw} and~\cite{Caratelli:2022llt}, which is to dope the LAr with a photosensitive dopant~\cite{psdopants,icDope}, which directly converts the isotropic scintillation signal to a directional ionization signal with high efficiency, with estimates up to 60\%~\cite{psdopants}.  Recently, the capabilities of LArTPCs with extended MeV-scale energy resolution have been explored, and it has been found that LArTPCs could, with a few R\&D advances, reach normal-ordering sensitivity to neutrinoless double-beta decay with xenon doping and the introduction of these photosensitive dopants~\cite{Mastbaum:2022rhw}.

\begin{figure}[t!]
\centering 
\includegraphics[width=0.9\textwidth]{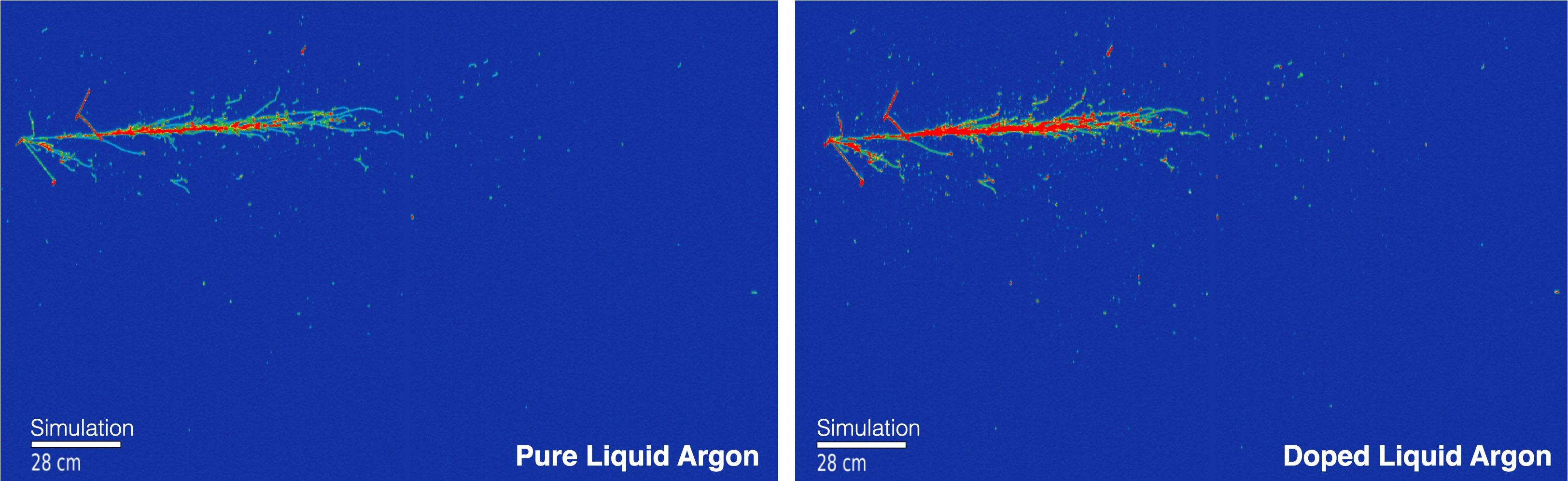}
\caption{A simulation of a GeV-scale electron antineutrino simulated in (left) pure LAr and (right) with a photosensitive dopant. This simulation utilizes the LArSoft~\cite{larsoft} simulation framework, and the photosensitive simulation converts the number of scintillation photons into ionization charge at the point of creation. This conversion leads to larger signals, a more linear detector response, and enables the visualization of a larger amount of soft deposits from neutron and low energy photon scatters.} 
\label{fig:photoionizing}
\end{figure}

The introduction of photosensitive dopants would lead to an enhanced ionization signal, especially where scintillation signals are large. This is expected to improve the energy reconstruction of low-energy signals~\cite{nesteres,Mastbaum:2022rhw}. It would also lead to a more linear detector response for highly quenched processes. Together these would lead to an improved reconstruction of alpha particles and low-energy protons. When looking toward GeV-scale neutrino interactions, these improvements could lead to improved neutrino energy reconstruction by improving vertex energy reconstruction, neutron and low energy photon reconstruction efficiency, and shower energy reconstruction, as well as reducing thresholds.  These gains in LArTPC GeV-scale performance coupled with the MeV-scale enhancements highlight the need for a robust R\&D program to enable this technique for use in large LArTPCs. The four primary areas of focus of such R\&D would be (1)~understanding the optimal doping strategy at the MeV and GeV scales, (2)~understanding the impact of dopants on long term detector stability, (3)~understanding the triggering and timing of a LArTPC with limited scintillation information, and (4)~studying if any light survives the photoconversion process~\cite{Nakajima:2015meb}. 

\subsubsection{Optical TPCs and the ARIADNE Program}

The ARIADNE program is dedicated to the development of optical readout of dual-phase LArTPCs, as a cost-effective and powerful alternative approach to the existing charge readout methodology. As first demonstrated in the 1-ton dual-phase ARIADNE detector, the secondary scintillation (S2) light produced in THGEM holes can be captured by fast TimePIX3 cameras to reconstruct in 3D the primary ionisation track.

\begin{figure}[ht]
\centering 
\includegraphics[width=0.9\textwidth]{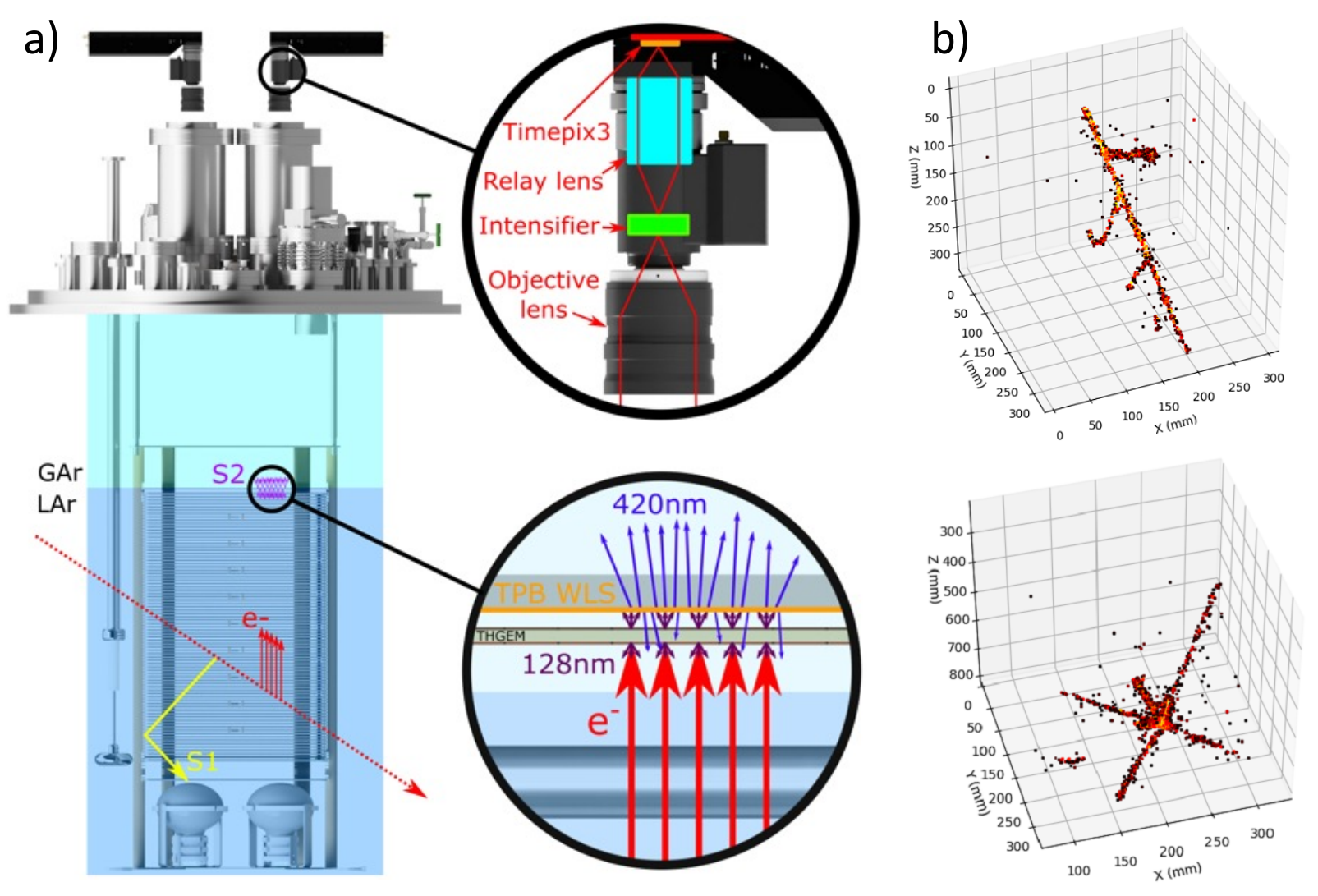}
\caption{a) Detection principle of dual phase optical TPC readout with TimePIX3 camera, first demonstrated in the 1-ton ARIADNE detector. b) LAr interactions from cosmics. Figures taken from~\cite{Hollywood_2020} and \cite{Roberts_2019}.} 
\label{fig:ariadne-1}
\end{figure}

The operation principle of a dual phase optical TPC read out with a TimePIX3 system is shown in Figure~\ref{fig:ariadne-1}. When a charged particle enters the LAr volume it causes prompt scintillation light (S1) and ionisation. The free ionisation electrons are drifted in a uniform electric field to the surface of the liquid. A higher field induced between an extraction grid and the bottom electrode of the THGEM extracts the electrons to the gas phase. Once in gas, the electrons are accelerated within the 500\,$\mu m$ holes of the THGEM at a field set between 22-31 kV/cm. As well as charge amplification, secondary scintillation (S2) VUV light is produced. The light is shifted with a TPB coated sheet to 430 nm and then detected by cameras mounted on optical viewports above the THGEM plane. 

Originally optical readout was tested with EMCCD cameras within the 1-ton ARIADNE detector at the T9 charged particle beamline at CERN~\cite{Hollywood_2020} and later was upgraded with fast TimePIX3. Within the TimePIX3 camera assembly, a lens coupled to a Photonis Cricket image intensifier ($\approx$ 33\% quantum efficiency at 430 nm) boosts the S2 light signal. The amplified photons hit light sensitive silicon, bump bonded to a TimePIX3 chip (55 µm, 256 × 256 pixel array). The camera measures simultaneously 10-bit Time over Threshold (ToT) and Time of Arrival (ToA); ToT allows accurate calorimetry and ToA gives accurate timing (1.6 ns resolution). The TimePIX3 chip then sends a packet containing 4 pieces of information: x and y pixel position, ToA (z) and ToT allowing for full 3D reconstruction using a single device. The camera provides a “Data driven readout” where pixels are read out asynchronously allowing for very efficient sparse readout; the maximum readout rate is 80 Mhits/s. The high readout rate, natively 3D raw data and low storage due to zero suppression make TimePIX3 ideal for optical TPC readout.

\begin{figure}[ht]
\centering 
\includegraphics[width=0.9\textwidth]{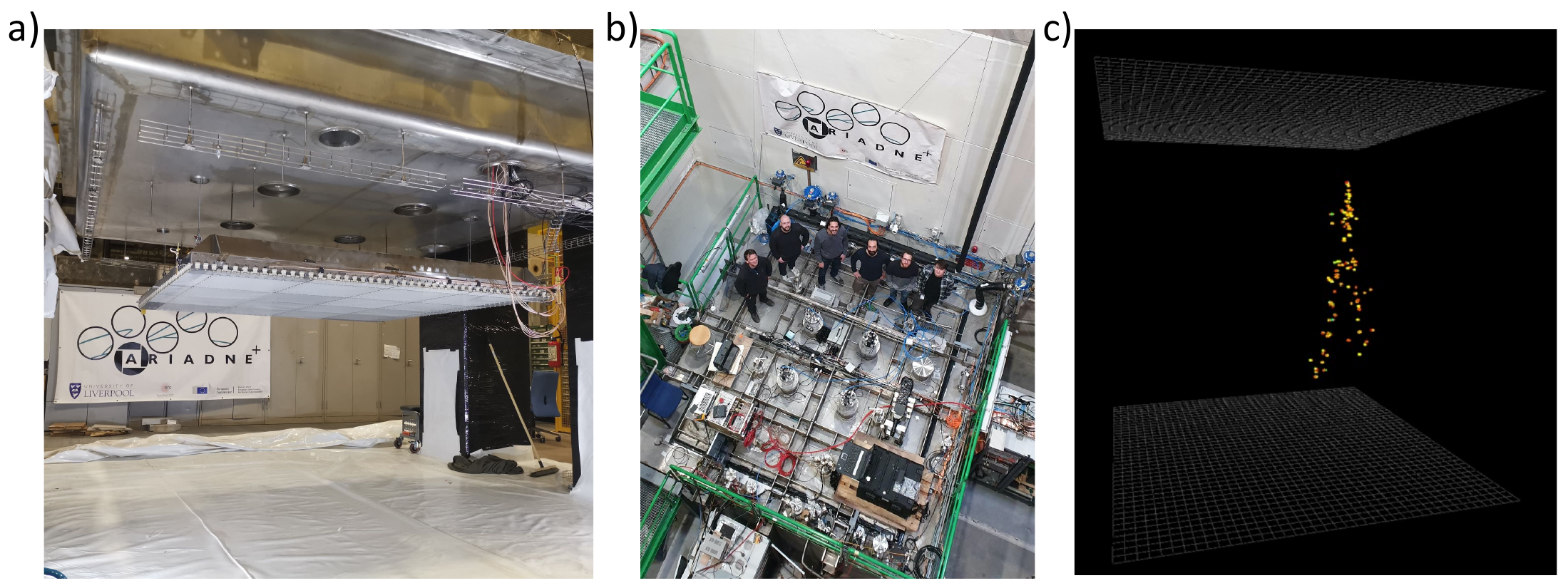}
\caption{a) The LRP under the cryostat lid. b) The ARIADNE+ team on top of the cryostat. c) A LAr interaction.} 
\label{fig:ariadne-2}
\end{figure}

The TPX3camera system was first tested in low pressure CF4 gas within the ARIADNE 40 l TPC prototype~\cite{Roberts_2019}; following this demonstration, a TimePIX3 camera was mounted on ARIADNE and particle tracks from cosmic showers were successfully imaged in 3D for first time (see Figure~\ref{fig:ariadne-1})~\cite{instruments4040035}. The cameras are shown to be sensitive even to pure electroluminescence light generated at the lower end of the THGEM field; this mitigates difficulties often faced when trying to operate THGEMs at a higher field where there can be issues with stability. Use of cameras has additional benefits such as ease of upgrade as they are externally mounted, they are decoupled from TPC and acoustic noise and as, large areas can be covered with one camera bringing cost and operational benefits.

In order to demonstrate this technology further and at a scale relevant to the 10 kton DUNE detector modules, a larger-scale test (ARIADNE+) was proposed~\cite{ariadne-loi} and recently performed at the CERN Neutrino Platform. Four cameras, each imaging a 1 m x 1 m field of view were employed of which one utilised a novel VUV image intensifier negating the need for wavelength shifter. The test also showcased a light readout plane (LRP) comprising of 16, 50 x 50 cm glass THGEMs supported within an Invar welded structure. The novel manufacturing process for the glass THGEMs allows for mass production at the large-scale~\cite{app11209450}. Successful, stable operation was achieved, and cosmic muon data were collected from both the visible and VUV intensifiers showcasing optical readout technology in combination with a novel glass THGEM array. An image of the detector setup is shown in Figure~\ref{fig:ariadne-2} and publication is pending.

\subsubsection{Scintillation Detection in LArTPCs}

The LAr is known to be an abundant scintillator: it emits about 40 photons per keV of energy deposited by minimum ionizing particles. The scintillation proceeds through the formation of the excited dimer Ar$_2^*$ and in particular the photons are emitted by the de-excitation of the lowest lying singlet and triplet states $^1\Sigma_u$ and  $^3\Sigma_u$. The transition from the $^1\Sigma_u$ state to the dissociative ground state (Ar + Ar) is very fast ($\tau_{fast} \sim$ 5 ns) while the transition from the  $^3\Sigma_u$ is much slower, being prohibited by the selection rules, and has a characteristic decay time of  $\tau_{slow} \sim$ 1300 ns \cite{Doke:1981eac}. The time evolution of LAr scintillation light can be described as the sum of two decaying exponential functions:
 
 \begin{equation}
 L(t) = \frac{A}{\tau_{fast}}e^{-\frac{t}{\tau_{fast}}}+\frac{B}{\tau_{slow}}e^{-\frac{t}{\tau_{slow}}}
 \end{equation}

\noindent where  A+B = 1. The relative abundance of the fast to slow components depends heavily on the type of particle that caused the excitation and is 0.3 for electrons, 1 for alphas and 3 
for neutrons. This property of LAr is responsible of its exceptional pulse shape discrimination capability that is extensively used by DM experiments to discard $\gamma$/e$^-$ and $\alpha$ background over neutron-like WIMPs (Weakly Interacting Massive Particles) signals.

The fast component is often used for triggering purposes and for determining the time of occurrence of the event (t$_0$ time) that is fundamental in LArTPC for position and 
calorimetric reconstruction. Scintillation light is also used to perform calorimetric measurements eventually in combination with charge detection.

Unfortunately LAr scintillation spectrum lies in the Vacuum Ultra Violet (VUV) region of the electromagnetic spectrum: it is a 10 nm band centered around 127 nm. This complicates 
its detection because common cryogenic photo-sensors have glass or fused silica windows, which are not transparent to these wavelength.

The most common paradigm for LAr scintillation light detection is based on the down conversion of VUV photons into visible ones by means of a wavelength shifter. Widely used shifters are TetraPhenyl Butadiene (TPB) and para-Terphenyl which absorb 127nm photons and re-emit around 430nm and 350nm, respectively. Down converted photons are them eventually collected with photo-collectors (ARAPUCA, light guides, optical fibers) and detected with cryogenic photo-sensors (PMT, SiPM, APD, CCD...)

    
The most commonly used cryogenic photo-sensors in LArTPCs in the last years are the photomultiplier (PMT), however, a new generation of photo-sensors based on silicon technology (APD, CCDs, and SiPMs, in particular) has been developed and is actually promising to surpass the PMTs in the next few years. This technology is growing fast, driven by applications in communications and medicine and by the competition among many producers. These devices are mainly constituted by a Si wafer, few electrical connections, an optical window and an almost inessential packing. The overall amount of material is on the order of grams and sensors can be manufactured with more or less any desired level of radio-purity. They do not need HV and have detection efficiencies comparable or better than those of the most performing PMTs. These are ideal devices to be coupled with cryogenic noble liquids because their performances strongly increase at low temperature (dark counts at the level of few Hz). SiPMs in particular have a gain per single photo-electron comparable to that of PMTs, and allow single photon counting in principle without amplification.
    

\subsubsection{The ARAPUCA Light Collector}
 
Detecting light in LAr and specially in a strong electric field like the one needed to operate a Time Projection Chamber (TPC), can be difficult with a standard PMT.  A new approach consist in the development of ARAPUCAs \cite{Machado_2016}, which are light traps built out of short pass dichroic filters and wavelength shifters. The dichroic filter is a multilayer film, that has the property of being highly transparent for wavelength below a certain cutoff and highly reflective above it. The device is a flattened box with highly reflective internal surfaces The open side hosts the dichroic filter that is the acceptance window. The filter is deposited with two shifters – one on each side. 

\begin{figure}[htp]
\centering 
\includegraphics[width=0.9\textwidth]{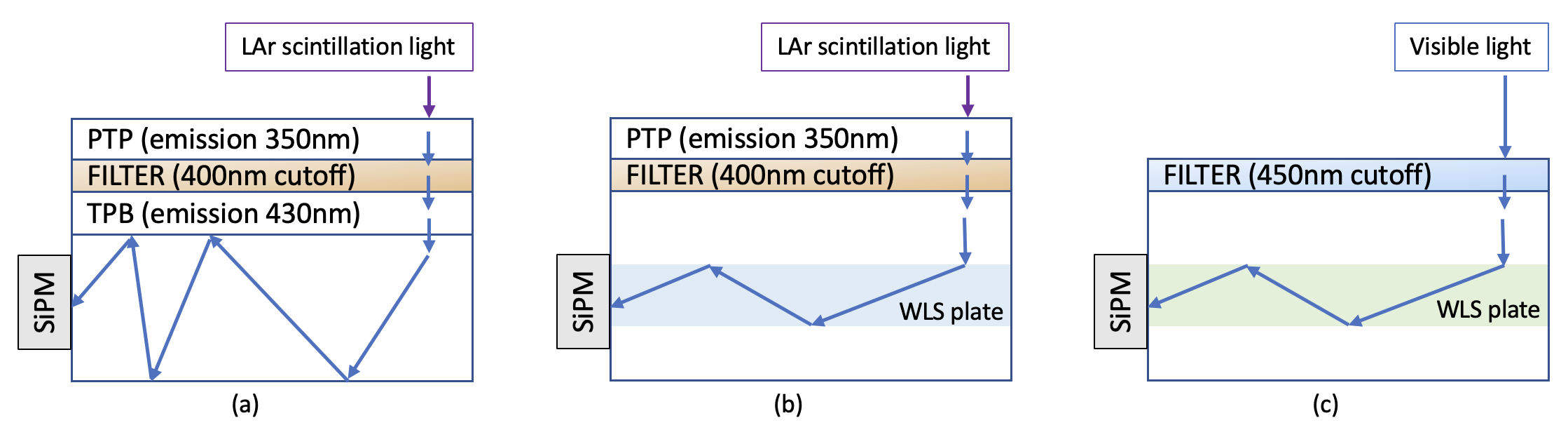}
\caption{a) Working principle of the ARAPUCA device. b) X-ARAPUCA for LAr scintilaltion light. c) X-ARAPUCA for visible light.} 
\label{fig:arpkfamily}
\end{figure}

LAr scintillation light is shifted by the external shifter (paraTerphenyl - PTP) from the ultraviolet (127nm) to 350nm (emission wavelength of PTP) and passes through the filter. On the internal side the second shifter (TetraPhenyl Butadiene - TPB, emission wavelength arounf 430nm) absorb 350nm photons and down-converts them to a wavelength above the filter cutoff. In this way the internal side of the filter becomes reflective to re-emitted photons and the photon is trapped inside the reflective cavity of the ARAPUCA. After some reflections the light will be eventually detected by the array of SiPMs installed on the internal surface of the ARAPUCA. The working principle of the ARAPUCA is illustred in Fig.~\ref{fig:arpkfamily}(a))

The coupling of the passive ARAPUCA photon trap with SIPMs has the advantage of increasing their effective area since trapped photons have more than one chance of hitting the SiPM sensitive surface. The ARAPUCA validation program has demonstrated an amplification factor of the effective SiPM area around 4.

The ARAPUCA technology was proposed to be used in the far detector of the DUNE experiment in 2016 and was extensively tested during the first run of protoDUNE (CERN2018-2020)\cite{prtdn_2020}. An evolution of the ARAPUCA device, the X-ARAPUCA \cite{Segreto_2018}, has been chosen as the baseline choice for the Photon Detection system of the DUNE far detector in 2019. The X-ARAPUCA also make part of the photon detection system of SBND experiment.
 
The X-ARAPUCA represents an optimization of the ARAPUCA design. In this case, the second wavelength shifter (deposited on the internal side of the filter) is replaced with a light guide in form of a plate, doped with a secondary shifter, see Fig.~\ref{fig:arpkfamily}(b). This design has the advantage of increasing the trapping efficiency with respect to an ARAPUCA, since a fraction of the photons converted inside the light guide suffers total internal reflection and is guided towards the edge of the plate, where the SiPM are installed. In addition, the construction of the device is simpler and the production is faster, since it is necessary to evaporate just one side of the dichroic filter.

Some experiments may want to detect the visible component of light as well. In this case, a type of X-ARAPUCA specific to this need can be developed. This is the case of the SBND experiment which, in addition to VUV photons, can also detect visible photons converted by TPB deposited on reflective foils installed on the cathode of the experiment. The X-ARAPUCA for visible light is showed in Fig.~\ref{fig:arpkfamily}(c)). 
   
The X-ARAPUCA has been demonstrated to have a detection efficiency 30\% higher than the standard ARAPUCA in several experimental tests \cite{Souza_2021} and \cite{Brizzolari_2021}.
    
\begin{figure}[htp]
\centering 
\includegraphics[width=0.85\textwidth]{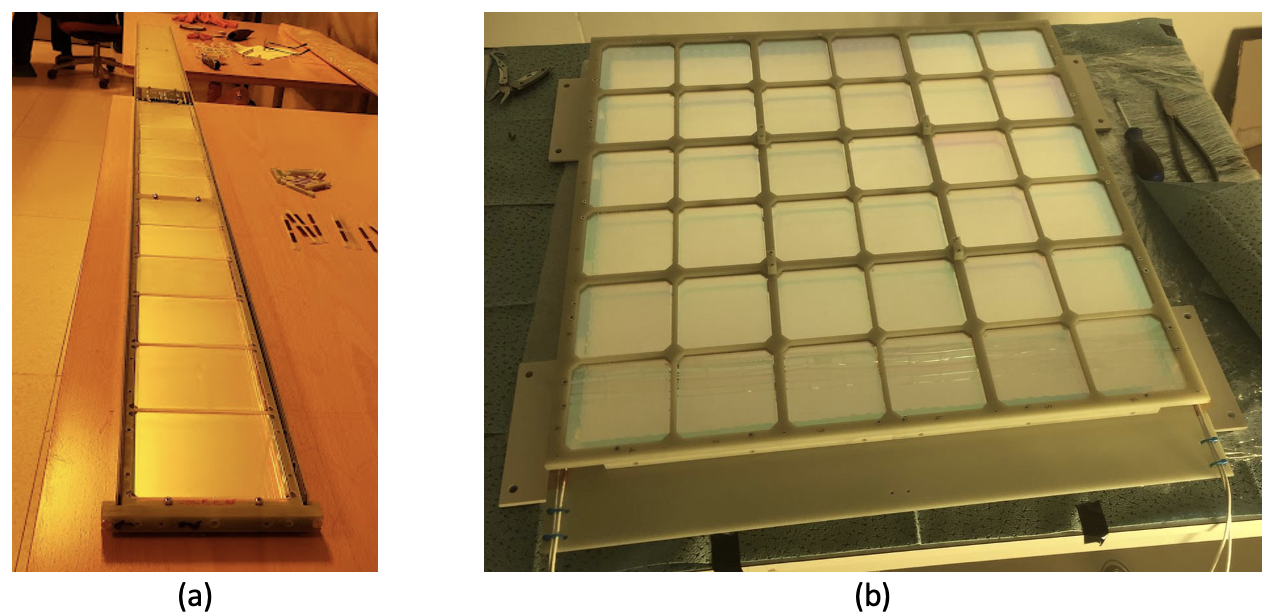}
\caption{a) X-ARAPUCA photon detection module of DUNE far detector one (horizontal drift). b) X-ARAPUCA photon detection module of DUNE far detector two (vertical drift).} 
\label{fig:xarpkdune}
\end{figure}

The X-ARAPUCA represents the baseline choice for the two modules of the DUNE far detector. The TPC of the two modules have different designs and use different charge read-out technologies: traditional wires for far detector one (also referred as horizontal drift) and printed PCB for far detector two (vertical drift). The differences in the designs of the TPC are also reflected on the geometrical shapes of the X-ARAPUCA modules: those for far detector one are in the form of bars (210 x 10 cm, see Fig.~\ref{fig:xarpkdune}(a)), while those for far detector two are squares (60 cm x 60 cm, see Fig.~\ref{fig:xarpkdune}(b)).
    
The ARAPUCA and X-ARAPUCA devices were proposed for large LArTPC detectors with the idea of improving the response to low energy events, such as supernova neutrino bursts and solar neutrinos. However, the use of the ARAPUCA technology is not limited to Liquid Argon detectors and can be extended to other areas, such as, for example, Cerenkov and Water Cerenkov detectors. A dedicated R\&D program is currently investigating the optimization of the optical components of the X-ARAPUCA for the detection of Cerenkov rdiation produced in water by cosmic particles. This program, called C-ARAPUCA, is developing specific dichroic filters, waveshifting plates and SiPMs for this application.

\subsubsection{Dopants to Increase Photon Yields}

LAr is a particularly bright scintillator, with around 40,000 photons/MeV with no electric field, and with excellent discrimination between $\beta$s and heavier particles. Because the scintillation light is very short, at 128~nm, it needs to be wavelength shifted to be detected by PMTs or SiPMS, and the wavelength-shifter of choice is usually TPB.  Adding Xe to the LAr, however, shifts the light in advance of getting to the TPB, overall increasing light yield. The DUNE Collaboration has performed a large-scale demonstration of the effect by adding Xe to the ProtoDUNE-SP prototype at CERN's Neutrino Platform~\cite{pdunexe}.  The fact that the shifted light tends to be the LAr triplet state also means that one might be able to do pulse-shape discrimination using spectral discrimination.



\subsubsection{Simulation and Calibration in LArTPC Neutrino Detectors}

Scintillation light detectors at LArTPC neutrino experiments are often used for timing measurements and triggering on non-beam events (e.g. proton decay). Additionally, they may provide better energy measurements when combined with ionization data from the TPC. In all of these applications, there must be precise understanding of initial scintillation light yields, effects of Rayleigh scattering, and light detector acceptance and quantum efficiency beyond what current measurements can provide.
The simulation of the optical signal is usually divided in a few steps.
The simulation of passage of particles through argon (GEANT4, FLUKA, ...)
yields the amount of energy deposited along the particle path.
Starting from it, a sequence of steps decides which fraction turns into
scintillation light and how many scintillation photons,
where they end their life (``propagation''), and if they convert into a signal
in a photodetector.
The simulation of their electronics signal follows.

The deposited energy contributes to both scintillation and ionization, and
these two processes compete against each other. Relevant physics parameters
includes the electric field and the electron-ion recombination model and
parameters. While simple models featuring their anticorrelation are already in
use, work is ongoing by the NEST collaboration 
to extend to liquid argon
the results already delivered for liquid xenon detectors. The standard
approach includes two scintillation times, 1.6~\textmu{}s and 6~ns, although
hints of a third scintillation time have appeared.

The propagation of scintillation light can in principle be performed via
detailed photon-by-photon simulation. Relevant physics parameters include the
scattering and absorption lengths in liquid argon, the refraction index and
group velocity of the light, and the optical properties of the surfaces in
contact with argon. While possible, this approach to simulation is deemed too
computing-intense to be performed individually for each of the million photons
(similarly to the ionized electrons drifting to the anode), and
parametrizations are commonly in use mapping each position in the detector to
the fraction of photons reaching each photodetector and their arrival time.
The first practical approach to the problem is to finely dice the detector
volume and to represent the mapping in a precomputed discrete table with one
cell per small volume. This soon hits computational limitations when large
detector volumes need to be covered. Conversely, simple physics-based
parametrizations are cheaper, but often inadequate in at least some regions.
Good promise is shown by hybrid approaches that attempt to decouple geometry
effects, simpler to describe analytically, from light interaction effects,
reducing the space where the latter need to be mapped. These
techniques fail to describe the visibility of regions with little or no line
of sight to the optical detectors, where physics effects dominate. Such
limitation is relevant in configurations with optical detectors outside the
active TPC volume, while they do not concern light-hermetic LArTPC modules.
Another alternative approach is being studied using machine learning
techniques[SIREN] that tune an existing map from simulation using actual
detector response to long tracks. This technique achieves a compact
description of the mapping, and also improves its quality.
The commonly used photodetectors are not sensitive to the argon scintillation
wavelength (128~nm), so somewhere in the path a wavelength shift needs to
happen. The relevant parameters, mainly the conversion efficiency, depend not
only on the wavelength shifting material, but also critically on its
deployment on the surfaces. This requires dedicated measurements specific to
each detector design. An aspect often neglected by the simulation is the 50\%
of visible light which is emitted back into argon by the wavelength shifter
coating a photodetector. In detectors using photomultipliers with curved
cathode there can be a direct optical path connecting their sensitive areas,
enhancing the chance of one illuminating the other.
If the photodetection system is sensitive to different wavelengths, multiple
maps will be needed, one describing each of them. Such setups can be chosen to
resolve light reflecting on the far surface of the detector (like in SBND) or
light scintillating from other materials (like in xenon doping).

The simulation of the photodetector response is very specific to the detector
itself. In general, in the signal formation both the amount of charge and its
time distribution need to be correctly simulated: seeking a time resolution at
nanosecond level requires a careful characterization of the setup which is
typically included as part of the detector calibration.

Simulation of readout electronics features most of the aspects common to the
one for the TPC charge. While the number of channels in the optical detector
readout is significantly lower than the TPC counterpart, the requirement of
nanosecond time resolution imposes a significantly higher sampling rate. As a
result, full readout for the optical detector is no more possible than for the
TPC charge, and data reduction techniques (selective readout, including lossy
compression) need to be applied, and simulated.
One special aspect arises in experiments whose the trigger is based on the
optical detection system, which is a common feature because of its prompt
response, and where the trigger signal is used to determine a selective
readout. For example, a large detector may decide to read out channels only in
an area close to where activity is being noticed. In such cases, the readout
simulation effectively requires a trigger simulation as well, and the two are
in fact intertwined.

The capability of the present and upcoming LAr-TPC based neutrino physics detectors in reaching their required precision in particle tracking and calorimetry is based on the performance and precision of the detector models for simulation and reconstruction, and therefore on the capabilities of calibration techniques. Due to the size of neutrino detectors and the long data-taking times, an essential component of a calibration program is the measurement and validation of spatial and time variations, both of low-level detector description parameters -- efficiency of ionization charge and scintillation light detector components, uniformity of electric field and drift velocity, electron diffusion, lifetime and recombination -- and high-level detector performance metrics -- trigger efficiency, track and cluster reconstruction efficiency and uncertainties, energy resolution and scale uncertainty, for different particle species.

 Natural sources, such as cosmic ray muons or natural radioactivity, are extensively used, for instance in MicroBooNE\cite{uB} and the DUNE prototype ProtoDUNE\cite{pdsp} at CERN. Samples of stopping charged particles, typically muons and protons, are used to measure the recombination model parameters and a work function scaling\cite{recomb}, both needed for the absolute energy scale. The calibration of the electromagnetic energy scale can typically be done with Michel electrons in the tens of MeV range or, at higher energies, with $\pi^0$ decay samples\cite{uBpi0}, whether from cosmic ray or beam events. Through-going muons can be used to map the detector efficiency, including the effect of electric field distortions from space-charge\cite{uBSCE} or to measure electron lifetime \cite{pdresults}. 
 
However, the use of cosmic ray muons presents limitations when the detectors are placed deep underground. For this reason, calibration methods based on UV laser ionization, external neutron pulsed, or the "classical" deployment of radioactive sources, are being actively pursued\cite{tdr4}.

Since several decades, argon gas TPCs have been calibrated with ionizing laser beams\cite{anderhub}, and more recently the technique has been further developed for use in liquid TPCs\cite{cao,badrhees,rossi}. The method is based on employing UV laser beams intense enough to ionize the LAr, directed at various locations in the TPC via steerable cold mirrors. It was first employed in a large detector in MicroBooNE\cite{ereditato2014}, where detailed electric field distortion maps were carried out\cite{uBlaser}. Due to the underground location, DUNE is also planning to use UV lasers to calibrate the far detectors \cite{tdr4}, and a prototype system will be installed at ProtoDUNE-II at CERN\cite{maneira}, where additional measurements, based on charge and not just position, will be tested.

 Other recent ideas for LAr TPC calibration include the use of external neutron generators creating pulses that propagate into the detector. Due to an anti-resonance\cite{artie} where the cross-section for neutron scattering in LAr is quite low and to the small energy loss per scatter, the travel range for neutrons in LAr is large. Therefore, only a few well-shielded commercial neutron DD generators installed on a few locations outside the cryostat are enough to cover a large fraction of a large detector such as DUNE. The capture of neutrons in $^{40}Ar$ leads to a 6.1 MeV (total) gamma ray cascade\cite{aced}, allowing the calibration of the energy range relevant for Supernova and solar neutrino physics. Initial encouraging tests were carried out at ProtoDUNE-I, but further development is upcoming in ProtoDUNE-II.


\subsubsection{Gaseous Argon TPCs}
In the DUNE near detector complex~\cite{ref:DUNEwp}, one of the detectors will be a high-pressure gaseous-argon time projection chamber (HPgTPC) surrounded by a calorimeter and a magnet (ND-GAr)~\cite{ref:ndgarwp}. In the direction of the neutrino beam, ND-GAr will be positioned downstream of a modular liquid-argon time projection chamber (ND-LAr). ND-GAr can complement ND-LAr by reconstructing the momentum and sign of charged particles ranging out of ND-LAr~\cite{ref:dune_cdr}. ND-GAr will also collect its own independent sample of neutrino interactions on argon, offering a unique cross-section and BSM physics program. In particular, due to its lower detection threshold than a liquid argon TPC, ND-GAr's HPgTPC will be able to reconstruct low energy pion and proton tracks and constrain uncertainties in the oscillation analysis by pinning down neutrino-argon interaction models~\cite{ref:ndgarwp}. HPgTPC is also less likely to confuse primary and secondary interactions since secondary interactions are rare in the lower density gas detecting medium~\cite{ref:ndgarwp}. This makes it possible to collect a neutrino event sample with less influence from the detector response and secondary interaction models~\cite{ref:ndgarwp}. Additionally, ND-GAr's excellent PID capabilities will allow it to constrain backgrounds to achieve a strong BSM reach~\cite{ref:ndgarwp}. While much of ND-GAr's baseline design takes advantage of the ALICE TPC and the CALICE calorimeter designs, there are some optimizations that need to be done to adapt the designs for the DUNE near detector environment~\cite{ref:ndgarwp}. Dedicated and novel R\&D efforts, including gas-mixture studies, light collection studies, charge readout, electronics and data acquisition development, high voltage studies, and detector calibration, are underway to optimize the designs and explore new technological developments~\cite{ref:ndgarwp}. These efforts will offer DUNE the opportunity to expand its capabilities.

\subsubsection{Low-Radioactivity Argon}

Argon from underground sources, depleted in $^{39}$Ar and $^{42}$Ar, has been important for the current and next-generation argon dark matter experiments~\cite{https://doi.org/10.48550/arxiv.2203.09734, https://doi.org/10.48550/arxiv.1901.10108}. Recent interest has been in using these underground sources within the neutrino frontier to fill next-generation detectors such as Coherent or a low background DUNE module [7,8] or to provide shielding for neutrinoless double beta decay experiments such as LEGEND. 
Experiments targeting sub-MeV signals are sensitive to both $^{39}$Ar and $^{42}$Ar, while MeV-scale experiments are sensitive primarily to the decay progeny of $^{42}$Ar. 

$^{39}$Ar decays primarily via a beta with a $Q_{\beta} = 565$~keV and a half-life of 269 years. This large rate of counts can overwhelm rare process searches and lead to challenges with multisite tagging due to pile-up. In a liquid argon detector filled with atmospheric argon, there is a 2\% chance that a $^{39}$Ar will decay within 32\,cm of any arbitrary point within the detector. $^{42}$Ar decays via a beta decay with a 599~keV endpoint and a half-life of 33 years. This decay also produces the radioactive isotope $^{42}$K, which beta decays with $Q_{\beta} = 3.5$~MeV and a cascade of de-excitation photons with a half-life of 12 hours. This long half-life means that coincidence tagging would be unreliable, and the high $Q_{\beta}$ falls across many critical regions of interest for MeV-scale physics searches. In the atmosphere, the $^{39}$Ar isotope is primarily produced through cosmogenic spallation, leading to 1 Bq/kg concentrations~\cite{https://doi.org/10.48550/arxiv.1901.10108}. $^{42}$Ar is less easily produced, requiring the fusion of two neutrons into the stable $^{40}$Ar~\cite{PEURRUNG1997425}, and the atmospheric argon level is 92 $\mu$Bq/kg~\cite{ASHITKOV1998179,https://doi.org/10.48550/arxiv.1901.10108}. 

DarkSide-50 has demonstrated a depleted underground argon source for a dark matter search. The argon was extracted from a commercial CO$_{2}$ gas stream, and after deployment in the detector, $^{39}$Ar levels were reduced by a factor of 1400. $^{42}$Ar was found to be below detection levels. Recent calculations accounting for the reduced cosmic flux level underground estimate this level to be greater than $10^{10}$ below atmospheric levels~\cite{SagarLRT2022}. Such reductions are crucial for the next-generation argon dark matter experiments. There is also evidence that the underground argon stream was contaminated by an air leak during processing, indicating that these measured $^{39}$Ar levels are due to residual atmospheric argon so that the ultimate level could be lower. A processing facility (URANIA) is being constructed to process the required amounts for DarkSide-20k and the proposed ARGO (~300 tons) next-generation experiments. The maximum production of such a facility is 300 kg/day, which though large enough for these dark matter experiments, is too small to fill a DUNE-scale (10+ ktons) module. 

A new facility will be required to provide underground argon at the scale needed for a DUNE module~\cite{https://doi.org/10.48550/arxiv.2203.09734}. This will require addressing a number of challenges. A major challenge is finding suitable sources of underground argon. Ideally, a commercial gas stream including a high concentration of chemically enriched underground argon should be identified. PNNL has identified such streams in discussion with a commercial supplier. The producer estimated that 10+ kilo-tonne-scale production might cost as little as a factor of three more than commercially available atmospheric underground sources. Another challenge to address will be monitoring the quality of the argon during production. Low background proportional counters can measure $^{39}$Ar levels to approximately 5\% of atmospheric levels. The DART liquid argon detector can assay 1 liter of argon to a factor 1000 below atmospheric $^{39}$Ar levels, which will be used to check batches of argon for DarkSide-20k production. Alternate indirect methods include searches for associated air contaminates or $^{36}$Ar:$^{40}$Ar ratios. A final challenge will be the storage of the large amounts of argon to prevent cosmogenic activation before the experiments begin. Calculations for the dark matter experiments show that significant activation requires several years of storage above ground~\cite{Zhang:2022dlg}.

\subsection{Novel Large Liquid Argon Detector Concepts}

Opportunities to deploy combinations of new enabling technologies in future large-scale liquid argon detectors are actively being considered by the neutrino physics community.  Low-energy physics in LArTPCs is a topic of great community interest, particularly as it would broaden the already exciting program of DUNE.  To do just about any physics in LArTPCs below about 6~MeV would require first, shielding for ambient neutrons in combination with extremely low-background materials for the cryostat (ensuring very low ($\alpha$,$n$) reactions).  In addition, a critically important enabling technology would be production of large quantities of low-background argon, likely through the extraction of underground argon at scales much larger than have been done to date.  Underground argon would significantly reduce the levels of $^{39}$Ar and $^{42}$Ar.  Below we describe several current ideas for optimizing large LArTPCs performance at low-energies and expanding the physics reach over first-generation LArTPC neutrino detector designs. 


\subsubsection{Solar Neutrinos in LAr (SoLAr)}

The SoLAr design (see Snowmass white paper~\cite{Para:2022gju}) is aimed at measuring physics at the MeV-scale in a large liquid argon neutrino detector, leveraging technological developments to expand the physics reach of next-generation devices to include solar neutrinos.  This novel concept will significantly improve the precision on solar neutrino mixing parameters and will enable observation of the ``hep branch'' of the proton-proton fusion chain.

The SoLAr technology will be based on the concept of monolithic light-charge pixel-based readout which addresses the main requirements for such a detector: a low energy threshold with excellent energy resolution (7\%) and background rejection through pulse-shape discrimination.
The SoLAr concept is also timely as a possible technology choice for the DUNE ``Module of Opportunity", which could serve as a next-generation multipurpose observatory for neutrinos from the MeV to the GeV range.

\begin{figure}[ht]
    \centering
    \includegraphics[height=0.25\textheight]{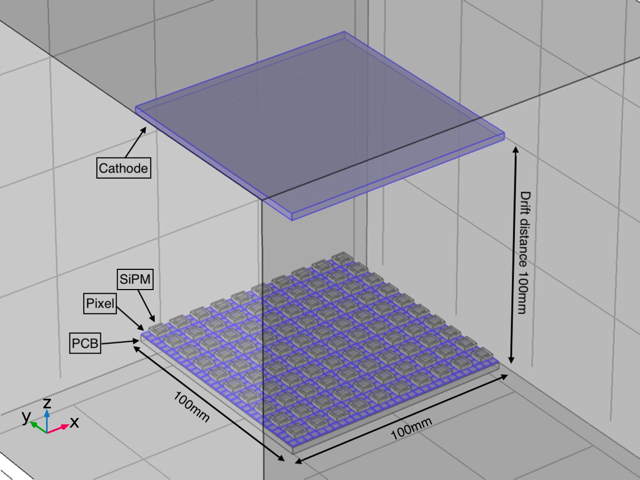}
    \includegraphics[height=0.25\textheight]{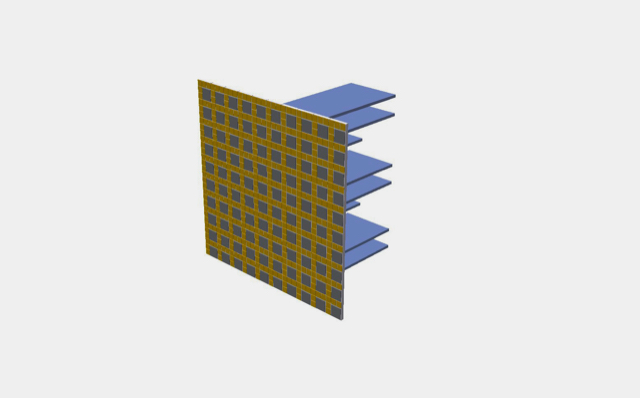}    
    \caption{SoLAr detector concept.}
    \label{fig:solar}
\end{figure}

The SoLAr readout unit will be a pixel tile in CMOS technology that embeds charge readout pads located at the focal point of the LArTPC field shaping system and collects VUV photons in thousands of microcells operated in Geiger mode. Each monolithic sensor produces analog signals corresponding to the charge of the electrons collected in each pixel and to the photons collected in the pixel frame, respectively.  SoLAr leverages the independent technology development of both VUV SiPMs and CMOS pixel readout, combining them to develop a detector technology which is scalable to the multi-kiloton scale.  

The goal of SoLAr is to observe solar neutrinos in a 10 ton-scale detector and to demonstrate that the required background suppression and energy resolution can be achieved. To achieve this, SoLAr is adopting a staged approach, building from the current table-top prototype (see figure~\ref{fig:solar}) whose design and assembly is ongoing, to a medium sized demonstrator module. 

\subsubsection{A SURF Low Background Module (SLoMo)}

\begin{figure}[ht]
    \centering
    \includegraphics[width=\textwidth]{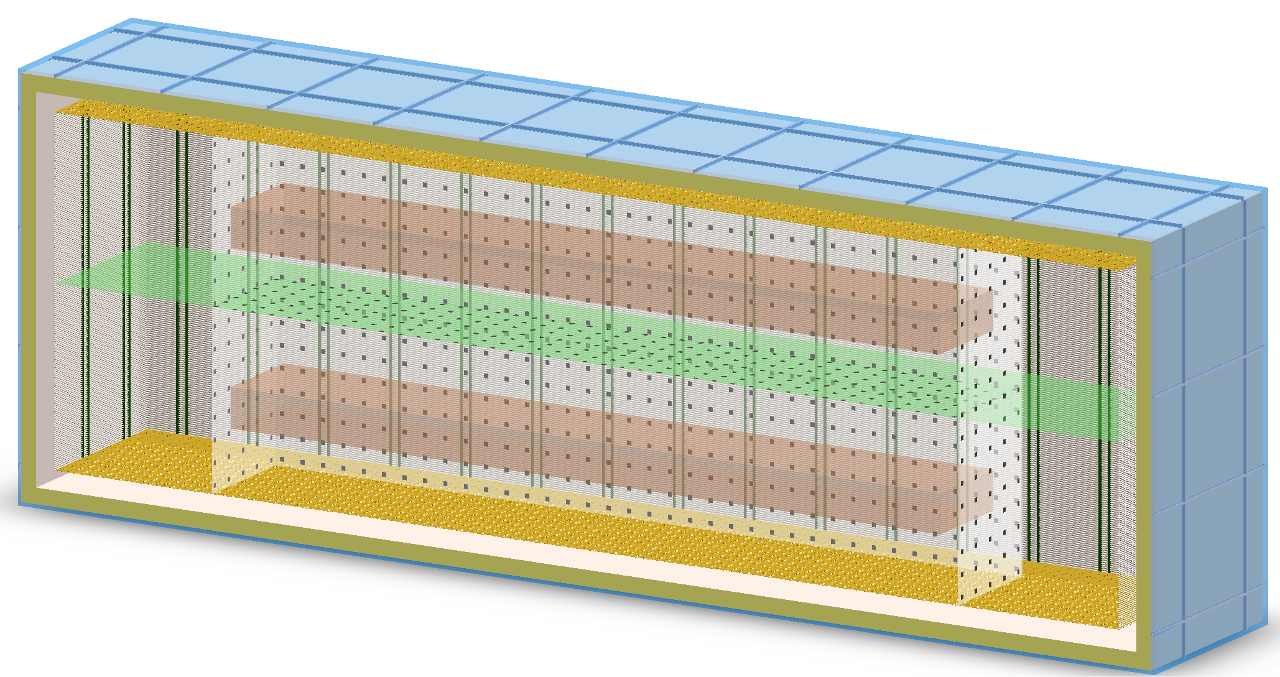}
    \caption{Shown is the baseline design for the proposed low background detector. Blue shows external water ``bricks". The top and bottom yellow planes are the Charge Readout Panels unchanged from the Vertical Detector design. The central cathode is in green. The white box of acrylic (full interior volume) is of dimensions 6x12x20 (12x12x60) m$^3$. The black points are SiPM modules shown here at a low coverage of 10\% for viewing's sake. A proposed fiducial volume totaling 2kT is shown in the two beige boxes.}
    \label{fig:SLoMo}
    \end{figure}

The possibility of building a low background kTon-scale liquid argon time projection chamber (the SURF Low Background Module -- SLoMo), that could be a design for the 3rd or 4th module of the Deep Underground Neutrino Experiment (DUNE), has been explored in~\cite{https://doi.org/10.48550/arxiv.2203.08821}. Such a module would allow the physics scope of that experiment to be increased with the addition of several low-energy physics topics, without disrupting the main oscillation physics program. This includes enhanced supernova and solar neutrino sensitivity, and even searches for neutrinoless double beta decay and Weakly Interacting Massive Particle (WIMP) dark matter. The detector design will take as a starting point the standard DUNE vertical drift single phase TPC design~\cite{snowmass_pof}. It will be modified with the addition of an optically isolated inner volume, where fiducialization allows significantly lower background levels, and photosensor coverage can be increased to improve energy resolution. The proposed design is shown in Figure~\ref{fig:SLoMo}.

Methods to control radioactive backgrounds have been explored. In particular, the large size of such a module ($\sim17$~kTons) allows the self-shielding properties of the argon to be exploited and fiducialization of a 1-3~kTon volume allows a radioactively clean central part of the detector to be utilized. Further techniques to control backgrounds that may reach the fiducial volume include:

\begin{itemize}
    \item External neutrons from the cavern. As first proposed in~\cite{PhysRevC.99.055810}\cite{PhysRevLett.123.131803}, a 40 cm layer of water shield will reduce external neutrons by a factor $10^{3}$. Such a water shield could be implemented around the current DUNE cryostat design, within the structural supports.
    \item Cryostat and detector background. An intensive materials assay campaign will allow selection of construction materials with radioactive backgrounds reduced by a factor $10^{3}$. Such material purity has been exceeded by a further two orders of magnitude in current dark matter experiments (e.g.~\cite{LZ_bkgs}). Additional internal shielding can also be added, including neutron absorbers such as boron, lithium or gadolinium loaded cryostat layers.
    \item Radon in the liquid argon. Active purification of the inner volume and an emanation measurement campaign will reduce radon levels by a factor $10^{3}$. This factor has been exceeded by current dark matter experiments~\cite{PhysRevD.100.022004}.
    \item Internal argon-42, argon-39, krypton-85. As demonstrated in DarkSide-50~\cite{UAr_DS50_39Ar}, underground sources of argon can have depleted amounts of critical background isotopes, with $^{39}$Ar reduced by a factor 1400. No $^{42}$Ar was measured in DarkSide, and reduction factors are expected to be of $10^8$ or more. To fill a DUNE-scale detector new sources must be found, and work is underway to identify viable commercial options. Discussion of a future underground argon facility can be found in~\cite{https://doi.org/10.48550/arxiv.2203.09734}.
\end{itemize}

The light collection will be enhanced in the inner volume with at least $\sim 10 \%$ coverage of SiPM tiles on walls and cathode, additional reflectors on the inner walls and anode collection planes, and additional argon purity requirements. This will allow energy resolution of $\sim 2\%$ at 1 MeV when combined with the TPC charge signal~\cite{nesteres} and allow a pulse shape discrimination measurement for background reduction.

A number of important physics topics can be explored with this detector. For supernova neutrino physics lower neutron levels allow the lowering of energy threshold for the search to 600 keV, permitting access to interesting low energy and late or early time information. The reduction in backgrounds allows improvements to the trigger sensitivity, and the detector will be fully sensitive to supernova from the Magellanic cloud. In addition, the increased light collection allows the supernova neutrino CE$\nu$NS interactions within the detector to be measured. For solar neutrino physics lowering the threshold and improving the energy resolution will increase sensitivity to $\Delta m^2_{21}$ allowing a precision measurement. This would also improve non-standard interaction constraints or explain the solar anomaly between reactor and solar measurements of $\Delta m_{21}^2$. A precision measurement of the CNO flux is also possible.

Loading the detector with a few-percent of xenon-136 allows a sensitive search for neutrinoless double beta decay beyond the coming ton-scale experiments. As first discussed in~\cite{DUNE-DM-PNNL}, the increased light coverage allows pulse shape discrimination to remove electron recoil backgrounds (primarily argon-39) to a WIMP dark matter search through nuclear recoils of at least 100 keV and perhaps just 50 keV. Such a search will be competitive with coming dark matter searches on a reduced timescale. Due to the unrivaled large mass of 3 ktons and a potentially very long DUNE operation of one decade (or even several), this concept can offer a unique seasonal variation detection at sufficient statistical significance for providing a smoking gun signature for the nature of WIMPs. This would be particularly of interest in the case upcoming generation-2 experiments such as LZ, XENONnT and/or DarkSide-20k have evidence for WIMPs near their sensitivity. It would be nearly impossible for the planned generation-3 experiments to make such a smoking gun detection proving the WIMP nature of dark matter. 

This concept takes full advantage of the planned DUNE far-detector facilities to create a true multipurpose detector to address important questions in high energy physics, astronomy, and nuclear physics.

\subsubsection{Neutrinoless Double Beta Decay with Xe+LArTPC}

With several of the exciting enabling technologies described above---underground argon for very low radioactivity, photo-ionizing dopants, and xenon doping in LArTPCs---there is the possibility of performing an extremely sensitive neutrinoless double beta decay search in a DUNE module, as described by Mastbaum, Pshihas, and Zennamo in Ref.~\cite{Mastbaum:2022rhw}.  The photo-ionizing dopants improve the energy resolution substantially over exclusive primary ionization and direct photon detection, and by re-using a DUNE module infrastructure costs would be dramatically lower than an entirely new detector.  
    
\subsubsection{Detecting Cherenkov and Scintillation Photons in LAr (ArCherS)}

Distinguishing Cherenkov and scintillation (C/S) photons is challenging in organic liquid scintillators because the short-wavelength end of the Cherenkov spectrum, where most of the photons reside, is buried by the more intense scintillation spectrum.  In contrast, liquid Ar (LAr) scintillates narrowly around 128 nm, leaving the broad Cherenkov spectrum isolated above this wavelength, which is readily detected by common devices such as PMTs.  The scintillation photons can be detected with the same devices upon coating their photosensitive surface with a wavelength shifter like TPB.  The narrow range of scintillation photons ensures that the uncoated PMTs see purely Cherenkov light.  This pure sample of Cherenkov photons can be used to cleanly reconstruct the position, direction, and energy of events from several MeV and above, depending on the PMT photocoverage and efficiency.  The photons detected by the TPB-coated PMTs would be primarily from scintillation due to its far greater yield and could be used to independently reconstruct the position and energy of events as low as a few MeV.  Using both photon samples together would provide more robust particle ID and reconstruction. 

Relative to liquid Ar time projection chambers like those proposed for DUNE, this type of detector would provide the same particle interaction channels while requiring simpler hardware (no charge readout) and event analysis (like a scintillator or Cherenkov detector).  Collaborators at Penn and Berkeley are studying the capabilities of a LAr detector to distinguish C/S photons.  GPU-accelerated simulations of a large liquid Ar detector with TPB-coated and -uncoated PMTs is performed using the Chroma package and pyrat wrapper.

\subsection{Liquid and Gaseous Xenon Detectors}

    Detectors that use liquid xenon for both scintillation and charge readout---typically as a TPC---have had great success in exploring low-energy physics, particularly for dark matter searches~\cite{Aprile:2020thb,Akerib:2016vxi,LZ} and neutrinoless double beta decay~\cite{EXO-200}.  
    
    For dark matter searches, LXe has the advantage of a high cross section and very low backgrounds (no real analog of $^{39}$Ar).  It is also a better neutron shield than LAr.  And liquid detectors lend themselves to large, monolithic designs, which help attenuate external backgrounds like $\gamma$ rays. 
    
    For neutrinoless double beta decay, LXe has similar advantages, although clearly cross section (other than the cross section for neutron scattering) is not relevant.  The high density of LXe is, however, important, as it helps to attenuate $\gamma$ rays even in small volumes.  In addition, the apparent anti-correlation of scintillation light and charge~\cite{EXO-200}, helps to provide better energy resolution than simple counting statistics would imply, a fact that is important for searches near the endpoint of the $2\nu\beta\beta$ spectrum, as well as for general background reduction.
    
    Clearly, there are synergies between LXe dark matter detectors and neutrinoless double beta decay detectors, although also significant differences.  Dark matter experiments have very low threshold requirements---to see the $\sim$ $10 {\rm keV}_{\rm ee}$ recoils from WIMPs, while neutrinoless double beta decay experiments have low-background (at much higher energies) requirements, which depend in part on requirements on energy resolution that are less critical for dark matter searches.  
    
    Nevertheless, it is likely there are many places where LXe detectors can be co-developed, including on low-background materials, veto systems, front-end electronics, etc.  This would seem to be a very good opportunity for the Cosmic Frontier and the Neutrino Frontier to work toward common goals.
    
    There are many new ideas for xenon-based detectors for both dark matter and neutrinoless double beta decay.  Future ideas for LXe dark matter detectors assume a two-phase detector, where electroluminescence converts drifting charge into photons at a liquid/gas transition layer. Like previous LXe dark matter experiments, the fast scintillation light is used to provide a $t_0$ that allows position reconstruction of events via timing as well as the pixelization of the electroluminescence photon sensors. Many details of this approach, including its physics program, can be found in Ref.~\cite{Aalbers:2022dzr}.  The two-phase LXe detector described there will also likely be able to measure solar neutrinos, down into the $pp$ energy regime.
    
    For neutrinoless double beta decay, the next-generation LXe detector planned, nEXO~\cite{nEXO:2017nam}, is a single-phase detector that detects both light and charge.  At the 5-tonne scale, nEXO is part of the Office of Nuclear Physics' ``tonne-scale'' double beta decay program.  
    
    Looking further ahead, the NEXT experiment plans to use high-pressure xenon gas which, like the ND-GAr detector described above, allows tracking of the two $\beta$s emitted in a double beta decay event.  The possibility of detecting the barium daughter ion via fluorescence is being actively investigated, as this would be a tremendous reduction in non-$\beta\beta$ backgrounds.

\cleardoublepage

\section{Photon-Based Neutrino Detectors}

 Between the last SNOWMASS and now, there has been an explosion of new ideas and new, enabling technologies that will significantly expand the capabilities of next-generation photon-based neutrino detectors.
Detectors that use photons as the primary carrier of interaction information have a long, rich history in neutrino physics, going as far back as the discovery of the neutrino itself.  Large-scale, monolithic detectors that use either Cherenkov or scintillation light have played major roles in nearly every discovery of neutrino oscillation phenomena~\cite{superk,sno,dayabay,kamland,t2k} or observation of astrophysical neutrinos~\cite{kii,sno,imb,borexino,borexinopep,icecube}.  New detectors at even larger scales are being built right now, including JUNO~\cite{juno}, Hyper-Kamiokande~\cite{hyperk}, and DUNE~\cite{dunearapucas}.
    Photon-based detectors have been so successful because they are inexpensive, remarkably versatile, and have dynamic ranges that reach all the way from tens of keV~\cite{snoplus} to PeV~\cite{icecube}.  
    
        The new photon-based technologies that NF10 has surveyed include neutrino physics and astrophysics programs of great breadth: from high-precision accelerator neutrino oscillation measurements, to detection of reactor and solar neutrinos, and even to neutrinoless double beta decay measurements that will probe the normal hiearachy regime.   They will also be valuable for neutrino applications, such as non-proliferation via reactor monitoring.
        
        Of particular community interest is the development of {\it hybrid} Cherenkov/scintillation detectors, which can simultaneously exploit the advantages of Cherenkov light---reconstruction of direction and related high-energy PID---and the advantages of scintillation light---high light-yield, low-threshold detection with low-energy PID. Hybrid Cherenkov/scintillation detectors could have an exceptionally broad dynamic range in a single experiment, allowing them to have both high-energy, accelerator-based sensitivity while also achieving a broad low-energy neutrino physics and astrophysics program.  Recently the Borexino Collaboration~\cite{borexinocherscint} has published results showing that even in a detector with standard scintillator and no special photon sensing or collecting, Cherenkov and scintillation light can be discriminated well enough on a statistical basis that a sub-MeV solar neutrino direction peak can be seen.  Thus the era of hybrid detectors has begun, and many of the enabling technologies described here will make full event-by-event direction reconstruction in such detectors possible.
        
        There were many relevant letters-of-intent 
        and several white papers including one aimed explicitly at these kinds of detectors and their technologies, and far more details can be found in that reference~\cite{photonwp}. 

\subsection{Enabling Technologies for Hybrid Cherenkov/Scintillation Detectors}

 The physics accessible in large Water Cherenkov (WC) detectors such as Super-Kamiokande (SK) is limited in many areas of interest by the inability to detect particles with energy below the Cherenkov threshold.  One example is the sensitivity to the Diffuse Supernova Neutrino Background (DSNB)~\cite{SK_DSNB:2021} due to backgrounds from low-energy atmospheric neutrino interactions, and reduced signal from the inability to detect positron annihilation, which enhances the prompt signal from the leading reaction $\overline{\nu}_e + p\rightarrow e^{+} + n$. Similarly for proton decay, the kaon from $p\rightarrow \overline{\nu}K^{+}$ is below Cherenkov threshold, and for solar neutrinos the $^{7}$Be and CNO neutrinos are practically undetectable as much of the energy from the neutrino electron scattering reaction is invisible.

Organic liquid scintillators (LS) have been used to enhance sensitivity for particles below Cherenkov threshold, and to provide high light yield and thus narrow energy resolution needed to see monoenergetic signals like the $0\nu$ peak in neutrinoless double beta decay. Liquid scintillators also typically allow excellent discrimination between heavy particles like $\alpha$s or protons and electrons or $\gamma$ rays, through the differences in the scintillation time profile. These detectors thus typically aim for a primarily low-energy program, like reactor antineutrinos (KamLAND, Daya Bay, and JUNO), solar neutrinos (Borexino and SNO+), or neutrinoless double beta decay (KamLAND-Zen and SNO+). 

    Yet the narrow energy resolution and low-energy particle ID of scintillation detectors comes at the cost of the high-energy particle ID possible with Cherenkov light, and the reconstruction of direction from Cherenkov rings which helps identify events from point sources like the Sun.  Most high-energy neutrino physics programs---from accelerator beams or atmospheric neutrinos (or nucleon decay)---use ring imaging and counting to eliminate backgrounds.

Hybrid neutrino detectors, which leverage both the unique topology of Cherenkov light and the high light yield of scintillation, have the potential to revolutionize the field of low- and high-energy neutrino detection, offering unprecedented event imaging capabilities and resulting background rejection.  Such detectors would be more than the sum of scintillation and Cherenkov physics programs, as the information from one source of light leverages background rejection for physics in the other.  One example of this is the rejection of radioactive backgrounds in a low-threshold scintillation measurement of solar neutrinos using direction~\cite{gdogrbonv}; another is the rejection of atmospheric neutrino backgrounds to DSNB neutrinos, by counting Cherenkov rings and using the Cherenkov/scintillation ratio to remove mostly hadronic final states~\cite{dsnb_psd}.
At high energies, it may be possible to use scintillation light to precisely reconstruct the neutrino interaction vertex in a charged-current event, thus helping to eliminate backgrounds from NC $\pi^0$ production.


Discrimination between ``chertons'' and ``scintons'' can be achieved in several ways, each of which has its own advantages and drawbacks, but which ultimately fall into four broad classes: 
\begin{itemize}
\item Make chertons easier to see by increasing the Cherenkov/scintillation ratio, typically through suppression of the scintillation light 
\item Separate chertons from scintons by timing, by using fast photons sensors, or by slowing down the scintillation time profile, or by having a large enough detector that dispersion between ``blue'' scintillation light and broadband Cherenkov light allows them to be resolved
\item Exploiting the differences in the spectra of Cherenkov and scintillation light to distinguish chertons from scintons, either by sorting the photons with devices like dichroicons~\cite{dichroicons}, or using filters on some subset of the photon sensors to exclusively see long-wavelength Cherenkov light.
\item Use the angular distribution of Cherenkov light to identify it, through analysis techniques.
\end{itemize}

    Which approach works best depends on many things, including the dominant backgrounds for the physics of interest, cost, development status, robustness, etc.  The tensions between them typically contrast the required number of detected Cherenkov photons against scintillation light yield; direction reconstruction against vertex resolution; and backgrounds and robustness of new scintillation cocktails against high costs of new photon sensing or collecting devices.  A program of both benchtop R\&D, and mid-scale prototypes is underway to help answer these questions.  In principle, all of the approaches above could be used in a single detector, as they are for the most part not mutually exclusive.
    
    Below we detail the various enabling technologies for hybrid Cherenkov/scintillation neutrino detectors.  Far more detail can be found in the related white paper~\cite{photonwp}.


\subsubsection{Water-based Liquid Scintillator}
\label{sec:wbls}

\begin{figure}[h!]
\centering
\includegraphics[width=0.4\textwidth]{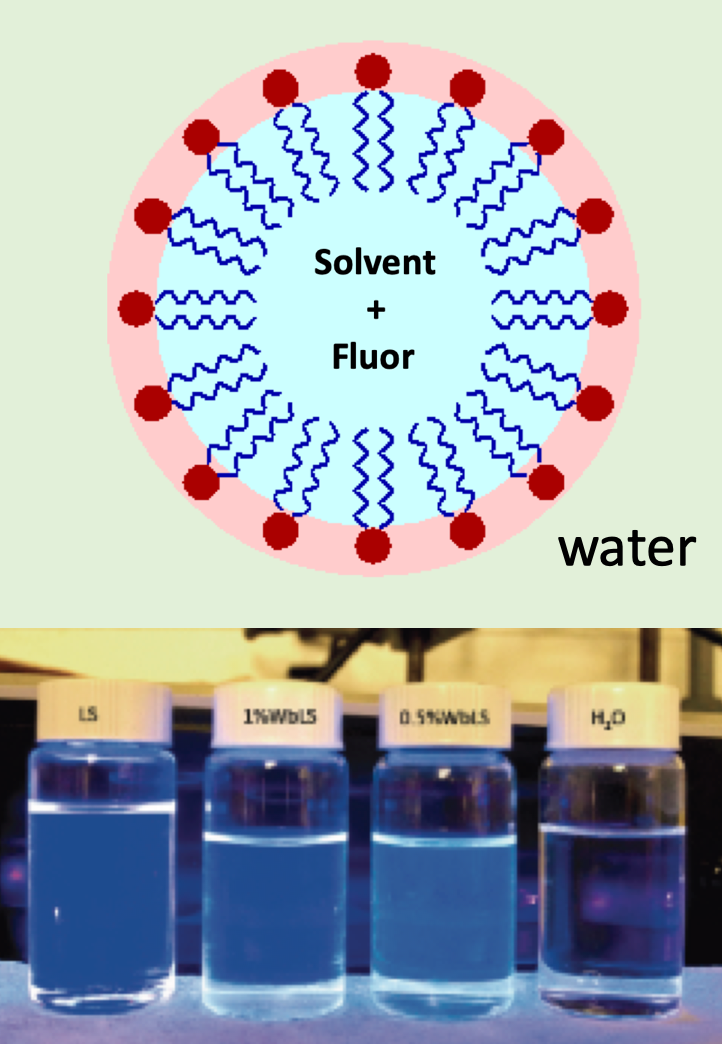}
\caption{Top: Liquid scintillator encapsulated in a surfactant micelles, which are typically around 10 nm in size. Bottom: From right to left: water, 0.5\% WbLS, 1\% WbLS, pure liquid scintillator.}
\label{fig:WbLS}
\end{figure}

The development of Water-based Liquid Scintillator (WbLS)~\cite{wbls} has the potential to signifcantly impact and enhance  hybrid particle detection capabilities.
WbLS is essentially liquid scintillator encapsulated in surfactant micelles that are thermodynamically stable in water (see Fig.~\ref{fig:WbLS}).
By introducing varying amounts (typically 1\%-10\%) of liquid scintillator into water, the liquid yield can be adjusted to allow detection of particles below Cherenkov threshold  while not sacrificing directional capability, cost, or environmental friendliness. First developed at Brookhaven National Lab (BNL), WbLS is a leading candidate for the main target medium for the proposed \theia detector, which would enhance the scientific program at the LBNF significantly, as described in the \theia White Paper ~\cite{theiawp}.

\subsubsection{Slow fluors}

The properties of four slow fluors have been studied in the context of LAB-based liquid scintillator mixtures to provide a means to effectively separate Cherenkov light in time from the scintillation signal with high efficiency. This allows for directional and particle ID information while also maintaining good energy resolution. Such an approach is highly economical (i.e. small compared to other experimental costs) and can be readily applied to existing and planned large-scale liquid scintillator instruments without the need of additional hardware development and installation. Using this technique,  Cherenkov/scintillation  separation has been demonstrated on a bench-top scale, showing clear directionality, for electron energies extending below 1 MeV. 


 While the use of slow fluors means that the vertex resolution may be worse than typical large-scale liquid scintillator detectors (but better than typical large-scale Cherenkov detectors), the balance between position resolution, Cherenkov separation purity and energy resolution can be tuned for a particular physics objective by modifying the fluor mixture. This balance is also affected by the presence of fluorescence quenchers, which may be naturally present in the case of loaded scintillator mixtures or could be purposely introduced to change the balance.

\subsubsection{Fast Photomultiplier Tubes and LAPPDs}
\label{sec:LAPPD}

Resolving Cherenkov and scintillation light via timing can be done even with ``standard'' scintillation cocktails, because the Cherenkov light is created promptly while scintillation light has both a risetime of hundreds of picoseconds and fall types that are in the multi-nanosecond regime, even without intentionally using a ``slow'' scintillator.
In addition, in large detectors, the dispersion of the broadband Cherenkov light means that many Cherenkov photons can arrive at a photon sensor a nanosecond or more before the scintillation light.

In the past decade there have been significant improvements in the time resolution of even standard photomultiple tubes (PMTs).  Hamamatsu, for example, has developed 8'' high-quantum efficiency PMTs that have transit time spreads (TTS) with FWHM of around 1.5~ns, or $\sigma\sim 700 ps$~\cite{tanner}.  With this kind of resolution, scintillator risetimes of 1~ns or so are slow enough to allow a prompt signal to be seen. Smaller PMTs
already have TTSs in the regime of 250~ps or so, and these can be ``ganged'' together to make larger-area arrays~\cite{chess}.  

The most aggressive approach to timing for photon sensors has been the development of
Large Area Picosecond Photodetectors (LAPPDs). LAPPDs are 20 cm x 20 cm microchannel plate photomultiplier tubes (MCP-PMTs)~\cite{Wiza:1979iia} now in use by the neutrino community and capable of millimeter-scale spatial resolutions, tens of picosecond sPE time resolutions, and gains exceeding $10^6$~\cite{LAPPDtiming}. 

\begin{figure}[ht!]
\centering
\includegraphics[width=0.4\textwidth]{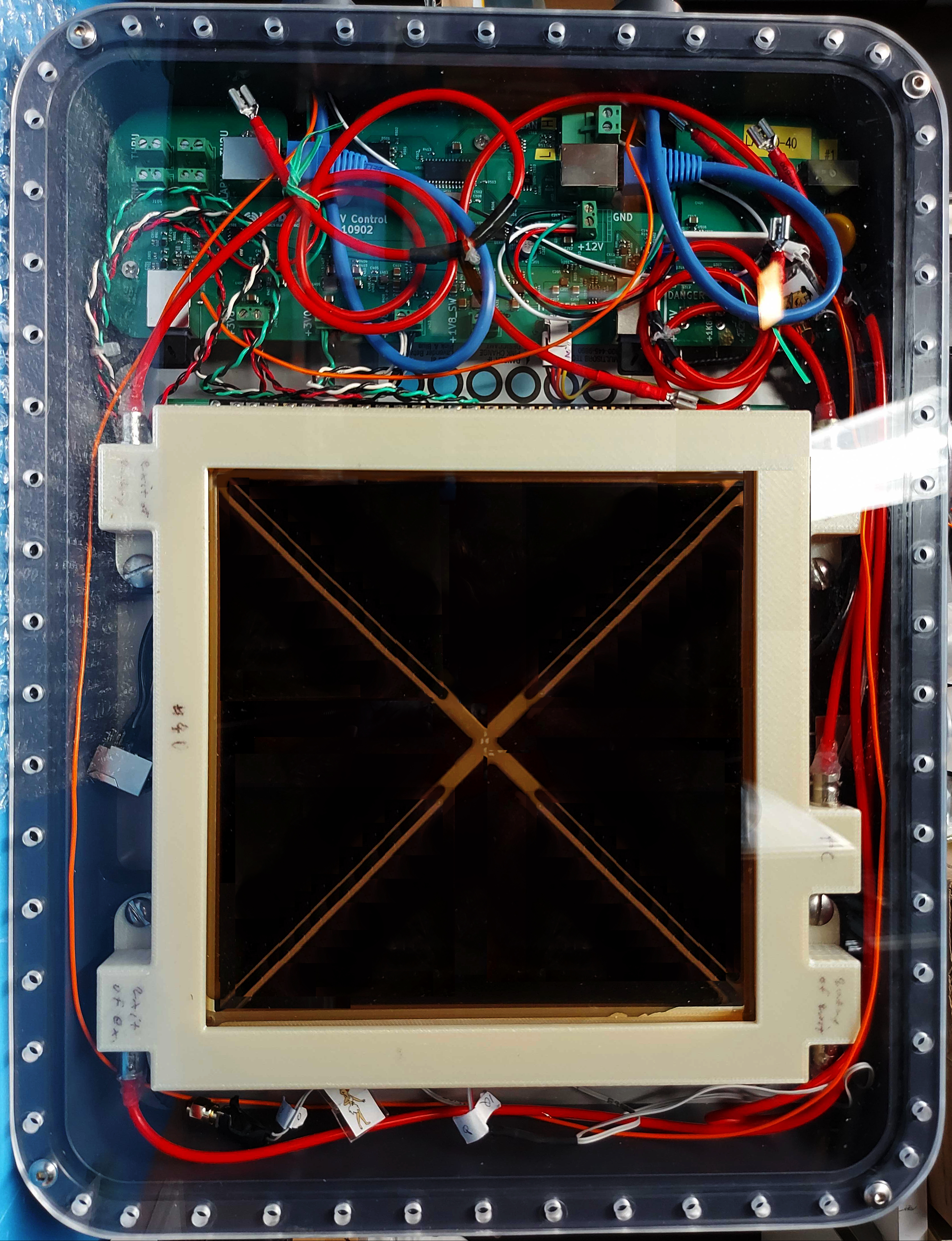}
\includegraphics[width=0.4\textwidth]{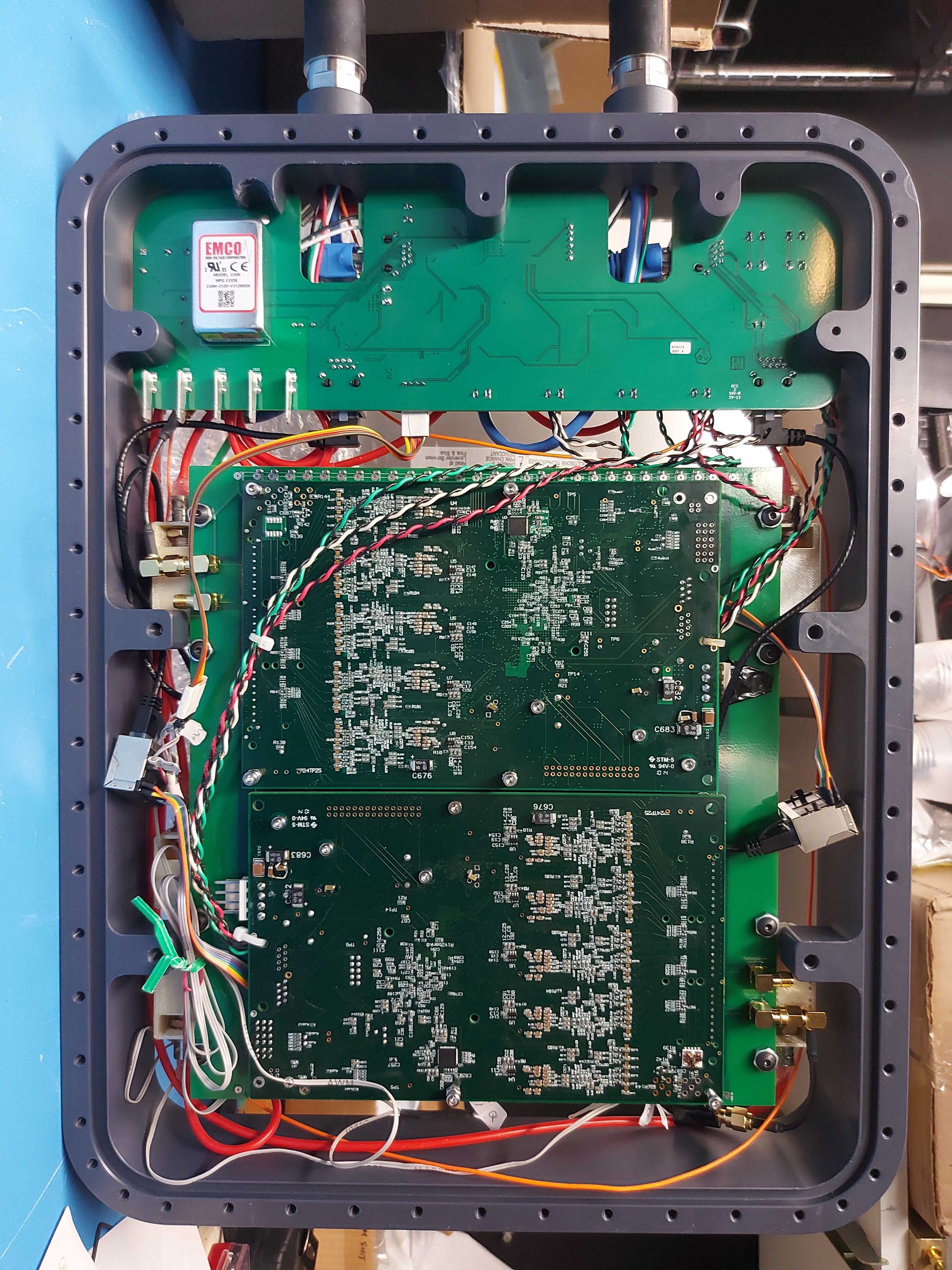}
\caption{Left: Front view of an ANNIE LAPPD Module. Right: The back of an LAPPD module with the back plate off.}
\label{fig:LAPPDmodule}
\end{figure}


\begin{figure}[ht!]
\centering
\includegraphics[width=0.5\textwidth]{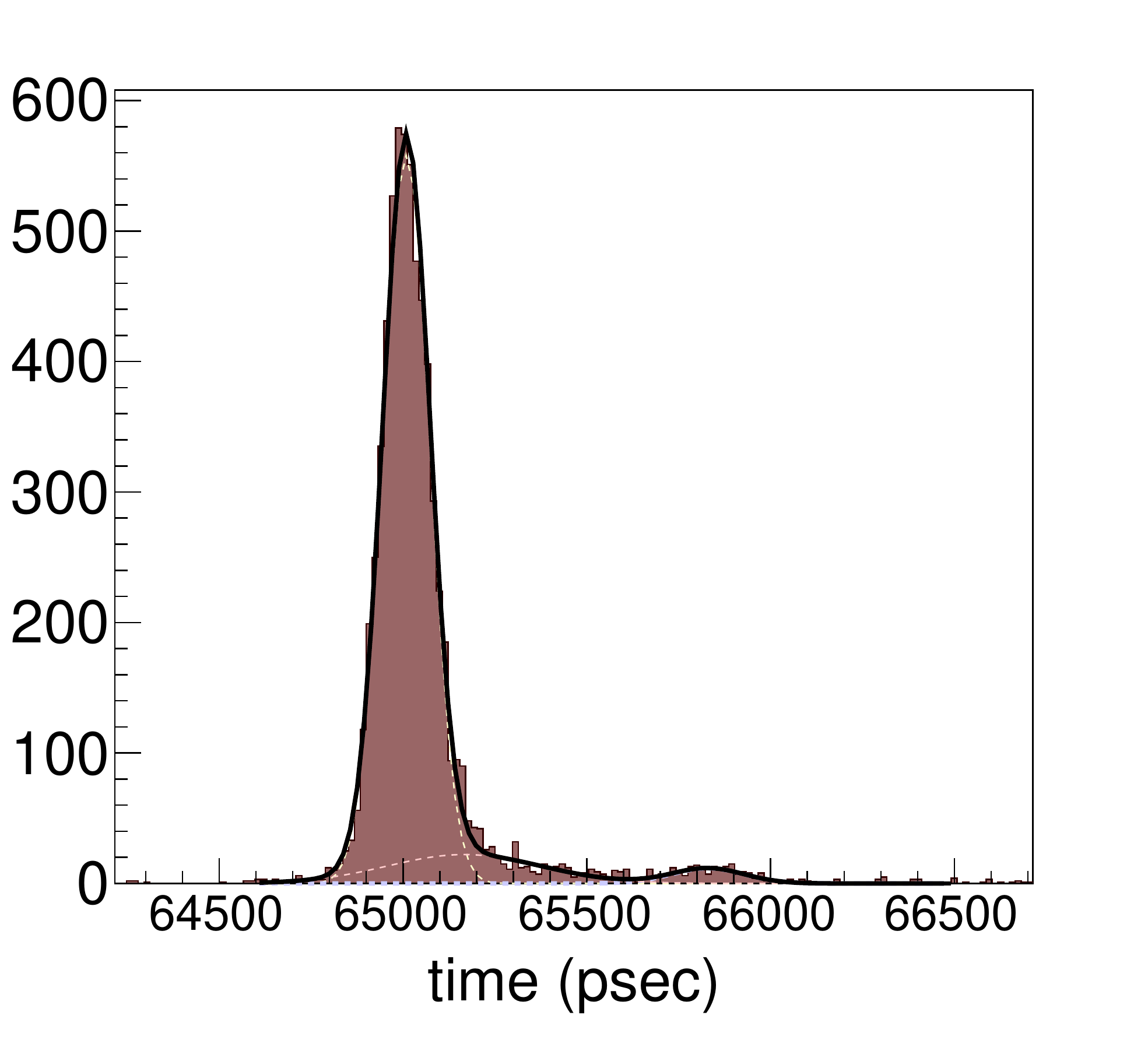}
\caption{The Transit Time Spread (TTS distribution) for ANNIE LAPPD-25. The time resolution for the prompt pulses is below 70ps, where roughly 30ps of smearing is due to the pulse duration of the diode laser used.}
\label{fig:LAPPDTTS}
\end{figure}

The combination of spatial and temporal information make LAPPDs ideal for Scintillation-Cherenkov separation. The $<$100 psec resolution of LAPPDs makes it possible to separate between the two components on the basis of timing alone. 

\subsubsection{Spectral Sorting and Dichroicons}

 One approach to separating Cherenkov and scintillation light is by discriminating photons by wavelength, as scintillation is typically within a narrow emission band, while Cherenkov is a broad spectrum of light, falling as roughly $1/\lambda^2$. 

A simple approach to doing this would be to add filters in front of some subset of the photon sensors in a detector, filtering out the blue scintillation light and allowing longer-wavelength Cherenkov light to pass.  This has the advantage of simplicity, and in a detector with low photocathode coverage would be reasonable, if the size of the ``red-sensitive'' PMTs were large enough to detect enough long-wavelength Cherenkov photons.  As the coverage increases however, using filters in front of a subset of the PMTs means that some of the scintillation light is lost, and for physics that requires narrow energy resolution, every photon is important.

Dichroicons, 
get around this problem by {\it sorting} photons by wavelength.
As shown in Figure~\ref{fig:dichroicon} below, the dichroicon can follow the off-axis parabolic
\begin{figure}[ht!]
\centering
\includegraphics[width=0.35\textwidth]{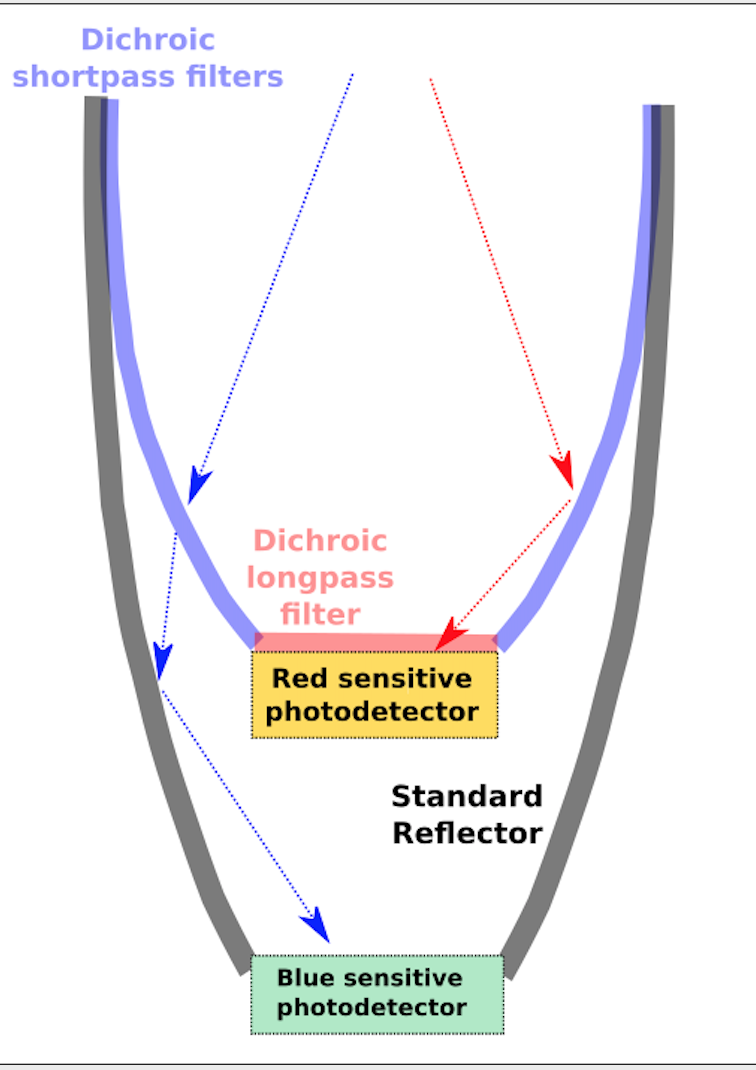}
\includegraphics[width=0.45\textwidth]{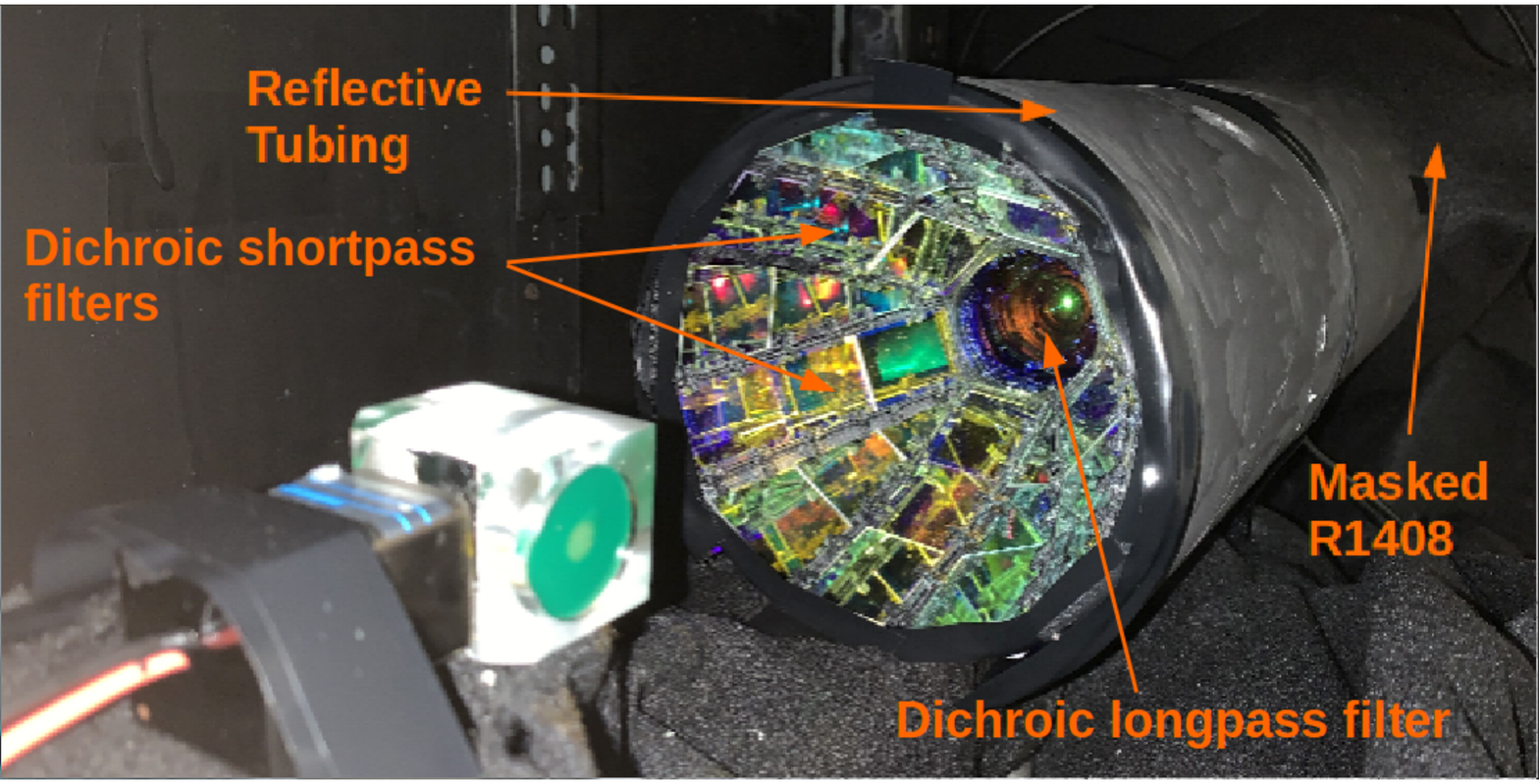}
\caption{Left: Schematic of a “nested” dichroicon configuration, with shortpass filters on the barrel and a longpass filter at the aperture. Right: Dichroicon in benchtop test setup, viewing $^{90}$Sr source embedded in an acrylic block containing LAB-PPO.}
\label{fig:dichroicon}
\end{figure}
design of an ideal Winston light concentrator but is built out of a tiled set of
dichroic filters. The filters are used to direct long-wavelength light towards
a central red-sensitive PMT, while transmitting the shorter wavelength light
through the ``barrel'' of the Winston cone to secondary photodetectors. This is
possible because of the remarkable property of the dichroic reflectors, which
reflect one passband of light (below or above a ‘cut-on’ wavelength) while
transmitting its complement, with very little absorption. As shown
schematically in one possible design in Figure~\ref{fig:dichroicon}, the
barrel of the dichroicon is built from shortpass dichroic filters (cutoff
wavelength near 480~nm) and a longpass dichroic filter is placed at the
aperture of the dichroicon. The shortpass filter passes short-wavelength light
while reflecting long-wavelength light; the longpass has the complementary
response. In the ``nested'' design shown, the back PMT detects the short
wavelength light.

There are many possible configurations of the dichroicon; the ones built to
date are not necessarily optimal, and different detectors may have different
needs. The nested photon sensor configuration of the design above requires
more than one photon sensor and is thus most useful when the available
detection area is limited (for example, when the desired coverage is $>50$\%, or
in a segmented detector where each segment is viewed by a single sensor).
Simpler designs could simply offset the dichroicon and its aperture PMT,
collecting the low-yield Cherenkov photons while allowing the scintillation
light to be detected by the rest of the PMT array. Using a pixelated photon
sensor, such as an LAPPD or an array of SiPMs, would also work, with the pixels
then mapping to different wavelength bands. A complementary design—--with
Cherenkov light passing through the barrel and scintillation light reflected
toward the aperture—--might be most useful when ring imaging is a high
priority.  To achieve more than two passbands, the ``nested'' design could be
extended using mutliple dichroicons.

\begin{figure}[ht!]
\centering
\includegraphics[width=0.35\textwidth]{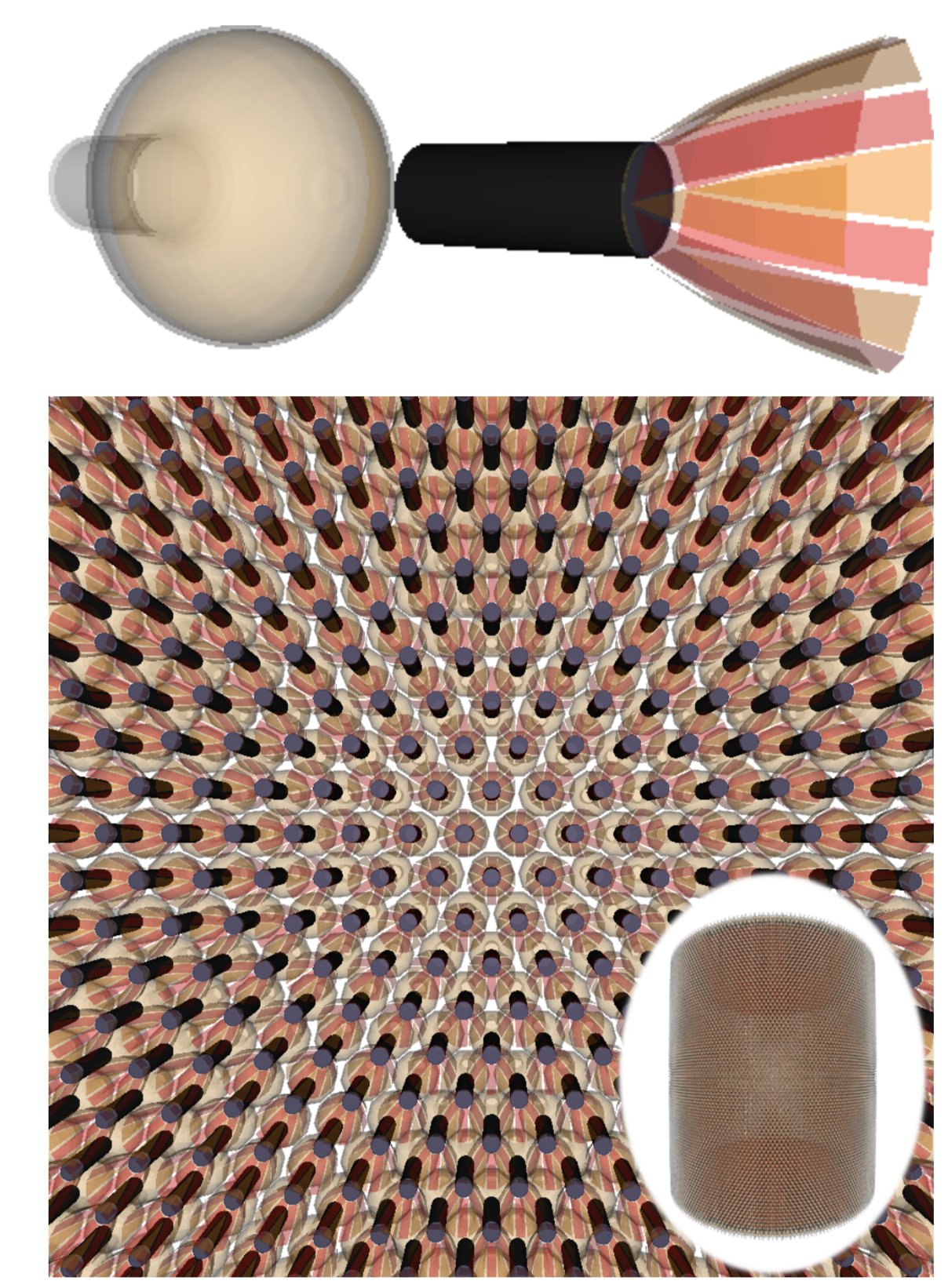}
\caption{A rendering of the {\it Chroma} geometry for (top)  a dichroicon unit with (left to right) 20” large area PMT, cylindrical 5” long-wavelength PMT, and dirchroic filter con- centrator, and (bottom) the dichroicon units tiled to create a 50 kt neutrino detector, viewed from inside the detector, with the full cylindrical geometry shown in the inset.}
\label{fig:chromadich}
\end{figure}

A simulation model of the dichroicon has been developed in {\it Chroma} (See Section~\ref{sec:chroma}) and calibrated against the measurements of a benchtop prototype, for use in simulating dichroicon performance large-scale detectors. This model was simplified and scaled up to use a 20" diameter large area PMT to collect photons passed by the short-pass dichroic filters, and a cylindrical 5” PMT to collect the long wavelength photons reflected by the dichroic filters shown at the top of Fig.~\ref{fig:chromadich}. This dichroicon unit was then tiled around a cylindrical volume to simulate a large-scale neutrino detector, seen from the inside at the bottom of Fig.~\ref{fig:chromadich}. In this model a 50 kt volume of the scintillator LAB with 2g/L of PPO [10] is surrounded by 90\% 
coverage of simplified dichroicons, which gives effectively 90\% coverage of both long and short wavelength photons, as the dichroic filters allow the long and short PMTs share the same solid angle. Using this model, we have begun to evaluate the impact of spectral photon sorting on future neutrino experiments.

A reconstruction algorithm developed in~\cite{wblsberk} has been modified such that it uses hit time information from all (short and long wavelength) PMTs to perform a position and time fit, and then uses the angular distribution of photons detected on long wavelength PMTs to perform a direction fit. 

\begin{figure}[ht!]
\centering
\includegraphics[width=0.35\textwidth]{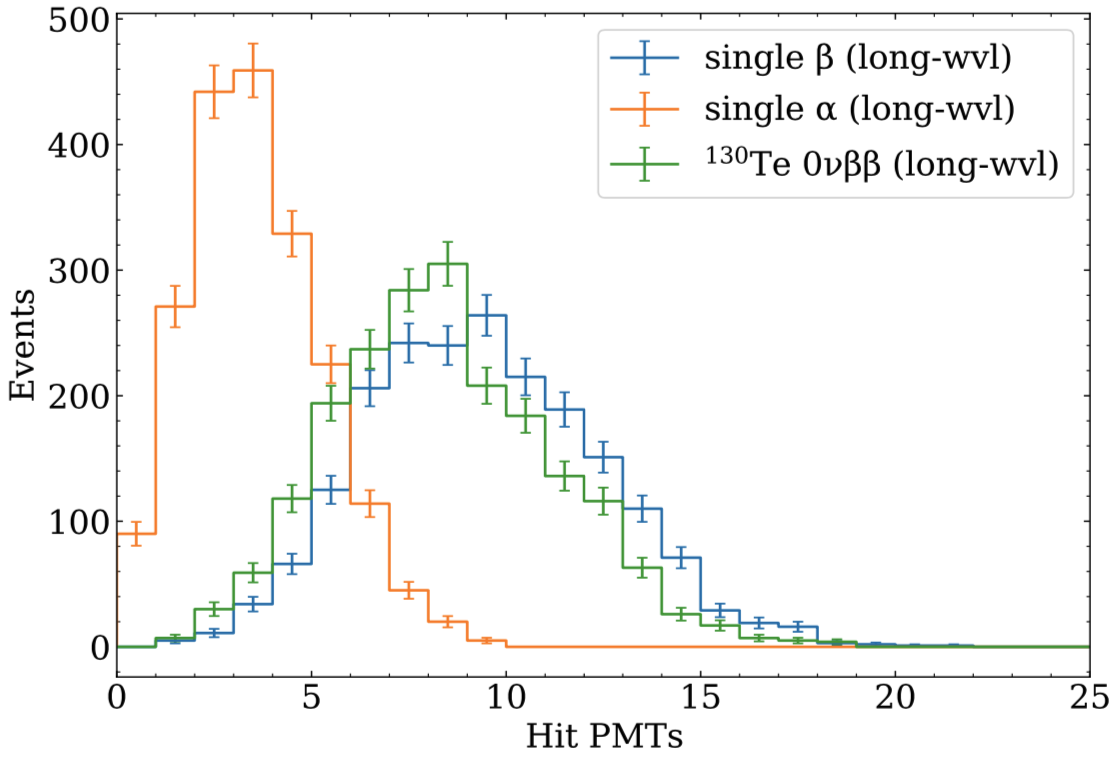}
\caption{Number of long-wavelength PMTs that detected photons for alpha and beta events with the same quenched energy as $^{130}$Te neutrinoless double beta decay, showing alpha/beta discrimination based on Cherenkov photon identification in a 50 kt liquid scintillator detector.}
\label{fig:dichpid}
\end{figure}

\begin{figure}[ht!]
\centering
\includegraphics[width=0.35\textwidth]{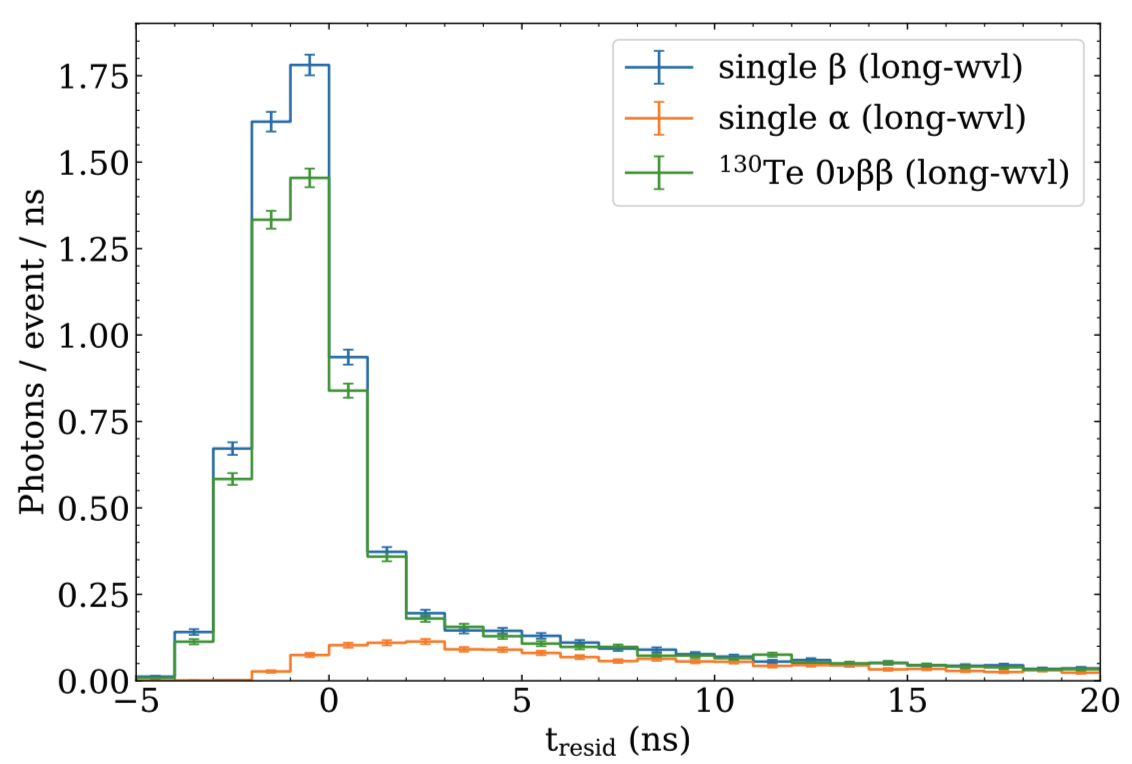}
\caption{The hit time residuals of long-wavelength PMTs for the events shown in Fig.~\ref{fig:dichpid}. The Cherenkov is clearly visible at early times for the beta events.}
\label{fig:dichtime}
\end{figure}

Alpha particle identification has been explored by simulating alpha and beta particles with the same quenched energy (number of scintillation photons) and inspecting the signal on the long-wavelength PMTs, which are pri- marily sensitive to Cherenkov photons. A quenched energy comparable to neutrinoless double beta decay in 130Te is chosen to highlight potential background rejection capabilities. The mean number of long-wavelength hits is indeed higher for betas than alphas as shown in Fig.~\ref{fig:dichpid} due to Cherenkov production with betas. The detected long wavelength photons from alphas are all scintillation, and indicate that the filters could be optimized for better rejection of scintillation. Fig.~\ref{fig:dichtime} shows a clear Cherenkov signal in the hit time residuals of betas at early times, which would allow for discrimination of alphas in a neutrinoless double beta decay region of interest. A similar method provides some discrimination between single and double beta events, as can be seen in the $^{130}$Te 0$\nu\beta\beta$ plots in Figs.~\ref{fig:dichpid} and~\ref{fig:dichtime}, which could further constrain backgrounds.

\subsubsection{Using Angular Information}

    The ring topology itself can be used to identify Cherenkon photons, essentially by reconstructing events using PDFs in time and angle.  The CHESS array~\cite{chess} has shown this works for cosmic-ray muons, and work with simulations~\cite{wblsberk} has shown that even at lower energies the separation can be good enough if the scintillation light yield is low (as one would see in a WbLS detector, or a detector with a low level of secondary fluor added). 

\subsubsection{Comparing Cherenkov/scintillation Separation Approaches}

\begin{figure}[ht!]
\centering
\includegraphics[width=0.25\textwidth]{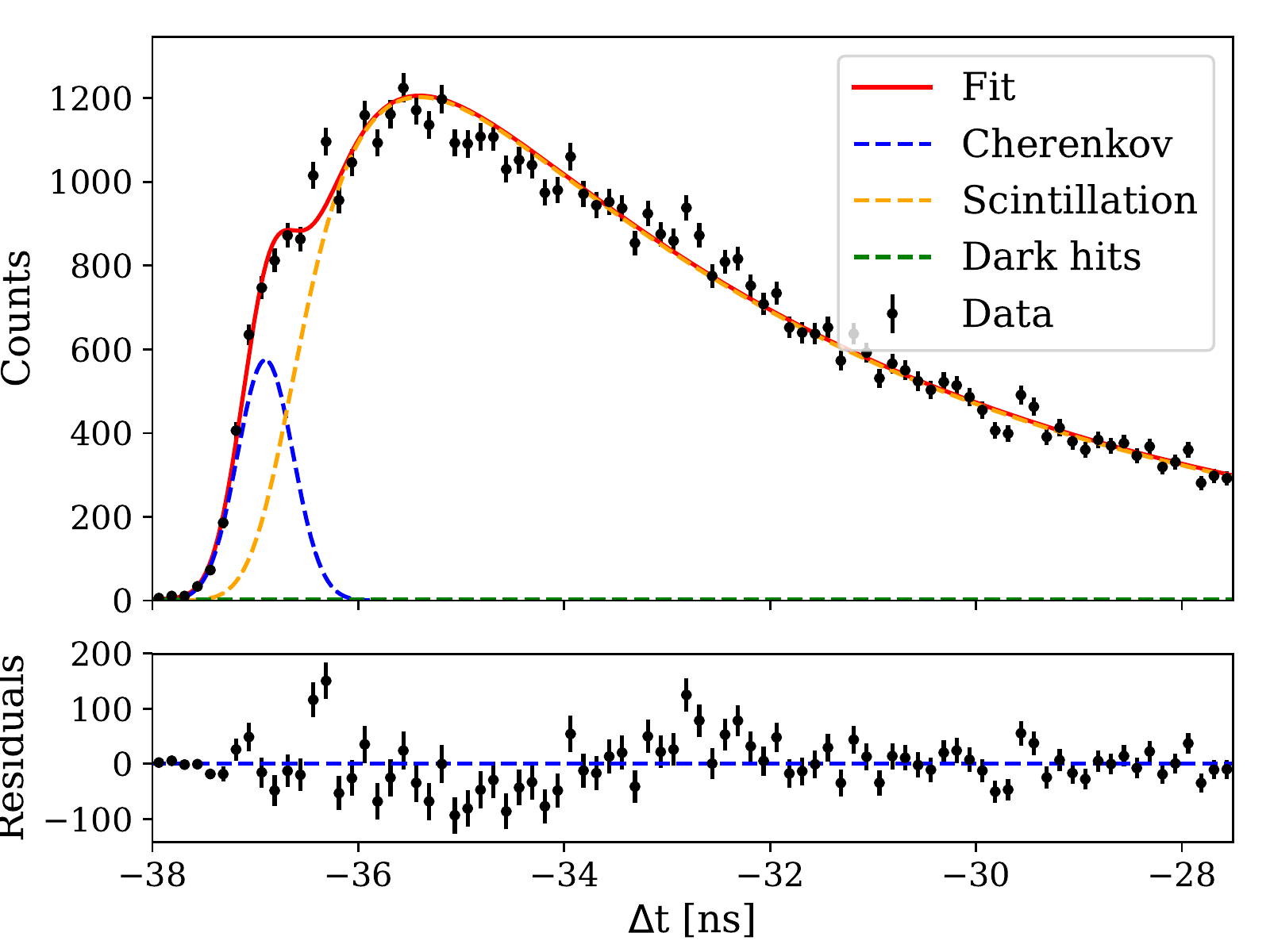} 
\includegraphics[width=0.25\textwidth]{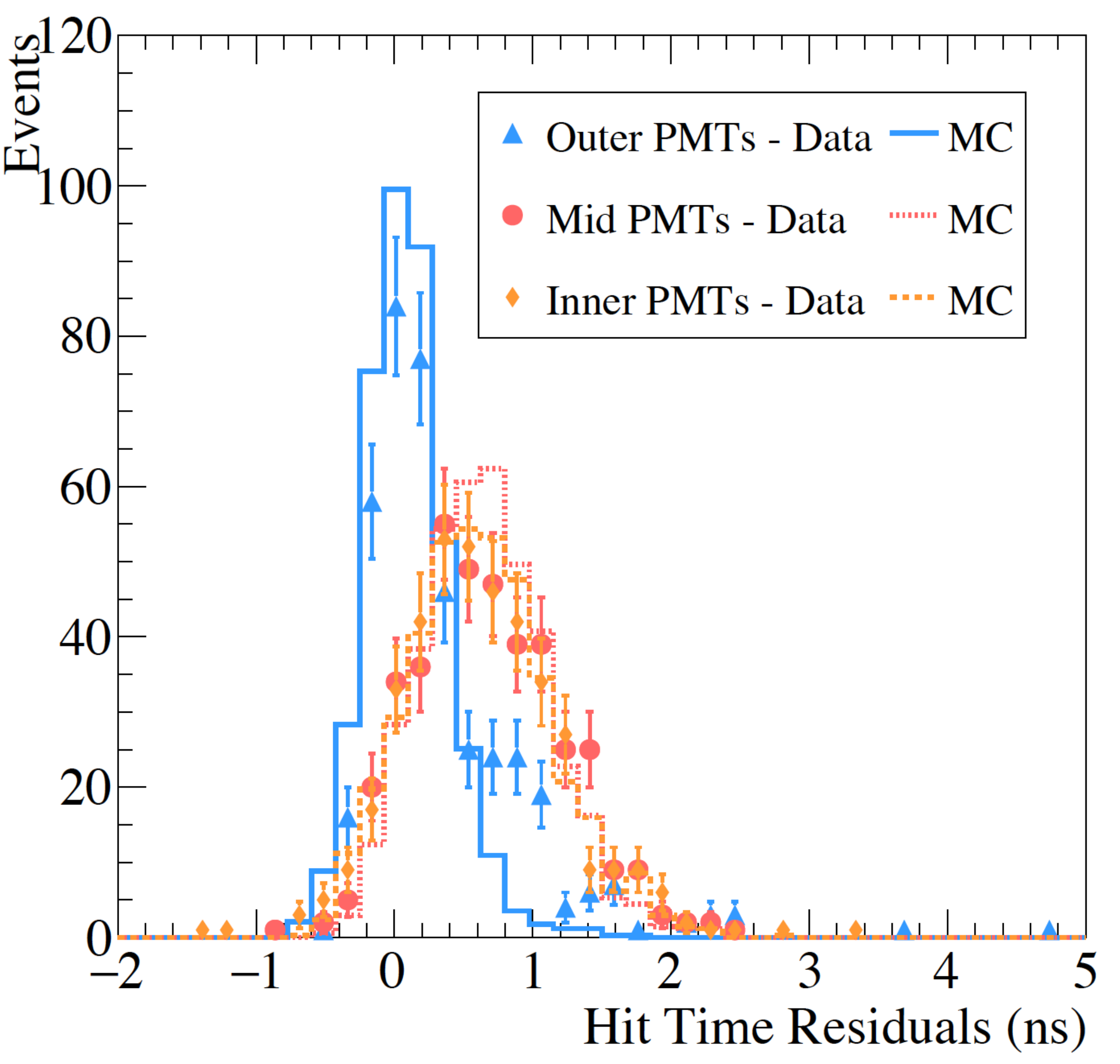} 
\includegraphics[width=0.25\textwidth]{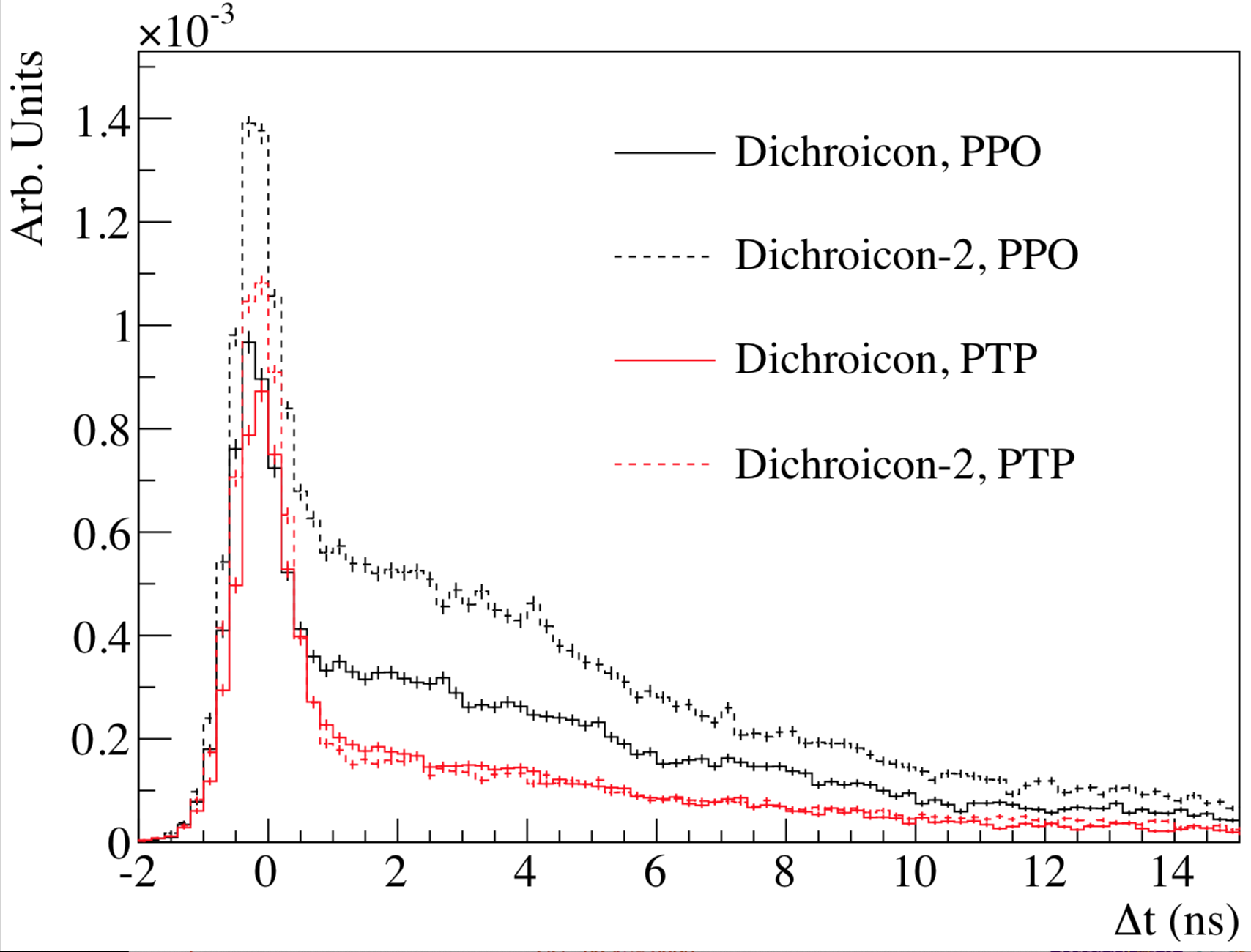}
\includegraphics[width=0.25\textwidth]{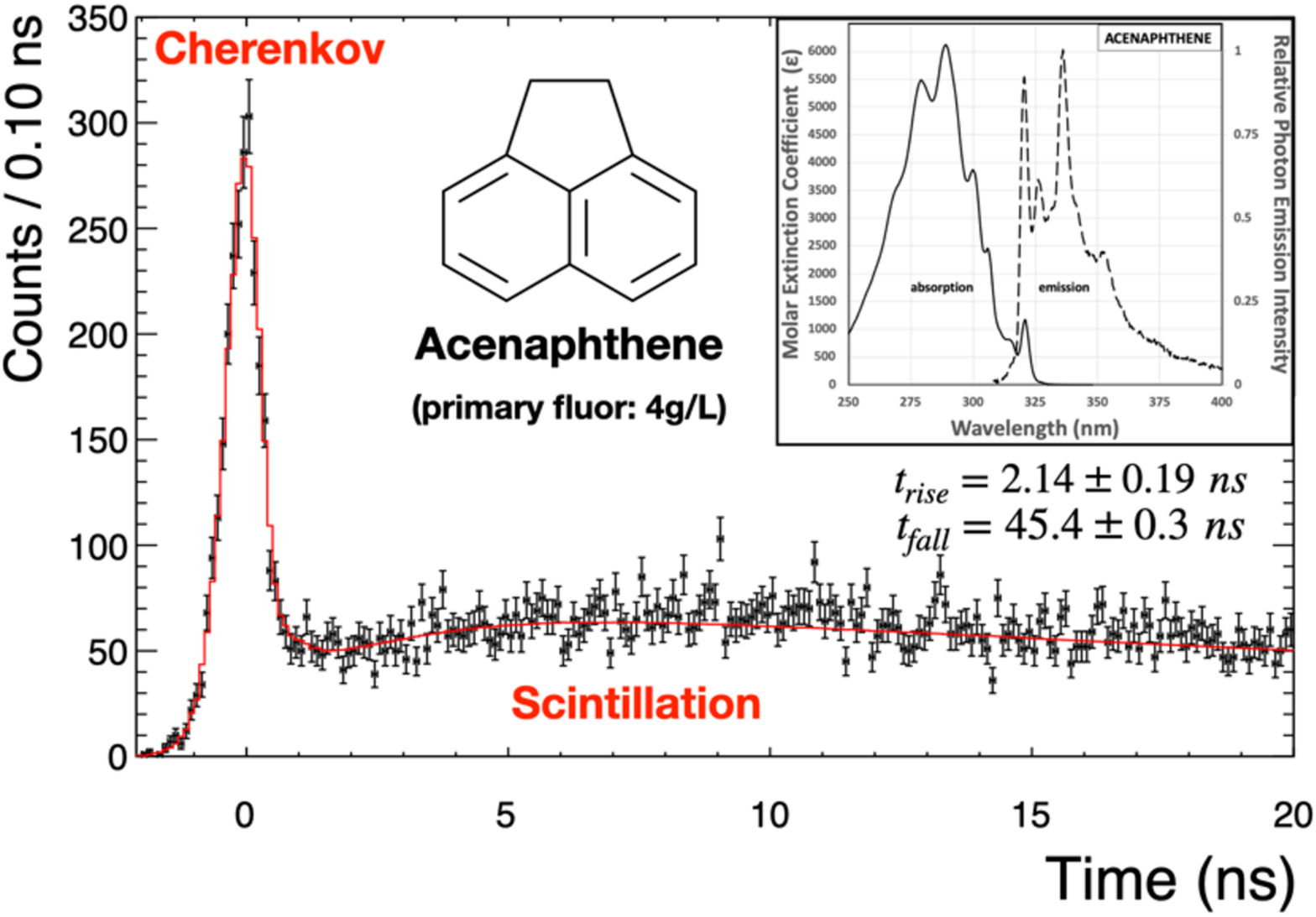} 
\includegraphics[width=0.25\textwidth]{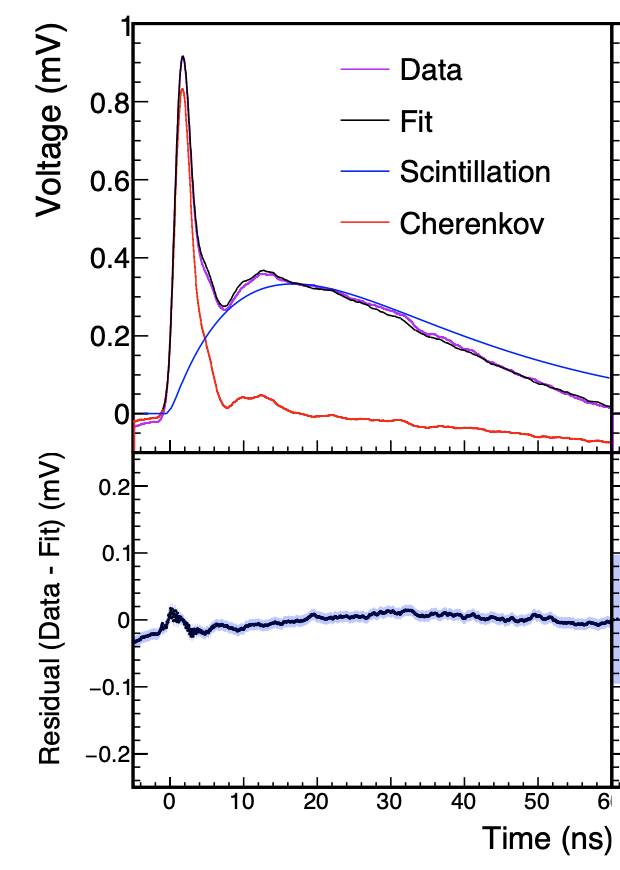} 
\includegraphics[width=0.25\textwidth]{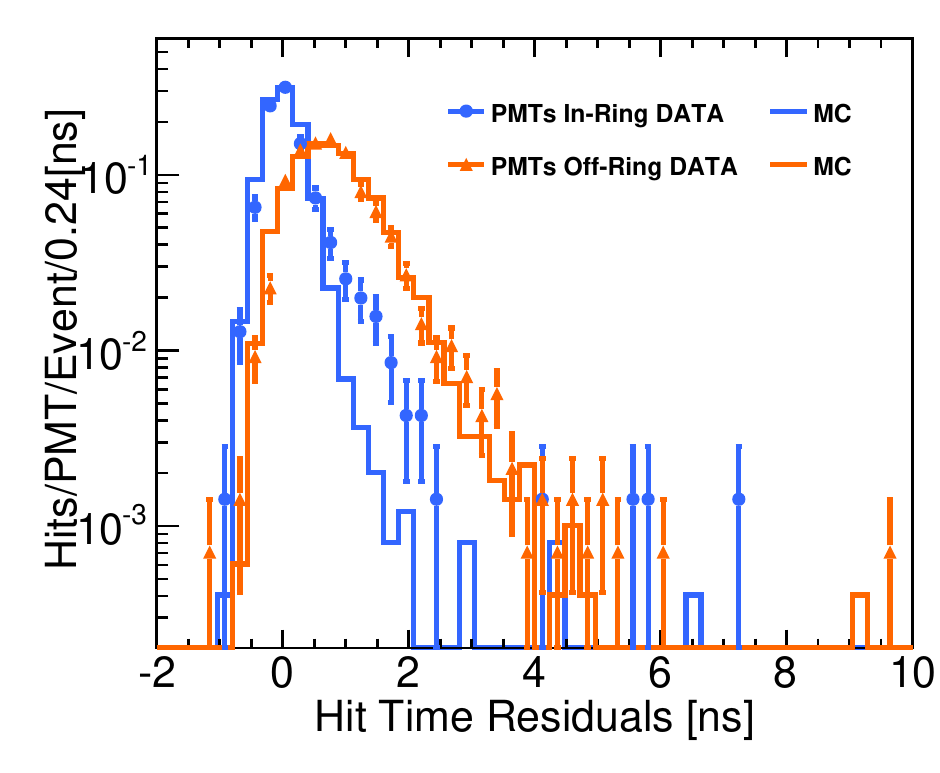} 
\caption{The hit time residuals for several different approaches to separating out Cherenkov and scintillation light. \label{fig:cherscintcomp}}
\end{figure}

    Any comparison between the approaches discussed above is dangerous because they have all been done on the benchtop, with different sources, photon sensors, analyses, and other conditions.  And because they will each scale differently in large detectors, depending on the target material and size of the detector itself. Nevertheless, it is interesting to see side-by-side so many different successful ways of doing this discrimination which, at the time of the last SNOWMASS, was mostly just an idea.
    Figure~\ref{fig:cherscintcomp} shows distributions from the various approaches side-by-side.

\subsubsection{Further Technological Needs}

    Despite the success of the development of these technologies, there are still may areas where additional development would have great benefit.  We describe these below:
    
    \begin{itemize}
    \item {\bf Lower-cost, large area, high-quantum efficiency photon sensors:}  While improvements in timing and quantum efficiency in traditional photomultipler tubes has been dramatic over the past several years, clearly the biggest single win for future large-scale detectors would be if these could b made less expensively, or with even higher photocathode efficiencies.
    \item {\bf Further advances in dichroic filters:} Dichroic filters are still expensive, and their production in large areas and on curve surfaces (like the dichroicon) has not yet been done.  In addition, improvements in the angular response of such filters, with sharp cut-on an cut-off tansmission and reflection curves as a function of angle would enable much better photon-sorting capabilities.
    \item {\bf Narrow-band fluors:} Any Cherenkov/scintillation separation using wavelength is hampered by the fact that most organic fluors are relatively broad-band compared to the sensitivity of typical photocathodes; they leave little room for ``clean'' long-wavelength Cherenkov detection.  A fluor that had an emission band from just 400-450 nm or something similar would make spectral sorting much more effective.
    \item {\bf High-yield scintillators with attenuation lengths $>$ 40~m:} Limitations on the size of big detectors using scintillation light come primarily from the short attenuation lengths these scintillators have.  WbLS is one approach to improving this, but it may come at the cost of lower light yield than some physics requires.  A high-light yield scintillator that had 40~m attenuation lengths or longer for the primary emission bands would make very large-scale (e.g., 250~ktonne) detectors possible.
    \item {\bf High light-yield ``slow'' ($\sim 10$ns or longer) fluors:} Existing slow fluors already look very good, but if the fluor is too slow, position reconstruction is poor, and particle ID may also be compromised.  A fluor that has a long risetime to allow identification of Cherenkov light, but maintains a light yield like some of the brightest scintillators while keeping good $\beta$-$\alpha$ discrimination would be an important step forward.
    \item {\bf New approaches to radiological background reduction:} Although not specific to hybrid detectors, for all Cherenkov and scintillation detectors that aim for a broad program that includes low-energy physics, continued reductions in radiological contamination is always critical.  This may be particularly important for new materials like WbLS or slow fluors.
    \end{itemize}


\subsection{Isotopic Loading Techniques for Low-energy Physics Programs}

    One of the great ways of expanding and improving the physics programs of photon-based detectors is through the loading of various isotopes.  These isotopes can be the source of the physics itself---as in neutrinoless double beta decay experiments---or leverage additional physics, such as the Gd loading of water to detect neutrons from supernova neutrinos, or $^6$Li to improve pulse-shape discrimination. We discuss some of the possibilities below.
    
    \subsubsection{Metal Loading in WbLS}
    \label{sec:mdwbls}
    
        Water-based liquid scintillator (WbLS) decribed above in Section~\ref{sec:wbls}, as an additional advantage in that loading of many isotopes is possible, using
        a mix of surface-active agents with hydrophilic as well as hydrophobic chelating groups to bridge aqueous (polar) and organic solvents (nonpolar). 
         WbLS allows inorganic metal salt to be first dissolved in water, followed by directly blended into any type of scintillator solvents, 
        regardless of the chemical property of each scintillator. WbLS has 100\%  metallic extraction efficiency and is particularly operative in extracting the hydrophilic elements (e.g. Boron, Lithium) into organic solvents.
       The metal-doped principal for oil-based WbLS ($>$80\% scintillator) has been successfully demonstrated by PROSPECT experiment, a $\sim$0.1\% 6Lithium-doped DIPN-based WbLS, with good light yield ($>$10,000 ph/MeV) and superior pulse shape discrimination.

A summary of applications and competences of metallic targets doped in WbLS, either being deployed or still under development, is presented in Table~\ref{tbl:wblsload}.
\begin{table}[hbt!]
\begin{centering}
\begin{tabular}{|l|l|l|}
\hline\hline
Target	&  Loading (mass) & Potential Applications\\
\hline
 Indium		 & $>$8\% In  &  Solar $\nu$ \\ 
\hline
 Tellurium 		 & $>$ 6\% Te  &  $0\nu\beta\beta$\\ 
\hline
 Lithium 		 & $0.1$\% $^6$Li  & Reactor $\bar{\nu}$; excellent PSD  \\ 
 &  $>$0.2\% $^6$Li  & Reactor $\bar{\nu}$; super PSD with improved optics \\ 
\hline
 Boron		 & $>$0.5\%   & Dark Matter veto, reator $\bar{\nu}$ \\ 
\hline
 Potassium 		 & $>$1\%  &  Calibration for LS detectors\\ 
\hline
 Iron, Strontium & ppm to 1\%  & Nuclear waste management, \\ 
 &   & enviromental tracers\\ 
\hline
 Gadolinium & 0.1\% Gd  & Dark matter veto \\ 
 &   & Reactor monitoring\\ 
 &   & Reactor $\bar{\nu}$ oscillations\\ 
\hline
\hline
 High-Z elements & 10-15\%Pb  & Solar $\nu$ \\ 
 &   & Calorimeters \\ 
 &   & Medical QA/AC \\ 
\hline\hline
\end{tabular}
\caption{Examples of metallic targets loaded in Liquid Scintillator and Water-based Liquid Scintillator with projected applications\label{tbl:wblsload}}
\end{centering}
\end{table}

\subsubsection{Te Loading in Liquid Scintillator for $0\nu\beta\beta$}

A method for loading tellurium into organic liquid scintillator has been developed based on the formation of soluble organic compounds derived from telluric acid (Te(OH)6, hereafter TeA) and 1,2-butanediol (BD) in conjunction with N,N-dimethyldodecylamine (DDA), which acts as a stabilisation/solubilisation agent. The chemicals involved can all be purified to high levels, have high flash points and are relatively safe to work with in underground environments. The loading process results in acceptable optical absorbance and light output in larger detectors for loading levels up to several percent Te by weight. Stability of the loading has been demonstrated to be at least on the timescale of years.

\begin{figure}[htp]
\centering 
\includegraphics[width=0.6\textwidth]{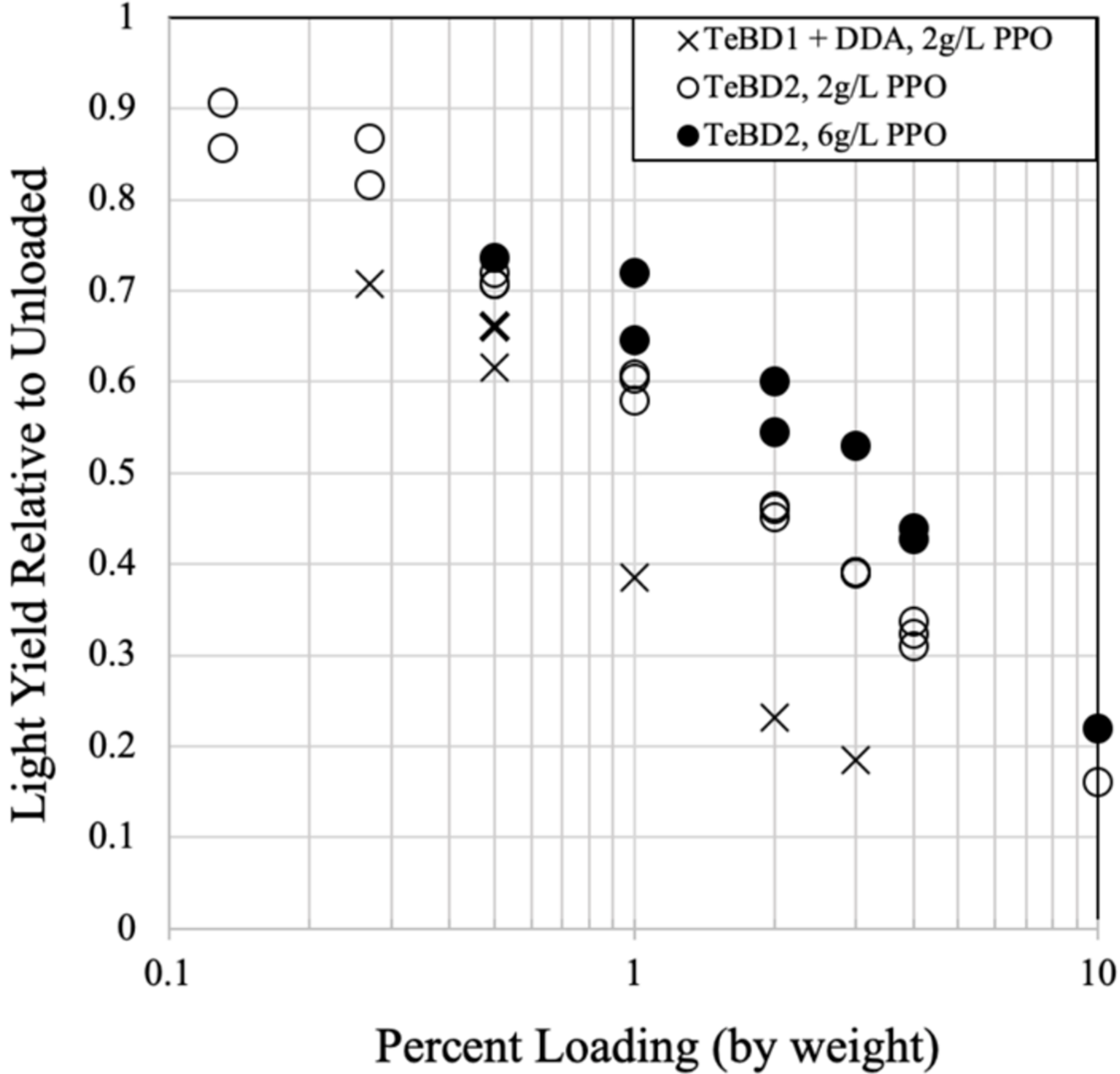}
\caption{Relative light levels as a function of percent Te loading for two variants of the technique: TeBD1 (heated solubilization) and TeBD2 (DDA-induced solubilization.} 
\label{fig:SlowFluor4}
\end{figure}

This is an important advance that opens the door to a highly scalable and economical approach for neutrinoless double beta decay. Further advances in Te loading and associated purification techniques could provide a practical path to realising sensitivity to the non-degenerate normal mass ordering.

\subsubsection{$^6$Li Loading in Plastic Scintillator}

\begin{figure}[h!]
     \centering
         \centering
         \includegraphics[width=\textwidth]{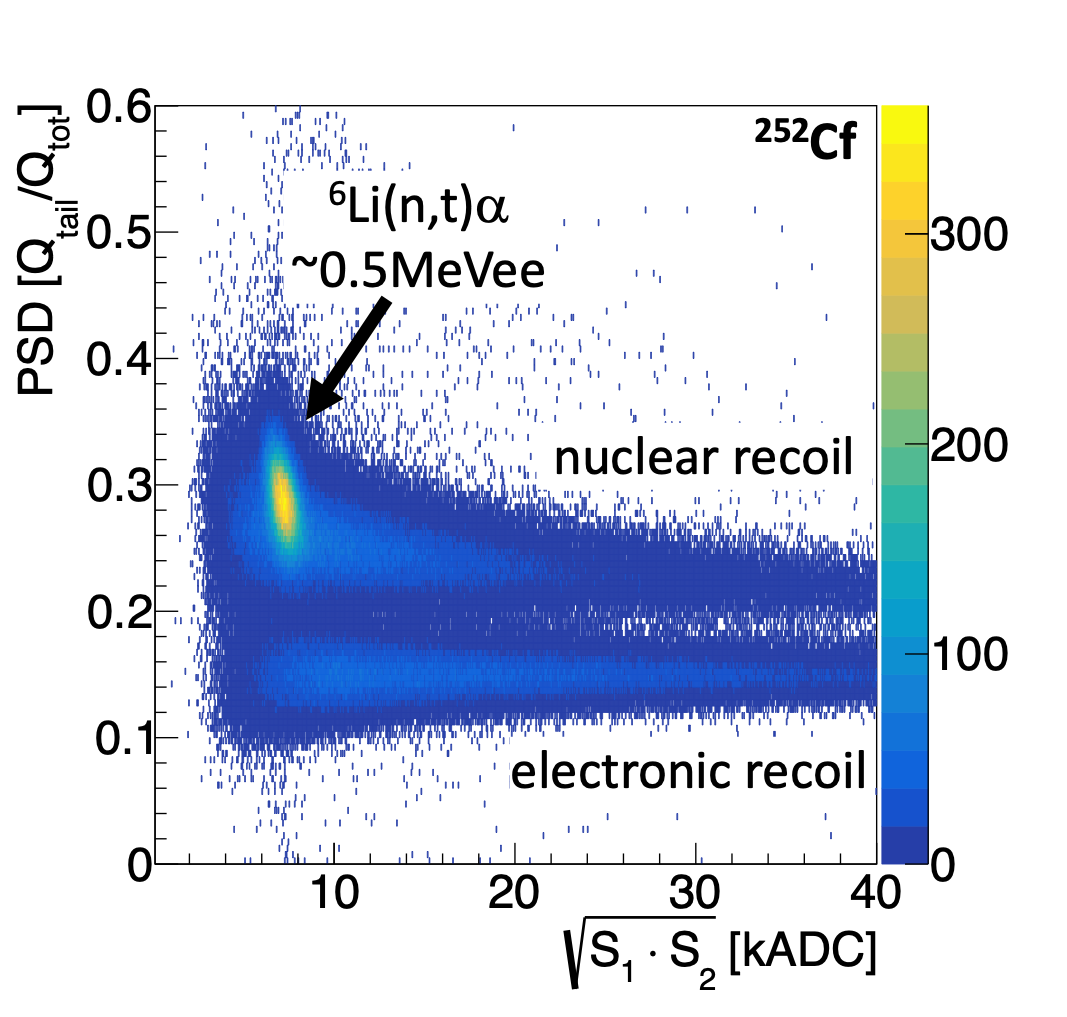}
         \caption{Pulse Shape Discrimination performance of the 5.5~cm $\times$ 5.5~cm $\times$ 50~cm $^6$Li-doped PSD plastic scintillator shown in panel (b) under $^{252}$Cf flood-field illumination , in terms of uncalibrated ADC units. A clear neutron capture feature is observed at about 0.5~MeV$_{ee}$, along with electronic and nuclear recoil separation.}
         \label{fig:Li6-PSD}
         \end{figure}

\begin{figure}[h!]
         \centering
         \includegraphics[width=\textwidth]{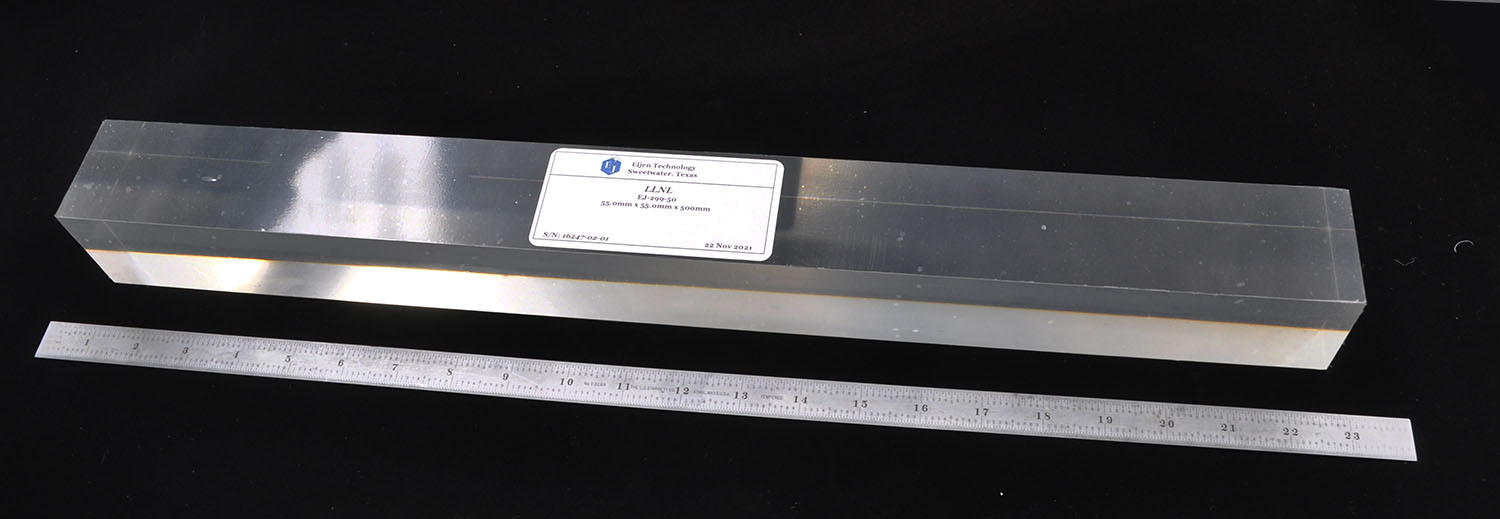}
         \caption{A 5.5~cm $\times$ 5.5~cm $\times$ 50~cm bar produced by Eljen Technology. \label{fig:Li6-PS}} 
\end{figure}

Plastic scintillators are comprised of dopants entrained in a polymer matrix. In some cases, the scintillation signal from these dopants can be used to discriminate between particle types, e.g. electromagnetic depositions vs those from heavy ions, via ``Pulse Shape Discrimination'' (PSD)~\cite{ZAITSEVA201288}. Further particle discrimination can be enabled by other dopants that have specific neutron capture reactions, which can also be identified via a combination of energy gating and PSD. For example, when loaded with lithium-6 and high amounts of a primary dye and secondary dye,
plastic scintillators can use PSD to distinguish between electrons or positrons, neutrons, and thermalized neutron captures~\cite{ ZAITSEVA2013747, MABE201680}. 
At the current stage of development these plastics have broadly similar performance in terms of light yield, optical attenuation, and PSD separation to liquids (e.g. Fig.~\ref{fig:Li6-PSD}). Elements have been produced with greater than 50-cm length and 7.5-cm width, and a commercial vendor is developing a production process (Fig.~\ref{fig:Li6-PS}). 

A key challenge to the production of these $^6$Li-doped PSD plastics is the balance between scintillation properties and solubility of the dopants in the polymer matrix and the polymer precursors (i.e., monomers). 
One option to improve the solubility of these lithium-6 salts is to incorporate a co-monomer with styrene that can dissolve the lithium-6 salts. 
Methacrylic acid is one possible co-monomer that has polar substituents that can solubilize the lithium-6 salts~\cite{osti_1490925,Frangville}. 
Another option for soluble lithium-6 salts is to use organic salts with aromatic organic groups that have good solubility in primary dyes like PPO. 
Overall, the balance of scintillation properties and solubility of dopants that enable multiple modes of PSD remains an interesting challenge to investigate and solve.

Plastic scintillators with PSD and $^6$Li-doping provide the same capabilities as the PSD $^6$Li-doped liquid scintillator used for  PROSPECT that is described in Sec.~\ref{sec:mdwbls}. 
As described in a recent LOI~\cite{Li6Organic:2021loi}, the new solid form makes self-supporting fine-grained segmentation on a large scale possible, enables light guiding approaches based on air gaps, and eliminates the need for containment vessels and supporting infrastructure. Several detector implementations are proposed and under development using $^6$Li-doped PSD plastics. The SANDD concept (Fig.~\ref{fig:Li6-PS}) has implemented mm-scale segmentation to study reactor antineutrino directionality and cosmogenic background rejection~\cite{Li:2019sof,Sutanto:2021xpo}. Segmentation at the 5--10-cm scale in 2D bar geometries is being developed for efficient, aboveground reactor antineutrino detection (e.g. ROADSTR~\cite{ROADSTR:2021loi}) with the goal of a mobile system appropriate for monitoring applications and measurements at  multiple reactors with correlated systematics.


\subsubsection{Quantum Dots}

In recent years new methods for the production of metal loaded LS in neutrino physics were studied involving semiconducting nanocrystals, which are known as ``quantum dots''. The optical and electrical properties of the quantum dots are directly proportional to their size, which is typically in the order of few nanometers. The emission band consists of a narrow resonance around the characteristic wavelength of the dot.  Since the dot size can be controlled to high precision in the synthesis, the absorption and re-emission spectrum of the dots can be tuned and optimized for a respective application. In some synthesis methods the quantum dots are already delivered in colloidal suspension with the aromatic solvent toluene at concentrations of several grams per liter.

The most commonly used quantum dot cores are binary alloys such as CdS, CdSe, CdTe, and ZnS. Alternatively, there are also phosphor-based rare-earth dots. Therefore, quantum dots provide a method to dope scintillator with various metals and rare-earth elements. For a cadmium (Cd) based LS there are two different applications in the field of neutrino physics: neutron-enhanced isotopes (113Cd) and $\beta\beta$-decay candidates (106,116Cd). Also Se, Te, and Zn, which are present in common quantum dot cores, have $\beta\beta$-decay candidates. There is also a possibility to tune the scintillator emission spectrum in a way that allows to separate scintillation from Cherenkov light by narrowing scintillation emission from the quantum dots separated from the Cerenkov contribution.

The basic limitations of quantum-dot-doped liquid scintillator in use for particle physics detectors are probably cost and availability in large quantity ($\sim$ ton). The stability tests on typical solutions of quantum dots in concentration at few g/l indicate that larger particles are formed by aggregation in the concentrated solutions over long time scales. This could explain the fact that filtering improves the attenuation length as well as the observation of transparency degradation after several weeks. However, there is room for optimization on the stability and optical performance by improvements in the loading process of quantum dots in scintillator solution. Instead of suspension, incorporating chelating agents or WbLS surface active agents into the mixing procedures could load the quantum dots in organic solvents homogeneously with lengthy stability.



\subsection{Improvements in Simulation and Analysis}

    The ever-increasing desire for more precise detectors models and better background rejection has led to development of many new approaches to simulation and analysis.
    There are far too many to include all of them here, but we have focused on a few that were represented in relevant LOIs to NF10.

\subsubsection{GPU-Accelerated Photon Ray Tracing ({\it Chroma})}
\label{sec:chroma}

As large-scale neutrino detectors become ever larger, photon coverage
becomes higher, and photon sensor technology becomes more complex with devices 
like the ARAPUCA~\cite{arapucas}, the dichroicon~\cite{dichroicons}, or 
distributed imaging~\cite{gratta_image}, a major bottleneck in both 
simulation and reconstruction of physics events is the propagation of photons 
through the detector geometry. 
Originally created for the water Cherenkov option for the LBNE experiment, a
fast photon ray-tracer was developed by Stan Seibert and Anthony
LaTorre~\cite{chroma} that improved photon simulation speeds by a factor of 200 over
what GEANT4 itself could do.  In order to achieve such high performance, {\it
Chroma} combines techniques from 3D rendering algorithms with the massively
parallel calculation hardware inside GPUs.  
A high-end GPU costs approximately \$5000
(twice the cost of a fast consumer-grade CPU), yet provides forty times the raw
floating point performance and ten times the memory bandwidth.  {\it Chroma}
uses the CUDA toolkit, provided by NVIDIA, to directly access the GPU resources
and perform all major calculations.  CUDA-compatible GPUs are being used more
and more in the construction of large supercomputing clusters, which will make
it easier in the future for the work on next-generation neutrino detectors such
as \textsc{Theia}~\cite{theiawp} to use {\it Chroma}.

Nearly all fast 3D rendering systems represent the world geometry using a mesh
of triangles.  Triangle meshes are very simple to represent, and can be used to
approximate any surface shape, limited by how much memory can be devoted to
triangle storage.  With only one surface primitive, there is only code path to
optimize.  In particular, we have adopted the Bounding Volume Hierarchy (BVH)
technique from the graphics world to speed up ray intersection tests with
triangle meshes.  A BVH is a data structure that organizes a spatial
arrangement of shapes (triangles in our case) into a tree of nested boxes.
Rather than test for ray intersection with every triangle in a geometry, the
photon propagator tests for intersection with boxes in the BVH.  If the ray
does not intersect the box, then all of the children of that box can be
skipped, leading to a large reduction in the number of intersection tests
required.  For example, a model of a large, 200 kton water Cherenkov detector
consists of 62 million triangles, but the BVH reduces a typical propagation
step for a photon to 130 box intersection tests and only 2 triangle
intersection tests.

\begin{figure}[h!]
\centering
\includegraphics[width=0.5\textwidth]{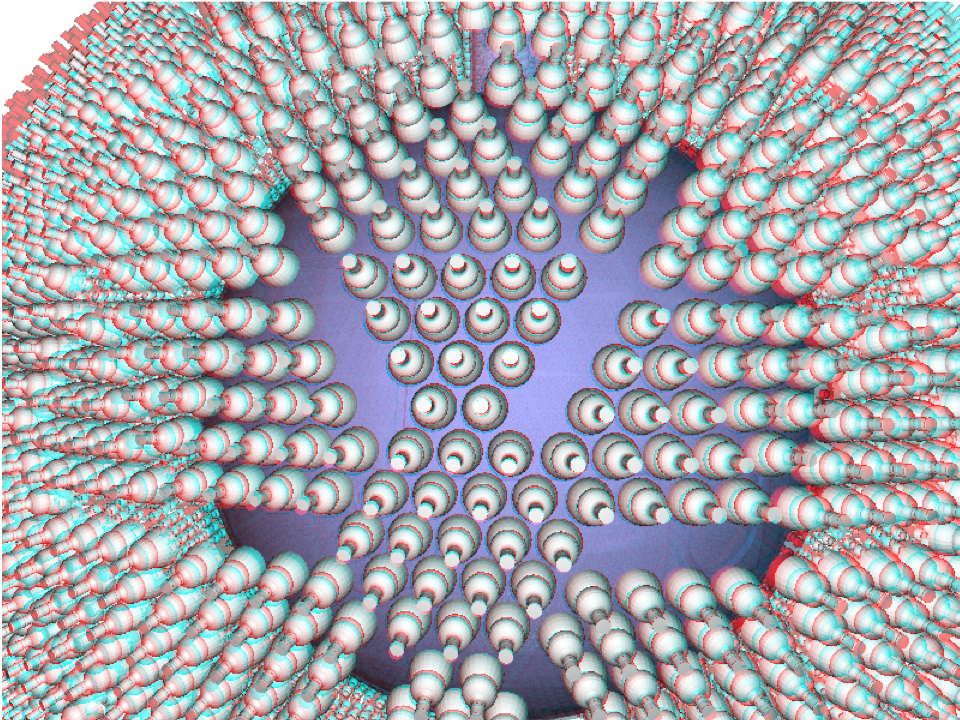}
\includegraphics[width=0.3\columnwidth]{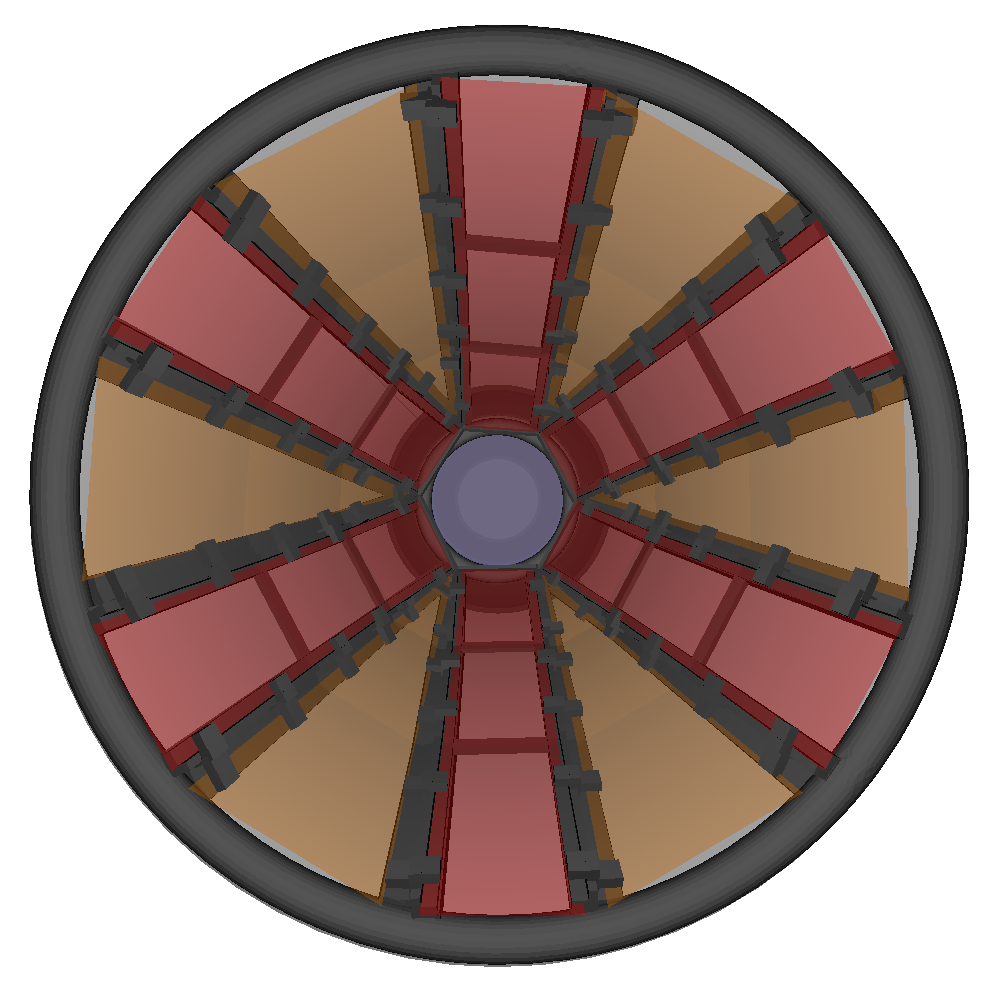}
\caption{Left: A screen capture from a realtime, three-dimensional ``movie'' of the
SNO+~\cite{snoplus} detector created with {\it Chroma}, which is best viewed with red-blue 3D glasses. The model shown is exactly the model
used by {\it Chroma}'s physics simulation, albeit with false-color optics for
the display rather than the complete physics the simulation uses.  The PMTs are
fully rendered, including their Winston-cone light
concentrators. Right: 
The CAD drawing for the 3D-printed dichroic filter holder for the dichroicon~\cite{dichroicons} was directly imported into {\it Chroma}, and used to accurately simulate the orientations of the dichroic filters (shown in red and orange) with respect to the benchtop experiment.\label{fig:chromapics}}.
\end{figure}

One collateral bonus of {\it Chroma's} speed is that it provides
remarkably beatiful, {\it realtime} detector displays. It is quite easy to
``fly through'' a detector and see it rendered in all of its detailed geometry,
exactly the way the photons themselves will see the detector.
Fig.~\ref{fig:chromapics} shows on the left a rendering of the SNO+~\cite{snoplus} detector which,
in a static pdf document like this proposal cannot be examined dynamically in
realtime, but was created from that realtime fly-through and simply captured
by screen-shot. The rendering is in fact a 3D image; with red-blue glasses one
can see the relief in the image.

\begin{figure}[h!]
\centering
\includegraphics[width=0.4\textwidth]{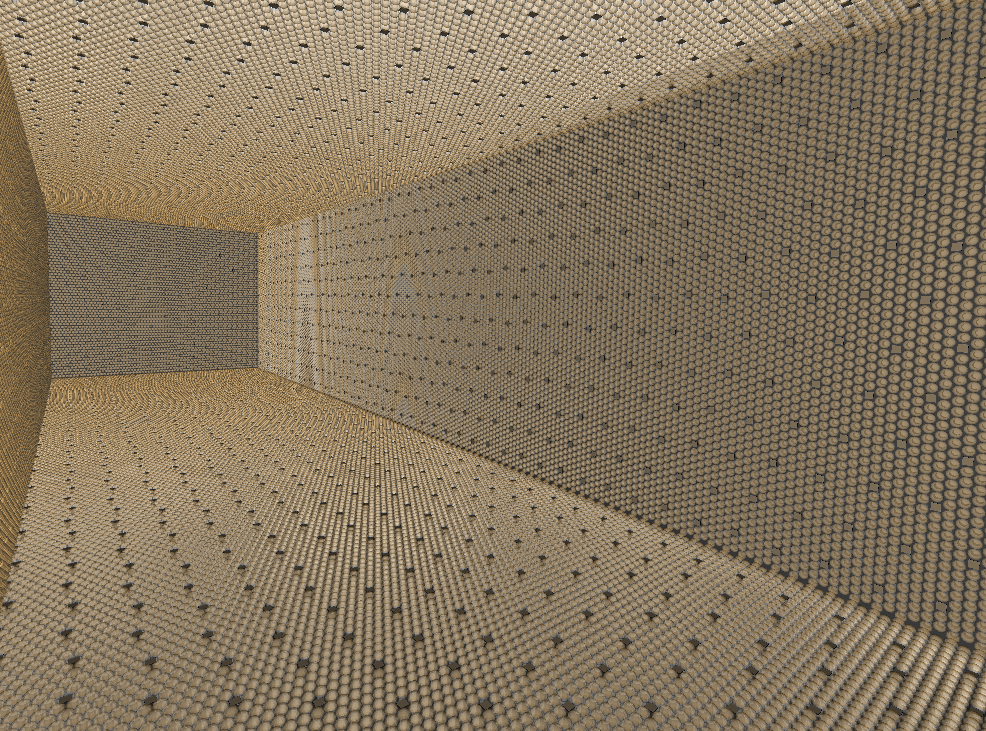}
\caption{\label{fig:theia25}
Next-generation photon detectors like LAPPDs~\cite{LAPPDtiming} are straightforward to simulate and include in larger geometries, as shown in this internal view of a \textsc{Theia}25~\cite{theiawp} model made with {\it Chroma}, which has mixed PMTs and LAPPDs for detecting photons.
The {\it Chroma} framework makes it quick and easy to explore new detector geometries and photon detecting devices.}
\end{figure}

The choice of representing geometries in {\it Chroma} as triangle mesh makes it straightforward to import CAD drawings into the optical simulation, as shown in Figure~\ref{fig:chromapics}.
All that is required is that the triangles on the mesh have optical properties assigned to them, which is a trivial operation in the case of a CAD model with uniform properties over its entire surface, but allows for fine-grained position dependent control of material properties as well. 
This allows anyone with CAD experience to quickly create an arbitrarily complicated simulation without having to learn a new way to represent geometries. 
The ability to rapidly prototype designs makes {\it Chroma} well suited for benchtop studies as well, and with easy scaling up from there to very large detectors,
as shown in \ref{fig:theia25}, which includes 3D Large Area Picosecond Photodetectors (LAPPD)~\cite{LAPPDtiming} mixed with standard PMTs in a {\it Chroma} model of the \textsc{Theia}25~\cite{theiawp} detector.

\subsubsection{GEANT4-based toolkits {\it RAT-PAC}}

RAT-PAC is an open-source GEANT4-based toolkit that offers both micro-physical simulation capabilities and analysis tools for high-precision event modeling, evaluation, and characterization, from benchtop test stands to large-scale detectors.

The RAT-PAC Monte Carlo simulation and analysis suite~\cite{ratpac} is a free and open-source version of the RAT toolkit.  RAT was first written for the Braidwood reactor experiment~\cite{braidwood},
and is now the official simulation and analysis package for  SNO+~\cite{snoplus}, DEAP, and MiniCLEAN experiments, thus benefiting from shared efforts in development and verification. A GEANT4-based package~\cite{geant4}, RAT-PAC (standing for ``RAT Plus Additional Code'') was branched off from the core RAT development some years ago, to form an open-source version of the code, available for public use.  RAT-PAC forms the basis of the official software for the \theia collaboration~\cite{theiawp}, the proposed third phase of ANNIE~\cite{annie-results}, and for the WATCHMAN collaboration, who are developing a design for the NEO detector to be located at the AIT facility in the UK~\cite{watchman}.

\begin{figure}[ht!]
    \centering
    \includegraphics[width=0.4\textwidth]{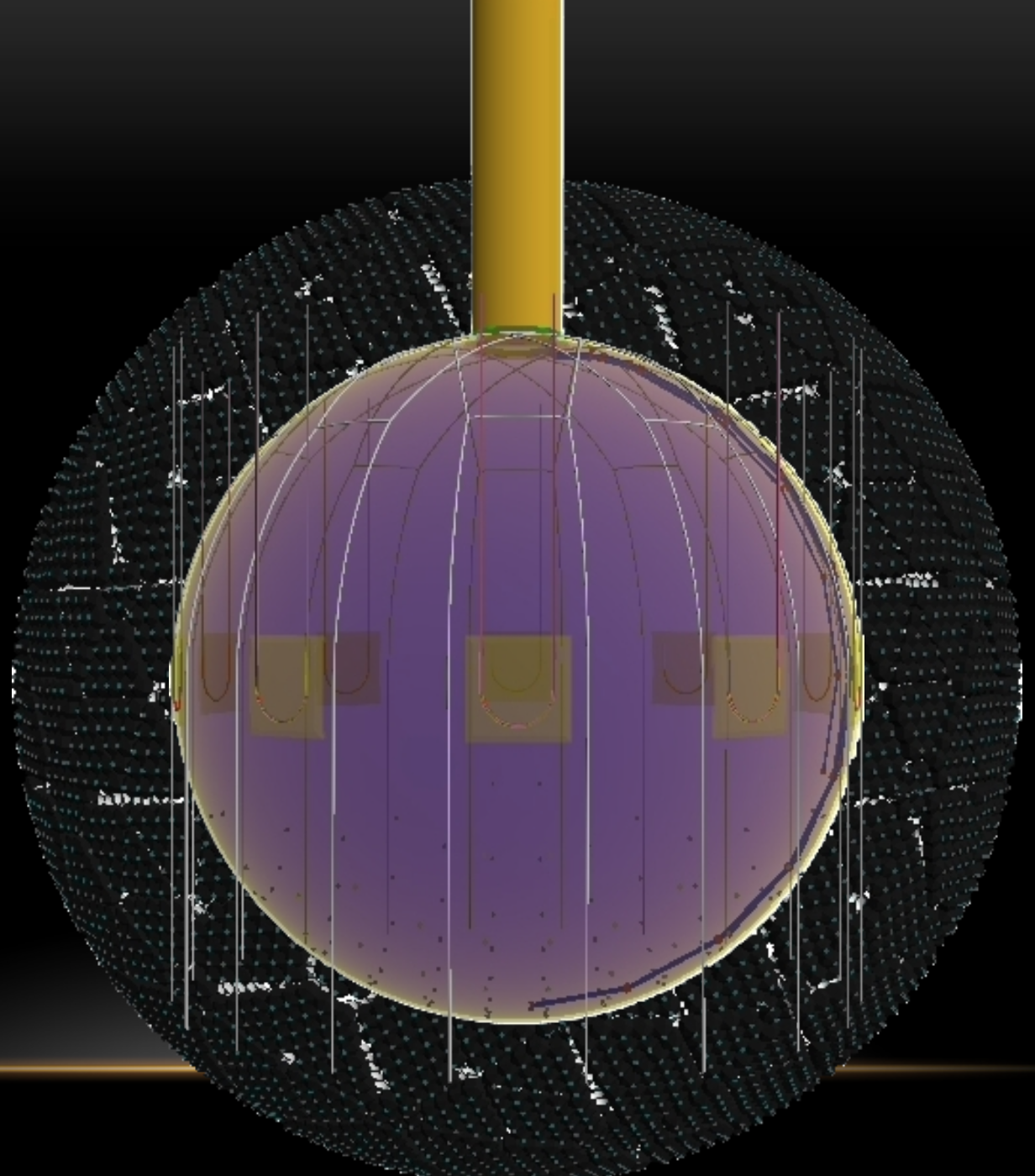} 
    \includegraphics[width=0.39\textwidth]{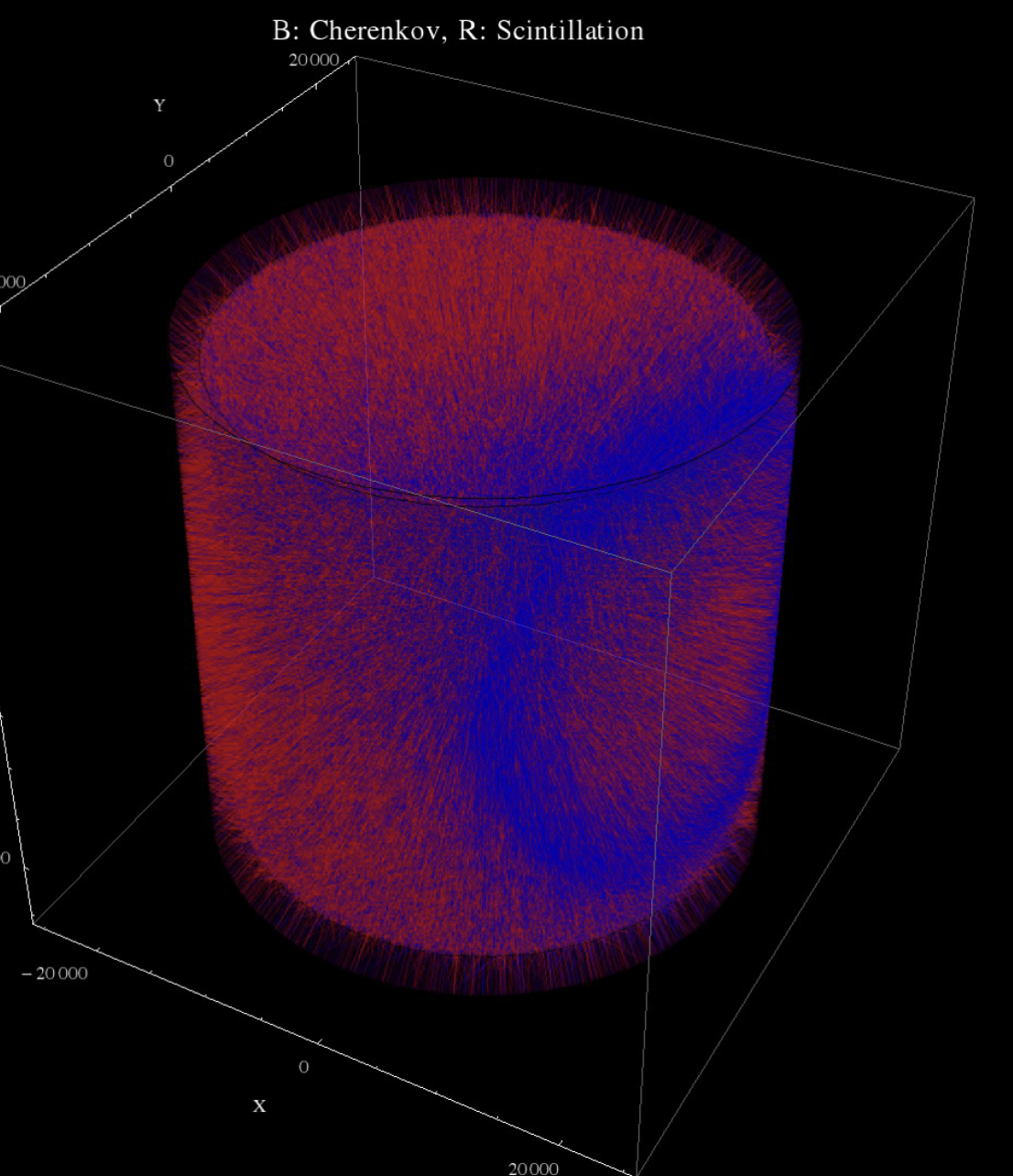} 
    \caption{(Left)  RAT-PAC generated image of the ktonne-scale SNO+ detector. (Right) RAT-PAC simulation of a high-energy (GeV) electron in the 50-ktonne \theia detector, including full photon tracking.  Blue shows Cherenkov photon track and red shows scintillation.}
    \label{fig:SNOp}
\end{figure}

\begin{figure}[ht!]
    \centering
    \includegraphics[width=0.5\textwidth]{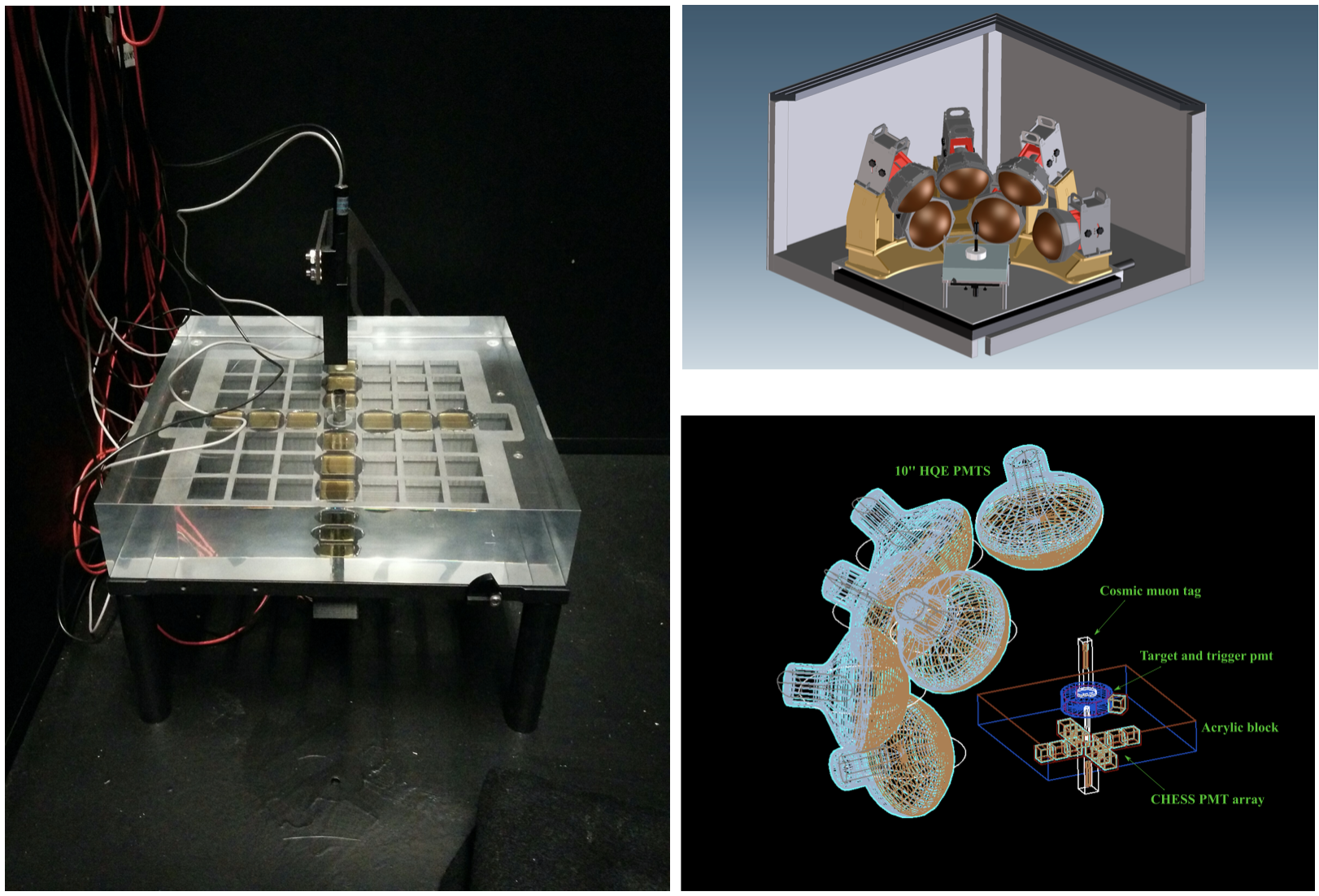} 
    \includegraphics[width=0.48\textwidth]{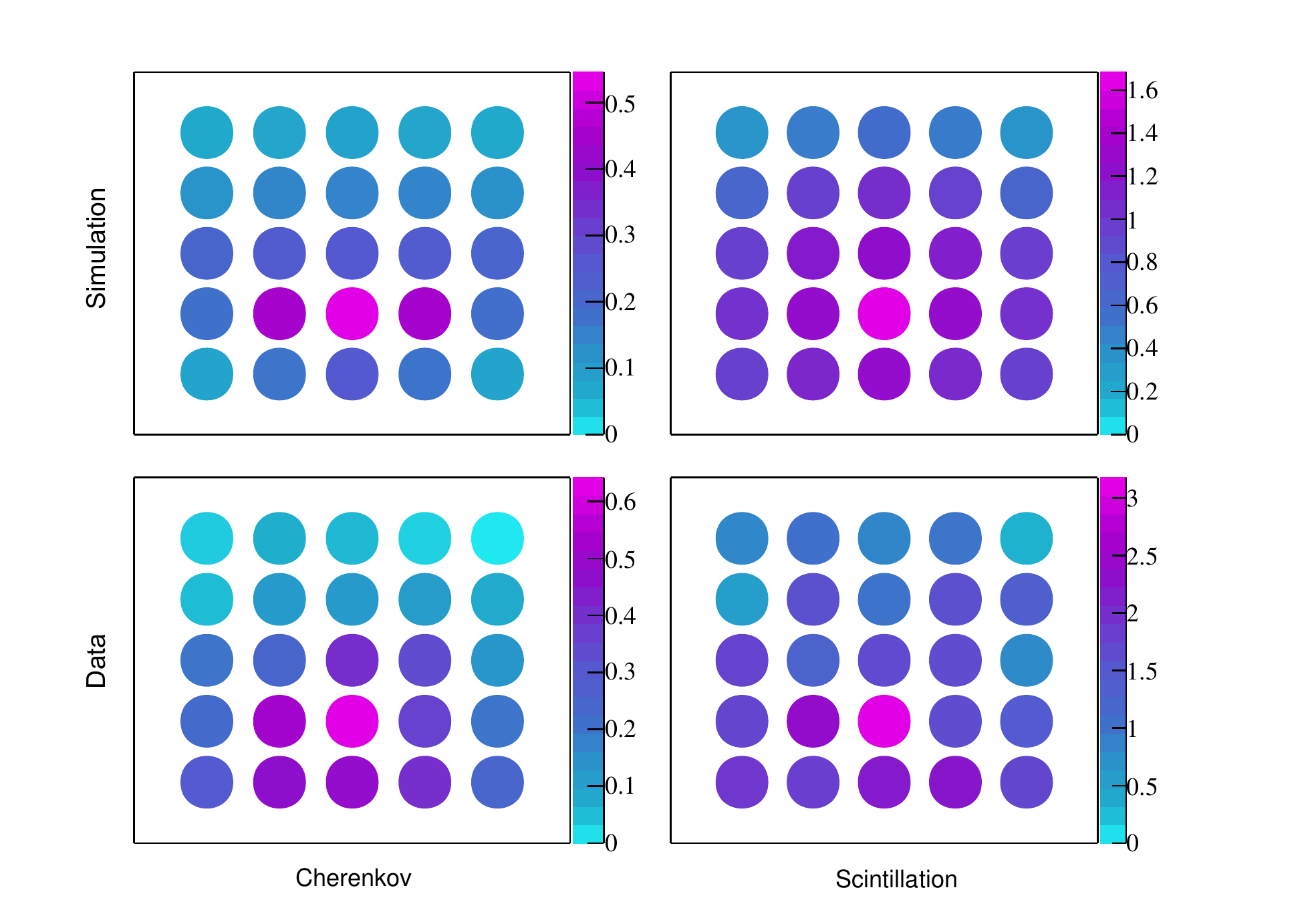}
    \caption{(Left) Photograph of the CHESS PMT array.  (Top centre) CAD image of the full CHESS detector.  (Bottom centre) RAT-PAC generated image of the full CHESS detector.  (Right) RAT-PAC simulations of both Cherenkov \textit{(left)} and scintillation \textit{(right)} signals show good agreement with data from the FlatDot experiment \textit{(bottom)}, up to a normalization factor reflecting the absolute light yield of the liquid scintillator.}
    \label{fig:CHESS}
    \end{figure}

One of the great advantages of the RAT-PAC approach is that its procedural geometry description allows the same code to be used to simulate or analyze data from a large-scale experiment and a small benchtop test-stand.  Figure~\ref{fig:SNOp} shows the detailed geometry of the full ktonne-scale SNO+ detector, and the even larger \theia detector, and Fig.~\ref{fig:CHESS} shows the much smaller CHESS detector at UC Berkeley/LBNL~\cite{chess}. In addition to the flexible geometry descriptions, RAT-PAC takes a micro-physical approach, relying on physical, rather than phenomenological models. For example, individual photons are simulated hitting photon sensors and the resulting timing and charge are evaluated photon-by-photon, rather than by application of a phenomenological risetime correction.  Therefore, simulating both benchtop test stands and large-scale detectors with the same micro-physical detail and the same code means that parameter measurements made by the benchtop are more easily translated into the larger-scale detector.  A measurement of, for example, the light yield of a scintillator cocktail performed in a small-scale setup can be straightforwardly propagated to predict performance in large detectors, complete with systematics associated with optical models or even data acquisition approaches. Comparisons between simulations of Cherenkov and scintillation light generated using RAT-PAC and data from test stands, such as at Penn, CHESS at LBNL, and FlatDot at MIT, show good agreement.  An example from FlatDot is shown in Fig.~\ref{fig:CHESS}~\cite{flatdot}. 

RAT(-PAC) is based on GEANT4.10~\cite{geant4} and the GLG4Sim package written by Glenn Horton-Smith, with custom code for scintillation and neutron absorption processes as well as a complete model of  PMT optics.
RAT(-PAC) handles all stages of event simulation: from the propagation of primary particles; production of optical photons via Cherenkov and scintillation processes; individual photon propagation, including a full optical model of all detector materials;  photon detection at the single PE level, including individual photon detector charge and timing response; and data acquisition including full
customizable simulation of  front end electronics, trigger systems, and event builders.  It also allows root-formatted data to be used as input, and provides simple analysis tools and ways to include many more, as well as a macro command structure for control. Lastly, RAT-PAC also includes the
ability to 
dynamically generate detector configurations via an external database.
Thus, RAT-PAC is a complete package that can be used with small modifications for entire experiments.

\subsubsection{Machine Learning Approaches}

    The applications of machine learning approaches to neutrino detectors in general, and to photon-based neutrino detectors in particular, is enormous.  For photon detectors, these approaches fall into two primary classes: those that are used to create fast simulations, and those that are used for particle ID and background rejection.
    
    

\subsection{Prototypes and Large-Scale R\&D Platforms}
\subsubsection{ANNIE}
\label{sec:ANNIE}

The ANNIE (Accelerator Neutrino Neutron Interaction Experiment)~\cite{annieLOI,annie-results,pershingdiss} is located in the Booster Neutrino Beam at Fermilab. This beam is about 93\% pure $\nu_{\mu}$ in neutrino mode and has a spectrum that peaks at about 700 MeV, an energy scale of great interest to current and future neutrino oscillation experiments. While ANNIE's current Phase 2 investigates the combination of a Gd-loaded water target read out by ultrafast LAPPDs, a first insertion of a small WbLS-filled vessel (SANDI) is planned for the upcoming summer break, adding scintillation to the Cherenkov signal at the neutrino interaction vertex. In an eventual Phase 3, the full detection volume is to be filled with WbLS, making ANNIE the first experiment exploring the benefits of the new hybrid detection technique in the reconstruction of GeV-scale beam neutrinos.

\begin{figure}[t]
    \centering
    \includegraphics[width=0.225\textwidth]{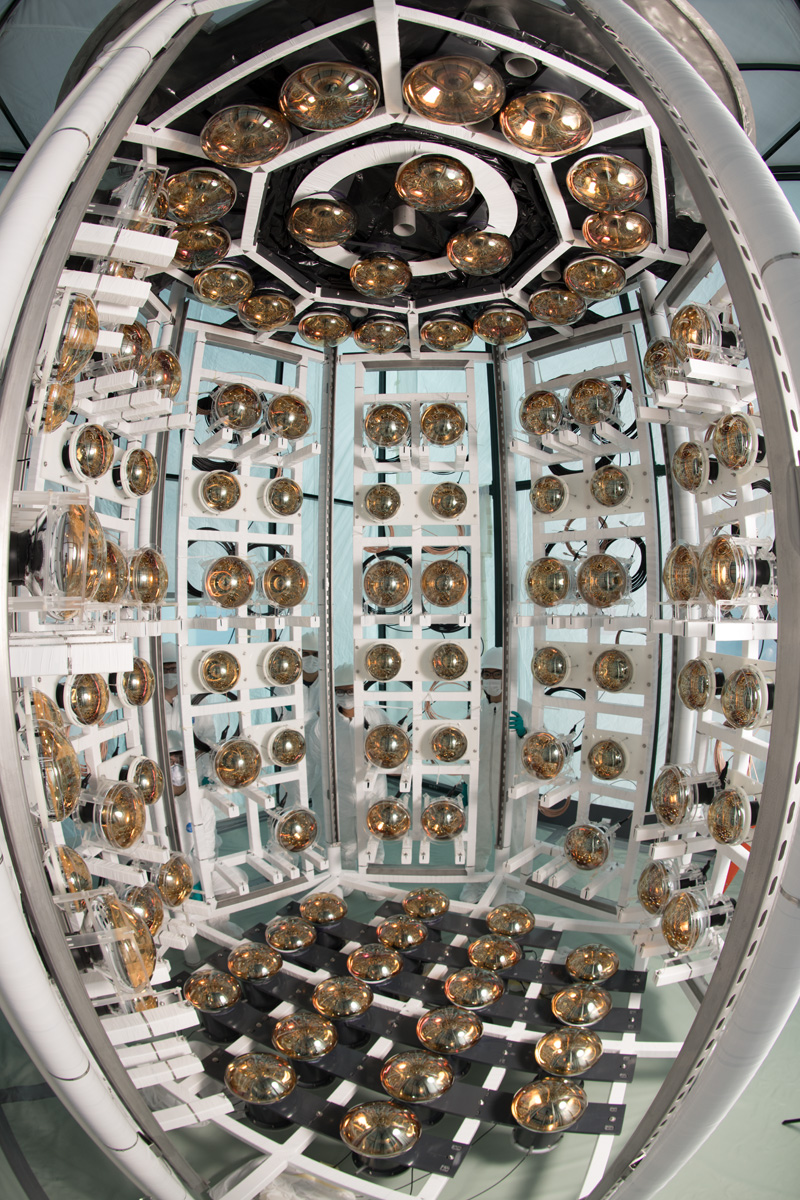}
    \hfill
    \includegraphics[width=0.43\textwidth]{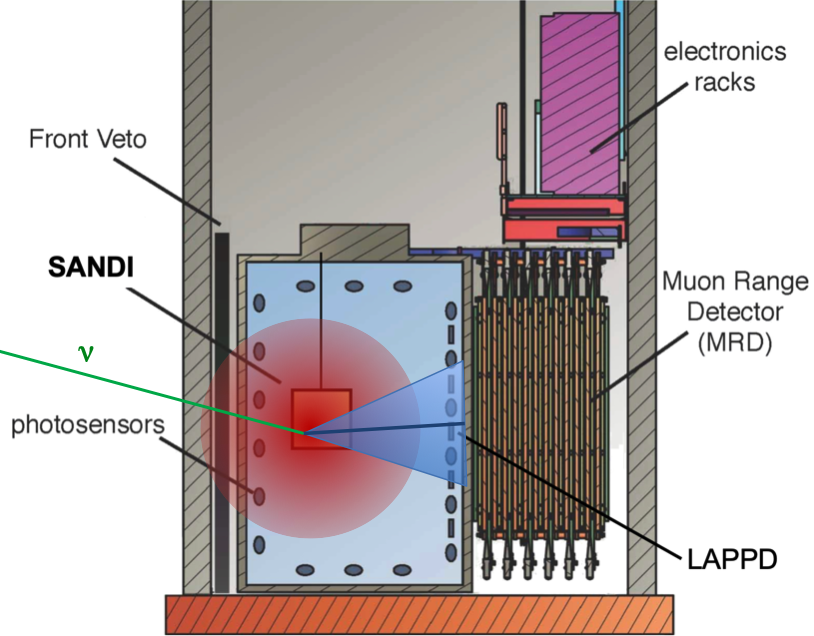}
    \hfill
    \includegraphics[width=0.255\textwidth]{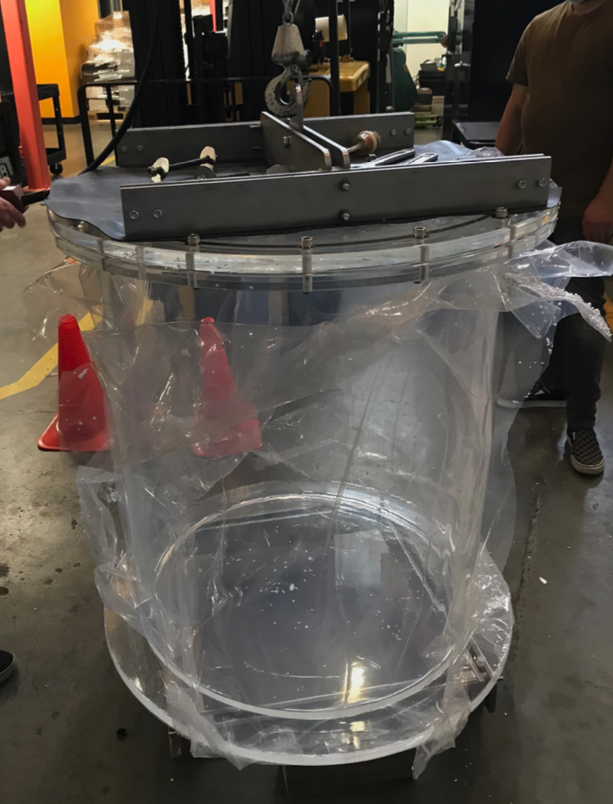}

    \caption{(left) NTT before installation. The rails allow LAPPDs to be inserted between the PMTs. (center) ANNIE detector setup with envisaged SANDI vessel containing 350\,kg of GdWbLS. (right) Test of SANDI acrylic vessel wrapped in protective plastics at UC Davis.} 
    \label{fig:ANNIE}
\end{figure}
\medskip

\medskip
\noindent{\bf ANNIE program with Gd-water target:} The two main goals of the current ANNIE Phase 2 are the precision measurements of the neutron yield from neutrino interactions at the 1\,GeV scale, and a test of the effectiveness of Large Area Picosecond PhotoDetectors (LAPPDs) in a realistic physics environment (Sec.~\ref{sec:LAPPD}). As illustrated by Figure~\ref{fig:ANNIE}, the current setup of the ANNIE detector comprises a 26-ton Neutrino Target Tank (NTT) equipped with 128 conventional PMTs, a Front Muon Veto (FMV) to reject ``dirt'' muons, and a Muon Range Detector (MRD) to track and range out muons. The water of the NTT was loaded with a mass fraction of 0.1\% of gadolinium (GdS) in early 2019. In January 2021, ANNIE started to take physics data in this basic configuration, while the deployment of first LAPPDs is foreseen for early 2022. The aim is to install in total five LAPPDs before the end of Phase 2 to evaluate their impact on the reconstruction of CC interaction final states as well as vertex reconstruction of the low-energy signals of neutron capture on gadolinium. The conventional PMT array provides a vertex reconstruction of $\Delta r=38\;$cm for CC final-state muons. The full configuration including the LAPPDs and their sub-nanosecond timing is expected to substantially improve this vertex resolution to about $\Delta r=12\;$cm.
\medskip\\
\noindent{\bf Scintillator for ANNIE Neutron Detection Improvement (SANDI):} From simulation studies, we know that $-$ even in the presence of LAPPDs $-$ the vertex resolution suffers from the need to image the ring edge with the conventional PMT array. There is a timing degeneracy in single track particles that makes the vertex location parallel to the track entirely dependent on ring edge imaging. One way to improve on this is to use a slightly scintillating material as neutrino target, resulting in a comparatively small amount of isotropic light emitted from the vertex that can be separated from Cherenkov light by timing and topology. 

To explore the potential benefit for event reconstruction, ANNIE plans to insert a 3'$\times$3' cylindrical vessel (SANDI) containing about 350\,kg of gadolinium-loaded Water-based Liquid Scintillator (GdWbLS) into the center of the NTT water volume (Figure~\ref{fig:ANNIE}). In addition, one or more of the five LAPPDs may be moved to the backward hemisphere. A deployment of several weeks will allow us to confirm models for track reconstruction and to identify additional scintillation light emitted by the hadronic recoil at the vertex. First simulation studies show that the inclusion of this signal will indeed lead to an improvement in reconstruction capability. While the muon energy reconstruction is dominated by the information from the (unchanged) MRD, the additional information from the GdWbLS volume improves the neutrino energy reconstruction from ($\Delta E/E\sim15\%$ to $11\%$).  
\medskip\\
\noindent {\bf Future full GdWbLS phase:} Following a successful SANDI test, ANNIE plans to propose a Phase 3 that features a complete replacement of the Gd-water with GdWbLS. Neutron captures in GdWbLS feature a substantially increased light output (factor $\sim$3) compared to Gd-water, greatly improving neutron detection efficiency ($\to\,\geq90\%$) and spatial resolution for the capture vertex ($\to\,\sim40\,$cm). This will enhance not only the precision of neutron multiplicity measurements but also bears the chance to reconstruct the neutron energy spectrum by their capture position relative to the production vertex. A full calorimetric measurement of the cross-section (including vertex hadronic energy) may also be possible. 

With these ambitious goals in mind, ANNIE is investigating the possibility to reconstruct the inner part of the NTT to make it compatible with WbLS, including encapsulation of PMTs and other components, in addition to the first-ever deployment of a WbLS liquid recirculation system. Systems based on nanofiltration and phase separation technology are being developed for Eos and Theia (Sects.~\ref{sec:Eos} and \ref{sec:Theia}). ANNIE Phase 3 will be a significant step towards realization of hybrid optical detectors in addition to making neutron-inclusive neutrino cross-section measurements of unprecedented scope and quality. 

\subsubsection{NuDOT}
\label{sec:NuDot}
NuDot is a mid-scale prototype designed to demonstrate timing-based Cherenkov/scintillation separation in a realistic experimental geometry, focusing on techniques applicable to searches for neutrinoless double-beta decay ($0\nu\beta\beta$). It builds on the successful demonstration of this approach in the FlatDot test stand~\cite{flatdot}. It is currently undergoing detector commissioning, with its initial physics data-taking campaign expected to begin by Summer 2022. 

\begin{figure}[]
    \centering
    \includegraphics[width=\textwidth]{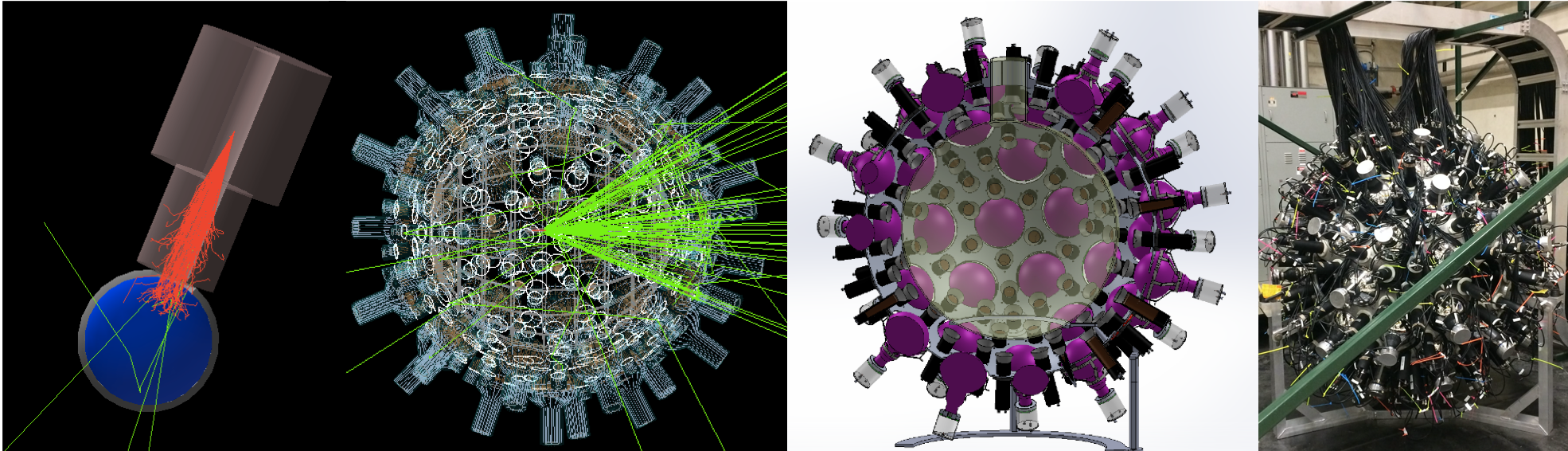} 
    \caption{\textit{From left to right:} A RAT-PAC simulation of electron tracks in the collimated $\beta$ source and coupled liquid scintillator cuvette used for ``dry" surface data-taking; a RAT-PAC simulation of Cherenkov light tracks for a 1\,MeV $\beta$ in NuDot; a rendering of a section view of the NuDot detector; a photo of NuDot, before the calibration system was deployed and the sphere was made light-tight for commissioning runs.}
    \label{fig:NuDot}
\end{figure}

The NuDot design, shown in Fig.~\ref{fig:NuDot}, relies on a combination of 59 8"-diameter PMTs (primarily a combination of EMI Model D642 and Hamamatsu Model R1408 PMTs previously used by the MACRO experiment) and 151 2"-diameter PMTs (Hamamatsu Module R13089, with an average transit time spread of 200\,ps) to achieve high light collection efficiency and fast-timing capability. These detectors surround a 36"-diameter acrylic sphere that will be filled with a LAB-based liquid scintillator cocktail, with the entire assembly immersed in a mineral oil tank to provide shielding and optical coupling. Detector calibration and the first two phases of data-taking will be conducted using a remotely-controlled $4\pi$ calibration system with 3 independent axes of motion. In addition to an LED source used for gain and pulse-shape calibrations, this system holds a collimated $\beta$-emitting $^{90}$Sr needle source that can be adjusted to shine on any point on the NuDot sphere from any point along the sphere diameter, allowing tests of the NuDot Cherenkov/scintillation separation capability as a function of position and direction. 

NuDot has 3 planned phases of operation, the first two of which are conducted prior to underground deployment:

\begin{enumerate}
    \item ``Dry" runs at Bates Research and Engineering Center: a small cuvette of liquid is coupled directly to the collimated source, allowing for calibration of the PMT timing response and pulse shape with water Cherenkov data, followed by C/s separation tests with a variety of liquid scintillator cocktails.
    \item Liquid scintillator- and mineral-oil-filled runs at Triangle Universities Nuclear Laboratories: following the completion of phase 1 data taking, NuDot will be moved to TUNL for any needed detector upgrades and liquid-filling. The goal of this phase is to demonstrate NuDot's C/s separation and energy reconstruction capability in the final operational configuration using the calibration system. 
    \item Underground proof-of-concept measurement: in phase 3, NuDot will be re-deployed underground for a proof-of-concept measurement. The current baseline is to conduct a $2\nu\beta\beta$ decay measurement, but alternatives like $\beta + $/EC searches are under consideration.
\end{enumerate}
The isotope and isotope-loading technique that will be used in phase 3 has not yet been selected. The NuDot Collaboration is currently conducting R\&D measurements of quantum-dot-loaded liquid scintillator cocktails that use perovskite wavelength shifters (see Ref.~\cite{perovskite}), but is also considering high-concentration Te loading and pressurized Xe loading options.


\subsubsection{Eos}
    \label{sec:Eos}

Successful detection of Cherenkov light from highly scintillating media has been achieved in CHESS, and complemented with an extensive characterisation program at LBNL, BNL, LLNL, and Davis.  These measurements inform large-scale Monte Carlo models, which are used to predict performance in kton-scale detectors. Ton-scale deployments of novel scintillators (ANNIE, BNL) are planned to test production, deployment, and recirculation.  

While bench-top measurements have been, and will continue to be used to measure microphysical properties of novel scintillators and photon detection technology, all demonstrations of event reconstruction and the resulting background rejection capabilities are purely Monte Carlo driven.  Data-driven demonstrations of the event imaging capabilities of hybrid detector technology is a critical step to realising a large detector for both fundamental physics and nonproliferation applications.

The proposed \eos prototype is a few-ton scale detector, designed to hold a range of novel scintillators, coupled with an array of photon detection options and the ability to deploy a range of low-energy calibration sources.  \eos will be constructed, calibrated and tested in Berkeley.  Assuming a successful surface deployment,  \eos could later be re-deployed underground, for example at SURF or SNOLAB, or at an alternative off-site location such as a reactor or test beam for further characterization of detector response to a range of particle interactions. \eos represents significant risk reduction for a large-scale deployment of (Wb)LS and novel detection technology.

\eos will be sufficiently large to use time-of-flight based reconstruction, and to fully contain a range of low-energy events ($\alpha$, $\beta$, $\gamma$, n) for detailed event-level characterisation.  \eos represents a balance of sufficient size for full event characterization, complemented with economy of scale, and flexibility to adapt for multiple target materials and photon detection options.

The primary goal of \eos is to validate performance predictions for large-scale hybrid neutrino detectors by performing a data-driven demonstration of low-energy event reconstruction leveraging both Cherenkov and scintillation light simultaneously.  By comparing data to model predictions, and allowing certain detector configuration parameters to vary -- such as the fraction of LS in a WbLS target cocktail, or by using PMTs with differing TTS, or deployment of dichroicons -- the predictive power of the model can be validated.  This validated microphysical model of hybrid neutrino detectors can then be used by the neutrino community for design optimization of next-generation hybrid detectors.

 \eos will provide an important test bed to the community, for testing alternative target media, photon detectors, and readout technology and methodology, and assessing the impact of these novel developments on detector performance.  
 
After the conclusion of operations at LBNL, \eos could be redeployed at an off-site location.  This would take place after the end of the primary project period, and successful completion of the project objective. Options include:
\begin{itemize}
\item Underground deployment, for example at SURF or SNOLAB, where scintillator handling procedures are well vetted and understood, and we have relationships of long standing.  This would allow more precise measurements in a low-background environment.
\item Deployment at a reactor for low-energy neutrino reconstruction.  This provides a near-field demonstration of the remote reactor monitoring concept.
\item Deployment at a test beam for hadronic event reconstruction, or a neutron source such as the Spallation Neutron Source at ORNL for neutron studies, useful for advanced event identification and background characterization.
\end{itemize}

\subsection{Large-Scale Detector Ideas}

\subsubsection{\theia}
\label{sec:Theia}
New developments in liquid scintillators, high-efficiency, fast photon detectors, and chromatic photon sorting have opened up the possibility for building a large-scale detector that can discriminate between Cherenkov and scintillation signals. Such a detector could exploit these two distinct signals to observe particle direction and species using Cherenkov light while also having the excellent energy resolution and low threshold of a scintillator detector. Situated in a deep underground laboratory, and utilizing new techniques in computing and reconstruction techniques, such a detector could achieve unprecedented levels of background rejection, thus enabling a rich physics program that would span topics in nuclear, high-energy, and astrophysics, and across a dynamic range from hundreds of keV to many GeV. The scientific program would include observations of low- and high-energy solar neutrinos, determination of neutrino mass ordering and measurement of the neutrino CP violating phase $\delta$, observations of diffuse supernova neutrinos and neutrinos from a supernova burst, sensitive searches for nucleon decay and, ultimately, a search for NeutrinoLess Double Beta Decay (NLDBD) with sensitivity reaching the normal ordering regime of neutrino mass phase space.

\theia is a detector concept that incorporates these new technologies in a practical and affordable way to accomplish the science goals described above. We consider two scenarios, one in which \theia would reside in a cavern the size and shape of the caverns intended to be excavated for the Deep Underground Neutrino Experiment (DUNE) which we call \theia-25, and a larger 100 ktonne version (\theia-100) that could achieve an even broader and more sensitive scientific program.

    \begin{figure}[htp!]
\centering
\includegraphics[width=0.95\textwidth]{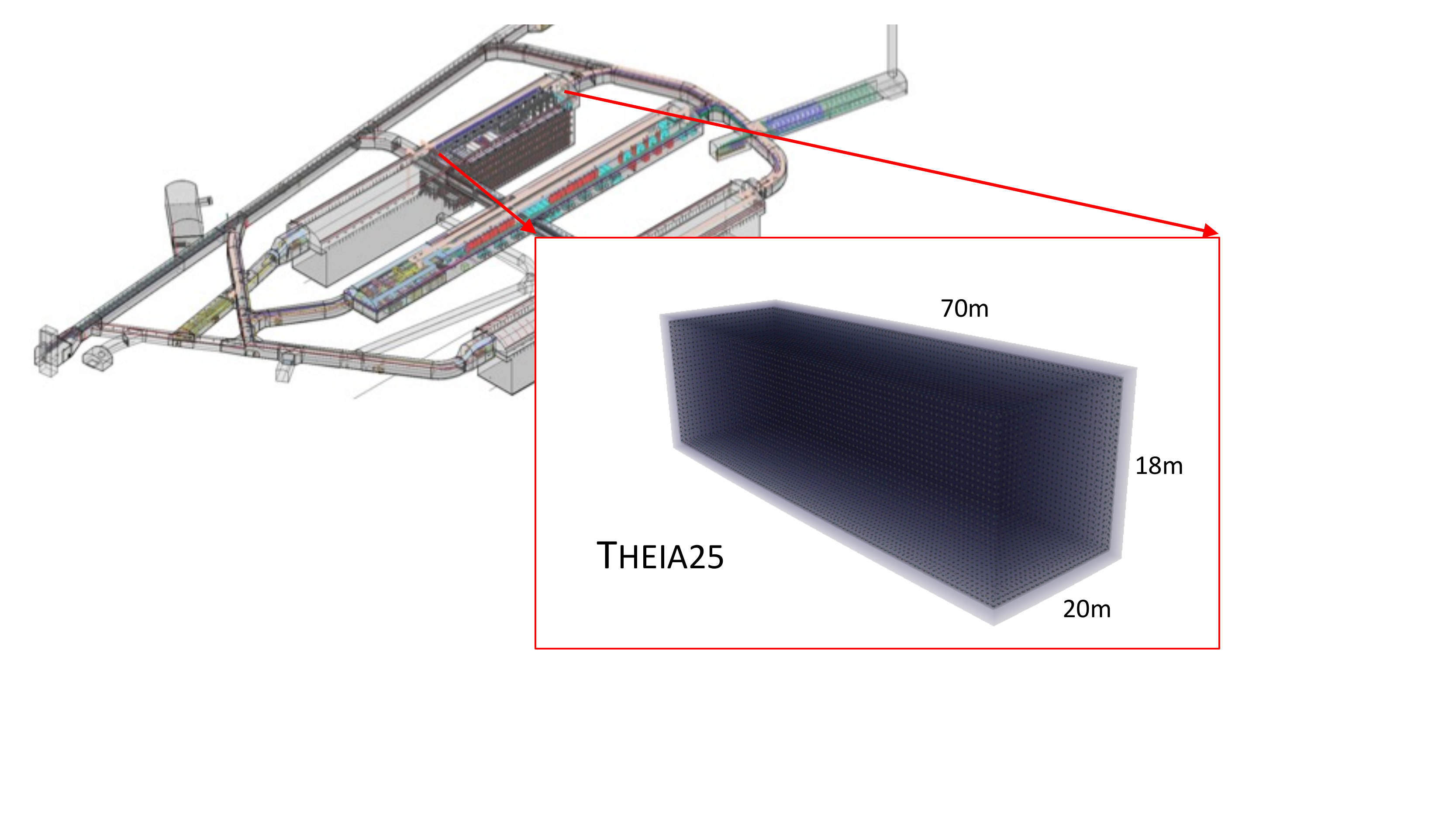}
\includegraphics[width=0.36\textwidth]{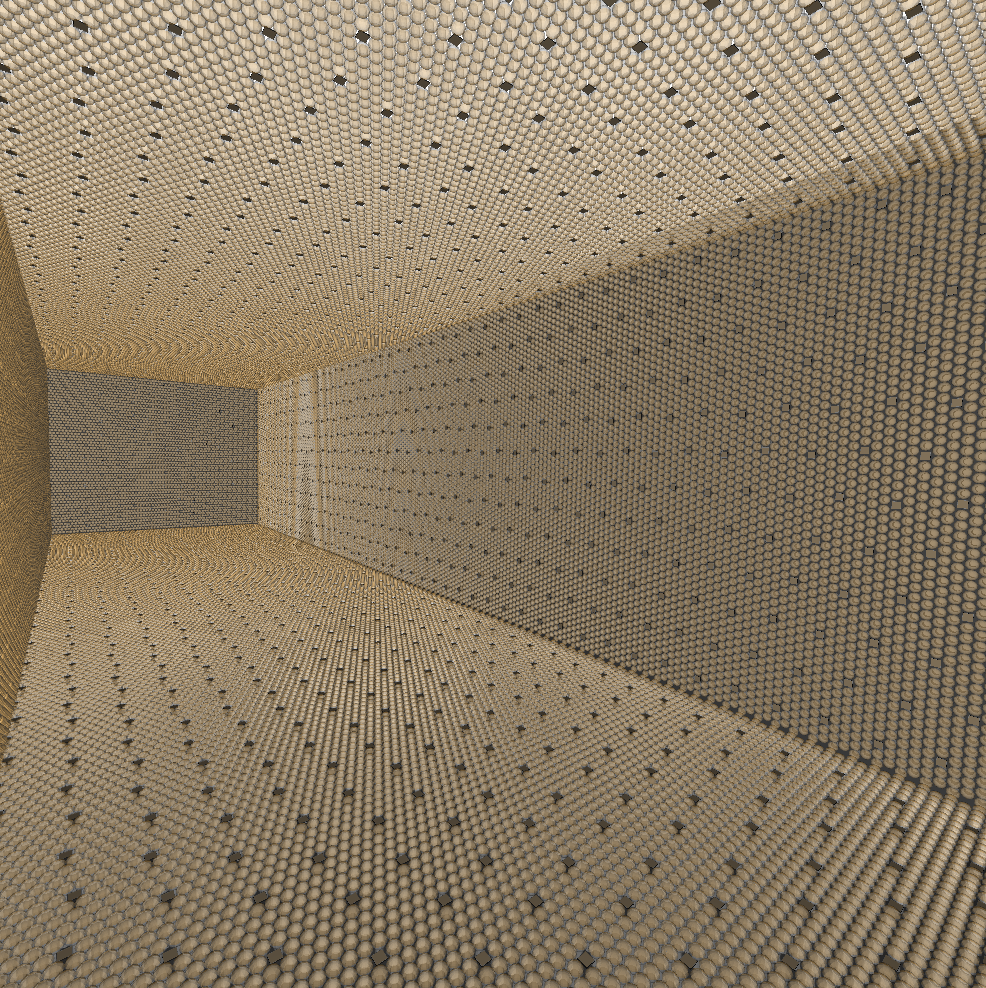}
\includegraphics[width=0.26\textwidth]{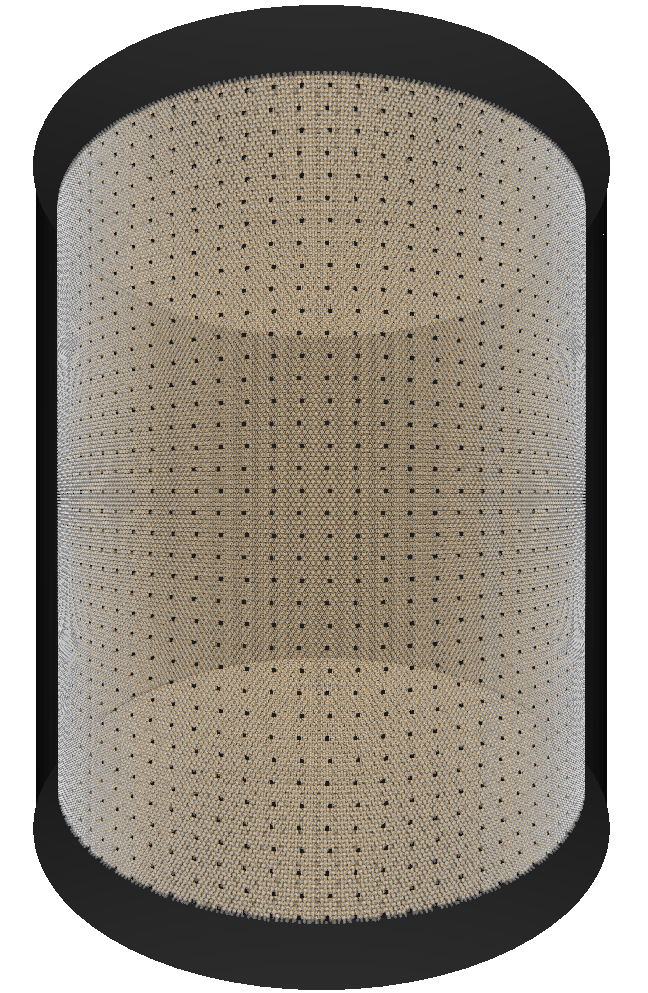}
\includegraphics[width=0.36\textwidth]{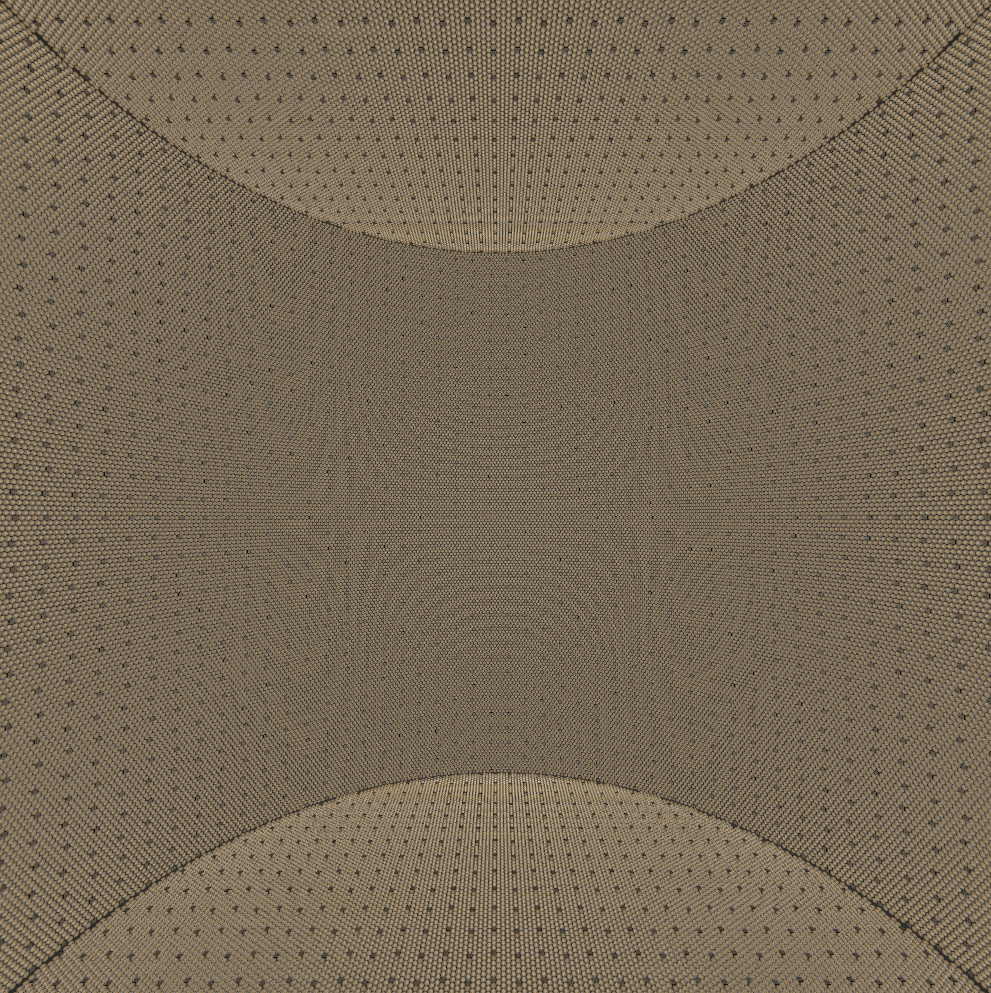}
\caption{The \theia detector.  
{\it Top  panel:} \theia-25 sited in the planned fourth DUNE cavern; 
{\it Lower left panel:} an interior view of \theia-25 modeled using the Chroma optical simulation package~\cite{chroma}; 
{\it Lower middle panel:} exterior view of \theia-100 in Chroma;
{\it Lower right panel:} an interior view of \theia-100 in Chroma.  In all cases, \theia has been modelled with 86\% coverage using standard 10-inch PMTs, and 4\% coverage with LAPPDs, uniformly distributed, for  illustrative purposes.  Taken from~\cite{theiawp}.}
\label{f:detector}
\end{figure}

In the \textsc{Theia} reference design, the target material would be water-based liquid scintillator (WbLS) described  in Section~\ref{sec:wbls}, which has an advantage for a big detector in its  long attenuation lengths. For much of the low-energy program of \textsc{Theia}, the WbLS would need to be made radiopure at levels not far from what can be done in an organic scintillator.   

The broadband neutrino beam being built for the Long Baseline Neutrino Facility (LBNF)
~\cite{lbnf} and the Deep Underground Neutrino Experiment (DUNE)~\cite{dunecdr} offers an opportunity for world-leading long-baseline neutrino oscillation measurements. Due to advances in Cherenkov ring reconstruction techniques, a \theia detector in the LBNF beam would have good sensitivity to neutrino oscillation parameters, including CP violation (CPV), with a relatively modestly-sized detector. 
In addition to this long-baseline neutrino program, \theia will also contribute to atmospheric neutrino measurements and searches for nucleon decay, particularly in the difficult $p\rightarrow K^+ + \overline{\nu}$ and $N\rightarrow 3\nu$ modes.

	\theia will also make a definitive measurement of the solar CNO neutrinos,
which have recently been detected exclusively by Borexino~\cite{bxcno}, but without enough precision to distinguish between competing models of the Sun's metallicity.
\theia will also provide a high-statistics, low-threshold ($\sim$ 3~MeV)
measurement of the shape of the $^8$B solar neutrinos and thus search for new
physics in the MSW-vacuum transition region~\cite{friedland2004, minakata2012}. Antineutrinos produced in the
crust and mantle of the Earth will be measured precisely by \theia with statistical uncertainly far exceeding all detectors to date. 

Should a supernova occur during
\theia  operations, a high-statistics detection of the $\bar{\nu}_e$ flux will be
made---literally complementary to the  detection of the $\nu_e$ flux in the DUNE liquid argon
detectors. The simultaneous detection of both messengers---and detection of an optical, x-ray, or gamma ray component will enable a great wealth of neutrino physics
and supernova astrophysics. With a very deep location and with the detection of a combination of scintillation and Cherenkov light, \theia will have world-leading sensitivity to make a detection of the Diffuse Supernova Neutrino Background (DSNB) antineutrino flux. The most ambitious goal,
which would likely come in a future phase, is a search for Neutrinoless double beta decay (NLDBD), with a total isotopic mass of 10 tonnes or more, and with decay lifetime sensitivity in excess of $10^{28}$ years~\cite{theiawp, biller_normal}.

\theia is able to achieve this broad range of physics by exploiting new
technologies to act simultaneously as a (low-energy) scintillation detector and
a (high-energy) Cherenkov detector. Scintillation light provides the energy
resolution necessary to get above the majority of radioactive backgrounds and
provides the ability to see slow-moving recoils; Cherenkov light enables event
direction reconstruction which provides particle ID at high energies and
background discrimination at low-energies. Thus, the scientific program benefits in many cases on
the ability of \theia to discriminate efficiently and precisely between the ``scintons'' (scintillator photons)
and ``chertons'' (Cherenkov photons). 

    A major advantage of \theia is that the target can be modified in a phased program to address the science priorities. In addition, since a major cost of \theia is expected to be photosensors, investments in \textsc{Theia25} instrumentation can be transferred directly over to \textsc{Theia100}. Thus, \theia can be realized in phases, with an initial phase consisting of lightly-doped scintillator and very fast photosensors, followed by a second phase with enhanced photon detection to enable a very low energy solar neutrino program, followed by a third phase that could include doping with a $0\nu\beta\beta$ isotope and perhaps an internal containment vessel. 

\subsubsection{LiquidO}

LiquidO is a new approach to detecting neutrinos that, in contrast with conventional liquid scintillator detectors, relies on using an opaque scintillator medium as the primary neutrino target~\cite{Cabrera:2019kxi}. The scintillators that can be best used by LiquidO have a short scattering length and a medium to long absorption length, an example of which has already been successfully produced~\cite{Buck:2019tsa}. In such a medium, the photons produced by the opaque scintillator undergo a random walk process near their creation point and are trapped in so-called {\it light balls} around each energy deposition point. The light is collected by a dense array of wavelength-shifting fibres that traverses the volume and that is readout by photo-sensors in the periphery. Silicon photomultipliers (SiPMs) are well-suited to this purpose given their affordable price, high efficiency, and excellent time resolution. 

\begin{figure}[!htbp]
\begin{center}
\includegraphics[width=0.90\textwidth]{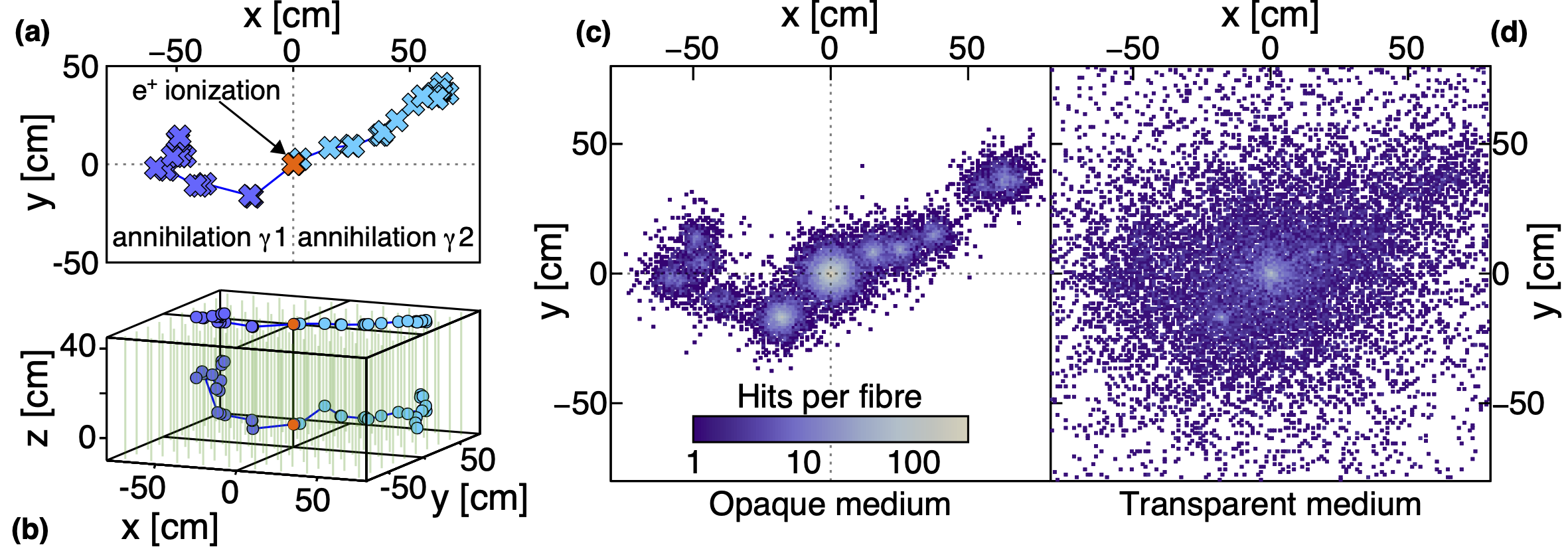}  
\caption{\label{fig:lozfig} Left: energy depositions of a simulated 1~MeV kinetic energy positron in a LiquidO detector with a regular 1~cm fibre pitch running along the $z$ direction. Panel (a) shows the $x$-$y$ projection, while panel (b) shows the full three-dimensional extent. The fibres are represented in green. Right: true number of photons hitting the fibres, each of which is represented by a pixel, in the opaque and transparent scintillator scenarios. In the former case, the scintillator is assumed to have a 5~mm scattering length and a 5~m absorption length. Figure obtained from Ref.~\cite{Cabrera:2019kxi}.}
\end{center}
\end{figure}

\begin{figure}
    \includegraphics[width=8cm]{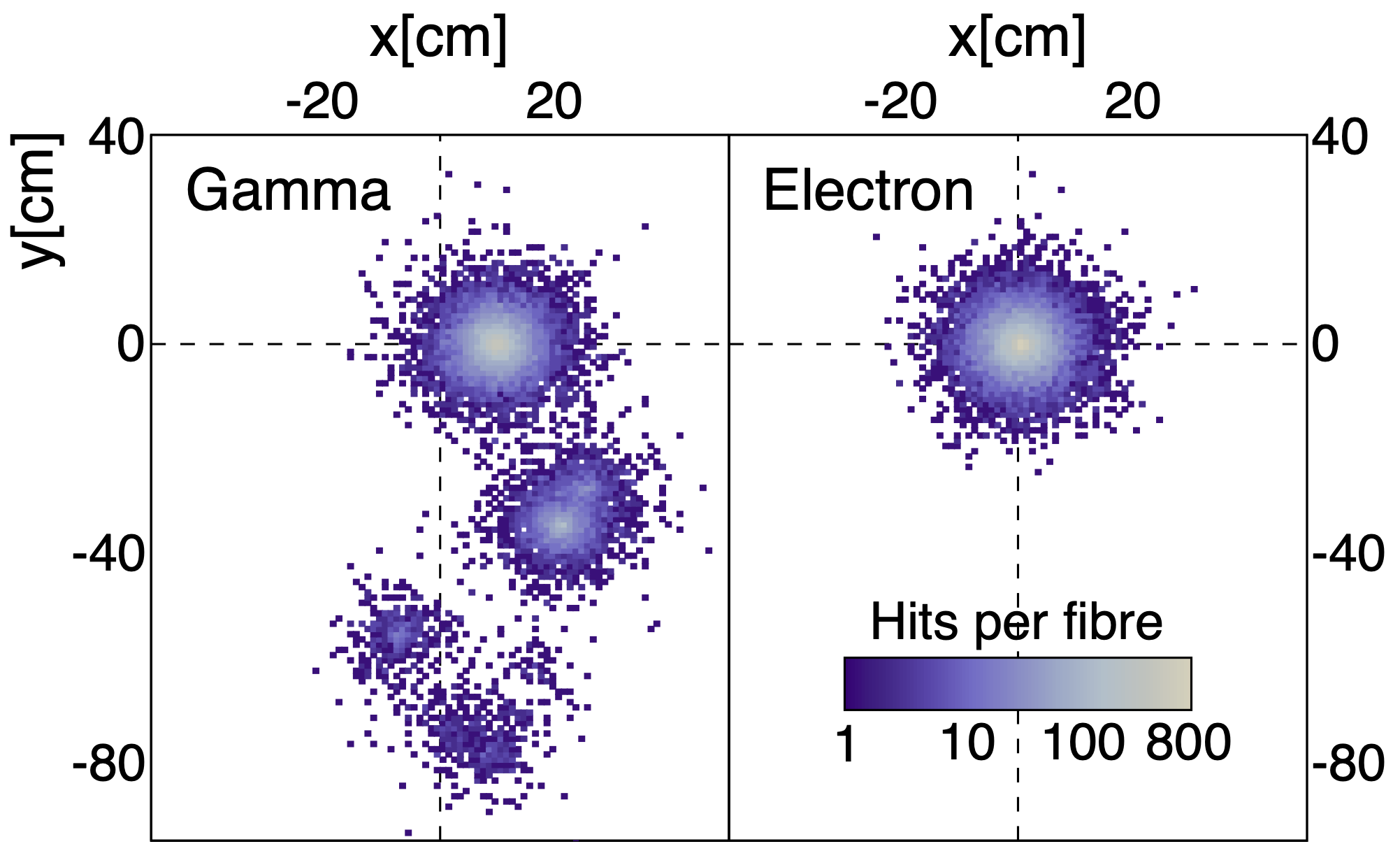}
    \caption{Simulated 2~MeV gamma (left) and electron (right) in the same detector configuration of Fig.~\ref{fig:lozfig}. Figure obtained from Ref.~\cite{Cabrera:2019kxi}.
    \label{fig:gammaelectron}}
\end{figure}

A full description of LiquidO, its expected performance, first experimental demonstration, and potential applications, can be found in Ref.~\cite{Cabrera:2019kxi}. Fig.~\ref{fig:lozfig} illustrates LiquidO's performance using a simulated positron with 1~MeV of kinetic energy. Here, the simplest configuration with fibres running only along one direction ($z$) is assumed. The true energy depositions of the positron are shown on the left panels, with (a) showing the $x$-$y$ projection and (b) the full three-dimensional extent. Panel (c) shows the number of true photons hitting each fibre in a 1-cm-pitch lattice when a scintillator with a 5~mm scattering length is used. The positron's loss of kinetic energy produces a light-ball at the vertex of the event. The two back-to-back gamma-rays resulting from its annihilation lose energy via Compton scattering, leaving two trails of smaller light balls that detach from the central one. A comparison is made in panel (d) of the light pattern collected by the same fibre array when using a transparent scintillator. Despite the use of fibres, the event topology is almost entirely washed out, illustrating the key role played by the scintillator's opacity in self-segmenting the detector.

The clear event topology of $\sim$MeV positrons in LiquidO stands in contrast with that of other events, as illustrated in Fig.~\ref{fig:gammaelectron}. At these energies, gammas lose their energy primarily via the Compton effect and produce trails of light balls over many tens of cm, whereas electrons produce single light balls. At higher energies (more than $\sim$10~MeV for electrons), charged particles have enough kinetic energy to travel several cm or more in the detector, producing sequences of point-like energy depositions that form clear tracks. As a result, many other interactions, from cosmic ray muons to charged and neutral current neutrino interactions of various energies, could also be precisely reconstructed in LiquidO. In this way, this detector technology combines some of the advantages of conventional liquid scintillator detectors with those of tracking detectors.

\subsubsection{SLIPs}

\begin{figure}[htp]
\centering 
\includegraphics[width=0.9\textwidth]{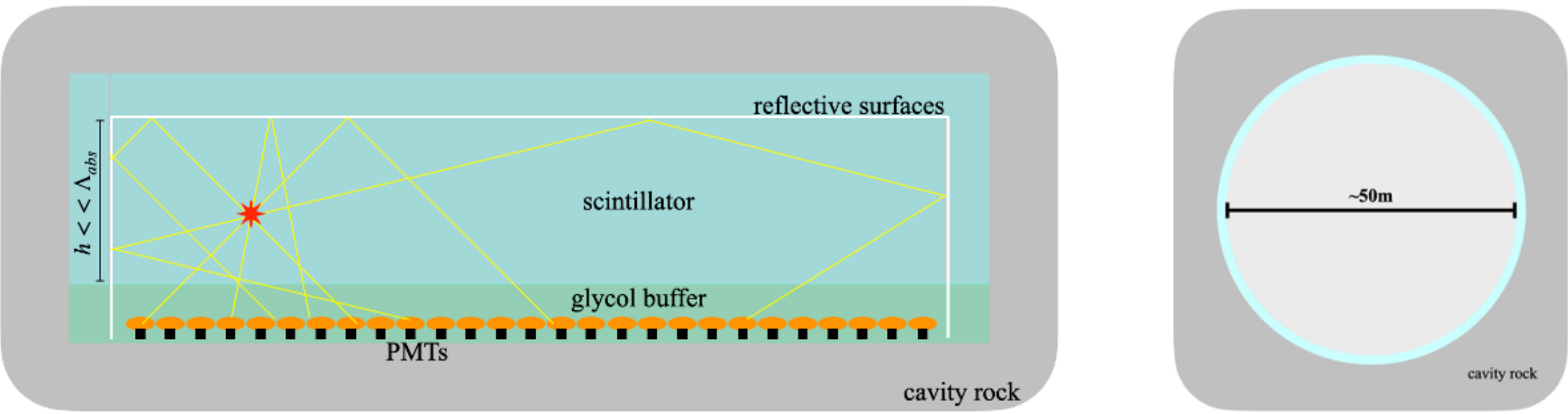} 
\caption{SLIPS design. Left - cross-sectional side view; right - top view.} 
\label{fig:SLIPS1}
\end{figure}

The construction of large-scale liquid scintillator detectors is complicated by the need to separate the scintillation region from photomultiplier tubes (PMTs) due to their intrinsic radioactivity. This is generally done using acrylic and/or nylon barriers, whose own intrinsic activity can also lead to substantial cuts to the fiducial detection volume for a number of low energy (~MeV) studies. Such barrier constructions also become increasingly difficult and expensive for larger detector volumes, with JUNO already pushing the boundaries of what might be achievable. The SLIPS concept is to do away with such physical barriers entirely by instead mounting PMTs on the bottom of a wide cavity and covering them with a distillable, lipophobic liquid, above which a less dense scintillator is layered. Liquids such as various ethylene glycols are good candidates for the bottom layer as they provide a good refractive index match to a number of liquid scintillator solvents. Thin, opaque and highly reflective (>90\%) surfaces are used near the top and side areas of the detector to provide a buffer region against radioactivity from the walls and to reflect scintillation light back to the bottom PMT array, where the time-separated reflected signals are used to reconstruct the 3D vertex position as well as the event energy. A conceptual design can be seen in Figure~\ref{fig:SLIPS1}. Initial simulation studies indicate that good position and energy reconstruction can be achieved with this approach. The notion is to use a shallow layer of scintillator relative to the cavity width, where the vertical depth of scintillator is chosen to be much less than the optical absorption length and can be optimised to balance fiducial volume, light level and reconstruction accuracy.

\begin{figure}[htp]
\centering 
\includegraphics[width=0.9\textwidth]{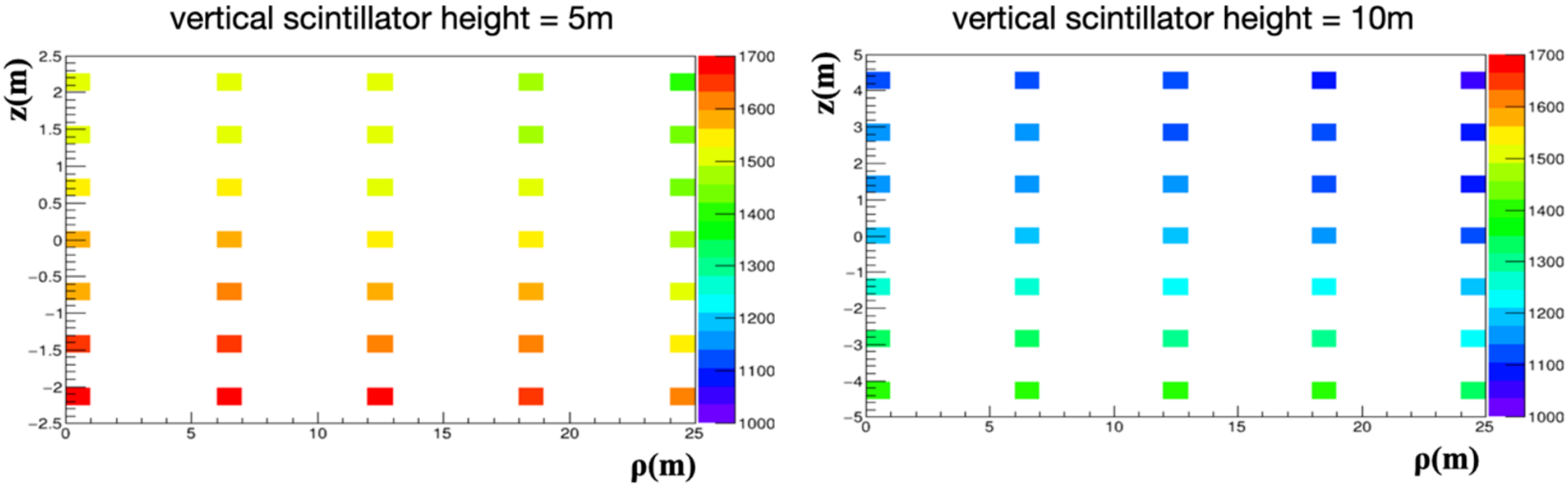} 
\caption{Detected number of photoelectrons for a 1 MeV electron for the simulated configuration as a function of $\rho$ (cylindrical radius from the centre of the detector) and $z$ (vertical height relative to the centre of the scintillator layer). The left plot is for a scintillator layer height of 5m and the right is for a height of 10m.} 
\label{fig:SLIPS3}
\end{figure}

Initial simulations have been carried out assuming a densely packed array of R5912-100 HQE PMTs in a pillbox-shaped detector with a diameter of 50m, an ethylene glycol layer extending 2m above the PMTs, a scintillator layer composed of LAB + 2g/L PPO, and 90\% specular reflective surfaces. Figure \ref{fig:SLIPS3} shows the resulting number of detected photons from a 1 MeV electron as a function of event position for vertical scintillator heights of both 5m and 10m. These correspond to fiducial volumes of $\sim$8kT and 16kT, respectively. 

\cleardoublepage

\section{Low-Threshold Neutrino Detectors}

 Low-threshold neutrino detectors are opening up new avenues in low-energy neutrino physics. This section summarizes the ongoing and planned projects as well as their enabling detector technologies. Recent breakthroughs in lowering the nuclear-recoil thresholds all the way down to the 10\,eV regime enabls the exploration of coherent elastic neutrino-nucleus scattering (CE$\nu$NS), which is a new portal to fundamental physics within and beyond the Standard Model of Particle Physics. The U.S.-based COHERENT experiment has set a milestone in 2017 with the first observation of CE$\nu$NS  and has triggered numerous experiments and  extensive  R\&D programs worldwide. Those have the potential to fully exploit CE$\nu$NS at spallation sources and at nuclear reactors within the next decade. The efforts will also lead to technological advances and potentially enable detector applications in science, industry and homeland security. Table~\ref{tab:7s} summarizes the many ongoing CE$\nu$NS detectors and their enabling technologies.

\begin{table}[h!]
    \begin{tabular}{|l|c|c|c|c|c|c|}
    \hline 
  Experiment &  Enabling technology  & $E_{th}$(NR)  & target material &  mass & status & $\nu$ source \\ 
  \hline  
  MINER   ~\cite{Agnolet:2016zir}      & phonon detectors &  $\mathcal{O}$(100eV) & Ge, Al$_2$O$_3$ &$\mathcal{O}$(100g) & D/C & R \\
    NUCLEUS~\cite{Strauss:2017cuu,Strauss:2017cam,Angloher:2017sxg}       & phonon detectors  &  $\mathcal{O}$(10eV) & CaWO$_4$,Al$_2$O$_3$    &$\mathcal{O}$(10g) & D/C & R \\
        RICOCHET ~\cite{Ricochet:2021rjo}       & phonon detectors&  $\mathcal{O}$(10eV) &Ge, Zn &$\mathcal{O}$(1kg) & D/C & R \\
   BULLKID ~\cite{colantoni:2020}      & phonon detectors &  $\mathcal{O}$(10eV) & Si  &$\mathcal{O}$(10g) & R\&D & - \\      
   
      CONNIE~\cite{Aguilar-Arevalo:2019jlr}        & CCD sensors &  $\mathcal{O}$(100eV) & Si  &$\mathcal{O}$(100g) & run & R \\  
      COHERENT~\cite{https://doi.org/10.48550/arxiv.2204.04575}      & cryog. scintillator &  $\mathcal{O}$(1keV) & CsI  &$\mathcal{O}$(10kg) & D/C & S \\  
        COHERENT~\cite{https://doi.org/10.48550/arxiv.2204.04575}      & HPGe &  $\mathcal{O}$(1keV) & Ge  &$\mathcal{O}$(10kg) & D/C & S \\  
  COHERENT~\cite{https://doi.org/10.48550/arxiv.2204.04575}      & scintillator &  $\mathcal{O}$(1keV) & NaI  &$\mathcal{O}$(1ton) & D/C & S \\

    CONUS~\cite{bonet2021constraints}      & HPGe &  $\mathcal{O}$(1keV) & Ge  &$\mathcal{O}$(1kg) & run & R \\  
    
    Dresden~\cite{Colaresi:2021kus}      & HPGe &  $\mathcal{O}$(1keV) & Ge  &$\mathcal{O}$(1kg) & run & R \\ 
       nuGen~\cite{nuGEN_mag7}      & HPGe &  $\mathcal{O}$(1keV) & Ge  &$\mathcal{O}$(1kg) & run & R \\ 
       
        TEXONO~\cite{Wong:2016lmb}      & HPGe &  $\mathcal{O}$(1keV) & Ge  &$\mathcal{O}$(1kg) & run & R \\ 
        
         NEON~\cite{https://doi.org/10.48550/arxiv.2204.06318}      & scintillator &  $\mathcal{O}$(1keV) & Ge  &$\mathcal{O}$(10kg) & run & R \\ 
         
          SBC~\cite{Alfonso-Pita:2022akn}      & bubble &  $\mathcal{O}$(100eV) & lAr(Xe)  &$\mathcal{O}$(10kg) & R\&D & R \\
           
            NEWS-G~\cite{Giomataris:2008ap}      & gaseous &  $\mathcal{O}$(100eV) & Ne,CH$_4$  &$\mathcal{O}$(100g) & run & -  \\
         COHERENT~\cite{PhysRevD.100.115020}      & liquid noble &  $\mathcal{O}$(1keV) & lAr  &$\mathcal{O}$(10kg) & run & S \\  
         
          RED100~\cite{Akimov:2017hee}      & liquid noble &  $\mathcal{O}$(1keV) & lXe  &$\mathcal{O}$(100kg) & run & R \\  
          
          CHILLAX~\cite{CHILLAX_M7_2021}      & liquid noble &  $\mathcal{O}$(100eV) & lAr,lXe  &$\mathcal{O}$(10kg) & R\&D & R/S \\  
          
            NUXE~\cite{Ni:2021mwa}      & liquid noble &  $\mathcal{O}$(100eV) & lAr,lXe  &$\mathcal{O}$(10kg) & R\&D & R \\

    \hline                      
    \end{tabular}
     \caption{Proposed detectors and enabling technologies for the exploration of CE$\nu$NS. The order of magnitude of the projected or demonstrated nuclear recoil energy threshold $E_{th}$(NR) and the envisioned total target mass (''mass'') are listed. The status is indicated as ''R\&D'' (R\&D phase), ''D/C'' (design or commissioning phase) and "run" (running experiment). If the project has a clear choice for the (anti)neutrino source, it is tagged by ''PR'' (power reactor), ''RR'' (research reactor) and ''S'' (spallation neutron source). }
    \label{tab:7s}
\end{table}
There are significant technological synergies of CE$\nu$NS detectors with high-resolution detectors for direct neutrino mass measurements \cite{PhysRevD.80.051301,velte,refId0} in particular for multiplexed readout, see e.g. \cite{doi:10.1063/1.4986222}. Furthermore, this section covers plenty of new ideas on future detector ideas to advance low-energy neutrino measurements by accessing new neutrino sources \cite{https://doi.org/10.48550/arxiv.1307.4738,Pattavina:2020cqc}, by entirely new detector concepts \cite{Baum:2018tfw,Alfonso:2022meh}, by exploiting new laboratories \cite{spacecraft} and by establishing directional sensitivity ~\cite{Vahsen:2020pzb,AristizabalSierra:2021uob}.

Furthermore, the striking mutual synergies of Dark Matter (DM) direct detection (Cosmic Frontier) and neutrino low-threshold detectors should be emphasized: on the one hand, the detector technology currently been used or considered for CE$\nu$NS measurements widely relies on the expertise gathered within the last decades in  direct DM detection --  both fields will highly profit from the currently blooming R\&D efforts for CE$\nu$NS. On the other hand, a laboratory measurement of the CE$\nu$NS cross-section will be essential to study solar neutrinos with future DM detector as well as to overcome the so-called neutrino-floor~\cite{Billard:2013qya}. Successful laboratory CE$\nu$NS experiments will further be an important tool to verify (''calibrate'') the DM detector technology by measuring a (fairly well) known neutrino signal.

\subsection{Common challenges of low-energy neutrino detectors}

To reach the ambitious physics goals of the next decade in low-energy neutrino physics, the enabling detector technology has to be continuously improved. The common technological and scientific challenges of low-threshold neutrino detectors can be summarized as follows: 

\begin{itemize}
     \item \textbf{Improving resolution and threshold}: one of the highest priority of current R\&D effort is improving the performance of the individual devices, since low thresholds in the 100\,eV (or even 10\,eV and below) and sub-eV resolutions are a prerequisite for CE$\nu$NS and direct neutrino mass measurements, respectively. What concerns cryogenic detectors there is extensive R\&D on the optimization of  transition-edge-sensors, the improvement of NTD phonon sensor readout and increasing the sensitivity of charge readout electronics. There has been enormous progress in lowering the thresholds of HPGe detectors and R\&D programs to further improve the charge collection and the low-noise readout electronics are ongoing. Scintillation detectors, liquid noble gas detectors and bubble chambers profit from advanced photon sensors and imaging techniques, as well as detector material with improved light yield and improved light collection efficiencies. 

  \item \textbf{Increasing detector mass}: Within the next decade low-threshold detector technology will enable precision physics at low energies, which requires larger detector masses. Typical phase1 CE$\nu$NS experiments will need to be scaled up in mass by factors of 10-100. This implies either increasing the target mass at same performance (requires improved sensor technology, see above), or increasing the number of detectors significantly. Usually, parallelizing the readout by ''simply'' multiplying the number of single electronic readout channels is not feasible, independent of the respective detector technology. However, the multiplexing technology has become mature in recent years, under leadership of U.S. institutes finds application in TES arrays e.g. in astronomy. This technology is a promising tool for next generation CE$\nu$NS experiments as well as for large scale direct neutrino mass experiments based on cryogenic detectors. 
  
  \item \textbf {Reduction of backgrounds}: Since most of the low-threshold CE$\nu$NS detectors for will be located at shallow depths, or even above ground as well as at radiogenically challenging environments (e.g. at reactors, accelerators), advanced background suppression and rejection techniques are needed. Promising techniques are: 1) Particle discrimination via 2-channel readout schemes (e.g. phonon-charge), pulse-shape discrimination, dual-phase TPCs or the different event topologies in bubble chambers and CCDs, 2) Passive shielding arrangements to moderate gamma and - in particular relevant at lowest energies - neutron radiation, 3) Active veto detectors that will act as large volume anti-coincidence detectors (e.g. HPGe detectors, liquid-scintillator) of instrumented detector holders that might veto stress-related events (in case of cryogenic detectors) , 4) Highly-efficient muon vetos  surrounding the setups, 5) Pulsed neutrino sources that provide a timing signal to increase signal-to-background,  and - ultimately - 6) Directional sensitivity to specially separate the sources for background and signal. In this context, it should be emphasized that the sensitivity of most of the CE$\nu$NS experiments is currently limited by the so-called low-energy Excess background~  \cite{https://doi.org/10.48550/arxiv.2202.05097}. A community-wide effort to understand and mitigate the Excess is absolutely required to reach the physics goals in the next decade. 
  
    \item \textbf {Understanding the low-energy detector response}: In recent years energy reach of neutrino detectors has been lowered by more than one order of magnitude, however the energy range of 10eV to 1KeV is lacking robust and precise calibration, in particular what concerns nuclear recoils. Several calibration methods have been proposed: X-ray fluorescence sources, LED calibration, identification of low-energetic Compton edges, as well as low-energy neutrino scattering and exploiting neutron-capture reaction. Fundamental measurements to understand the detector signals at lowest energies are absolutely required and should be conducted as a community-wide effort with high priority.  
  
  \item \textbf {Increase level of automatization} of experimental equipment for future applications of neutrino detectors in science, industry, and for society (e.g. reactor monitoring for nuclear proliferation via CE$\nu$NS).

\end{itemize}

\subsection{The eV frontier of neutrino detectors}
\subsubsection{Cryogenic particle detectors for CEvNS}
\bigskip
\noindent 
\underline{MINER}: The Mitchell Institute Neutrino Experiment at Reactor (MI$\nu$ER) experiment was launched to use cryogenic germanium and silicon detectors with a low nuclear recoil energy thresholds to register nuclear recoils of coherent elastic neutrino-nucleus scattering (CE$\nu$NS) at a TRIGA research nuclear reactor at the Texas A\&M University~\cite{Agnolet:2016zir}. This reactor has a movable core (1\,m to 10\,m) that will allow precision studies of very short baseline neutrino oscillation by comparing rates as a function of distance and largely eliminating reactor flux uncertainties. Close proximity of the detector to the reactor core, combined with multiple low threshold detectors with event-by-event discrimination between the dominant electromagnetic background and the nuclear recoil signal provides sensitivity to BSM physics, sterile neutrinos that oscillate away on a few-meter scale, and above all a highly sensitive probe for applied reactor monitoring for safeguards and non-proliferation. Planned deployment of the MIN$\nu$ER experimental set up at the South Texas Project (3 GW) power reactor will provide significant further improvement in measurement sensitivity.

\bigskip
\noindent 
\underline{NUCLEUS}: The NUCLEUS experiment aims for a detection of CE$\nu$NS using CaWO$_4$ cryogenic calorimeters with nuclear recoil thresholds around 20~eV. This unique feature demonstrated in an early prototype~\cite{Strauss:2017cuu,Strauss:2017cam,Angloher:2017sxg} will allow observation of the majority of tungsten recoils induced by reactor antineutrinos by accessing unprecedentedly low energies, taking full advantage of the coherent cross-section boost. The first experimental phase will deploy a 10~g cryogenic target~\cite{Rothe:2019aii} composed of approximately 6~g CaWO$_4$ and 4~g Al$_2$O$_3$ in a new experimental location~\cite{Angloher:2019flc} at the Chooz nuclear power plant in France. The two target materials feature widely different CE$\nu$NS cross-sections but a comparable neutron response, useful for in-situ measurement of potentially dangerous nuclear-recoil backgrounds.\\ 
The experimental setup will consist of a dry dilution refrigerator, a compact passive shielding made of neutron moderators and lead, active muon~\cite{Wagner:2022iqf} and gamma anticoincidence veto detectors as well as an integrated LED-based calibration system. The setup is under construction for commissioning in Munich in 2022 before deployment at the reactor site planned from 2023 on. 

\bigskip
\noindent 
\underline{Ricochet}: The Ricochet neutrino experiment aims to measure neutrinos produced from nuclear reactors by using cryogenic bolometers to identify the signature nuclear recoil from CE$\nu$NS~\cite{Ricochet:2021rjo}.  In order to overcome the high level of electromagnetic background present at low energies, Ricochet will make use of particle identification in order to discriminate between electron and nuclear recoils.  The future Ricochet experiment will be deployed at the ILL-H7 site in Grenoble, France. The H7 site starts at about 8~m from the ILL reactor core that provides a nominal nuclear power of 58.3~MW, leading to a neutrino flux at the Ricochet{} detectors 8.8~m from the reactor core of about 1.2$\times$10$^{12}$~cm$^{-2}$s$^{-1}$. 
The experiment will make use of two detector technologies/targets.
The CryoCube will consist of an array of 27 ($3\times3\times3$) high purity germanium crystal detectors, encapsulated in a radio-pure infrared-tight copper box suspended below a lead shield inside the crysotat~\cite{RICOCHET:2021gkf}. 
Q-Array -- the complimentary detector array within Ricochet -- will consist of 9 cubes of superconducting zinc cubes, each with a mass of about 35\,g, as its target. Using superconductors as the primary detector is a novel technology which is expected to provide detection sensitivity theoretically down to the Cooper pair binding energy.  The expected discrimination mechanism begins with the different efficiency of quasiparticle (QP) production (breaking Cooper pairs) by electron recoils (higher QP production) vs.~nuclear recoils (lower QP production). Ricochet is currently scheduled to see ``first light" in 2023.

\bigskip
\noindent 
\underline{BULLKID} is a R\&D project on cryogenic detectors for CE$\nu$NS and light Dark Matter~\cite{colantoni:2020}. By exploiting the high multiplexing levels of kinetic inductance detectors,  goal of BULLKID is to create a monolithic and highly-segmented array of silicon targets with energy threshold on nuclear recoils around 100~eV  and total mass of 30-60~g. In future experiments several arrays would be produced and stacked to obtain target masses exceeding 1~kg.

\noindent 
\underline{Cryogenic Carbon Detectors} have various advantages: They have good isotopic purity, outstanding radiation hardness, and excellent phonon properties and dynamics due to the particularly strong carbon bonds. In their tetrahedral forms have the highest sound speeds and most energetic optical phonons of any known crystals. These properties are expected to yield a significant improvement in energy sensitivity relative to more conventional semiconductors, thus enabling detection thresholds in the meV to eV energy range \cite{PhysRevD.103.075002,PhysRevD.99.123005}. Finally, much of the thin-film phonon-readout technology developed for germanium and silicon can be applied to diamond with minimal modification. Initial work on diamond and SiC particle detectors is promising, and a Transition Edge Sensor (TES) readout on a diamond sample was recently demonstrated with comparable performance to sapphire \cite{lucia}.

\noindent 
\underline{Phonon-mediated KIDs}
 Athermal (meV-scale) phonons in sub-Kelvin target materials can be produced by such small energy depositions, and superconducting kinetic inductance detectors (KIDs) can sense these meV phonons, with long-term potential for even single-phonon sensitivity. The inherent multiplexability of KIDs also makes them ideal for recovering position information from phonons. Different architectures are appropriate for different applications. In the case of low-mass dark matter (sub-GeV nuclear- and electron-scattering particles and sub-eV dark photons and axion-like particles (ALPs)), thresholds at the meV scale are eventually needed, motivating small individual detectors with a single KID sensor, exploiting the energy resolution potential of KIDs. For coherent elastic neutrino-nucleus scattering of solar and reactor neutrinos, information about recoil type (nucleus or electron) and position are as important as energy resolution, motivating larger individual detectors instrumented with many KIDs.

\subsubsection{CCD-based detectors for CEvNS}

\bigskip
\noindent 
\underline{CONNIE}: The Coherent Neutrino-Nucleus Interaction Experiment (CONNIE) uses low-noise fully depleted charge-coupled devices (CCDs) with the goal of measuring low-energy recoils from CE$\nu$NS of reactor antineutrinos with silicon nuclei~\cite{Aguilar-Arevalo:2019jlr}. The CCD detectors  can operate at a electron recoil threshold of approximately 30 eV, where the conversion from electron equivalent to silicon recoil energy is given by the so-called quenching
factor, from which measurements at low energies ($\approx 0.7$ keV) and new theoretical approaches are used \cite{QFmeasurementsSi, QFtheorynew, Sarkis2021}. CONNIE has reported results from its analysis with a detector array of 8 CCDs with a fiducial mass of 36.2 g, and a total exposure of 2.2 kg-days. In an analysis of the difference between the reactor-on and reactor-off spectra, no excess events are found at low energies, yielding upper limits at 95\% confidence level on the~CE$\nu$NS~rate. In the lowest-energy range analyzed by CONNIE, $50-180$ eV, the expected limit is 34 (39) times the standard model prediction, depending on the assumed quenching factor. 

\bigskip
\noindent 
\underline{$\nu$IOLETA} is a future experiment \cite{webpage_violetta} that will measure reactor antineutrinos and their physical properties through two channels: neutrino-electronic interaction and Elastic Coherent Neutrino Nucleus interaction. This experiment will use the new Skipper-CCD technology that achieves sub-electronic readout noise and a detection threshold in the eV range. Two positions were identified to place vIOLETA within the Atucha II Nuclear Power Plant in Argentina, 8 and 12 meters away from the center of the reactor core.

\subsubsection{Detectors for neutrino mass}
\noindent
\underline{Project8} is an experiment designed to measure the absolute neutrino mass using cyclotron radiation emission spectroscopy (CRES) \cite{PhysRevD.80.051301}. In addition to probing the electron-weighted neutrino mass through endpoint investigations, beta-decay spectrum measurements can also be used to study other neutrino properties that affect the spectrum shape. The Project 8 experiment plans to improve on the existing tritium beta-decay measurements in two different ways: by measuring beta-decay of atomic tritium and by using the CRES technique for spectrum measurement. The former reduces the systematic effects associated with the decay of molecular tritium [2], while the latter allows for a high-statistics differential spectrum measurement with excellent energy resolu- tion over a range of energies. Leveraging these unique capabilities, Project 8 is projected to achieve $m_\beta$ sensitivity of $\sim0.040$\,eV (90\% C.L.). The advantages of the CRES method and the use of atomic tritium generally also extend to the rest of the measured spectrum, equipping Project 8 with unique access to other secondary physics capabilities.

\bigskip
\noindent 
\underline{Future large-scale $^{163}$Ho experiment}:
The electron capture decay of  $^{163}$Ho provides an attractive system for kinematic measurements of the electron neutrino mass. When $^{163}$Ho is embedded in a calorimetric sensor, each decay deposits energy in the sensor equal to the Q-value of the reaction minus the energy of the departing neutrino. The rest mass of the neutrino is manifested as a deficit of events in the region of the decay spectrum near the Q-value. $^{^163}$Ho is particularly attractive because of its low Q-value of 2.833 keV and its reasonable half-life of 4570 years \cite{PhysRevLett.115.062501}. The ECHO collaboration has demonstrated decay spectra with 275,000 counts using a small array of magnetic microcalorimeters \cite{velte} and is presently analyzing data corresponding to a spectrum with about 10$^8$ counts. The HOLMES collaboration is developing multiplexed transition-edge-sensors (TESs) for this purpose \cite{refId0}. An international collaboration between groups in the US and Europe is an attractive path to execute a large-scale 163Ho experiment. While work on ${^163}$Ho in Europe is presently more mature than in the U.S., the U.S. community is leading on multiplexed readout \cite{doi:10.1063/1.4986222} and has excellent facilities for fabricating cryogenic detectors and SQUID multiplexers. This technology has large synergies with the CEvNS R\&D program and will enable an up-scaling of the detector mass of cryogenic detectors using TES sensors.

\subsubsection{Future detectors for relic neutrinos}
\bigskip
\noindent 
\underline{PTOLEMY} is designed to be the first instrument to detect neutrinos created in the early moments of the Big Bang based on the concept of neutrino capture on $\beta$-decay nuclei as a detection method for the Cosmic Neutrino Background (CNB). An experimental realization of the concept for CNB detection was proposed based on PTOLEMY \cite{https://doi.org/10.48550/arxiv.1307.4738} in 2013. The parameters for a relic neutrino experiment require 100 grams of weakly-bound atomic tritium, sub-eV energy resolution commensurate with the most massive neutrinos with electron-flavor content, and below microHertz of background rate in a narrow energy window above the tritium endpoint. The PTOLEMY experiment  aims to achieve these goals through a combination of a large area surface-deposition tritium target, MAC-E filter methods, cryogenic calorimetry, and RF tracking and time-of-flight systems.

\subsubsection{New detector concepts}
\noindent 
\underline{Paleo-detectors}
 use the nuclear damage tracks recorded in natural minerals over geological time-scales to detect weakly-interacting particles over exposures much larger than what is feasible in conventional terrestrial detectors \cite{Baum:2018tfw,Drukier:2018pdy}. Unlike conventional experiments which measure nuclear recoils in real time, paleo-detectors measure the number of events integrated over the age of the mineral, reaching up to a billion years for minerals routinely found on Earth. The sources of the keV-scale nuclear recoils which can be recovered as tracks is rich, including atmospheric neutrinos \cite{Jordan:2020gxx}, solar neutrinos \cite{Tapia-Arellano:2021cml}, supernova neutrinos \cite{Baum:2019fqm}, as well as dark matter \cite{Baum:2021jak,Baum:2021chx}. Low-energy neutrino and dark matter tracks are initiated dominantly via CE$\nu$NS (quasi-elastic charged-current interactions are more applicable for high-energy neutrinos). 
 
\noindent 
\underline{PALEOCCENE}: This concept~\cite{Cogswell:2021qlq,Alfonso:2022meh} aims to exploit the crystal defects caused by nuclear recoil by using an optical readout scheme based on the imaging of individual color centers. The resulting detectors would be room-temperature, passive devices with recoil thresholds close to the threshold damage energy of the detector material of 100\,eV or less. The project is in the early stages of R\&D.

\noindent 
\underline{Neutrino Spacecraft}:
The idea of new Science using a neutrino detector spacecraft \cite{spacecraft}, which is of interest to NASA and DOE elementary particle physics, is discussed in the context of two physics scenarios: the detection of solar neutrinos at close distance to the Sun, and the search for dark matter at the outskirt of the Solar system. It is possible to detect far more solar neutrinos by going closer to the Sun and at 7 solar radii distance the solar neutrino flux would be 1000x that on Earth and at 3 solar radii 10,000x that on Earth. This would result in the need for a smaller detector to do the same things that large detectors on earth do and a 1 ton detector need only be 1 kg at 7 solar radii. Other science that can be done with a spacecraft capable of detecting neutrinos in space are that solar neutrino backgrounds in dark matter searches go down as the detector goes away from the Sun and at Jupiter a 10x reduction is expected and 100x reduction at Uranus. Furthermore, there are ideas to study the gravitational focus of the sun for neutrinos to image exoplanets. The feasibility of the physics scenarios mentioned above depend crucially on the ability to operate detectors in unshielded space. 

\noindent 
\underline{Snowball detectors}:
A new detector technology proposed, the so-called “Snowball Chamber,”  is based on the phase transition (of liquid to solid) for metastable fluids. A water-based supercooled detector \cite{D1CP01083B} has the potential to move past the Neutrino Floor, and extend the reach of direct detection dark matter experiments and has also applications within CEvNS.

\subsection{Optimized conventional detectors to exploit CEvNS}
\subsubsection{HPGe detectors }

\bigskip
\noindent 
\underline{COHERENT-HPGe}:
Continued development over the past decade of P-type Point-Contact (PPC) High-Purity Germanium (HPGe) detectors has resulted in devices with masses in excess of 2 kg each and sub-keV energy resolution and thresholds \cite{COOPER201125}. Combined with the well-understood systematics and measured quenching factors, the resolution allowed by PPC detectors enables precision measurements of spectral shape distortions due to nuclear form factors or new physics. The drawbacks of these detectors are primarily in their relatively high cost to manufacture, and in the case of SNS operation, the limited timing resolution afforded due to the wide range of drift times in larger mass detectors (up to $\sim$ 2$\mu$sec).
Currently, the COHERENT Collaboration is deploying an array of 8 PPC germanium detectors with a total mass in excess of 16 kg. To enable a next-generation effort, we propose the development of new classes of PPC detectors with larger masses and a narrower range of drift lengths, while maintaining reasonable depletion/operating voltages and the low-noise performance that enables excellent energy resolution and low thresholds.

\bigskip
\noindent 
\underline{CONUS}: The CONUS (COherent elastic Neutrino nUcleus Scattering) experiment employs four 1\,kg low energy threshold high-purity point-contact Germanium detectors to look for CE$\nu$NS. The experiment is located at the commercial nuclear power plant of Brokdorf, Germany, in a distance of 17.1\,m from the reactor core with a maximum thermal power of 3.9\,GW. With the electrically cooled spectrometers an energy resolution of 150-160eV (ionization energy, full width at half maximum ionization energy) at the 10.4\,keV K-shell X-ray of $^{71}$Ge/$^{68}$Ge is achieved \cite{bonet2021large}. The energy threshold is $\leq$1.875\,keV for nuclear recoils. With the onion-like shield consisting of layers of lead, borated polyethylene and a muon anti-coincidence veto a background level of $\sim$10\,counts/kg/d below 1\,keV ionization energy is achieved at reactor site \cite{bonet2021full}.
With the data collected in 2018 and 2019 an upper limit on CE$\nu$NS was derived in dependence of the quenching factor (ratio between detected ionization energy and recoil energy) in germanium \cite{bonet2021constraints}. For a quenching factor of k=0.16, this corresponds to a limit of 0.34\,kgd$^{-1}$ at 90\% confidence level (factor 17 above the standard model prediction). 
The quenching factor for germanium at low recoil energies is not well known and large discrepancies between the existing measurements persist. Recently, the CONUS collaboration carried out an own quenching measurement \cite{bonhomme2022direct} to significantly reduce this major systematic uncertainty.

\bigskip
\noindent 
\underline{Dresden}: The experiment consists of a low-noise 3 kg p-type point contact germanium detector which has been installed and operated at the Dresden-II power reactor, near Chicago, at about 10 meters from its 2.96 GW$_{th}$ core~\cite{Colaresi:2021kus}. The detector are enclosed by a compact shielding made of several layers of active vetos and passive shielding material. The results from an upgraded setup suggest a preference for a CE$\nu$NS component in the data, when being compared to a background only model~\cite{Colaresi:2022obx}.

\bigskip
\noindent 
\underline{nuGen}: The nuGEN experiment aims at the detection of CE$\nu$NS with low-background, low-threshold HPGe detectors installed at a distance of $\sim10$\,m to one of the 3.1\,GW$_{th}$ reactors at the Kalinin Nuclear Power Plant (KNPP) in Russia. The detectors are surrounded by a compact passive shielding and an active muon veto, which are installed on a movable platform to modify the distance to the reactor core. nuGEN has been installed on-site in 2019 and preliminary data from first science runs have been presented~\cite{nuGEN_mag7}. 

\bigskip
\noindent 
\underline{TEXONO}: 
The TEXONO collaboration has been studying neutrino physics with sub-keV
germanium detectors at the Kuo-Sheng Reactor Neutrino Laboratory (KSNL) in Taiwan~\cite{Wong:2016lmb}. There is a national policy of de-commissioning nuclear power in Taiwan, and the Kuo-Sheng Reactor will be phased out by 2023. As a result, there are no plans on expansion or new projects to the KSNL program. The collaboration would seek to continue the studies via collaboration with other existing reactor laboratories. Data taking and R\&D program are conducted with electro-cooled (EC) point-contact germanium detectors (PCGe). As of summer 2020, detector mass up to 1.43 kg
are built and threshold as low as 200 eV ee is achieved. The data would also bring improved sensitivities to the searches on various Beyond Standard Model (BSM) physics channels, such as neutrino magnetic moments~\cite{Wong:2006nx} and milli-charged neutrinos~\cite{Chen:2014dsa}. Active theory program is being pursued in parallel, with focuses on atomic corrections to $\nu$N (and $\chi$N for dark matter) cross-sections~\cite{Chen:2014dsa}, as well as BSM searches.

\subsubsection{Photon-based low-threshold detectors }
\noindent 
\underline{COHERENT-CsI}:
The COHERENT collaboration is performing R\&D towards optimizing low threshold cryogenic scintillators. The most serious limitation in reducing the energy threshold of the COHERENT CsI(Na) detector was the Cherenkov radiation originated from its PMT quartz window by natural radiation and cosmic rays, which can be  eliminated by replacing PMTs with SiPM arrays. Cryogenic operation is needed to reduce the dark count rate of SiPM arrays, which also calls for the replacement of doped crystals with undoped ones due to much higher intrinsic light yield of the latter. The light yield of such a combination is expected to be at least 4 times higher than that of the CsI(Na) detector, and the energy threshold would be at least three times lower. With such a low threshold, even a $\sim$ 10 kg prototype can detect a thousand CEvNS events annually.

\noindent 
\underline{NEON}: The Neutrino Elastic-scattering Observation on NaI(Tl) (NEON) experiment uses high-light yield NaI(Tl) crystals to observe the CE$\nu$NS events at a distance of 24\,m from the core of the Hanbit nuclear reactor in Korea. 
This experiment utilized the previous experiences of the NaI(Tl) crystal detectors for the COSINE-100 dark matter search experiments~\cite{Adhikari:2018ljm,Adhikari:2019off}.  
The NaI(Tl) detector used in the COSINE-100 experiments showed a light yield of 15-photoelectrons/keV~\cite{Adhikari:2017esn}, and a multivariate machine larning technique was used to effectively remove the noise event caused by PMT to reach a low energy threshold of 1~keV~\cite{COSINE-100:2020wrv}. Preliminary studies to lower the energy threshold achieved 0.5~keV energy event access with 80\% selection efficiency and 25\% noise contamination level.  
In addition, a novel encapsulation method of the NaI(Tl) crystals with improved light collection efficiency up to 22 photoelectrons/keV~\cite{Choi:2020qcj} has been developed. Energy thresholds of less than 0.3\,keV can be achieved with these detectors. 
The first phase NEON experiment~(NEON-pilot) was built with a $2\times3$ array of 6 detectors with a total mass of 15~kg using the commercial quality crystals while the next phase experiment~(NEON-1) may use up to 100\,kg of the low-background NaI(Tl) crystals~\cite{Park:2020fsq}. 
The NEON-pilot crystals were immersed in an 800-L liquid scintillator. It was shielded with 10-cm-thick leads and 30-cm-thick polyethylene. The shields and DAQ systems closely follows the COSINE-100 dark matter experiment~\cite{Adhikari:2017esn,Adhikari:2018fpo}.

\noindent 
\underline{COHERENT-NaI}:
The COHERENT collaboration is currently constructing a multi-tonne modular array of re-purposed thallium-doped, 7.7 kg NaI crystals \cite{https://doi.org/10.48550/arxiv.2204.04575}. Each module of 63 crystals provides 485 kg of detector mass. The current phase consists of the sectional deployment of 5 modules for a 2.4 tonne detector designed using dual-gain bases on the photomultiplier tubes to measure the low-energy CEvNS signal ($\sim$3-25 keV$_{ee}$) simultaneously with the high-energy (10-50 MeV) charged-current signal on 137I. Background studies with the NaIvE-185 detector array of 24 crystals deployed in Neutrino Alley since 2016, indicates that environmental and intrinsic backgrounds are sufficiently low for a successful CEvNS measurement. Recent quenching-factor measurements and a calibration scheme will address nonlinearity issues for low-energy signals. Initial charged-current studies with the NaIvE-185 detector will inform the analysis of the NaI Tonne-scale Experiment (NaIvETe) neutrino scattering measurement on $^{127}$I. The future deployment of an additional two modules will bring the NaIvETe mass total to 3.4 tonnes.

\noindent 
\underline{COHERENT-D2O}:
One of the dominant systematics of the COHERENT measurements are neutrino flux uncertainties. To benchmark the neutrino flux, COHERENT plans to construct a 1300-kg D$_2$O detector in two modules. Its operation is based on detecting Cherenkov light from CC electron neutrino reactions with d. Each module will consist of an upright cylinder, with D2O contained inside a central acrylic cylinder, contained inside a steel tank with 10 cm of H2O tail catcher. Twelve PMTs view the volume from above. With the theoretical cross section known at the 2-3\% level \cite{PhysRevC.75.044610}, it has the potential to significantly improve the overall systematic uncertainties of the experiment. More than 500 CC neutrino-deuterium events per module per SNS beam-year are anticipated.

\subsubsection{Bubble chambers}


\underline{SBC}: The SBC (Scintillating Bubble Chamber) Collaboration is developing novel liquid-argon bubble chambers for GeV-scale dark matter and CE$\nu$NS physics~\cite{Alfonso-Pita:2022akn,Giampa:2021wte}. The first detector, SBC-Fermilab, with 10-kg active mass, is currently under construction for characterization and calibration in the NuMI tunnel at Fermilab, aiming to reach a threshold for nuclear recoils of 100~eV. The detector consists of a quartz jar filled with superheated liquid argon, which is spiked with ppm levels of xenon acting as a wavelength shifter. Cameras are used to image bubbles, silicon photo-multipliers detect scintillation, and piezo-acoustic sensors listen for bubble formation. A duplicate detector constructed with low-background components, SBC-SNOLAB, will follow for a search for 0.7--7~GeV dark matter. Following this initial program, a deployment of one of these detectors after $\sim$2024 at a nuclear reactor could make a high signal-to-background measurement of reactor neutrino CE$\nu$NS. Sensitivity to the weak mixing angle, neutrino magnetic moment, and a Z’ gauge boson mediator have been calculated for deployments at both a 1~MW research reactor and a 2~GW power reactor~\cite{SBC:2021yal}. Background characterizations are ongoing at the 1~MW TRIGA Mark III research reactor located at the National Institute for Nuclear Research (ININ) near Mexico City for a potential first reactor deployment 3~m from the core.

\subsubsection{Gaseous detectors }
\bigskip
\noindent 
\underline{NEWS-G} is a direct dark matter detection experiment, sensitive to  light Dark Matter (DM) between 0.1 and \SI{10}{GeV/c^2}. NEWS-G uses a spherical gaseous detector, the Spherical Proportional Counter (SPC)~\cite{Giomataris:2008ap}. The SPC's features that make it ideal for light DM searches make it also appealing for the detection of neutrinos through CE$\nu$NS. The study of CE$\nu$NS with a sub-keV energy threshold detector, like the SPC, allows for a rich physics program and opens a window to physics beyond the Standard Model that will appear as deviations from the expected recoil spectra predicted by SM interactions~\cite{Scholberg:2005qs}.
The detector is composed of a spherical shell made of radio-pure copper, acting as the cathode, and a read-out sensor at the center with either a single anode~\cite{Katsioulas:2018pyh} or multiple anodes~\cite{Giomataris:2020rna, Giganon:2017isb}. The detector is versatile and can operate with a wide variety of gas mixtures.  The SPC exhibits several features, such as:
  simple, few component build;
   very low energy threshold (single ionization electron level) independent of detector size~\cite{Bougamont:2010mj}; and
  fiducialization and event discrimination through pulse shape analysis. NEWS-G plans to construct a \SI{60}{cm} diameter sphere made of ultra-pure copper for CE$\nu$NS studies. The detector will be encased in a shielding inspired from the GIOVE~\cite{giove} and CONUS~\cite{conus} experiments. A muon veto will complete the shielding apparatus. The experimental setup will be installed at Queen's University, Kingston, Canada, to assess the environmental and cosmogenic backgrounds and to establish if any alterations are required for operations in the proximity of a reactor. The construction of the shield will start in fall 2020 and commissioning at Queen's University is expected to take place by summer 2021

\bigskip
\noindent 
\underline{ArgonTPC}: 
This R\&D program  intends to study the feasibility of, and help achieve the measurement of nuclear recoil (NR) directionality by tracking their ionization signatures in gaseous argon TPCs. The goal is to develop detectors capable of tracking the O(10-100) $\mu$m ionization tracks produced by O(10-100) keV NRs from CEvNS interactions in gaseous argon employing high-granularity GEM-based GAr TPCs. Such a detector would enable a broad range of physics measurements in focused neutrino beams of O(100) MeV neutrinos as well as high-intensity stopped pion neutrino sources which are currently available or will be operational in the coming years at facilities such as the SNS at Oak Ridge National Lab and in Fermilab’s next-generation neutrino beamline. A particularly interesting scenario would envision augmenting, through this effort, a DUNE-like GAr TPC detector in order to measure CEvNS interactions from sub-100 MeV neutrinos in a future underground near-detector experimental hall at Fermilab.

\subsubsection{Liquid noble gas detectors}

\bigskip
\noindent 
\underline{COHERENT-LAr}:
The COHERENT collaboration is developing Liquid argon (LAr) detectors, with focus on reducing energy thresholds and understanding the LAr detector response to make the technology ready for CEvNS measurements. COHERENT  has performed detailed studies of the 24 kg fiducial mass CENNS-10 LAr scintillator detector \cite{PhysRevD.100.115020} and performed the first detection of CEvNS in a light nucleus, advancing the understanding of neutrino-nucleus interactions and constraining non-standard interactions. Detailed background studies and lowering the scintillation threshold were crucial for these results, laying the groundwork for the 610 kg fiducial mass CENNS-750 detector \cite{https://doi.org/10.48550/arxiv.1803.09183}. The CENNS-750 detector will provide precision measurements of the CEvNS cross section on argon. Novel techniques are persued, including the  study of high-efficiency PMTs and SiPMs to reduce thresholds, and explore highly-segmented photodetector geometries to improve event selection and particle ID.

\bigskip
\noindent 
\underline{RED-100}: RED-100 is a two-phase xenon emission detector built to observe the coherent elastic scattering of reactor electron antineutrinos off xenon atomic nuclei~\cite{Akimov:2017hee}. The mass of the detector medium is 160 kg in the sensitive volume, and about 100 kg in the fiducial volume --- the largest value among detectors developed for CE$\nu$NS observations at reactors. The capability of detector full-scale operation in the background conditions caused by cosmic radiation has been demonstrated in a ground-level laboratory~\cite{Akimov:2019ogx}. The detector was deployed at the Kalinin NPP in 2021 at 19~m distance under the 3~GW reactor core (50~m.w.e. overburden) and acquired both reactor on and off data in January and February of 2022 with an energy threshold of about 1~keV$_{nr}$.

\noindent
\underline{CHILLAX}: The CoHerent Ionization Limit in Liquid Argon and Xenon (CHILLAX) project is an experimental effort to develop a xenon-doped argon ionization detector that can enjoy the benefits of both argon and xenon~\cite{CHILLAX_M7_2021}. Thanks to the relatively small atomic mass, an argon atom can pick up more kinetic energy from neutrino scatters than heavier elements can, and by doping it with xenon -- which has lower excitation/ionization energy and faster scintillation than argon does  -- the detector can be more efficient in producing ionization electrons from CEvNS  interactions and also generate detectable light signals with long wavelength and fast decays. Combining an argon target with a xenon detector-like performance, CHILLAX aims to develop the ideal large-mass noble liquid CEvNS detector. With an expected energy threshold of 200-300eV, CHILLAX may detect a few times less CEvNS interaction signal per kilogram than what is possible in eV-threshold detectors, but thanks to the scalability of the noble liquid technology CHILLAX can easily achieve an active mass of tens of kilogram and be a leading competitor in rate-oriented CEvNS  applications. 
CHILLAX is currently focusing on developing the first generation prototype detector. Once the xenon-doping benefits and low-energy sensitivity are experimentally demonstrated, we plan to build a $\sim$50kg detector to deploy either at the SNS (for BSM physics studies) or near a reactor (for sterile neutrino search and reactor monitoring demonstrations). 

\bigskip
\noindent 
\underline{NUXE}: The NUXE experiment will use a liquid xenon detector to observe reactor neutrino CE$\nu$NS events down to single ionization electron signals~\cite{Ni:2021mwa}. The experiment is currently under development at UC San Diego with a 30-kg liquid xenon target in an electron counting chamber (ECC). Major effort is reducing the background down to the single electrons, corresponding to a nuclear recoil energy threshold of $\sim$300~eV~\cite{Lenardo:2019fcn}. 

\subsection{Dark Matter detectors for next-generation neutrino measurements}

Neutrinos from the astrophysical sources like the Sun and supernovae as well as from the atmospheric cosmic-ray-showers produce nuclear recoils via CE$\nu$NS, leading to signals in future dark matter detectors (see \cite{https://doi.org/10.48550/arxiv.2203.07361}). 
The eventual presence of an unshieldable background in multiton-scale dark matter detectors has been anticipated for some time~\cite{Cabrera:1984rr,Drukier:1986tm,Monroe:2007xp,Vergados:2008jp,Strigari:2009bq,Gutlein:2010tq,Gelmini:2018ogy}, and is thought to present a major obstacle for improving the sensitivity of these experiments. The central problem is that for many of the most commonly sought-after dark matter models produce nuclear recoil signals that look remarkably similar to the CE$\nu$NS recoil energy spectra generated by natural neutrino sources. Due to the finite systematic uncertainty on the fluxes of those sources eventually a feeble-enough DM signal could disappear under the expected variation in the neutrino event rate, and because its signal would not be distinct enough from the background no positive identification of dark matter would be possible. Naively this implies that there is a ``floor'' to the sensitivity of direct detection experiments~\cite{Billard:2013qya}. 

Since the dark matter and CE$\nu$NS signals are not exactly identical--- it is not impossible to search for dark matter in the presence of a CE$\nu$NS background, but it does entail significant reduction in experiment's sensitivity. This observation has been called ``neutrino fog''~\cite{OHare:2021utq}. Neutrino floor considerations play an important role when considering the feasibility of future experiments, as decreased sensitivity to DM signals in vicinity of neutrino background also hinder the increase in experimental sensitivity with scaling in detector exposure~(e.g.~\cite{Dent:2016wor,Gelmini:2018ogy}). There are several approaches for circumventing the neutrino floor if an experiment can access additional information to further discriminate the dark matter signal and neutrino background, including the use of annual modulation~\cite{Davis:2014ama}, target complementarity~\cite{Ruppin:2014bra,Gelmini:2018ogy,Gaspert:2021gyj,OHare:2020lva}, and directionality~\cite{OHare:2015utx,Grothaus:2014hja,Mayet:2016zxu,OHare:2017rag,Franarin:2016ppr,Vahsen:2020pzb,Vahsen:2021gnb,Sassi:2021umf}.

\subsubsection{Multi-ton liquid noble detectors}

\bigskip
\noindent 
\underline{XENON}: 
The XENON1T experiment operated a dual-phase time projection chamber (TPC) filled with 3.2 tonnes of ultra-pure liquid xenon (LXe). The TPC contained 2.0 tonnes of LXe that is sensitive to ionization electrons (S2) and scintillation photons (S1) produced by interactions therein. In a fiducial volume with 1.0 tonnes of LXe, a background level down to $(76\pm2)\,\mathrm{events}/(\mathrm{tonne}\times\mathrm{year}\times\mathrm{keV})$ has been achieved. A WIMP dark matter search lasted from December 2017 to February 2018. Using this data, the XENON collaboration has performed a first sensitive search for solar $^{8}$B neutrinos through nuclear recoils from the CE$\nu$NS process~\cite{Aprile:2020thb}. $^8$B CE$\nu$NS leads to an average nuclear recoil energy of $\sim$\,1\,keV, requiring unprecedented low energy thresholds in identifying scintillation and ionization signals. In this analysis, the threshold in ionization and scintillation signals was lowered down to 4\,electrons and 2\,photon-electrons, respectively. With an exposure of 0.6\,tonne$\times$year, the expected CE$\nu$NS signal is 2.1 events with a background expectation of 5.4 events, dominated by the accidental pileup of isolated-S1 and isolated-S2 signals. The mean discovery power of $^8$B CE$\nu$NS is $2\,\sigma$, limited by the exposure of this experiment. No significant excess from $^8$B CE$\nu$NS is found in XENON1T. 

\bigskip
\noindent 
\underline{LUX-ZEPLIN}: 
LUX-ZEPLIN (LZ) is a dual-phase xenon TPC with a 7-tonne active mass located 1 mile underground at the Sanford Underground Research Facility (SURF) in Lead, South Dakota. 
The LZ detector was designed to search for interactions of particle dark matter in the mass range from 1 GeV/c$^2$ to 10 TeV/c$^2$.
Because the CEvNS process can produce low-energy nuclear recoil signals  similar to those produced by low-mass WIMPs, LZ will also be sensitive to astrophysical sources of neutrinos such as solar $^8$B neutrinos and atmospheric neutrinos. Over the full 15.34 tonne-year exposure, LZ is expected to observe 0.65 events from atmospheric neutrinos and 36 events from $^8$B neutrino 
in its WIMP search campaigns~\cite{akerib2020projected}. A positive detection of $^8$B CEvNS signals will be an unambiguous confirmation of LZ's low-mass WIMP sensitivity. As the CEvNS mechanism is insensitive to neutrino flavors, LZ's measured flux provides a  data point complementary to large solar neutrino experiments that rely on charge-current interactions. The observable rate of $^8$B CEvNS in LZ  depends strongly on the detector energy threshold. LZ relies on the detection of scintillation and ionization signals produced by particle interactions in the active liquid volume. 

\bigskip
\noindent 
\underline{DARWIN}: 
DARWIN (DARk matter WImp search with liquid xenoN) is a proposed next-generation dark matter experiment that will operate 50\,t (40\,t active) of xenon in a cylindrical time projection chamber with 2.6\,m in diameter and height~\cite{Aalbers:2022dzr}. The TPC will be placed in a double-walled cryostat vessel surrounded by neutron and muon vetoes. While DARWIN's primary goal is to observe particle dark matter in the $\sim$1\,GeV-100\,TeV mass range, it will also be able to measure the solar $^{8}$B neutrino flux, as well as atmospheric and supernovae neutrinos via CE$\nu$NS. The expected $^{8}$B neutrino rate is $\sim$90 events/(t\,yr)~\cite{Baudis:2013qla,Aalbers:2016jon}, depending on the achieved energy threshold. The measurement of the atmospheric neutrino flux requires a large exposure of about 700\,t\,yr~\cite{Newstead:2020fie}. A DARWIN-like detector would be able to observe astrophysical neutrinos of all flavours from core-collapse~\cite{Lang:2016zhv}, as well as failed core-collapse and thermonuclear runaway fusion~\cite{Raj:2019sci} supernovae. Typically, about 100 events are expected  from a core-collapse supernova at a distance of 10\,kpc and a 27\,M$_{\odot}$ progenitor mass~\cite{Lang:2016zhv}. The detection of neutrinos from failed core-collapse supernovae would deliver the time when the proto-neutron star collapses into a black hole~\cite{Raj:2019sci} and, in conjunction with the detection of gravitational waves, would identify the progenitor of a failed supernova. Finally, a large xenon detector such as DARWIN would dramatically improve the sensitivity to the diffuse supernova neutrino background (DSNB) in the $\nu_x$ channel, where $\nu_x \subset (\nu_{\mu}, \nu_{\tau}. \bar{\nu}_{\mu}, \bar{\nu}_{\tau})$~\cite{Suliga:2021hek}. While there are strong upper limits on the $\bar{\nu}_{e}$ flux from Super-Kamiokande, of 2.7\,cm$^{-2}$s$^{-1}$, the limits on $\nu_x$ are about three orders of magnitude weaker. DARWIN would be able to reach a sensitivity of $\sim$10\,cm$^{-2}$s$^{-1}$ per flavour. While this is not sufficient for a detection, such a constraint would exclude many DSNB scenarios with new astrophysics or physics~\cite{Suliga:2021hek}. 

\subsubsection{Large-scale cryogenic detectors}

\bigskip
\noindent 
\underline{SuperCDMS SNOLAB}: 
SuperCDMS SNOLAB is a dark matter search focused on the 0.5--5~GeV mass range~\cite{supercdms_sensitivity_2017, supercdms_snowmass_2022}.  It will use two kinds of cryogenic solid-state detectors. The first type, iZIP detectors, will have the capacity to discriminate nuclear recoils from electron recoils via measurement of athermal phonons and ionization production down to 1--2~keV recoil energy.  The second type, HV detectors, will use a high drift field to transduce the charge signal into athermal phonons, provide a recoil energy threshold about 10$\times$ lower, though without the ability to discriminate nuclear recoils.  These thresholds should enable the detection of $^8$B neutrino CEvNS in a solid-state detector for the first time, albeit with low statistics.  The experiment is currently under construction at SNOLAB and anticipates beginning to acquire data in late 2023.



Going forward, the SuperCDMS Collaboration anticipates extending its scientific reach for dark matter with the SNOLAB facility primarily through detector improvements that will provide access to much lower energy recoils~\cite{supercdms_snowmass_2022}.  (Modest background upgrades will also be implemented.)  This work will provide sufficiently low threshold to detect CEvNS of solar neutrinos from the CNO, $pep$, $^7$Be, and even $pp$ reaction chains, going well past the $^8$B neutrinos detectable in SuperCDMS SNOLAB.  In particular, with the 0.5~eV threshold anticipated for 25-gram Si detectors operated with phonon-only readout at 0V bias (i.e., neither iZIP nor HV, and smaller in mass than the kg-scale SuperCDMS SNOLAB detectors), the rate of solar neutrino events will be roughly 0.01/kg-day.  With anticipated exposures of 12--240 kg-yr, there will be the potential to detect tens to hundreds of $pp$ chain CEvNS events.  These events will compete with a background of coherent photonuclear scattering, which can be modeled well based on measurements of Compton scattering at high energies, and with environmental backgrounds such as vibrations, RF noise, infrared and blackbody radiation, etc., which will be explored and better understood during SuperCDMS SNOLAB.

\bigskip
\noindent 
\underline{RES-NOVA}: 
RES-NOVA is a newly proposed experiment for the detection of neutrinos from astrophysical sources~\cite{Pattavina:2020cqc}. RES-NOVA will employ an array of archaeological Pb-based cryogenic detectors sensitive to SN neutrino emission from the entire Milky Way Galaxy. Its modular design will be suited for the detection of nearby SN explosions ($<$3~kpc)~\cite{RES-NOVA:2021gqp}.

\subsubsection{Directional detectors}

\bigskip
\noindent 
\underline{DRIFT}: 
The goal of the Directional Recoil Identification From Tracks (DRIFT) collaboration was the detection of a directional signal from Weakly Interacting Massive Particle (WIMP), halo, dark matter~\cite{PhysRevD.61.101301}. In order to accomplish this goal a unique, low-pressure, Negative Ion Time Projection Chamber (NITPC) technology was developed. The negative-ion drift allowed DRIFT NITPCs to have the lowest energy threshold and best inherent directional sensitivity of any limit-setting, directional dark matter detector. With its unique directional and background rejection capabilities, the DRIFT NITPC technology is ideally suited to search for nuclear recoils in beam dump experiments (BDX-DRIFT). A 1 m$^3$ $\nu$BDX-DRIFT detector run for one year in the DUNE Near Detector Complex is estimated to detect several CE$\nu$NS events. In the near term a 1 m$^3$ $\nu$BDX-DRIFT detector is available to be deployed in the NuMI beam at Fermilab on a year or two timescale \cite{AristizabalSierra:2021uob}. 

\bigskip
\noindent 
\underline{CYGNUS} is a proposed modular and multi-site network of large-scale gas time projection chambers~\cite{Vahsen:2020pzb}. The primary goal of the CYGNUS experiment is to perform a direction-dependent search for dark matter, which has been shown to be one of the only ways to convincingly prove the galactic origin of a detected signal~\cite{Mayet:2016zxu,Vahsen:2021gnb}. Directionality is also the best means of circumventing the neutrino fog~\cite{Grothaus:2014hja,OHare:2015utx}, but requires that good performance can be achieved at the sub-10-keV nuclear recoil energies where the majority of CE$\nu$NS events coming from solar neutrinos would lie. 
A nuclear recoil threshold of 8 keV has already been shown to be feasible in the 755:5 He:SF$_6$ atmospheric pressure gas mixture suggested by Ref.~\cite{Vahsen:2020pzb}. This could be lowered to 3--5 keV with further gas/readout optimisation, and the development of specialized track-fitting techniques to improve particle identification at low energies. This would enable a CYGNUS-1000 m$^3$ detector to see between 30--50 CE$\nu$NS events over a few years.

\cleardoublepage

\section{High- and Ultra-High-Energy Neutrino Detectors}

    Neutrinos at energies of TeV and beyond are interesting in part because they come to us from astrophysical and cosmic sources, such as active galactic nuclei or perhaps even neutron-star mergers.  They are also interesting because as they pass through the Earth their propagation is very sensitive to non-standard interactions.  Precision measurements of standard model neutrino interactions at the TeV scale will provide a new opportunity to search for physics beyond the standard model, while interactions in the PeV-ZeV regime probe nucleon structure in a way that cannot be done at terrestrial accelerators.  The detectors at these ultra-high energies vary widely; neutrino telescopes need very large masses to deal with the rarity of such

\subsection{Advanced High- and Ultra-High-Energy Neutrino Telescopes}

      Neutrino telescopes aimed at looking at neutrinos from TeV scales and beyond have been extremely successful in both uncovering new astrophysical sources of neutrinos, and on searching  for neutrino physics beyond the standard model.   Detection techniques are 
      heavily dependent on neutrino energy.  In the TeV-PeV (``high-energy'') regime, Earth's opacity is high but neutrinos may still travel thousands of kilometers; beyond this regime detected neutrinos are either in regions surrounding large detectors, in dense media such as the lunar regolith, or by looking at air showers created by Earth-skimming $\nu_{\tau}$s.
      
      A detailed SNOWMASS white paper~\cite{uhe_wp} provides a comprehensive view of this very exciting and challenging area.  In keeping with the NF10 scope, we focus here primarily on future ideas that are not already at the technical design stage, with an emphasis on new enabling technologies. Interestingly, in many cases the ``enabling technology'' is a particular piece of geography, whether it is polar ice or a mountain range.
      
\begin{figure*}[h!]
  \centering
  \includegraphics[width=\textwidth]{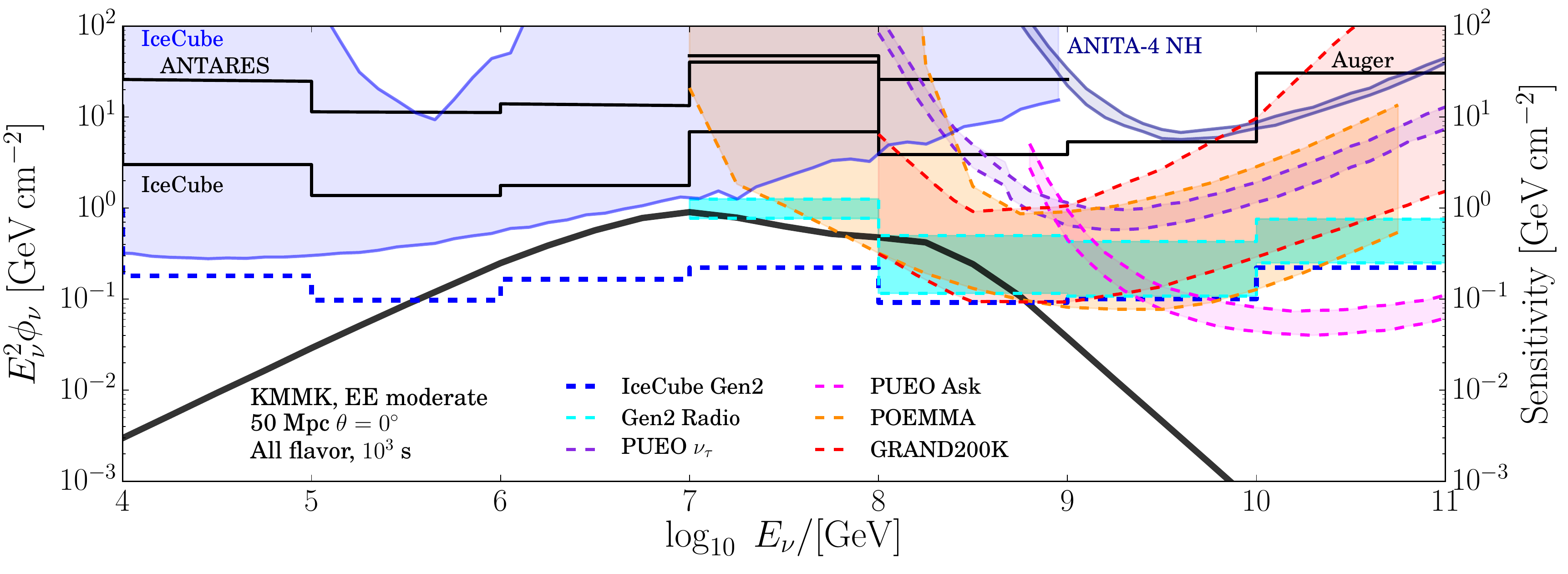}
  \caption{\label{fig:uhe_bursts}
  Sensitivity to short (1000 s) burst all-flavor neutrino plus antineutrino spectral fluence for some current and future detectors. The ANTARES, IceCube and Auger limits are 90\% confidence level limits in a $\pm$ 500 s window around the gravitational wave event from GW170817
  \cite{ANTARES:2017bia}. The dashed blue histogram shows IceCube-Gen2's projected sensitivity for such an event \cite{IceCube-Gen2:2020qha}, including IceCube-Gen2 radio \cite{IceCube-Gen2:2021rkf} (shown for a range of declination with cyan). 
  The blue shaded band comes from IceCube's
   all-sky  point-source  effective  area  values tabulated for 2012 with 86 strings \cite{IceCubePointSource,IceCube:2016tpw}.
  The ANITA-4 NH limit is shown with the solid purple curves, and
  projected all-flavor sensitivities for PUEO \cite{PUEO:2020bnn} from $\nu_\tau$-sourced and Askaryan signals are shown with purple and magenta dashed curves,
  for POEMMA (orange)\cite{Venters:2019xwi} and for GRAND200K (red) \cite{GRAND:2018iaj}. The Kimura et al.~\cite{Kimura:2017kan} extended emission short gamma ray burst fluence for on-axis viewing ($\theta=0^\circ$)\cite{Kimura:2017kan} from 50 Mpc is shown with the solid black curve. Taken from Ref.~\cite{uhe_wp}.
  }
\end{figure*}

\begin{figure}[h!]
    \centering
    \includegraphics[width=\textwidth]{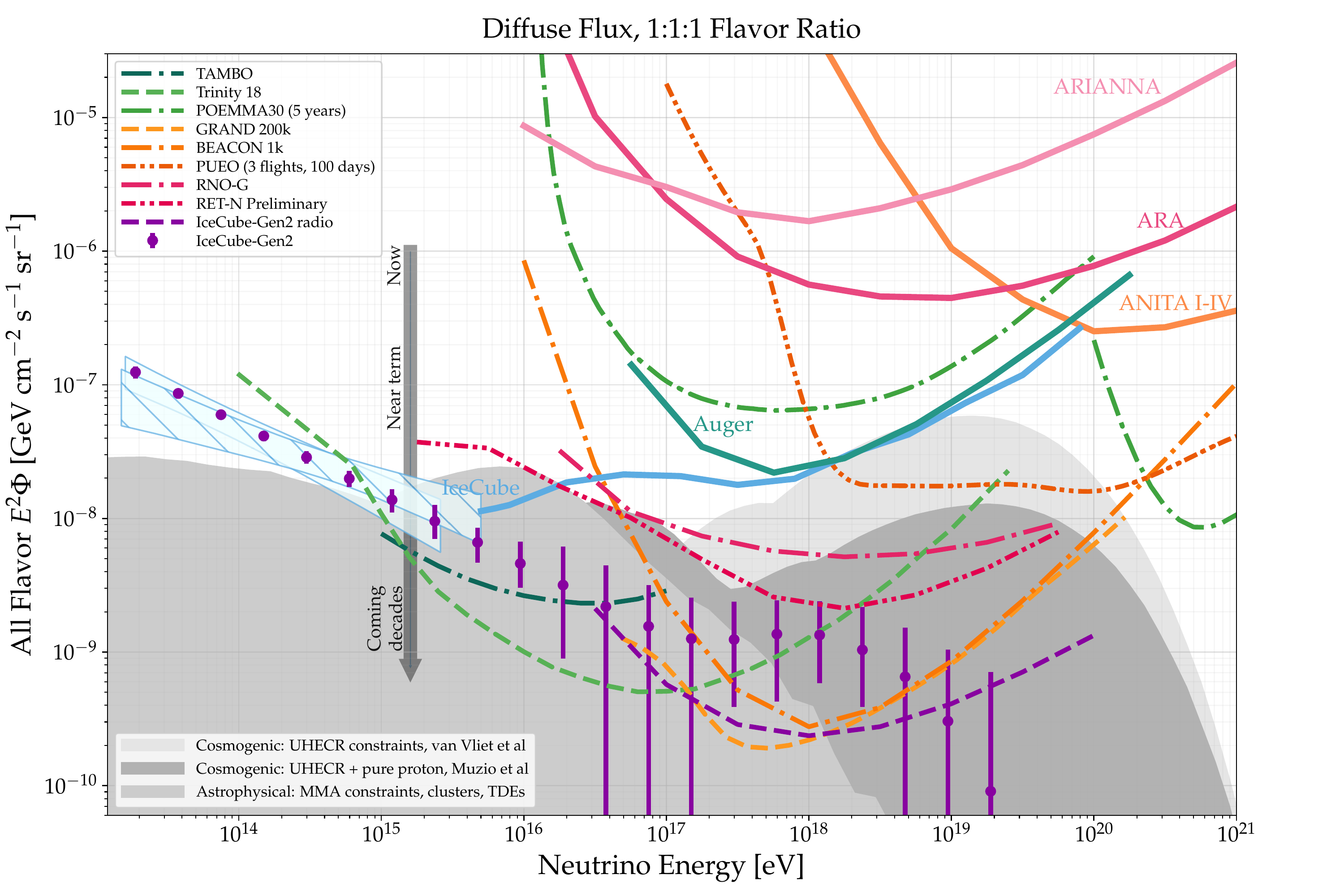}
    \caption{ The expected differential 90\%~C.L. sensitivities for a variety of experiments to an all-flavor diffuse neutrino flux computed in decade-wide energy bins and assuming a ten-yelr integration (unless otherwise noted in the legend). The measurements and sensitivities are compared with a range of cosmogenic neutrino models~\cite{vanVliet:2019nse, Muzio:2021zud} and astrophysical neutrino models~\cite{Muzio:2021zud, Fang:2017zjf, Biehl:2017hnb}.  The blue bordered bands show the astrophysical neutrino flux measured by IceCube using tracks ($\nu_\mu$~\cite{IceCube:2021uhz}) in hatch and using cascade-like events ($\nu_e$ and $\nu_\tau$~\cite{IceCube:2020acn}) in solid band. The solid lines show experimental upper limits at higher energies from the Pierre Auger Observatory~\cite{PierreAuger:2019ens}, ARA~\cite{ARA}, ARIANNA~\cite{Anker:2019rzo}, ANITA I-IV~\cite{Gorham:2019guw}, and IceCube~\cite{IceCube:2016zyt}. The dashed lines show the sensitivities of a selection of proposed experiments currently in various design and prototyping stages (GRAND with 200,000 stations~\cite{GRAND:2018iaj}, BEACON with 1000 stations~\cite{Wissel:2020sec}, TAMBO with 22,000 detectors~\cite{Romero-Wolf:2020pzh}, Trinity with 18 stations~\cite{Otte:2019aaf}, RET-N with 10 stations ~\cite{Prohira:2019glh}, POEMMA30~\cite{POEMMA:2020ykm}) and under construction (RNO-G~\cite{RNO-G:2021hfx}, PUEO~\cite{PUEO:2020bnn}). Experiments using the same detection technique are grouped into similar colors (orange, Earth-skimming radio (GRAND, BEACON); dark teal, particle showers (TAMBO); light green, Earth-skimming optical Cherenkov and fluorescence (Trinity, POEMMA30); pink, in-ice radio (ARIANNA, ARA, PUEO, RNO-G, RET-N (radar); blue, in-ice optical Cherenkov (IceCube)). Sensitivity from IceCube-Gen2 (dashed purple) is computed using radio and PUEO (dashed orange) uses both in-ice and Earth-skimming radio techniques. Auger (teal) uses particle showers and fluorescence and its upgrade, AugerPrime, will employ radio. The expected measurement of the diffuse astrophysical neutrino flux by IceCube-Gen2 in 10 years is shown with the purple points assuming a continuous single-power-law spectrum with an additional cosmogenic flux at the highest energies~\cite{IceCube-Gen2:2020qha}. The grey downward-pointing arrow is a reminder that experimental sensitivities improve not only as exposure increases with time, but also as new experimental techniques and analysis methods are both demonstrated and scaled to larger detection volumes. Taken from Ref.~\cite{uhe_wp}.}
    \label{fig:uhe_diffuse}
\end{figure}

      A summary of sensitivities of some current and proposed experiments to 1000 s bursts and to diffuse high-energy neutrino sources can be found in Figures~~\ref{fig:uhe_bursts} and~\ref{fig:uhe_diffuse}, respectively, taken from Ref.~\cite{uhe_wp}.
      
  Table~\ref{tab:uhe_technologies} summarizes the many ideas and their enabling technologies for high-energy and ultra-high energy natural-source neutrino detection.
\begin{table}[h!]
    \begin{tabular}{|l||c|c|c|}
    \hline 
  Proposed Detector &  Basic Approach  & Energy Range & Enabling Technology  \\ 
  \hline  
    P-ONE           & Water Cherenkov &  $>$ TeV & Scale, \\
                    &                 &   & telecomm fibers   \\ 
    ICECUBE-Gen2 (optical)   &  Ice Cherenkov & TeV-PeV & Scale, \\
                             &                &   &  multi-PMTs    \\ 
    Trinity   &  Air-shower Cherenkov &  10-1000 PeV & $60^\circ$ FOV optics \\ 
    RET   &  Radar reflection off ionization &  $>$ 10 PeV &  In-ice radar reflection \\ 
    TAMBO   & (Air-shower) water Cherenkov tanks &  1-100 PeV &  Mountain/valley geography \\ 
    RNO-G   & Askaryan emission in ice &  $>$ PeV &  Greenland ice, \\
            &                          &           & solar+wind power \\
            &                          &           & autonomous radio detectors \\ 
    ICECUBE-Gen2 (radio)   & Radio array &   PeV-EeV &  Omni-directional \\
                           &             &           &  cylindrical antennas \\ 
    BEACON  & Radio air-shower detection & 100 PeV--EeV  &  Mountain geometry, \\
            &                            &               & interferometric \\
            &                            &               & phased arrays \\ 
    GRAND & Geomagnetic air-shower radio & $>$ PeV & Very large-scale radio array \\
    POEMMA & Space-based optical air-shower & 10 PeV-40 EeV & Wide-FOV \\
           &                                &                &    Schmidt telescopes, \\
           &                                &                & Cherenkov camera \\
    PUEO & Balloon-based radio in ice and air & $>$ EeV & Realtime \\
         &                                    &         &  interfermetric beamforming, \\
         &                                    &          & Xilinx RFSoC \\
    GCOS & Nested water Cherenkov tanks & $>10$ EeV &  TBD \\
         &               +radio         &           &  \\
    \hline                      
    \end{tabular}
     \caption{Proposed detectors and enabling technologies for high-energy and ultra-high-energy neutrino detection. Only those experiments not yet past the technical design phase are included.}
    \label{tab:uhe_technologies}
\end{table}

\subsubsection{Detector Requirements}

\begin{figure*}[h!]
  \centering
  \includegraphics[width=\textwidth]{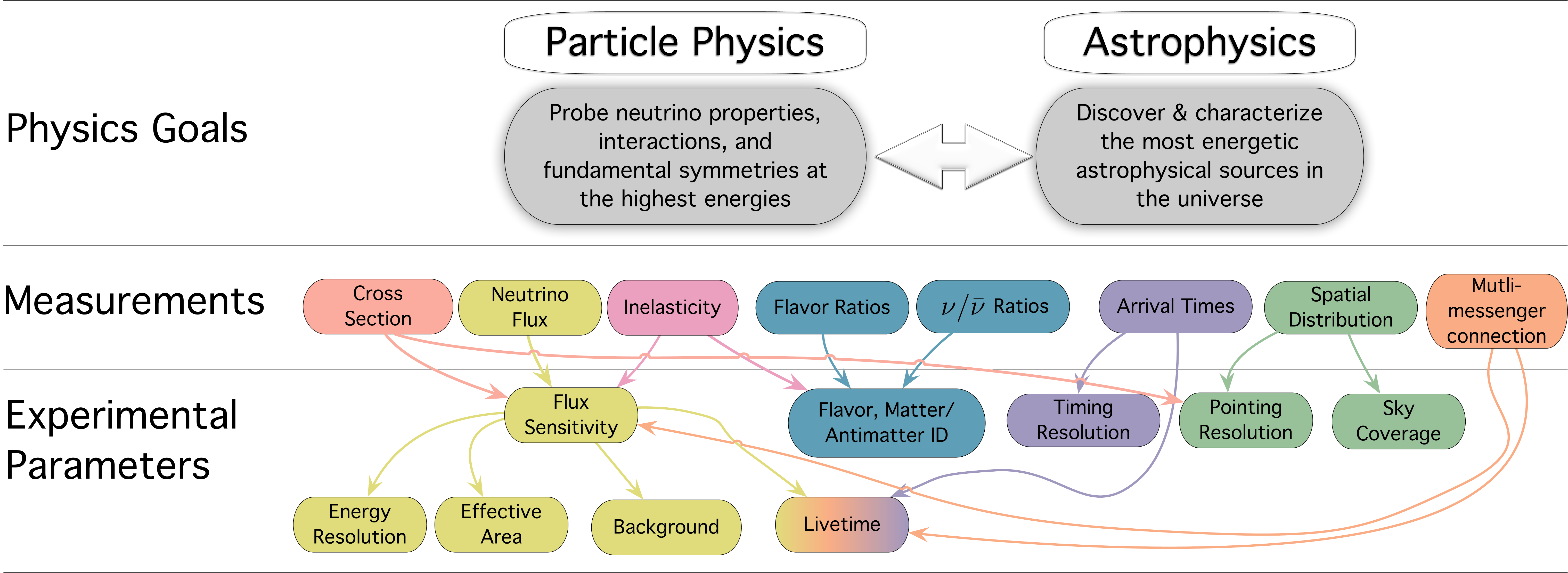}
  \caption{\label{fig:uhe_reqs}\small Flow of several neutrino telescope physics and astrophysica goals into detector requirements. Taken from Ref.~\cite{uhe_wp}.
  }
\end{figure*}

There are four important observables measured by neutrino telescopes to search for new physics and probe astrophysics: energy spectrum, distribution of arrival directions, flavor composition, and arrival times.  
A summary of how detector requirements flow from physics goals can be found in Figure~\ref{fig:uhe_reqs}, taken from Ref.~\cite{uhe_wp}.
Next-generation detectors aim for ever larger sizes to be sensitive to the very small fluxes of possible ultra-high energy sources, with goals of an order of magnitude in flux for HE sources and two orders of magnitude for UHE sources. Energy resolutions of 0.1 in $\log_{10}(E/{\rm GeV})$ at the HE scale, and in half-decade energy binning, are sufficient to resolve bumps and dips that may indicate new physics~\cite{Hooper:2007jr, Ioka:2014kca, Ng:2014pca, Dhuria:2017ihq, Bustamante:2020mep, Creque-Sarbinowski:2020qhz, Esteban:2021tub} while also distinguishing among models of astrophysical neutrino production~\cite{IceCube:2015gsk, IceCube:2020wum}.
In EeV energy scale, next-generation detectors will need tens of events to measure the cross section to within an order of magnitude.

Measuring the arrival times of neutrinos is important for both time-domain transient and multi-messenger astrophysics, but also to search for evidence for new physics that would cause photons, neutrinos, and gravitational waves to arrive at Earth at different times~\cite{Huang:2019etr, Jacob:2006gn, Addazi:2021xuf}. Being able to capture the transient behavior of sources requires the ability for instruments to send and respond to real-time alerts. Continuous operation is ideal for detecting transient events and improving overall flux sensitivity.

Flavor and $\nu/\bar{\nu}$ ratios provide complementary probes of new neutrino physics and neutrino production mechanisms. 
Large event statistics and complementary flavor-specific detection techniques are needed to identify flavor-specific signals and to measure the flavor composition statistically in a sample of collected events. 
Futrure directions should explore new techniques to improve flavor separation, like muon and neutron echoes~\cite{Li:2016kra}.  In the EeV range, some instruments will be sensitive only to certain flavors, while others will be sensitive to all flavors. A comprehensive approach may allow flavor information across multiple experiments to be combined.

Sub-degree pointing resolution is needed to resolve the neutrino sky~\cite{Murase:2016gly,Ahlers:2014ioa,Fang:2016hop,Bartos:2016wud,Bartos:2021tok} while also reducing the systematic uncertainties on cross section~\cite{IceCube:2021jhz,Bustamante:2017xuy,IceCube:2017roe, Denton:2020jft,Connolly:2011vc} and inelasticity measurements~\cite{IceCube:2018pgc}.  Resolving the neutrino sky will be important to search for BSM physics that causes anisotropies. 


\subsubsection{Optical Cherenkov Approaches}
    
  Experiments of this type typically look for tracks of secondary leptons produced by charged-current neutrino interactions or cascades produced by  neutral-current and charged-current neutrino interactions. 
  Track events have excellent angular resolution due to the long lever arm left by a final-state muon track, while cascade events have superior energy resolution. Typically, astrophysical purity increases with energy as the backgrounds from atmospheric neutrinos and muons fall more steeply with energy than the astrophysical neutrino flux.
  At UHE energies, these detectors are also sensitive to neutrinos that cascaded from EeV $\nu_\tau$ to PeV energies~\cite{Safa:2019ege}. 
  To get detectors large enough to see the very small fluxes, either polar ice or sea water typically need to be instrumented.  
Future plans to improve HE sensitivity and broaden energy coverage are to expand detector volumes of optical arrays in the Northern Hemisphere, e.g., KM3NeT\ \cite{KM3Net:2016zxf}, Baikal-GVD\ \cite{Avrorin:2019dli}, and P-ONE\ \cite{P-ONE:2020ljt}.  The primary enabling technology for these experiments is {\it scale} and accessibility of the target volume (e.g., availability of clear water or ice).
  
\subsubsection{Radio Detection in Ice}

  In dense media like ice, compact electromagnetic showers generated after UHE neutrino interactions emit coherent Askaryan radiation at the Cherenkov angle. Askaryan radiation---fast, coherent radio-frequency impulses---is due to the excess negative charge in the shower~\cite{Askaryan:1961pfb}. The long attenuation length at radio frequencies allows the signal to propagate over kilometer-long distances. In-ice radio experiments are sensitive to all three flavors~\cite{Anchordoqui:2019omw,Garcia-Fernandez:2020dhb}, and may have the power to discriminate flavors based on different event topologies~\cite{Garcia-Fernandez:2020dhb,Lai:2013kja}, such as the stretching of electromagnetic showers due to the Landau-Pomeranchuk-Migdal (LPM) effect~\cite{Gerhardt:2010bj}. In ARA \cite{Allison:2014kha} and ARIANNA~\cite{ARIANNA:2019scz, Anker:2020lre}, radio antennas are buried in the Antarctic ice. RNO-G, based on a similar concept, is a radio detector under construction in Greenland~\cite{Aguilar:2019jay}. The experience gained in these experiments will directly feed into the design of the expansive, sparse radio array of IceCube-Gen2~\cite{IceCube-Gen2:2020qha}. Radar signals reflecting off in-ice showers is also being explored as a detection method~ \cite{deVries:2013qwa, Prohira:2017nyr, Prohira:2019glh}.
  
  \subsubsection{Air-shower detection techniques for UHE $\nu_{\tau}$s}
  
 CC interactions of $\nu_\tau$s produce tau leptons that at UHE energies typically decay over tens to hundreds of kilometers. If the geometry of the experimental setting is right, the neutrino interacts inside the Earth and the tau emerges and decays in the atmosphere. The decay initiates an extensive air shower, which can be detected with particle, air-shower imaging, or radio detectors~\cite{Fargion:2000iz}. But even in cases where a $\nu_\tau$ interacts deep underground and the tau decays before reaching the surface, a new $\nu_\tau$ is produced in the decay which can again interact in the Earth and generate a tau emerging from the surface.  This ``$\nu_\tau$ regeneration" increases the chances of Earth-skimming $\nu_\tau$ reaching the detector.  
 Earth-skimming neutrino fluxes have the added advantage of being unaffected by atmospheric neutrinos and only mildly affected by atmospheric muons~\cite{Garcia-Fernandez:2020dhb}.

The Pierre Auger Observatory (Auger) is a long-running large-scale array of surface water tanks that detect the Cherenkov light from air-shower particles passing through them.  Auger is designed to detect UHECRs, but it has been used to search for horizontal showers initiated by UHE neutrinos in the atmosphere~\cite{PierreAuger:2019ens}.  The Telescope Array (TA)~\cite{Abbasi:2019fmh} and  HAWC~\cite{Vargas:2016hcp} experiments have used a similar detection strategy, but have a more limited sensitivity.  TAMBO is a planned array of water tanks to be located on one side of an Andean canyon, designed to detect the showers initiated by UHE taus emerging from the opposite side~\cite{Romero-Wolf:2020pzh}.
  \subsubsection{Air-shower radio detection}  Radio-detection of Earth-skimming tau neutrinos is a promising method due to the long attenuation lengths of radio waves in air. As with in-ice radio detection, sparse arrays can be used to instrument large areas.  Radio emission is generated in air showers initiated by tau decays, via the geomagnetic effect, due to charge separation in the magnetic field of Earth as the air showers progress through the atmosphere. Moreover, the narrow Cherenkov cone and fast radio imaging enables sub-degree angular resolution. BEACON~\cite{Wissel:2020sec}, in its prototype phase, TAROGE~\cite{Nam:2016cib}, and TAROGE-M~\cite{Nam:2020hng} are compact antenna arrays in elevated locations that aim to detect UHE $\nu_\tau$ emerging upwards via the radio emission of the air showers that they trigger.  ANITA~\cite{Deaconu:2019rdx} and PUEO~\cite{Deaconu:2019rdx} are also sensitive to upgoing $\nu_\tau$, from a higher elevation. GRAND~\cite{GRAND:2018iaj} is a planned experiment that will cover large areas with a sparse antenna array to detect the radio emission from air showers triggered by UHE $\nu_\tau$, cosmic rays, and gamma rays.
  \subsubsection{Air-shower imaging} Several air-shower imaging instruments, although optimized for cosmic-ray and gamma-ray detection, have demonstrated that the imaging of air showers via the Cherenkov and fluorescence light radiated by shower particles is a viable detection method of UHE $\nu_\tau$~\cite{Aramo:2004pr,  Gora:2014lya, Gora:2016mmy, Gora:2016mmy, MAGIC:2018gza, Fiorillo:2020xst}. Air-shower imaging allows the reconstruction of the air-shower arrival direction with arcminute resolution and the shower energy within a few tens of percent of uncertainty. These excellent reconstruction characteristics are why the very-high-energy gamma-ray and UHECR communities have been using air-shower imaging for quite some time~\cite{Weekes:1989tc}. Two planned instruments optimized for the detection of UHE neutrinos from the ground are Trinity~\cite{Otte:2018uxj} and Ashra NTA~\cite{Sasaki:2014mwa}.
  POEMMA is designed to detect the Cherenkov light of UHE $\nu_\tau$-initiated showers from a satellite. A unique feature of POEMMA is its ability to rapidly reposition to target transient multi-messenger events~\cite{Venters:2019xwi, POEMMA:2020ykm}. EUSO-SPB2 is a telescope mounted on a super-pressure balloon which will fly at high altitudes and serve as a pathfinder for POEMMA~\cite{Adams:2017fjh}.

\subsection{Detectors for the Forward Physics Facility at the LHC}

The Forward Physics Facility (FPF) will be a new underground cavern at the Large Hadron Collider (LHC) to host a suite of far-forward experiments during the High-Luminosity LHC era. The existing large LHC detectors do not cover particles beyond pseudorapidity range of $|\eta| \leq 4.5$ \cite{Aad:2008zzm},  and so they miss the physics opportunities provided by the enormous flux of particles produced in the far-forward direction.  The FPF will realize this physics potential. In the following we will briefly summarize the physics considerations for FPF detectors, the  technical challenges, and novel approaches to these detectors.  


\begin{figure}[bp]
\centering
\includegraphics[width=0.99\textwidth]{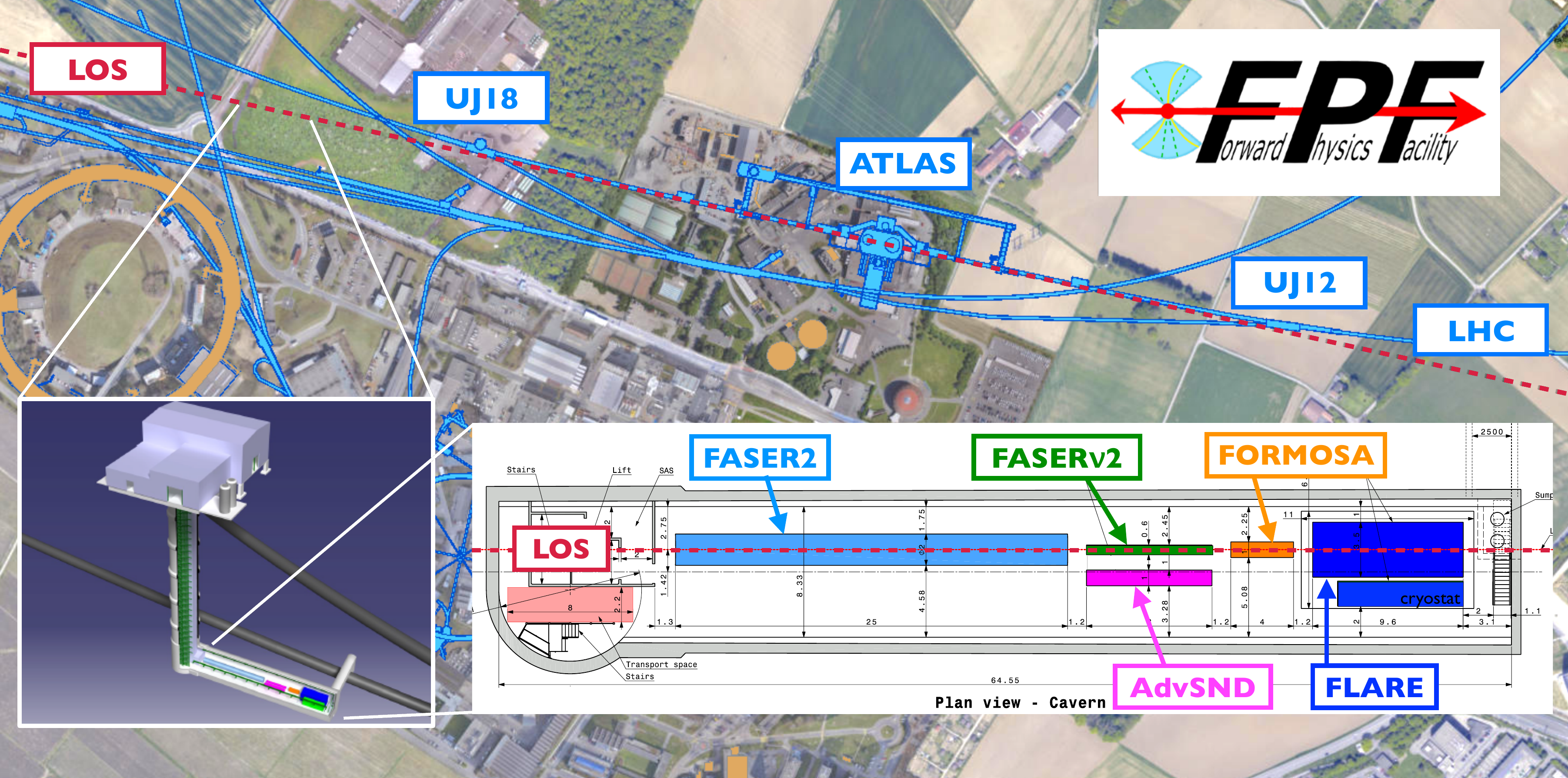}
\caption{The preferred location for the Forward Physics Facility, a proposed new cavern for the High-Luminosity era.  The FPF will be 65~m-long and 8.5~m-wide and will house a diverse set of experiments
along the line of sight (LOS) to explore the many physics opportunities in the far-forward region. 
}
\label{fig:ExecutiveSummaryMap}
\end{figure}

\subsubsection{Detector Requirements}  
The FPF will  need a diverse set of experiments, each optimized for particular physics goals. 
Details of the physics reach can be obtained in  \cite{Anchordoqui:2021ghd, Feng:2022inv}.

FASER2, a magnetic spectrometer and tracker, will search for light and weakly-interacting states, including long-lived particles, new force carriers, axion-like particles, light neutralinos, and dark sector particles.  FASER2 will increase sensitivity by many orders of magnitude beyond the FASER experiment \cite{FASER:2018bac, snomassfaserdoc}  by increasing the acceptance and the length of the magnetic spectrometer.  
FASER$\nu$2 and Advanced SND are  proposed emulsion and electronic neutrino detectors, respectively. They are both in the several ton scale. FLArE is a proposed 10-tonne-scale noble liquid detector.
These detectors will detect neutrinos and also search for light dark matter.  They will each 
detect $\sim 10^6$ neutrinos and anti-neutrinos at energies ranging from 100 GeV to few TeV.  Each is expected to detect  $\sim 10^3$ tau neutrinos also at very high energies. Laboratory experiments have not detected neutrinos in this high energy range in the past. The detection, reconstruction, energy measurement, and particle identification present unique challenges to existing instrumentation.   Detection of tau particles either by explicit identification of decays near the vertex (in emulsion)  or by kinematics  (in electronic detectors)  in a high energy charged current tau neutrino  interaction particularly  presents unique challenges. Collecting a large sample of well identified tau neutrino interactions is a unique capability of the FPF. This capability will be fully utilized by multiple detectors with very different technical approaches. 
Lastly, it is of great interest that these same detectors have low thresholds ($< $ few hundred MeV) for detection of light dark matter scattering. 

Finally, the FORMOSA  detector is  composed of scintillating bars.  These bars are arranged to measure charge deposition from straight-through  millicharged particles and other very weakly-interacting particles across a large range of masses.


\subsection{Technical Considerations} 

The most important technical issue is finding a location along the line of sight (LOS) at an appropriate distance and with features that include access, safety, and the ability to have a large space to house multiple detectors and their infrastructure. During 2021-2022, in a series of workshops \cite{fpfworkshops}  informed  by the CERN civil engineering team and accelerator advisory body, the location 617-682 m west of the ATLAS interaction point (IP) has been selected. This location is about $\sim 80$ m deep and  shielded from the ATLAS IP by over 200 m of concrete and rock, providing an ideal location to search for rare processes and very weakly-interacting particles. A dedicated shaft will provide access to the underground hall, and the construction of both the facility and detectors can be substantially decoupled from HL-LHC construction and activities; this will make planning, scheduling, as well as safety aspects easier for the FPF.  

Despite the shielding from the IP due to LHC magnets, concrete and rock, there is a substantial flux of high energy $\mu^- / \mu^+$ particles directly from the IP as well as produced in the various LHC accelerator structures. Studies are progressing to reduce this flux to $< 1$ Hz/cm$^2$ during the HL-LHC running (with nominal luminosity of $\sim5 \times 10^{34}{\rm /cm^2/sec}$) using sweeping magnets with momentum kick of $\sim 7$ Tesla-meter in a location $\sim 350$ m from the IP.  
Given the many uncertainties including the maximum luminosity in the planned long running of the HL-LHC (until 2040) detectors at the FPF should be  designed to operate with an expected high energy muon flux in the range of $\sim 1$ Hz/cm$^2$. For comparison, the neutrino event rate will be approximately 50 charged current events per day per ton of detector with 45 $\nu_\mu$ CC, 5 $\nu_e$ CC, and $\sim$0.2 $\nu_\tau$ CC. 

\subsubsection{Experimental Technologies} 

The proposed experiments vary in their technological readiness. While they  are conceptually similar to previous efforts, many of them require scaling up of hardware and software  technologies to a new level.  And some will require focused R\&D efforts to tackle specific issues.  Two such efforts will be  the ability to trigger the detectors in the presence of considerable muon backgrounds, and the ability to track and reconstruct neutrino vertices with high multiplicities due to the very high energies of the events. 

{\bf FASER2:} The experiment proposes a large superconducting spectrometer  magnet of   2 to 3 meter diameter and decay volume length of 10-20 meters. The very high energies of expected tracks from  decays of long lived particles implies that for reasonably  large magnetic fields (1 T), the spectrometer will require charged particle resolution exceeding $\sim $1 mm in large area tracking detectors.  The detector will also need to include a fine grained EM calorimeter and  a muon identification system.  A sophisticated trigger system to select LLP decays in the spectrometer volume in the presence of background muons will be needed.   

{\bf FASERnu2:} This will be a $\times$10 scaled up version of the current FASERnu (1.2 ton) detector composed of emulsion and tungsten stacks.  A pilot run of this technology  has already yielded 6 candidate neutrino events from the LHC \cite{FASER:2021mtu}. The scaled up version for the FPF  presents two issues that need to be solved. The HL-LHC presents much high muon rate for this kind of detector which has no trigger capability; the emulsion detectors need to be changed regularly because of the muon radiation. The increased size and frequency of replacement will mean over two orders of magnitude greater production and handling of emulsion films with correspondingly larger analysis effort for FASERnu2.  The experience from FASERnu, however, will prove to be extremely valuable to automate much of this process.

{\bf Electronic Neutrino Detectors:} 

\underline{Advanced-SND} is conceived to be an electronic fine grained 
detector with a magnetic muon spectrometer. The neutrino target is nominally expected to be a few tons.  The technology is still under development with several choices including a detector with silicon tracking detectors sandwiched with passive neutrino target material.  
Such a silicon readout will require further development.  

\underline{FLArE} will be a noble liquid  tracking time projection chamber 
with a hadronic calorimeter to capture  particles escaping from the downstream end, as well as a muon tagger. The default fill is considered to be liquid argon, although liquid krypton is also under consideration.  The nominal size of the detector is approximately 2m  wide, 2m high, and 7m long for the TPC and an additional few meters for the hadronic calorimeter and muon tagging system.  The key technical challenges for the detector are the 
installation of the cryostat and cryogenic systems, the TPC design to obtain the highest spatial resolution for tracks near the vertex, and the scintillation photon system to trigger on neutrino and dark matter events with sufficiently low threshold. This detector benefits enormously from the last decade of R\&D investment into DUNE\cite{DUNE:2018tke} technologies, however the needed tracking resolution and trigger capability is unique for FLArE and will require dedicated R\&D. 

{\bf FORMOSA:} The technologies for FORMOSA are standard with scintillation counters and readout of photo-multiplier pulses, however care has be exercised in limiting backgrounds from muons and instrumental effects.  Much will  be learned from a prototype experiment, expected to be built and run during the upcoming Run-3 of the LHC. 

All detectors will require intelligent flexible trigger systems that can capture interesting events and associate them across detector systems.  Requirements for such triggers to cooperate across detector systems as well as with the collider detector at the IP are in consideration.  

\begin{table}[]
    \begin{tabular}{|l||c|c|c|c|c|}
    \hline  
  Technology      &  ~FASER2~  & ~FASERnu2~  & ~Adv-SND~ & ~FLArE~  & ~FORMOSA~  \\ 
  \hline  
    Large aperture SC magnet             &    x      &         &           &        &  \\ 
    High resolution tracking           &    x      &         &  x         &    x    &  \\ 
    Large scale emulsion            &        & x        &           &        &  \\ 
    Silicon tracking    &  & & x & &  \\ 
    High purity noble liquids   & &    &  & x &  \\        
    Low noise cold electronics &  & & & x &  \\ 
    Scintillation  &  & & & x & x  \\ 
    Optical materials & & & & x &  x  \\ 
    Cold SiPM  &  & &  & x &  \\ 
    Picosec synchronization & & & x & x & x \\ 
    Intelligent Trigger & x & & x& x & x \\ 
    \hline                      

    \end{tabular}
     \caption{Enabling technologies for the detectors and systems of the far forward physics facility.}
    \label{tab:fpf-technologies}
\end{table}

\cleardoublepage

\section{Additional Detector Ideas}

\subsection{Hydrogen and Deuterium Detectors}

Understanding neutrino interactions at the nucleon-level will require a new generation of precision experiments.  These measurements are simplest on the lightest of nuclei---hydrogen and deuterium. Yet the available neutrino data on hydrogen
or light nuclei come from bubble-chamber experiments of the 1970s
and 1980s~\cite{Mann:1973pr,Barish:1977qk,Miller:1982qi,Baker:1981su,Kitagaki:1983px}.   These data have served the community well but they have
essential shortcomings.  They have poor statistical precision, coming
from the relatively low-intensity neutrino beams of an earlier era, and they have
poorly-constrained systematic uncertainties associated with
hand-scanning of events and poorly known fluxes.  Moreover, in most
cases, event-level data have been lost, and information exists only as
one-dimensional projections in publications.  Using the published bubble-chamber
data in subsequent work involves making assumptions about the details of the
analysis, such as whether or not efficiency corrections have been applied to
event rate distributions~\cite{Wilkinson:2014yfa}.

While these experiments were pioneering in their age, and probed
qualitative features of neutrino interactions that helped establish
our modern Standard Model (SM) of the strong and electroweak forces, 
they were not designed to underpin the ambitious neutrino oscillation
experiments of the current precision era.
It is clear that new and better data are needed.  

Detector options to make these measurements include using DUNE's SAND detector's hydrocarbon (CH$_2$)$^n$ radiators as a source of hydrogen.  Pure carbon radiators and precise primary vertex location will allow the separate measurement of neutrino interactions on the hydrocarbon and carbon.  Subtraction of the carbon measurements from the hydrocarbon measurements in the same beam yields estimates of the interaction rates on pure hydrogen.  The use of transverse kinematic imbalance (TKI) variables further improves the signal-to-background ratio before the subtraction and allows for sideband control of important backgrounds.  Unfortunately, the measurements cannot be extended to deuterium in this way as deuterated plastics are prohibitively expensive.

Another option is to include a hydrogen-rich gas in the ND-GAr subdetector. ND-GAr, with its full solid-angle acceptance and MeV-level proton tracking threshold, is ideal for performing exclusive final-state measurements. 
One unique advantage of the ND-GAr, compared to other DUNE ND components, is its flexibility to
use various gas mixtures as interaction targets. Possible hydrogen-rich mixtures include binary systems such as He-Alkane. For example, in terms of the ratio between the numbers of neutrino-interacting free protons and bound protons, a mixture of 90\% He + 10\% CH$_{4}$ equals to polystyrene (CH), and  75\% He + 25\% CH$_4$ equals to polypropylene (CH$_2$); at 50\%, CH$_{4}$ provides the same target mass as pure hydrogen gas (H$_2$)---higher concentration will not only further improve the event purity, but also increase the free-proton target mass.

With the help of the superb tracking of ND-GAr, a successful search for a safe and hydrogen-rich gas mixture would enable measurements of
event-by-event neutrino-hydrogen interactions~\cite{Lu:2015hea,Hamacher-Baumann:2020ogq}. The idea was that beamed neutrinos  interacting on hydrogen could be selected out of those on the heavier targets via Transverse Kinematic Imbalance (TKI) once sufficient momentum resolution is achieved: with perfect tracking, interactions on hydrogen would have balanced final-state transverse momenta (that is, zero TKI), while the TKI on heavier nuclei is irreducibly wide due to nuclear effects such as Fermi motion and final-state interactions. 

A separate hall can be constructed upstream of the DUNE Near Detector hall to house  large, hydrogen or deuterium target detectors.
Two choices exist for an active-target liquid hydrogen detector -- a TPC that collects drifting electrons from ionized hydrogen molecules and a liquid hydrogen bubble chamber that is read out optically on each spill.  Both a TPC and bubble chamber would need auxiliary devices and detectors to perform high-precision cross section studies.
Both detectors would require a strong magnet for charge separation as well as the measurement of charged particle momentum.
Depending on their size and shape, both detectors could contribute complementary calorimetric measurements for short-lived particles, but heavier particles and those with higher lifetimes would require calorimetry and a muon chamber.

 A liquid hydrogen TPC presents several challenges in design, construction and operation.  The very slow electron drift velocity in liquid hydrogen~\cite{SAKAI198289,HARRISON1971418} presents pileup and background issues, which will need to be evaluated in the context of detector design.   Furthermore, extremely long free-electron lifetimes (of order seconds) will be required in order to collect charge after a long drift distance.  A time-projection chamber can be built with crossed wire planes similar to the single-phase liquid-argon TPC design of the DUNE Far Detector, or it could have pixel readout, similar to the LArTPC near detector design. This would yield a resolution of around 4~mm \cite{DUNE:2021tad}. The presence of high-voltage electrodes in contact with hydrogen will require careful attention to safety, such as the removal of all oxidizers in the hall.

While bubble chambers have many similar safety concerns, they offer several advantages over a similarly-sized hydrogen TPC and have a strong history of successful use in particle physics.
Spatial resolutions of about 100 microns and momentum resolutions of 2\% were standard for historic devices using traditional cameras and relatively weak magnets \cite{Bradner:1960iok}. 
Also, the use of special holographic optical techniques can reduce spatial resolutions down to less than 10 microns \cite{Dykes:1981dvb}. 

The prospect of measuring neutrino cross sections on a polarized target is very exciting.  It has never been done before, and the inherent polarization of neutrinos in the Standard Model and the parity-violating nature of the weak interactions result in very large polarization asymmetries.  A polarized target would have to be scaled up in size relative to existing ones in order to provide event rates high enough to make useful measurements.  A polarized target would require very low temperatures, of order 1~K, a strong magnetic field, and RF applied so that dynamical nuclear polarization can be used.
The very intense LBNF neutrino beam provides opportunities for even small targets to produce enough data over time to make an impact in this previously unexplored domain.

The requirement for such detectors of having both high
nuclear target polarization and high heat load simultaneously is achievable through dynamic nuclear polarization (DNP) of solid-state targets.  These systems contain a high-powered microwave generator, a superconducting magnet ($\sim$5T) to produce the Zeeman splitting, and a high-cooling-power evaporation refrigerator used to hold the target at 1~K despite the heat load from the beam, the microwaves, the integrated particle detection, and external heat leaks~\cite{Crabb:1997cy,Dael:1992xh}. The technology of these targets is quite mature and it is now possible to polarize nucleons and other nuclei in any orientation required for beams of photons, muons, electrons, protons, and mesons.

Due to the negligible heat load from high-intensity neutrino beams, either an evaporation or a frozen spin target could be used.  Both of these systems can also be scaled up longitudinally along the beam line to increase the exposure.  There are, however, limits to the width of the target as both require a strong polarizing magnetic field with high field homogeneity.  It is also possible to increase the target's overall diameter as well but the limiting factor is the size of bore of the magnet~\cite{Berryhill:2019gan} and the expense of building a polarizing magnet on the desired scale (more than a meter).  Polarizing solenoids can be built to have much larger warm bores but to keep the desired homogeneity usually the field strength will need to decrease to keep the magnet construction expenses reasonable. Using a 1.5~T magnet to polarize materials such as NH$_3$ is still practical assuming that the nuclei are polarized at a much lower temperature. The DNP process will then take significantly longer to reach optimal polarization enhancement.

The detector must be integrated with the target due to the low energies of the outgoing production particles and the very small interaction cross section of neutrinos on nuclei.  Historically this type of constraint has been addressed using active targets where the target material also acts as a detector of the low-energy decay products.  For nitroxyl radicals such as TEMPO, DTBN and oxo-TEMPO, the unpaired electron is localized predominately in the N-O bond and is surrounded and shielded by four methyl groups \cite{BUNYATOVA200422}.  This molecular structure allows the dopant to be chemically mixed without losing the free electron.  These free radicals can be combined with materials like polymethyl methacrylate (PMMA) and polystyrene which can also be used as scintillating detectors.  Such approaches provide active targets for experiments that require a polarized target combined with particle detection very close to the polarized target nucleus \cite{BUNYATOVA200422,vandenBrandt:2002ab,vandenBrandt:2002rs}.

An initial design consists of a scintillator mesh with 1~mm strips running both longitudinally and transversely to provide pixel and cluster reconstruction information.  These strips are connected to wavelength shifting fiber inside the mixing chamber volume of a large scale $^3$He/$^4$He dilution refrigerator. In the insulating vacuum space, these fibers interface and optically connect to a set of Multi-Pixel Photon Counter (MPPC), also known as silicon photomultiplier (SiPM), a solid state photomultiplier comprised of a high density matrix of Geiger-mode-operated avalanche photodiodes.  Routing the optical fibers outside of the mK volume distances the heat source of the SiPMs and associated electronics from the cold target.

The suggested system could be built to house a target with diameter of 25~cm with a length of several meters.  The system could be polarized with a long solenoid magnet (2.5~T) that could move back and forth to polarize different sections and the two different spin states of the long sections of target.  A separate large Helmholtz holding coil (0.3~T) could be positioned outside of the cryostat on the section not being actively polarized with the solenoid. This holding field would preserve the polarization during the frozen spin state.  In this system, the dilution refrigerator is static but the magnets can move on a track in the longitudinal direction to polarize different sections of the target.



\textbf{Acknowledgements}

We thank the editors of the Snowmass Whitepaper 'Coherent elastic neutrino-nucleus scattering: Terrestrial and astrophysical applications', L. Strigari, P. Barbeau and R. Strauss for important input to this report. We also acknowledge the many authors of the Snowmass Whitepaper, ``Future Advances in Photon-based Neutrino Detectors,'' along with the many participants in the Photon-based Neutrino Detector Workshop; we have relied heavily in several sections on these.  We also thank the many authors of the Snowmass Whitepaper, ``High-Energy and Ultra-High Energy Neutrinos,'' which we have used as the basis for many of the relevant sections in this report.  We also thank the authors of the Snowmass Whitepaper, ``Neutrino Scattering Measurements on Hydrogen and Deuterium: A Snowmass Whitepaper.'' We would also like to thank Diego Garcia Gamez, Ettore Segreto, and Andrzej Szelc for very helpful input.
\cleardoublepage

\bibliographystyle{hieeetr}
\bibliography{refs}

\begin{thebibliography}{100}

\bibitem{darkside}
P.~Agnes {\em et~al.}, ``{Low-Mass Dark Matter Search with the DarkSide-50
  Experiment},'' {\em Phys. Rev. Lett.}, vol.~121, no.~8, p.~081307, 2018.

\bibitem{DEAP}
DEAP-3600 Collaboration, P.~A. Amaudruz {\em et~al.}, ``{Design and
  Construction of the DEAP-3600 Dark Matter Detector},'' {\em Astropart.
  Phys.}, vol.~108, pp.~1--23, 2019.

\bibitem{LZ}
LZ Collaboration, D.~S. Akerib {\em et~al.}, ``{LUX-ZEPLIN (LZ) Conceptual
  Design Report},'' 9 2015.

\bibitem{EXO-200}
EXO-200 Collaboration, M.~Auger {\em et~al.}, ``{Search for Neutrinoless
  Double-Beta Decay in $^{136}$Xe with EXO-200},'' {\em Phys. Rev. Lett.},
  vol.~109, p.~032505, 2012.

\bibitem{MicroBooNE:2015bmn}
MicroBooNE, LAr1-ND, ICARUS-WA104 Collaboration, M.~Antonello {\em et~al.},
  ``{A Proposal for a Three Detector Short-Baseline Neutrino Oscillation
  Program in the Fermilab Booster Neutrino Beam},'' arXiv:1503.01520
  [physics.ins-det].

\bibitem{DUNE:2020lwj}
DUNE Collaboration, B.~Abi {\em et~al.}, ``{Deep Underground Neutrino
  Experiment (DUNE), Far Detector Technical Design Report, Volume I
  Introduction to DUNE},'' {\em JINST}, vol.~15, no.~08, p.~T08008,
  arXiv:2002.02967 [physics.ins-det].

\bibitem{nEXO:2017nam}
nEXO Collaboration, J.~B. Albert {\em et~al.}, ``{Sensitivity and Discovery
  Potential of nEXO to Neutrinoless Double Beta Decay},'' {\em Phys. Rev. C},
  vol.~97, no.~6, p.~065503, 2018.

\bibitem{verticaldrift:2020loi}
{X. Qian and others}, ``Development of lartpc vertical drift solutions with pcb
  anode readouts for dune,'' 2020.
\newblock {Snowmass 2022 Letter of Interest}.

\bibitem{Dwyer_2018}
D.~Dwyer, M.~Garcia-Sciveres, D.~Gnani, C.~Grace, S.~Kohn, M.~Kramer,
  A.~Krieger, C.~Lin, K.~Luk, P.~Madigan, C.~Marshall, H.~Steiner, and
  T.~Stezelberger, ``{LArPix}: demonstration of low-power 3d pixelated charge
  readout for liquid argon time projection chambers,'' {\em Journal of
  Instrumentation}, vol.~13, pp.~P10007--P10007, oct 2018.

\bibitem{Nygren:2018rbl}
D.~Nygren and Y.~Mei, ``{Q-Pix: Pixel-scale Signal Capture for Kiloton Liquid
  Argon TPC Detectors: Time-to-Charge Waveform Capture, Local Clocks, Dynamic
  Networks},'' arXiv:1809.10213 [physics.ins-det].

\bibitem{Adams_2020}
C.~Adams, M.~D. Tutto, J.~Asaadi, M.~Bernstein, E.~Church, R.~Guenette,
  J.~Rojas, H.~Sullivan, and A.~Tripathi, ``Enhancing neutrino event
  reconstruction with pixel-based 3d readout for liquid argon time projection
  chambers,'' {\em Journal of Instrumentation}, vol.~15, pp.~P04009--P04009,
  apr 2020.

\bibitem{Q-Pix:2022zjm}
Q-Pix Collaboration, S.~Kubota {\em et~al.}, ``{Enhanced Low-Energy Supernova
  Burst Detection in Large Liquid Argon Time Projection Chambers Enabled by
  Q-Pix},'' arXiv:2203.12109 [hep-ex].

\bibitem{pixels:2020loi}
{D. Dwyer and others}, ``An r\&d collaboration for scalable pixelated detector
  systems,'' 2020.
\newblock {Snowmass 2022 Letter of Interest}.

\bibitem{multimodalpixels:2020loi}
{E. Gramellini and others}, ``Multi-modal pixels for noble element time
  projection chambers,'' 2020.
\newblock {Snowmass 2022 Letter of Interest}.

\bibitem{dual-readout:2020loi}
{J. Asaadi, E. Gramellini, B. Jones, K. Kelly, P. Machado}, ``Dual-readout time
  projection chamber: exploring sub-millimeter pitch for directional dark
  matter and tau identification in $\nu_\tau$ cc interactions,'' 2020.
\newblock {Snowmass 2022 Letter of Interest}.

\bibitem{nesteres}
M.~Szydagis {\em et~al.}, ``{A Review of Basic Energy Reconstruction Techniques
  in Liquid Xenon and Argon Detectors for Dark Matter and Neutrino Physics
  Using NEST},'' {\em Instruments}, vol.~5, p.~13, arXiv:2102.10209 [hep-ex].

\bibitem{Mastbaum:2022rhw}
A.~Mastbaum, F.~Psihas, and J.~Zennamo, ``{Xenon-Doped Liquid Argon TPCs as a
  Neutrinoless Double Beta Decay Platform},'' arXiv:2203.14700
  [physics.ins-det].

\bibitem{Caratelli:2022llt}
D.~Caratelli {\em et~al.}, ``{Low-Energy Physics in Neutrino LArTPCs},''
  arXiv:2203.00740 [physics.ins-det].

\bibitem{psdopants}
D.~F. Anderson, ``{New Photosensitive Dopants for Liquid Argon},'' {\em Nucl.
  Instrum. Meth. A}, vol.~245, p.~361, 1986.

\bibitem{icDope}
P.~Cennini {\em et~al.}, ``Improving the performance of the liquid argon tpc by
  doping with tetra-methyl-germanium,'' {\em Nuclear Instruments and Methods in
  Physics Research Section A: Accelerators, Spectrometers, Detectors and
  Associated Equipment}, vol.~355, no.~2, pp.~660 -- 662, 1995.

\bibitem{larsoft}
E.~D. Church, ``{LArSoft: A Software Package for Liquid Argon Time Projection
  Drift Chambers},'' arXiv:1311.6774 [physics.ins-det].

\bibitem{Nakajima:2015meb}
Y.~Nakajima, A.~Goldshmidt, H.~S. Matis, T.~Miller, D.~R. Nygren, C.~A.~B.
  Oliveira, and J.~Renner, ``{Measurement of scintillation and ionization yield
  with high-pressure gaseous mixtures of Xe and TMA for improved neutrinoless
  double beta decay and dark matter searches},'' {\em JINST}, vol.~11, no.~03,
  p.~C03041, arXiv:1511.02257 [physics.ins-det].

\bibitem{Hollywood_2020}
D.~Hollywood, K.~Majumdar, K.~Mavrokoridis, K.~McCormick, B.~Philippou,
  S.~Powell, A.~Roberts, N.~Smith, G.~Stavrakis, C.~Touramanis, and J.~Vann,
  ``{ARIADNE}{\textemdash}a novel optical {LArTPC}: technical design report and
  initial characterisation using a secondary beam from the {CERN} {PS} and
  cosmic muons,'' {\em Journal of Instrumentation}, vol.~15,
  pp.~P03003--P03003, mar 2020.

\bibitem{Roberts_2019}
A.~Roberts, P.~Svihra, A.~Al-Refaie, H.~Graafsma, J.~Küpper, K.~Majumdar,
  K.~Mavrokoridis, A.~Nomerotski, D.~Pennicard, B.~Philippou, S.~Trippel,
  C.~Touramanis, and J.~Vann, ``First demonstration of 3d optical readout of a
  {TPC} using a single photon sensitive timepix3 based camera,'' {\em Journal
  of Instrumentation}, vol.~14, pp.~P06001--P06001, jun 2019.

\bibitem{instruments4040035}
A.~Lowe, K.~Majumdar, K.~Mavrokoridis, B.~Philippou, A.~Roberts, C.~Touramanis,
  and J.~Vann, ``Optical readout of the ariadne lartpc using a timepix3-based
  camera,'' {\em Instruments}, vol.~4, no.~4, 2020.

\bibitem{ariadne-loi}
{P. Amedo and others}, ``Letter of intent: Large-scale demonstration of the
  ariadne lartpc optical readout system at the cern neutrino platform,'' 2020.
\newblock {CERN Letter of Interest}.

\bibitem{app11209450}
A.~Lowe, K.~Majumdar, K.~Mavrokoridis, B.~Philippou, A.~Roberts, and
  C.~Touramanis, ``A novel manufacturing process for glass thgems and first
  characterisation in an optical gaseous argon tpc,'' {\em Applied Sciences},
  vol.~11, no.~20, 2021.

\bibitem{Doke:1981eac}
T.~Doke, ``{Fundamental Properties of Liquid Argon, Krypton and Xenon as
  Radiation Detector Media},'' {\em Portugal. Phys.}, vol.~12, pp.~9--48, 1981.

\bibitem{Machado_2016}
A.~Machado and E.~Segreto, ``{ARAPUCA} a new device for liquid argon
  scintillation light detection,'' {\em Journal of Instrumentation}, vol.~11,
  pp.~C02004--C02004, feb 2016.

\bibitem{prtdn_2020}
D.~Collaboration, ``First results on {ProtoDUNE}-{SP} liquid argon time
  projection chamber performance from a beam test at the {CERN} neutrino
  platform,'' {\em Journal of Instrumentation}, vol.~15, pp.~P12004--P12004,
  dec 2020.

\bibitem{Segreto_2018}
 vol.~13, pp.~C04026--C04026, apr 2018.

\bibitem{Souza_2021}
H.~Souza, E.~Segreto, A.~Machado, R.~Sarmento, M.~Bazetto, L.~Paulucci,
  F.~Marinho, V.~Pimentel, F.~Demolin, G.~de~Souza, A.~Fauth, and
  M.~Ayala-Torres, ``Liquid argon characterization of the x-{ARAPUCA} with
  alpha particles, gamma rays and cosmic muons,'' {\em Journal of
  Instrumentation}, vol.~16, p.~P11002, nov 2021.

\bibitem{Brizzolari_2021}
C.~Brizzolari, S.~Brovelli, F.~Bruni, P.~Carniti, C.~Cattadori, A.~Falcone,
  C.~Gotti, A.~Machado, F.~Meinardi, G.~Pessina, E.~Segreto, H.~Souza,
  M.~Spanu, F.~Terranova, and M.~Torti, ``Enhancement of the x-arapuca photon
  detection device for the {DUNE} experiment,'' {\em Journal of
  Instrumentation}, vol.~16, p.~P09027, sep 2021.

\bibitem{pdunexe}
DUNE Collaboration, N.~Gallice, ``{Xenon doping of liquid argon in ProtoDUNE
  single phase},'' {\em JINST}, vol.~17, no.~01, p.~C01034, arXiv:2111.00347
  [physics.ins-det].

\bibitem{uB}
M.~Collaboration, ``Design and construction of the {MicroBooNE} detector,''
  {\em Journal of Instrumentation}, vol.~12, pp.~P02017--P02017, feb 2017.

\bibitem{pdsp}
D.~Collaboration, ``Design, construction and operation of the {ProtoDUNE}-{SP}
  liquid argon {TPC},'' {\em Journal of Instrumentation}, vol.~17, p.~P01005,
  jan 2022.

\bibitem{recomb}
I.~Collaboration, ``Study of electron recombination in liquid argon with the
  icarus tpc,'' {\em Nuclear Instruments and Methods in Physics Research
  Section A: Accelerators, Spectrometers, Detectors and Associated Equipment},
  vol.~523, no.~3, pp.~275--286, 2004.

\bibitem{uBpi0}
M.~Collaboration, ``Reconstruction and measurement of o(100) {MeV} energy
  electromagnetic activity from $\pi^0\rightarrow \gamma\gamma$ decays in the
  {MicroBooNE} {LArTPC},'' {\em Journal of Instrumentation}, vol.~15,
  pp.~P02007--P02007, feb 2020.

\bibitem{uBSCE}
M.~Collaboration, ``Study of space-charge effects in microboone,'' {\em Public
  Note 1018}, 2016.

\bibitem{pdresults}
D.~Collaboration, ``First results on {ProtoDUNE}-{SP} liquid argon time
  projection chamber performance from a beam test at the {CERN} neutrino
  platform,'' {\em Journal of Instrumentation}, vol.~15, pp.~P12004--P12004,
  dec 2020.

\bibitem{tdr4}
D.~Collaboration, ``Volume {IV}. the {DUNE} far detector single-phase
  technology,'' {\em Journal of Instrumentation}, vol.~15, pp.~T08010--T08010,
  aug 2020.

\bibitem{anderhub}
H.~Anderhub, M.~Devereux, and P.-G. Seiler, ``On a new method for testing and
  calibrating ionizing particle detectors,'' {\em Nuclear Instruments and
  Methods}, vol.~166, no.~3, pp.~581--582, 1979.

\bibitem{cao}
J.~Sun, D.~Cao, and J.~Dimmock, ``Investigating laser-induced ionization of
  purified liquid argon in a time projection chamber,'' {\em Nuclear
  Instruments and Methods in Physics Research Section A: Accelerators,
  Spectrometers, Detectors and Associated Equipment}, vol.~370, no.~2,
  pp.~372--376, 1996.

\bibitem{badrhees}
I.~Badhrees, A.~Ereditato, I.~Kreslo, M.~Messina, U.~Moser, B.~Rossi, M.~S.
  Weber, M.~Zeller, C.~Altucci, S.~Amoruso, R.~Bruzzese, and R.~Velotta,
  ``Measurement of the two-photon absorption cross-section of liquid argon with
  a time projection chamber,'' {\em New Journal of Physics}, vol.~12,
  p.~113024, nov 2010.

\bibitem{rossi}
B.~Rossi, I.~Badhress, A.~Ereditato, S.~Haug, R.~Hänni, M.~Hess,
  S.~Jano{\^{s}}, F.~Juget, I.~Kreslo, S.~Lehmann, P.~Lutz, R.~Mathieu,
  M.~Messina, U.~Moser, F.~Nydegger, H.~U. Schütz, M.~S. Weber, and M.~Zeller,
  ``A prototype liquid argon time projection chamber for the study of {UV}
  laser multi-photonic ionization,'' {\em Journal of Instrumentation}, vol.~4,
  pp.~P07011--P07011, jul 2009.

\bibitem{ereditato2014}
A.~Ereditato, I.~Kreslo, M.~Lüthi, C.~R. von Rohr, M.~Schenk, T.~Strauss,
  M.~Weber, and M.~Zeller, ``A steerable {UV} laser system for the calibration
  of liquid argon time projection chambers,'' {\em Journal of Instrumentation},
  vol.~9, pp.~T11007--T11007, nov 2014.

\bibitem{uBlaser}
M.~Collaboration, ``A method to determine the electric field of liquid argon
  time projection chambers using a {UV} laser system and its application in
  {MicroBooNE},'' {\em Journal of Instrumentation}, vol.~15,
  pp.~P07010--P07010, jul 2020.

\bibitem{maneira}
J.~Maneira, ``Techniques for tpc calibration: Application to liquid ar-tpcs,''
  {\em Particles}, vol.~5, no.~1, pp.~74--83, 2022.

\bibitem{artie}
ARTIE Collaboration Collaboration, V.~Fischer, L.~Pagani, L.~Pickard,
  A.~Couture, S.~Gardiner, C.~Grant, J.~He, T.~Johnson, E.~Pantic, C.~Prokop,
  R.~Svoboda, J.~Ullmann, and J.~Wang, ``Artie final results,'' {\em Oral
  communication at the PANIC2021 conference}, 2021.

\bibitem{aced}
ACED Collaboration Collaboration, V.~Fischer, L.~Pagani, L.~Pickard,
  A.~Couture, S.~Gardiner, C.~Grant, J.~He, T.~Johnson, E.~Pantic, C.~Prokop,
  R.~Svoboda, J.~Ullmann, and J.~Wang, ``Measurement of the neutron capture
  cross section on argon,'' {\em Phys. Rev. D}, vol.~99, p.~103021, May 2019.

\bibitem{ref:DUNEwp}
A.~A. Abud {\em et~al.}, ``{Snowmass Neutrino Frontier: DUNE Physics
  Summary},'' in {\em {2022 Snowmass Summer Study}}, 3 2022, 2203.06100.

\bibitem{ref:ndgarwp}
A.~A. Abud {\em et~al.}, ``{A Gaseous Argon-Based Near Detector to Enhance the
  Physics Capabilities of DUNE},'' in {\em {2022 Snowmass Summer Study}}, 3
  2022, 2203.06281.

\bibitem{ref:dune_cdr}
DUNE Collaboration, A.~Abed~Abud {\em et~al.}, ``{Deep Underground Neutrino
  Experiment (DUNE) Near Detector Conceptual Design Report},'' {\em
  Instruments}, vol.~5, no.~4, p.~31, arXiv:2103.13910 [physics.ins-det].

\bibitem{https://doi.org/10.48550/arxiv.2203.09734}
H.~O. Back, W.~Bonivento, M.~Boulay, E.~Church, S.~R. Elliott, F.~Gabriele,
  C.~Galbiati, G.~K. Giovanetti, C.~Jackson, A.~McDonald, A.~Renshaw,
  R.~Santorelli, K.~Scholberg, M.~Simeone, R.~Tayloe, and R.~Van~de Water, ``{A
  Facility for Low-Radioactivity Underground Argon},'' 2022.
\newblock https://arxiv.org/abs/2203.09734.

\bibitem{https://doi.org/10.48550/arxiv.1901.10108}
T.~Alexander, H.~O. Back, W.~Bonivento, M.~Boulay, P.~Collon, Z.~Feng, M.~Foxe,
  P.~G. Abia, P.~Giampa, C.~Jackson, C.~Johnson, E.~Mace, P.~Mueller,
  L.~Palcsu, W.~Pettus, R.~Purtschert, A.~Renshaw, R.~Saldanha, K.~Scholberg,
  M.~Simeone, O.~Šrámek, R.~Tayloe, W.~TeGrotenhuis, S.~White, and
  R.~Williams, ``The low-radioactivity underground argon workshop: A workshop
  synopsis,'' 2019.

\bibitem{PEURRUNG1997425}
A.~Peurrung, T.~Bowyer, R.~Craig, and P.~Reeder, ``Expected atmospheric
  concentration of 42ar,'' {\em Nuclear Instruments and Methods in Physics
  Research Section A: Accelerators, Spectrometers, Detectors and Associated
  Equipment}, vol.~396, no.~3, pp.~425--426, 1997.

\bibitem{ASHITKOV1998179}
V.~Ashitkov, A.~Barabash, S.~Belogurov, G.~Carugno, S.~Konovalov, I.~Pilugin,
  G.~Puglierin, R.~Saakyan, V.~Stekhanov, V.~Umatov, and I.~Vanushin, ``New
  experimental limit on the 42ar content in the earth’s atmosphere,'' {\em
  Nuclear Instruments and Methods in Physics Research Section A: Accelerators,
  Spectrometers, Detectors and Associated Equipment}, vol.~416, no.~1,
  pp.~179--181, 1998.

\bibitem{SagarLRT2022}
S.~S. Poudel, ``{Low radioactivity argon for rare event searches},'' Presented
  at Workshop on Low Radioactivity Technique (LRT 2022) (AAP 2018), 2022.

\bibitem{Zhang:2022dlg}
C.~Zhang and D.~Mei, ``{Evaluation of cosmogenic production of ~39Ar and ~42Ar
  for rare-event physics using underground argon},'' {\em Astropart. Phys.},
  vol.~142, p.~102733, arXiv:2202.06403 [physics.ins-det].

\bibitem{Para:2022gju}
S.~Para {\em et~al.}, ``{SoLAr: Solar Neutrinos in Liquid Argon},'' in {\em
  {2022 Snowmass Summer Study}}, 3 2022, 2203.07501.

\bibitem{https://doi.org/10.48550/arxiv.2203.08821}
A.~Avasthi, T.~Bezerra, A.~Borkum, E.~Church, J.~Genovesi, J.~Haiston, C.~M.
  Jackson, I.~Lazanu, B.~Monreal, S.~Munson, C.~Ortiz, M.~Parvu, S.~J.~M.
  Peeters, D.~Pershey, S.~S. Poudel, J.~Reichenbacher, R.~Saldanha,
  K.~Scholberg, G.~Sinev, J.~Zennamo, H.~O. Back, J.~F. Beacom, F.~Capozzi,
  C.~Cuesta, Z.~Djurcic, A.~C. Ezeribe, I.~Gil-Botella, S.~W. Li, M.~Mooney,
  M.~Sore, and S.~Westerdale, ``{Low Background kTon-Scale Liquid Argon Time
  Projection Chambers},'' 2022.
\newblock https://arxiv.org/abs/2203.08821.

\bibitem{snowmass_pof}
{Xin Qian, et al.}, ``{Snowmass 2021 Letter of Interest: ``Development of
  LArTPC Vertical Drift Solutions with PCB Anode Readouts for DUNE''}.''

\bibitem{PhysRevC.99.055810}
G.~Zhu, S.~W. Li, and J.~F. Beacom, ``{Developing the MeV potential of DUNE:
  Detailed considerations of muon-induced spallation and other backgrounds},''
  {\em Phys. Rev. C}, vol.~99, p.~055810, May 2019.

\bibitem{PhysRevLett.123.131803}
F.~Capozzi, S.~W. Li, G.~Zhu, and J.~F. Beacom, ``{DUNE as the Next-Generation
  Solar Neutrino Experiment},'' {\em Phys. Rev. Lett.}, vol.~123, p.~131803,
  Sep 2019.

\bibitem{LZ_bkgs}
D.~S. Akerib, C.~W. Akerlof, D.~Y. Akimov, A.~Alquahtani, S.~K. Alsum, T.~J.
  Anderson, N.~Angelides, H.~M. Ara{\'u}jo, A.~Arbuckle, J.~E. Armstrong,
  M.~Arthurs, H.~Auyeung, S.~Aviles, X.~Bai, A.~J. Bailey, J.~Balajthy,
  S.~Balashov, J.~Bang, M.~J. Barry, D.~Bauer, P.~Bauer, A.~Baxter, J.~Belle,
  P.~Beltrame, J.~Bensinger, T.~Benson, E.~P. Bernard, A.~Bernstein, A.~Bhatti,
  A.~Biekert, T.~P. Biesiadzinski, H.~J. Birch, B.~Birrittella, K.~E. Boast,
  A.~I. Bolozdynya, E.~M. Boulton, B.~Boxer, R.~Bramante, S.~Branson,
  P.~Br{\'a}s, M.~Breidenbach, C.~A.~J. Brew, J.~H. Buckley, V.~V. Bugaev,
  R.~Bunker, S.~Burdin, J.~K. Busenitz, R.~Cabrita, J.~S. Campbell, C.~Carels,
  D.~L. Carlsmith, B.~Carlson, M.~C. Carmona-Benitez, M.~Cascella, C.~Chan,
  J.~J. Cherwinka, A.~A. Chiller, C.~Chiller, N.~I. Chott, A.~Cole, J.~Coleman,
  D.~Colling, R.~A. Conley, A.~Cottle, R.~Coughlen, G.~Cox, W.~W. Craddock,
  D.~Curran, A.~Currie, J.~E. Cutter, J.~P. da~Cunha, C.~E. Dahl, S.~Dardin,
  S.~Dasu, J.~Davis, T.~J.~R. Davison, L.~de~Viveiros, N.~Decheine, A.~Dobi,
  J.~E.~Y. Dobson, E.~Druszkiewicz, A.~Dushkin, T.~K. Edberg, W.~R. Edwards,
  B.~N. Edwards, J.~Edwards, M.~M. Elnimr, W.~T. Emmet, S.~R. Eriksen, C.~H.
  Faham, A.~Fan, S.~Fayer, S.~Fiorucci, H.~Flaecher, I.~M.~F. Florang, P.~Ford,
  V.~B. Francis, E.~D. Fraser, F.~Froborg, T.~Fruth, R.~J. Gaitskell, N.~J.
  Gantos, D.~Garcia, V.~M. Gehman, R.~Gelfand, J.~Genovesi, R.~M. Gerhard,
  C.~Ghag, E.~Gibson, M.~G.~D. Gilchriese, S.~Gokhale, B.~Gomber, T.~G. Gonda,
  A.~Greenall, S.~Greenwood, G.~Gregerson, M.~G.~D. van~der Grinten, C.~B.
  Gwilliam, C.~R. Hall, D.~Hamilton, S.~Hans, K.~Hanzel, T.~Harrington,
  A.~Harrison, J.~Harrison, C.~Hasselkus, S.~J. Haselschwardt, D.~Hemer, S.~A.
  Hertel, J.~Heise, S.~Hillbrand, O.~Hitchcock, C.~Hjemfelt, M.~D. Hoff,
  B.~Holbrook, E.~Holtom, J.~Y.-K. Hor, M.~Horn, D.~Q. Huang, T.~W. Hurteau,
  C.~M. Ignarra, M.~N. Irving, R.~G. Jacobsen, O.~Jahangir, S.~N. Jeffery,
  W.~Ji, M.~Johnson, J.~Johnson, P.~Johnson, W.~G. Jones, A.~C. Kaboth,
  A.~Kamaha, K.~Kamdin, V.~Kasey, K.~Kazkaz, J.~Keefner, D.~Khaitan,
  M.~Khaleeq, A.~Khazov, A.~V. Khromov, I.~Khurana, Y.~D. Kim, W.~T. Kim, C.~D.
  Kocher, D.~Kodroff, A.~M. Konovalov, L.~Korley, E.~V. Korolkova, M.~Koyuncu,
  J.~Kras, H.~Kraus, S.~W. Kravitz, H.~J. Krebs, L.~Kreczko, B.~Krikler, V.~A.
  Kudryavtsev, A.~V. Kumpan, S.~Kyre, A.~R. Lambert, B.~Landerud, N.~A. Larsen,
  A.~Laundrie, E.~A. Leason, H.~S. Lee, J.~Lee, C.~Lee, B.~G. Lenardo, D.~S.
  Leonard, R.~Leonard, K.~T. Lesko, C.~Levy, J.~Li, Y.~Liu, J.~Liao, F.~T.
  Liao, J.~Lin, A.~Lindote, R.~Linehan, W.~H. Lippincott, R.~Liu, X.~Liu,
  C.~Loniewski, M.~I. Lopes, E.~Lopez-Asamar, B.~L. Paredes, W.~Lorenzon,
  D.~Lucero, S.~Luitz, J.~M. Lyle, C.~Lynch, P.~A. Majewski, J.~Makkinje, D.~C.
  Malling, A.~Manalaysay, L.~Manenti, R.~L. Mannino, N.~Marangou, D.~J.
  Markley, P.~MarrLaundrie, T.~J. Martin, M.~F. Marzioni, C.~Maupin, C.~T.
  McConnell, D.~N. McKinsey, J.~McLaughlin, D.~M. Mei, Y.~Meng, E.~H. Miller,
  Z.~J. Minaker, E.~Mizrachi, J.~Mock, D.~Molash, A.~Monte, M.~E. Monzani,
  J.~A. Morad, E.~Morrison, B.~J. Mount, A.~S.~J. Murphy, D.~Naim, A.~Naylor,
  C.~Nedlik, C.~Nehrkorn, H.~N. Nelson, J.~Nesbit, F.~Neves, J.~A. Nikkel,
  J.~A. Nikoleyczik, A.~Nilima, J.~O'Dell, H.~Oh, F.~G. O'Neill, K.~O'Sullivan,
  I.~Olcina, M.~A. Olevitch, K.~C. Oliver-Mallory, L.~Oxborough, A.~Pagac,
  D.~Pagenkopf, S.~Pal, K.~J. Palladino, V.~M. Palmaccio, J.~Palmer,
  M.~Pangilinan, N.~Parveen, S.~J. Patton, E.~K. Pease, B.~P. Penning,
  G.~Pereira, C.~Pereira, I.~B. Peterson, A.~Piepke, S.~Pierson, S.~Powell,
  R.~M. Preece, K.~Pushkin, Y.~Qie, M.~Racine, B.~N. Ratcliff,
  J.~Reichenbacher, L.~Reichhart, C.~A. Rhyne, A.~Richards, Q.~Riffard,
  G.~R.~C. Rischbieter, J.~P. Rodrigues, H.~J. Rose, R.~Rosero, P.~Rossiter,
  R.~Rucinski, G.~Rutherford, J.~S. Saba, L.~Sabarots, D.~Santone, M.~Sarychev,
  A.~B. M.~R. Sazzad, R.~W. Schnee, M.~Schubnell, P.~R. Scovell, M.~Severson,
  and D.~Seymour, ``{The LUX-ZEPLIN (LZ) radioactivity and cleanliness control
  programs},'' {\em The European Physical Journal C}, vol.~80, no.~11, p.~1044,
  2020.

\bibitem{PhysRevD.100.022004}
DEAP Collaboration Collaboration, R.~Ajaj, P.-A. Amaudruz, G.~R. Araujo,
  M.~Baldwin, M.~Batygov, B.~Beltran, C.~E. Bina, J.~Bonatt, M.~G. Boulay,
  B.~Broerman, J.~F. Bueno, P.~M. Burghardt, A.~Butcher, B.~Cai, S.~Cavuoti,
  M.~Chen, Y.~Chen, B.~T. Cleveland, D.~Cranshaw, K.~Dering, J.~DiGioseffo,
  L.~Doria, F.~A. Duncan, M.~Dunford, A.~Erlandson, N.~Fatemighomi,
  G.~Fiorillo, S.~Florian, A.~Flower, R.~J. Ford, R.~Gagnon, D.~Gallacher,
  E.~A. Garc\'es, S.~Garg, P.~Giampa, D.~Goeldi, V.~V. Golovko, P.~Gorel,
  K.~Graham, D.~R. Grant, A.~L. Hallin, M.~Hamstra, P.~J. Harvey, C.~Hearns,
  A.~Joy, C.~J. Jillings, O.~Kamaev, G.~Kaur, A.~Kemp, I.~Kochanek,
  M.~Ku\ifmmode~\acute{z}\else \'{z}\fi{}niak, S.~Langrock, F.~La~Zia,
  B.~Lehnert, X.~Li, J.~Lidgard, T.~Lindner, O.~Litvinov, J.~Lock, G.~Longo,
  P.~Majewski, A.~B. McDonald, T.~McElroy, T.~McGinn, J.~B. McLaughlin,
  R.~Mehdiyev, C.~Mielnichuk, J.~Monroe, P.~Nadeau, C.~Nantais, C.~Ng, A.~J.
  Noble, E.~O'Dwyer, C.~Ouellet, P.~Pasuthip, S.~J.~M. Peeters, M.-C. Piro,
  T.~R. Pollmann, E.~T. Rand, C.~Rethmeier, F.~Reti\`ere, N.~Seeburn,
  K.~Singhrao, P.~Skensved, B.~Smith, N.~J.~T. Smith, T.~Sonley, J.~Soukup,
  R.~Stainforth, C.~Stone, V.~Strickland, B.~Sur, J.~Tang,
  E.~V\'azquez-J\'auregui, L.~Veloce, S.~Viel, J.~Walding, M.~Waqar, M.~Ward,
  S.~Westerdale, J.~Willis, and A.~Zu\~niga Reyes, ``{Search for dark matter
  with a 231-day exposure of liquid argon using DEAP-3600 at SNOLAB},'' {\em
  Phys. Rev. D}, vol.~100, p.~022004, Jul 2019.

\bibitem{UAr_DS50_39Ar}
DarkSide Collaboration Collaboration, P.~Agnes, L.~Agostino, I.~F.~M.
  Albuquerque, T.~Alexander, A.~K. Alton, K.~Arisaka, H.~O. Back, B.~Baldin,
  K.~Biery, G.~Bonfini, M.~Bossa, B.~Bottino, A.~Brigatti, J.~Brodsky,
  F.~Budano, S.~Bussino, M.~Cadeddu, L.~Cadonati, M.~Cadoni, F.~Calaprice,
  N.~Canci, A.~Candela, H.~Cao, M.~Cariello, M.~Carlini, S.~Catalanotti,
  P.~Cavalcante, A.~Chepurnov, A.~G. Cocco, G.~Covone, L.~Crippa, D.~D'Angelo,
  M.~D'Incecco, S.~Davini, S.~De~Cecco, M.~De~Deo, M.~De~Vincenzi, A.~Derbin,
  A.~Devoto, F.~Di~Eusanio, G.~Di~Pietro, E.~Edkins, A.~Empl, A.~Fan,
  G.~Fiorillo, K.~Fomenko, G.~Forster, D.~Franco, F.~Gabriele, C.~Galbiati,
  C.~Giganti, A.~M. Goretti, F.~Granato, L.~Grandi, M.~Gromov, M.~Guan,
  Y.~Guardincerri, B.~R. Hackett, J.~Hall, K.~Herner, P.~H. Humble, E.~V.
  Hungerford, A.~Ianni, A.~Ianni, I.~James, C.~Jollet, K.~Keeter, C.~L.
  Kendziora, V.~Kobychev, G.~Koh, D.~Korablev, G.~Korga, A.~Kubankin, X.~Li,
  M.~Lissia, P.~Lombardi, S.~Luitz, Y.~Ma, I.~N. Machulin, A.~Mandarano, S.~M.
  Mari, J.~Maricic, L.~Marini, C.~J. Martoff, A.~Meregaglia, P.~D. Meyers,
  T.~Miletic, R.~Milincic, D.~Montanari, A.~Monte, M.~Montuschi, M.~Monzani,
  P.~Mosteiro, B.~J. Mount, V.~N. Muratova, P.~Musico, J.~Napolitano,
  A.~Nelson, S.~Odrowski, M.~Orsini, F.~Ortica, L.~Pagani, M.~Pallavicini,
  E.~Pantic, S.~Parmeggiano, K.~Pelczar, N.~Pelliccia, S.~Perasso, A.~Pocar,
  S.~Pordes, D.~A. Pugachev, H.~Qian, K.~Randle, G.~Ranucci, A.~Razeto,
  B.~Reinhold, A.~L. Renshaw, A.~Romani, B.~Rossi, N.~Rossi, D.~Rountree,
  D.~Sablone, P.~Saggese, R.~Saldanha, W.~Sands, S.~Sangiorgio, C.~Savarese,
  E.~Segreto, D.~A. Semenov, E.~Shields, P.~N. Singh, M.~D. Skorokhvatov,
  O.~Smirnov, A.~Sotnikov, C.~Stanford, Y.~Suvorov, R.~Tartaglia,
  J.~Tatarowicz, G.~Testera, A.~Tonazzo, P.~Trinchese, E.~V. Unzhakov,
  A.~Vishneva, B.~Vogelaar, M.~Wada, S.~Walker, H.~Wang, Y.~Wang, A.~W. Watson,
  S.~Westerdale, J.~Wilhelmi, M.~M. Wojcik, X.~Xiang, J.~Xu, C.~Yang, J.~Yoo,
  S.~Zavatarelli, A.~Zec, W.~Zhong, C.~Zhu, and G.~Zuzel, ``{Results from the
  first use of low radioactivity argon in a dark matter search},'' {\em Phys.
  Rev. D}, vol.~93, p.~081101, Apr 2016.

\bibitem{DUNE-DM-PNNL}
E.~Church, C.~Jackson, and R.~Saldanha, ``{Dark matter detection capabilities
  of a large multipurpose Liquid Argon Time Projection Chamber},'' {\em Journal
  of Instrumentation}, vol.~15, p.~P09026, Sep 2020.

\bibitem{Aprile:2020thb}
XENON Collaboration, E.~Aprile {\em et~al.}, ``{Search for Coherent Elastic
  Scattering of Solar $^8$B Neutrinos in the XENON1T Dark Matter Experiment},''
  {\em Phys. Rev. Lett.}, vol.~126, p.~091301, arXiv:2012.02846 [hep-ex].

\bibitem{Akerib:2016vxi}
LUX Collaboration, D.~S. Akerib {\em et~al.}, ``{Results from a search for dark
  matter in the complete LUX exposure},'' {\em Phys. Rev. Lett.}, vol.~118,
  no.~2, p.~021303, arXiv:1608.07648 [astro-ph.CO].

\bibitem{Aalbers:2022dzr}
J.~Aalbers {\em et~al.}, ``{A Next-Generation Liquid Xenon Observatory for Dark
  Matter and Neutrino Physics},'' 3 2022.

\bibitem{superk}
Super-Kamiokande Collaboration, Y.~Fukuda {\em et~al.}, ``{Evidence for
  oscillation of atmospheric neutrinos},'' {\em Phys. Rev. Lett.}, vol.~81,
  pp.~1562--1567, 1998.

\bibitem{sno}
SNO Collaboration, Q.~R. Ahmad {\em et~al.}, ``{Direct evidence for neutrino
  flavor transformation from neutral current interactions in the Sudbury
  Neutrino Observatory},'' {\em Phys. Rev. Lett.}, vol.~89, p.~011301, 2002.

\bibitem{dayabay}
Daya Bay Collaboration, F.~P. An {\em et~al.}, ``{Observation of
  electron-antineutrino disappearance at Daya Bay},'' {\em Phys. Rev. Lett.},
  vol.~108, p.~171803, 2012.

\bibitem{kamland}
KamLAND Collaboration, K.~Eguchi {\em et~al.}, ``{First results from KamLAND:
  Evidence for reactor anti-neutrino disappearance},'' {\em Phys. Rev. Lett.},
  vol.~90, p.~021802, 2003.

\bibitem{t2k}
T2K Collaboration, K.~Abe {\em et~al.}, ``{Indication of Electron Neutrino
  Appearance from an Accelerator-produced Off-axis Muon Neutrino Beam},'' {\em
  Phys. Rev. Lett.}, vol.~107, p.~041801, 2011.

\bibitem{kii}
Kamiokande-II Collaboration, K.~Hirata {\em et~al.}, ``{Observation of a
  Neutrino Burst from the Supernova SN 1987a},'' {\em Phys. Rev. Lett.},
  vol.~58, pp.~1490--1493, 1987.

\bibitem{imb}
IMB Collaboration, C.~B. Bratton {\em et~al.}, ``{Angular Distribution of
  Events From Sn1987a},'' {\em Phys. Rev. D}, vol.~37, p.~3361, 1988.

\bibitem{borexino}
Borexino Collaboration, C.~Arpesella {\em et~al.}, ``{Direct Measurement of the
  Be-7 Solar Neutrino Flux with 192 Days of Borexino Data},'' {\em Phys. Rev.
  Lett.}, vol.~101, p.~091302, 2008.

\bibitem{borexinopep}
Borexino Collaboration, G.~Bellini {\em et~al.}, ``{First evidence of pep solar
  neutrinos by direct detection in Borexino},'' {\em Phys. Rev. Lett.},
  vol.~108, p.~051302, 2012.

\bibitem{icecube}
IceCube Collaboration, M.~G. Aartsen {\em et~al.}, ``{Evidence for High-Energy
  Extraterrestrial Neutrinos at the IceCube Detector},'' {\em Science},
  vol.~342, p.~1242856, 2013.

\bibitem{juno}
JUNO Collaboration, A.~Abusleme {\em et~al.}, ``{JUNO Physics and Detector},''
  4 2021.

\bibitem{hyperk}
Hyper-Kamiokande Collaboration, K.~Abe {\em et~al.}, ``{Hyper-Kamiokande Design
  Report},'' 5 2018.

\bibitem{dunearapucas}
DUNE Collaboration, D.~Totani {\em et~al.}, ``{A measurement of absolute
  efficiency of the ARAPUCA photon detector in liquid argon},'' {\em JINST},
  vol.~15, no.~06, p.~T06003, 2020.

\bibitem{snoplus}
SNO+ Collaboration, V.~Albanese {\em et~al.}, ``{The SNO+ experiment},'' {\em
  JINST}, vol.~16, no.~08, p.~P08059, arXiv:2104.11687 [physics.ins-det].

\bibitem{borexinocherscint}
BOREXINO Collaboration, M.~Agostini {\em et~al.}, ``{First Directional
  Measurement of Sub-MeV Solar Neutrinos with Borexino},'' {\em Phys. Rev.
  Lett.}, vol.~128, no.~9, p.~091803, 2022.

\bibitem{photonwp}
J.~R. Klein {\em et~al.}, ``{Future Advances in Photon-Based Neutrino
  Detectors: A SNOWMASS White Paper},'' arXiv:2203.07479 [physics.ins-det].

\bibitem{SK_DSNB:2021}
Super-Kamiokande Collaboration, K.~Abe {\em et~al.}, ``{Diffuse supernova
  neutrino background search at Super-Kamiokande},'' {\em Phys. Rev. D},
  vol.~104, no.~12, p.~122002, 2021.

\bibitem{gdogrbonv}
R.~Bonventre and G.~D. Orebi~Gann, ``{Sensitivity of a low threshold
  directional detector to CNO-cycle solar neutrinos},'' {\em Eur. Phys. J. C},
  vol.~78, no.~6, p.~435, 2018.

\bibitem{dsnb_psd}
J.~Sawatzki, M.~Wurm, and D.~Kresse, ``{Detecting the Diffuse Supernova
  Neutrino Background in the future Water-based Liquid Scintillator Detector
  Theia},'' {\em Phys. Rev. D}, vol.~103, no.~2, p.~023021, 2021.

\bibitem{dichroicons}
T.~Kaptanoglu, M.~Luo, B.~Land, A.~Bacon, and J.~Klein, ``{Spectral Photon
  Sorting For Large-Scale Cherenkov and Scintillation Detectors},'' {\em Phys.
  Rev. D}, vol.~101, no.~7, p.~072002, 2020.

\bibitem{wbls}
M.~Yeh, S.~Hans, W.~Beriguete, R.~Rosero, L.~Hu, R.~L. Hahn, M.~V. Diwan, D.~E.
  Jaffe, S.~H. Kettell, and L.~Littenberg, ``{A new water-based liquid
  scintillator and potential applications},'' {\em Nucl. Instrum. Meth. A},
  vol.~660, pp.~51--56, 2011.

\bibitem{theiawp}
Theia Collaboration, M.~Askins {\em et~al.}, ``{THEIA: an advanced optical
  neutrino detector},'' {\em Eur. Phys. J. C}, vol.~80, no.~5, p.~416, 2020.

\bibitem{tanner}
T.~Kaptanoglu, ``{Characterization of the Hamamatsu 8'' R5912-MOD
  Photomultiplier Tube},'' {\em Nucl. Instrum. Meth. A}, vol.~889, pp.~69--77,
  arXiv:1710.03334 [physics.ins-det].

\bibitem{chess}
J.~Caravaca, F.~B. Descamps, B.~J. Land, J.~Wallig, M.~Yeh, and G.~D.
  Orebi~Gann, ``{Experiment to demonstrate separation of Cherenkov and
  scintillation signals},'' {\em Phys. Rev. C}, vol.~95, no.~5, p.~055801,
  2017.

\bibitem{Wiza:1979iia}
J.~L. Wiza, ``{Microchannel plate detectors},'' {\em Nucl. Instrum. Meth.},
  vol.~162, no.~1-3, pp.~587--601, 1979.

\bibitem{LAPPDtiming}
{Adams, B. W. and Elagin, A. and Frisch, H. J. and Obaid, R. and Oberla, E. and
  Vostrikov, A. and Wagner, R. G. and Wang, J. and Wetstein, M.}, ``{Timing
  characteristics of Large Area Picosecond Photodetectors},'' {\em Nuclear
  Instruments and Methods in Physics Research. Section A, Accelerators,
  Spectrometers, Detectors and Associated Equipment}, vol.~795, 9 2015.

\bibitem{wblsberk}
B.~J. Land, Z.~Bagdasarian, J.~Caravaca, M.~Smiley, M.~Yeh, and G.~D.
  Orebi~Gann, ``{MeV-scale performance of water-based and pure liquid
  scintillator detectors},'' {\em Phys. Rev. D}, vol.~103, no.~5, p.~052004,
  2021.

\bibitem{ZAITSEVA201288}
N.~Zaitseva {\em et~al.}, ``{Plastic scintillators with efficient neutron/gamma
  pulse shape discrimination},'' {\em Nucl. Instrum. Meth.}, vol.~A668, pp.~88
  -- 93, 2012.

\bibitem{ZAITSEVA2013747}
N.~Zaitseva {\em et~al.}, ``{Pulse shape discrimination with lithium-containing
  organic scintillators},'' {\em Nucl. Instrum. Meth.}, vol.~A729, pp.~747 --
  754, 2013.

\bibitem{MABE201680}
A.~N. Mabe {\em et~al.},
  ``\href{https://doi.org/10.1016/j.nima.2015.09.111}{Transparent plastic
  scintillators for neutron detection based on lithium salicylate},'' {\em
  Nucl. Instrum. Meth.}, vol.~A806, pp.~80 -- 86, 2016.

\bibitem{osti_1490925}
A.~N. Mabe {\em et~al.}, ``Plastic scintillator materials development at
  llnl,'' {\em Contribution to the Workshop on Applied Antineutrino Physics
  (AAP 2018)}, 12 2018.

\bibitem{Frangville}
C.~Frangville {\em et~al.}, ``{Large solubility of lithium carboxylates
  reaching high rates of 6Li incorporation in polystyrene-based plastic
  scintillators for fast/thermal neutron and gamma ray detection},'' {\em
  Mater. Chem. Front.,}, vol.~3, p.~1626, 2019.

\bibitem{Li6Organic:2021loi}
{N. S. Bowden and H. P Mumm}, ``Neutrino physics and nuclear security
  motivations for the continued development of organic scintillators with pulse
  shape discrimination capability and $^6$li-doping,'' 2020.
\newblock {Snowmass 2022 Letter of Interest}.

\bibitem{Li:2019sof}
V.~A. Li {\em et~al.}, ``{A prototype for SANDD: A highly-segmented
  pulse-shape-sensitive plastic scintillator detector incorporating silicon
  photomultiplier arrays},'' {\em Nucl. Instrum. Meth. A}, vol.~942, p.~162334,
  arXiv:1903.11668 [physics.ins-det].

\bibitem{Sutanto:2021xpo}
F.~Sutanto {\em et~al.}, ``{SANDD: A directional antineutrino detector with
  segmented $^6$Li-doped pulse-shape-sensitive plastic scintillator},'' {\em
  Nucl. Instrum. Meth. A}, vol.~1006, p.~165409, arXiv:2105.00083
  [physics.ins-det].

\bibitem{ROADSTR:2021loi}
{ROADSTR Near-Field Working Group}, ``Roadstr: a mobile antineutrino detector
  platform for enabling multi-reactor spectrum, oscillation, and application
  measurements,'' 2020.
\newblock {Snowmass 2022 Letter of Interest}.

\bibitem{arapucas}
G.~Cancelo, F.~Cavanna, C.~O. Escobar, E.~Kemp, A.~A. Machado, A.~Para,
  E.~Segreto, D.~Totani, and D.~Warner, ``{Increasing the efficiency of photon
  collection in LArTPCs: the ARAPUCA light trap},'' {\em JINST}, vol.~13,
  no.~03, p.~C03040, arXiv:1802.09726 [physics.ins-det].

\bibitem{gratta_image}
J.~Dalmasson, G.~Gratta, A.~Jamil, S.~Kravitz, M.~Malek, K.~Wells, J.~Bentley,
  S.~Steven, and J.~Su, ``{Distributed Imaging for Liquid Scintillation
  Detectors},'' {\em Phys. Rev. D}, vol.~97, no.~5, p.~052006, arXiv:1711.09851
  [physics.ins-det].

\bibitem{chroma}
S.~Seibert and A.~LaTorre, ``{Fast Optical Monte Carlo Simulation with
  Surface-based Geometries Using {\it Chroma}},'' {\em Semantic Scholar}, 2011.

\bibitem{ratpac}
{S. Seibert~{\it et al.}}, ``{RAT-PAC analysic package},''

\bibitem{braidwood}
T.~Bolton, ``{The Braidwood reactor antineutrino experiment},'' {\em Nucl.
  Phys. B Proc. Suppl.}, vol.~149, pp.~166--169, 2005.

\bibitem{geant4}


\bibitem{annie-results}
{\it et al.}.~A.R.~Back, ``{Accelerator Neutrino Neutron Interaction Experiment
  (ANNIE): Preliminary Results and Physics Phase Proposal},'' {\em FNAL Report
  No. P-1063}, arXiv:1707.08222 [physics.ins-det].

\bibitem{watchman}
WATCHMAN Collaboration, M.~Askins {\em et~al.}, ``{The Physics and Nuclear
  Nonproliferation Goals of WATCHMAN: A WAter CHerenkov Monitor for
  ANtineutrinos},'' 2 2015.

\bibitem{flatdot}
J.~Gruszko, B.~Naranjo, B.~Daniel, A.~Elagin, D.~Gooding, C.~Grant, J.~Ouellet,
  and L.~Winslow, ``{Detecting Cherenkov light from 1\textendash{}2 MeV
  electrons in linear alkylbenzene},'' {\em JINST}, vol.~14, no.~02, p.~P02005,
  arXiv:1811.11144 [physics.ins-det].

\bibitem{annieLOI}
{\it et al.}.~I.~Anghel, ``{LETTER OF INTENT: The Accelerator Neutrino Neutron
  Interaction Experiment (ANNIE)},'' {\em FNAL Report No. P-1063}, 2015.

\bibitem{pershingdiss}
T.~J. Pershing, {\em {The Accelerator Neutrino-Neutron Interaction
  Experiment}}.
\newblock PhD thesis, University of California, Davis, 2020.

\bibitem{perovskite}
E.~Graham, D.~Gooding, J.~Gruszko, C.~Grant, B.~Naranjo, and L.~Winslow,
  ``{Light Yield of Perovskite Nanocrystal-Doped Liquid Scintillator},''
  arXiv:1908.03564 [physics.ins-det].

\bibitem{lbnf}
DUNE Collaboration, J.~Strait {\em et~al.}, ``{Long-Baseline Neutrino Facility
  (LBNF) and Deep Underground Neutrino Experiment (DUNE)}: {Conceptual Design
  Report, Volume 3: Long-Baseline Neutrino Facility for DUNE June 24, 2015},''
  arXiv:1601.05823 [physics.ins-det].

\bibitem{dunecdr}
DUNE Collaboration, R.~Acciarri {\em et~al.}, ``{Long-Baseline Neutrino
  Facility (LBNF) and Deep Underground Neutrino Experiment (DUNE)}: {Conceptual
  Design Report, Volume 1: The LBNF and DUNE Projects},'' arXiv:1601.05471
  [physics.ins-det].

\bibitem{bxcno}
BOREXINO Collaboration, M.~Agostini {\em et~al.}, ``{Experimental evidence of
  neutrinos produced in the CNO fusion cycle in the Sun},'' {\em Nature},
  vol.~587, pp.~577--582, arXiv:2006.15115 [hep-ex].

\bibitem{friedland2004}
A.~Friedland, C.~Lunardini, and C.~Pena-Garay, ``{Solar neutrinos as probes of
  neutrino matter interactions},'' {\em Phys. Lett. B}, vol.~594, p.~347, 2004.

\bibitem{minakata2012}
H.~Minakata and C.~Pena-Garay, ``{Solar Neutrino Observables Sensitive to
  Matter Effects},'' {\em Adv. High Energy Phys.}, vol.~2012, p.~349686,
  arXiv:1009.4869 [hep-ph].

\bibitem{biller_normal}
S.~D. Biller, ``{Probing Majorana neutrinos in the regime of the normal mass
  hierarchy},'' {\em Phys. Rev. D}, vol.~87, no.~7, p.~071301, arXiv:1306.5654
  [physics.ins-det].

\bibitem{Cabrera:2019kxi}
A.~Cabrera {\em et~al.}, ``{Neutrino Physics with an Opaque Detector},'' {\em
  Commun. Phys.}, vol.~4, p.~273, 2021.

\bibitem{Buck:2019tsa}
C.~Buck, B.~Gramlich, and S.~Schoppmann, ``{Novel Opaque Scintillator for
  Neutrino Detection},'' {\em JINST}, vol.~14, no.~11, p.~P11007, 2019.

\bibitem{Agnolet:2016zir}
MINER Collaboration, G.~Agnolet {\em et~al.}, ``{Background Studies for the
  MINER Coherent Neutrino Scattering Reactor Experiment},'' {\em Nucl. Instrum.
  Meth. A}, vol.~853, pp.~53--60, arXiv:1609.02066 [physics.ins-det].

\bibitem{Strauss:2017cuu}
R.~Strauss {\em et~al.}, ``{The $\nu$-cleus experiment: A gram-scale
  fiducial-volume cryogenic detector for the first detection of coherent
  neutrino-nucleus scattering},'' {\em Eur. Phys. J. C}, vol.~77, p.~506,
  arXiv:1704.04320 [physics.ins-det].

\bibitem{Strauss:2017cam}
R.~Strauss {\em et~al.}, ``{Gram-scale cryogenic calorimeters for rare-event
  searches},'' {\em Phys. Rev. D}, vol.~96, no.~2, p.~022009, arXiv:1704.04317
  [physics.ins-det].

\bibitem{Angloher:2017sxg}
CRESST Collaboration, G.~Angloher {\em et~al.}, ``{Results on MeV-scale dark
  matter from a gram-scale cryogenic calorimeter operated above ground},'' {\em
  Eur. Phys. J. C}, vol.~77, no.~9, p.~637, arXiv:1707.06749 [astro-ph.CO].

\bibitem{Ricochet:2021rjo}
C.~Augier {\em et~al.}, ``{Ricochet Progress and Status},'' in {\em {19th
  International Workshop on Low Temperature Detectors}}, 11 2021, 2111.06745.

\bibitem{colantoni:2020}
I.~{Colantoni}, C.~{Bellenghi}, M.~{Calvo}, R.~{Camattari}, L.~{Cardani},
  N.~{Casali}, A.~{Cruciani}, S.~{Di Domizio}, J.~{Goupy}, V.~{Guidi}, H.~{Le
  Sueur}, M.~{Martinez}, A.~{Mazzolari}, A.~{Monfardini}, V.~{Pettinacci},
  G.~{Pettinari}, M.~{Romagnoni}, and M.~{Vignati}, ``{BULLKID: BULky and
  Low-Threshold Kinetic Inductance Detectors},'' {\em Journal of Low
  Temperature Physics}, vol.~199, pp.~593--597, Feb. 2020.

\bibitem{Aguilar-Arevalo:2019jlr}
CONNIE Collaboration, A.~Aguilar-Arevalo {\em et~al.}, ``{Exploring low-energy
  neutrino physics with the Coherent Neutrino Nucleus Interaction
  Experiment},'' {\em Phys. Rev. D}, vol.~100, no.~9, p.~092005,
  arXiv:1906.02200 [physics.ins-det].

\bibitem{https://doi.org/10.48550/arxiv.2204.04575}
D.~Akimov, S.~Alawabdeh, P.~An, A.~Arteaga, C.~Awe, P.~S. Barbeau, C.~Barry,
  B.~Becker, V.~Belov, I.~Bernardi, M.~A. Blackston, L.~Blokland, C.~Bock,
  B.~Bodur, A.~Bolozdynya, R.~Bouabid, A.~Bracho, J.~Browning,
  B.~Cabrera-Palmer, N.~Chen, D.~Chernyak, E.~Conley, J.~Daughhetee,
  J.~Daughtry, E.~Day, M.~d.~V. Coello, J.~Detwiler, K.~Ding, M.~R. Durand,
  Y.~Efremenko, S.~R. Elliott, L.~Fabris, M.~Febbraro, W.~Fox, J.~Galambos,
  A.~G. Rosso, A.~Galindo-Uribarri, C.~Gilbert, M.~P. Green, K.~R. Hansen,
  B.~Harris, M.~R. Heath, S.~Hedges, R.~Henderson, D.~Hoang, C.~Hughes,
  M.~Hughes, E.~Iverson, P.~Jairam, B.~A. Johnson, T.~Johnson, L.~Kaufman,
  A.~Khromov, A.~Konovalov, J.~Koros, E.~Kozlova, A.~Kumpan, L.~Li, J.~T.
  Librande, J.~M. Link, J.~Liu, A.~Major, K.~Mann, D.~M. Markoff,
  J.~Mastroberti, J.~Mattingly, O.~McGoldrick, M.~McIntyre, Y.~A. Melikyan,
  M.~Mishra, P.~E. Mueller, J.~Newby, D.~S. Parno, A.~Penne, S.~I. Penttila,
  D.~Pershey, C.~Prior, D.~Radford, F.~Rahman, R.~Rapp, H.~Ray, J.~Raybern,
  O.~Razuvaeva, D.~Reyna, G.~C. Rich, D.~Rimal, J.~Ross, A.~Rouzky, D.~Rudik,
  J.~Runge, D.~J. Salvat, A.~M. Salyapongse, J.~Sander, K.~Scholberg,
  P.~Siehien, A.~Shakirov, G.~Simakov, G.~Sinev, W.~M. Snow, V.~Sosnovstsev,
  J.~Steele, A.~S. Hjelmstad, T.~Subedi, B.~Suh, R.~Tayloe,
  K.~Tellez-Giron-Flores, R.~T. Thornton, I.~Tolstukhin, S.~Trotter, F.~Tsai,
  Y.~T. Tsai, E.~Ujah, J.~Vanderwerp, E.~van Nieuwenhuizen, R.~L. Varner,
  S.~Vasquez, C.~J. Virtue, G.~Visser, K.~Walkup, J.~Wang, E.~M. Ward,
  C.~Wiseman, T.~Wongjirad, D.~Wu, J.~Yang, Y.~Yang, Y.~R. Yen, J.~Yoo, C.~H.
  Yu, J.~Zettlemoyer, and S.~Zhang, ``The coherent experimental program,''
  2022.

\bibitem{bonet2021constraints}
CONUS Collaboration, H.~Bonet {\em et~al.}, ``{Constraints on Elastic Neutrino
  Nucleus Scattering in the Fully Coherent Regime from the CONUS Experiment},''
  {\em Phys. Rev. Lett.}, vol.~126, no.~4, p.~041804, arXiv:2011.00210
  [hep-ex].

\bibitem{Colaresi:2021kus}
J.~Colaresi, J.~I. Collar, T.~W. Hossbach, A.~R.~L. Kavner, C.~M. Lewis, A.~E.
  Robinson, and K.~M. Yocum, ``{First results from a search for coherent
  elastic neutrino-nucleus scattering at a reactor site},'' {\em Phys. Rev. D},
  vol.~104, no.~7, p.~072003, arXiv:2108.02880 [hep-ex].

\bibitem{nuGEN_mag7}
Presentation at the Magnificent CEvNS workshop, Nov 2021,
  https://indico.cern.ch/event/1075677/contributions/4556660/.

\bibitem{Wong:2016lmb}
H.~T.-K. Wong, ``{Taiwan EXperiment On NeutrinO --- History and Prospects},''
  {\em The Universe}, vol.~3, no.~4, pp.~22--37, arXiv:1608.00306 [hep-ex].

\bibitem{https://doi.org/10.48550/arxiv.2204.06318}
J.~J. Choi, E.~J. Jeon, J.~Y. Kim, K.~W. Kim, S.~H. Kim, S.~K. Kim, Y.~D. Kim,
  Y.~J. Ko, B.~C. Koh, C.~Ha, B.~J. Park, S.~H. Lee, I.~S. Lee, H.~Lee, H.~S.
  Lee, J.~Lee, Y.~M. Oh, and S.~L. Olsen, ``Exploring coherent elastic
  neutrino-nucleus scattering using reactor electron antineutrinos in the neon
  experiment,'' 2022.

\bibitem{Alfonso-Pita:2022akn}
E.~Alfonso-Pita {\em et~al.}, ``{Snowmass 2021 Scintillating Bubble Chambers:
  Liquid-noble Bubble Chambers for Dark Matter and CE$\nu$NS Detection},'' in
  {\em {2022 Snowmass Summer Study}}, 7 2022, 2207.12400.

\bibitem{Giomataris:2008ap}
I.~Giomataris {\em et~al.}, ``{A Novel large-volume Spherical Detector with
  Proportional Amplification read-out},'' {\em JINST}, vol.~3, p.~P09007,
  arXiv:0807.2802 [physics.ins-det].

\bibitem{PhysRevD.100.115020}
D.~Akimov {\em et~al.}, ``First constraint on coherent elastic neutrino-nucleus
  scattering in argon,'' {\em Phys. Rev. D}, vol.~100, p.~115020, Dec 2019.

\bibitem{Akimov:2017hee}
D.~Y. Akimov {\em et~al.}, ``{Status of the RED-100 experiment},'' {\em JINST},
  vol.~12, no.~06, p.~C06018, 2017.

\bibitem{CHILLAX_M7_2021}
J.~Xu, ``"status of the chillax detector development".'' Magnificent CEvNS
  2021, 2021.

\bibitem{Ni:2021mwa}
K.~Ni, J.~Qi, E.~Shockley, and Y.~Wei, ``{Sensitivity of a Liquid Xenon
  Detector to Neutrino\textendash{}Nucleus Coherent Scattering and Neutrino
  Magnetic Moment from Reactor Neutrinos},'' {\em Universe}, vol.~7, no.~3,
  p.~54, 2021.

\bibitem{PhysRevD.80.051301}
B.~Monreal and J.~A. Formaggio, ``Relativistic cyclotron radiation detection of
  tritium decay electrons as a new technique for measuring the neutrino mass,''
  {\em Phys. Rev. D}, vol.~80, p.~051301, Sep 2009.

\bibitem{velte}
C.~Velte, F.~Ahrens, A.~Barth, K.~Blaum, M.~Bra{\ss}, M.~Door, H.~Dorrer, C.~E.
  D{\"u}llmann, S.~Eliseev, C.~Enss, P.~Filianin, A.~Fleischmann, L.~Gastaldo,
  A.~Goeggelmann, T.~D. Goodacre, M.~W. Haverkort, D.~Hengstler, J.~Jochum,
  K.~Johnston, M.~Keller, S.~Kempf, T.~Kieck, C.~M. K{\"o}nig, U.~K{\"o}ster,
  K.~Kromer, F.~Mantegazzini, B.~Marsh, Y.~N. Novikov, F.~Piquemal, C.~Riccio,
  D.~Richter, A.~Rischka, S.~Rothe, R.~X. Sch{\"u}ssler, C.~Schweiger,
  T.~Stora, M.~Wegner, K.~Wendt, M.~Zampaolo, and K.~Zuber, ``High-resolution
  and low-background {\$}{\$}\^{}{\{}163{\}}{\$}{\$}ho spectrum: interpretation
  of the resonance tails,'' {\em The European Physical Journal C}, vol.~79,
  no.~12, p.~1026, 2019.

\bibitem{refId0}
{Nucciotti, Angelo}, ``Statistical sensitivity of ho electron capture neutrino
  mass experiments,'' {\em Eur. Phys. J. C}, vol.~74, no.~11, p.~3161, 2014.

\bibitem{doi:10.1063/1.4986222}
J.~A.~B. Mates, D.~T. Becker, D.~A. Bennett, B.~J. Dober, J.~D. Gard, J.~P.
  Hays-Wehle, J.~W. Fowler, G.~C. Hilton, C.~D. Reintsema, D.~R. Schmidt, D.~S.
  Swetz, L.~R. Vale, and J.~N. Ullom, ``Simultaneous readout of 128 x-ray and
  gamma-ray transition-edge microcalorimeters using microwave squid
  multiplexing,'' {\em Applied Physics Letters}, vol.~111, no.~6, p.~062601,
  2017.

\bibitem{https://doi.org/10.48550/arxiv.1307.4738}
S.~Betts {\em et~al.}, ``Development of a relic neutrino detection experiment
  at ptolemy: Princeton tritium observatory for light, early-universe,
  massive-neutrino yield,'' 2013.

\bibitem{Pattavina:2020cqc}
L.~Pattavina, N.~Ferreiro~Iachellini, and I.~Tamborra, ``{Neutrino observatory
  based on archaeological lead},'' {\em Phys. Rev. D}, vol.~102, no.~6,
  p.~063001, arXiv:2004.06936 [astro-ph.HE].

\bibitem{Baum:2018tfw}
S.~Baum, A.~K. Drukier, K.~Freese, M.~G\'orski, and P.~Stengel, ``{Searching
  for Dark Matter with Paleo-Detectors},'' {\em Phys. Lett. B}, vol.~803,
  p.~135325, arXiv:1806.05991 [astro-ph.CO].

\bibitem{Alfonso:2022meh}
K.~Alfonso {\em et~al.}, ``{Passive low energy nuclear recoil detection with
  color centers -- PALEOCCENE},'' in {\em {2022 Snowmass Summer Study}}, 3
  2022, 2203.05525.

\bibitem{spacecraft}
N.~Solomey {\em et~al.}, ``Science mission of a neutrino space-craft.''
  https://www.nasa.gov/directorates/spacetech/niac/.

\bibitem{Vahsen:2020pzb}
S.~Vahsen {\em et~al.}, ``{CYGNUS: Feasibility of a nuclear recoil observatory
  with directional sensitivity to dark matter and neutrinos},''
  arXiv:2008.12587 [physics.ins-det].

\bibitem{AristizabalSierra:2021uob}
D.~Aristizabal~Sierra, B.~Dutta, D.~Kim, D.~Snowden-Ifft, and L.~E. Strigari,
  ``{Coherent elastic neutrino-nucleus scattering with the
  \ensuremath{\nu}BDX-DRIFT directional detector at next generation neutrino
  facilities},'' {\em Phys. Rev. D}, vol.~104, no.~3, p.~033004,
  arXiv:2103.10857 [hep-ph].

\bibitem{Billard:2013qya}
J.~Billard, L.~Strigari, and E.~Figueroa-Feliciano, ``{Implication of neutrino
  backgrounds on the reach of next generation dark matter direct detection
  experiments},'' {\em Phys. Rev. D}, vol.~89, no.~2, p.~023524,
  arXiv:1307.5458 [hep-ph].

\bibitem{https://doi.org/10.48550/arxiv.2202.05097}
P.~Adari {\em et~al.}, ``Excess workshop: Descriptions of rising low-energy
  spectra,'' 2022.

\bibitem{Rothe:2019aii}
NUCLEUS Collaboration, J.~Rothe {\em et~al.}, ``{NUCLEUS: Exploring Coherent
  Neutrino-Nucleus Scattering with Cryogenic Detectors},'' {\em J. Low Temp.
  Phys.}, vol.~199, no.~1-2, pp.~433--440, 2019.

\bibitem{Angloher:2019flc}
NUCLEUS Collaboration, G.~Angloher {\em et~al.}, ``{Exploring $\hbox {CE}\nu
  \hbox {NS}$ with NUCLEUS at the Chooz nuclear power plant},'' {\em Eur. Phys.
  J. C}, vol.~79, no.~12, p.~1018, arXiv:1905.10258 [physics.ins-det].

\bibitem{Wagner:2022iqf}
V.~Wagner {\em et~al.}, ``{Development of a compact muon veto for the NUCLEUS
  experiment},'' arXiv:2202.03991 [physics.ins-det].

\bibitem{RICOCHET:2021gkf}
T.~Salagnac {\em et~al.}, ``{Optimization and performance of the CryoCube
  detector for the future RICOCHET low-energy neutrino experiment},'' in {\em
  {19th International Workshop on Low Temperature Detectors}}, 11 2021,
  2111.12438.

\bibitem{PhysRevD.103.075002}
S.~M. Griffin, Y.~Hochberg, K.~Inzani, N.~Kurinsky, T.~Lin, and T.~C. Yu,
  ``Silicon carbide detectors for sub-gev dark matter,'' {\em Phys. Rev. D},
  vol.~103, p.~075002, Apr 2021.

\bibitem{PhysRevD.99.123005}
N.~Kurinsky, T.~C. Yu, Y.~Hochberg, and B.~Cabrera, ``Diamond detectors for
  direct detection of sub-gev dark matter,'' {\em Phys. Rev. D}, vol.~99,
  p.~123005, Jun 2019.

\bibitem{lucia}
L.~Canonica, A.~H. Abdelhameed, P.~Bauer, A.~Bento, E.~Bertoldo,
  N.~Ferreiro~Iachellini, D.~Fuchs, D.~Hauff, M.~Mancuso, F.~Petricca,
  F.~Pr{\"o}bst, and J.~Rothe, ``Operation of a diamond cryogenic detector for
  low-mass dark matter searches,'' {\em Journal of Low Temperature Physics},
  vol.~199, no.~3, pp.~606--613, 2020.

\bibitem{QFmeasurementsSi}
A.~E. Chavarria {\em et~al.}, ``{Measurement of the ionization produced by
  sub-keV silicon nuclear recoils in a CCD dark matter detector},'' {\em Phys.
  Rev. D}, vol.~94, no.~8, p.~082007, arXiv:1608.00957 [astro-ph.IM].

\bibitem{QFtheorynew}
Y.~Sarkis, A.~Aguilar-Arevalo, and J.~C. D'Olivo, ``{Study of the ionization
  efficiency for nuclear recoils in pure crystals},'' {\em Phys. Rev. D},
  vol.~101, no.~10, p.~102001, arXiv:2001.06503 [hep-ph].

\bibitem{Sarkis2021}
Y.~Sarkis, A.~Aguilar-Arevalo, and J.~C. D\textquoteright{}Olivo, ``{A Study of
  the Ionization Efficiency for Nuclear Recoils in Pure Crystals},'' {\em Phys.
  At. Nucl.}, vol.~84, no.~4, pp.~590--594, 2021.

\bibitem{webpage_violetta}
https://www.violetaexperiment.com.

\bibitem{PhysRevLett.115.062501}
S.~Eliseev, K.~Blaum, M.~Block, S.~Chenmarev, H.~Dorrer, C.~E. D\"ullmann,
  C.~Enss, P.~E. Filianin, L.~Gastaldo, M.~Goncharov, U.~K\"oster,
  F.~Lautenschl\"ager, Y.~N. Novikov, A.~Rischka, R.~X. Sch\"ussler,
  L.~Schweikhard, and A.~T\"urler, ``Direct measurement of the mass difference
  of $^{163}\mathrm{Ho}$ and $^{163}\mathrm{Dy}$ solves the $q$-value puzzle
  for the neutrino mass determination,'' {\em Phys. Rev. Lett.}, vol.~115,
  p.~062501, Aug 2015.

\bibitem{Drukier:2018pdy}
A.~K. Drukier, S.~Baum, K.~Freese, M.~G\'orski, and P.~Stengel,
  ``{Paleo-detectors: Searching for Dark Matter with Ancient Minerals},'' {\em
  Phys. Rev. D}, vol.~99, no.~4, p.~043014, arXiv:1811.06844 [astro-ph.CO].

\bibitem{Jordan:2020gxx}
J.~R. Jordan, S.~Baum, P.~Stengel, A.~Ferrari, M.~C. Morone, P.~Sala, and
  J.~Spitz, ``{Measuring Changes in the Atmospheric Neutrino Rate Over Gigayear
  Timescales},'' {\em Phys. Rev. Lett.}, vol.~125, no.~23, p.~231802,
  arXiv:2004.08394 [hep-ph].

\bibitem{Tapia-Arellano:2021cml}
N.~Tapia-Arellano and S.~Horiuchi, ``{Measuring solar neutrinos over gigayear
  timescales with paleo detectors},'' {\em Phys. Rev. D}, vol.~103, no.~12,
  p.~123016, arXiv:2102.01755 [hep-ph].

\bibitem{Baum:2019fqm}
S.~Baum, T.~D.~P. Edwards, B.~J. Kavanagh, P.~Stengel, A.~K. Drukier,
  K.~Freese, M.~G\'orski, and C.~Weniger, ``{Paleodetectors for Galactic
  supernova neutrinos},'' {\em Phys. Rev. D}, vol.~101, no.~10, p.~103017,
  arXiv:1906.05800 [astro-ph.GA].

\bibitem{Baum:2021jak}
S.~Baum, T.~D.~P. Edwards, K.~Freese, and P.~Stengel, ``{New Projections for
  Dark Matter Searches with Paleo-Detectors},'' {\em Instruments}, vol.~5,
  no.~2, p.~21, arXiv:2106.06559 [astro-ph.CO].

\bibitem{Baum:2021chx}
S.~Baum, W.~DeRocco, T.~D.~P. Edwards, and S.~Kalia, ``{Galactic geology:
  Probing time-varying dark matter signals with paleodetectors},'' {\em Phys.
  Rev. D}, vol.~104, no.~12, p.~123015, arXiv:2107.02812 [astro-ph.GA].

\bibitem{Cogswell:2021qlq}
B.~K. Cogswell, A.~Goel, and P.~Huber, ``{Passive Low-Energy Nuclear-Recoil
  Detection with Color Centers},'' {\em Phys. Rev. Applied}, vol.~16, no.~6,
  p.~064060, arXiv:2104.13926 [physics.ins-det].

\bibitem{D1CP01083B}
M.~Szydagis, C.~Levy, Y.~Huang, A.~C. Kamaha, C.~C. Knight, G.~R.~C.
  Rischbieter, and P.~W. Wilson, ``Demonstration of neutron radiation-induced
  nucleation of supercooled water,'' {\em Phys. Chem. Chem. Phys.}, vol.~23,
  pp.~13440--13446, 2021.

\bibitem{COOPER201125}
R.~Cooper, D.~Radford, P.~Hausladen, and K.~Lagergren, ``A novel hpge detector
  for gamma-ray tracking and imaging,'' {\em Nuclear Instruments and Methods in
  Physics Research Section A: Accelerators, Spectrometers, Detectors and
  Associated Equipment}, vol.~665, pp.~25--32, 2011.

\bibitem{bonet2021large}
H.~Bonet {\em et~al.}, ``{Large-size sub-keV sensitive germanium detectors for
  the CONUS experiment},'' {\em Eur. Phys. J. C}, vol.~81, no.~3, p.~267,
  arXiv:2010.11241 [physics.ins-det].

\bibitem{bonet2021full}
H.~Bonet {\em et~al.}, ``Full background decomposition of the conus
  experiment,'' arXiv:2112.09585 [physics.ins-det].

\bibitem{bonhomme2022direct}
A.~Bonhomme {\em et~al.}, ``Direct measurement of the ionization quenching
  factor of nuclear recoils in germanium in the kev energy range,''
  arXiv:2202.03754 [physics.ins-det].

\bibitem{Colaresi:2022obx}
J.~Colaresi, J.~I. Collar, T.~W. Hossbach, C.~M. Lewis, and K.~M. Yocum,
  ``{Suggestive evidence for Coherent Elastic Neutrino-Nucleus Scattering from
  reactor antineutrinos},'' arXiv:2202.09672 [hep-ex].

\bibitem{Wong:2006nx}
TEXONO Collaboration, H.~Wong {\em et~al.}, ``{A Search of Neutrino Magnetic
  Moments with a High-Purity Germanium Detector at the Kuo-Sheng Nuclear Power
  Station},'' {\em Phys. Rev. D}, vol.~75, p.~012001, 2007.

\bibitem{Chen:2014dsa}
J.-W. Chen, H.-C. Chi, H.-B. Li, C.~P. Liu, L.~Singh, H.~T. Wong, C.-L. Wu, and
  C.-P. Wu, ``{Constraints on millicharged neutrinos via analysis of data from
  atomic ionizations with germanium detectors at sub-keV sensitivities},'' {\em
  Phys. Rev. D}, vol.~90, no.~1, p.~011301, arXiv:1405.7168 [hep-ph].

\bibitem{Adhikari:2018ljm}
COSINE-100 Collaboration, G.~Adhikari {\em et~al.}, ``{An experiment to search
  for dark-matter interactions using sodium iodide detectors},'' {\em Nature},
  vol.~564, no.~7734, pp.~83--86, arXiv:1906.01791 [astro-ph.IM].

\bibitem{Adhikari:2019off}
COSINE-100 Collaboration, G.~Adhikari {\em et~al.}, ``{Search for a Dark
  Matter-Induced Annual Modulation Signal in NaI(Tl) with the COSINE-100
  Experiment},'' {\em Phys. Rev. Lett.}, vol.~123, no.~3, p.~031302,
  arXiv:1903.10098 [astro-ph.IM].

\bibitem{Adhikari:2017esn}
COSINE-100 Collaboration, G.~Adhikari {\em et~al.}, ``{Initial Performance of
  the COSINE-100 Experiment},'' {\em Eur. Phys. J. C}, vol.~78, no.~2, p.~107,
  arXiv:1710.05299 [physics.ins-det].

\bibitem{COSINE-100:2020wrv}
COSINE-100 Collaboration, G.~Adhikari {\em et~al.}, ``{Lowering the energy
  threshold in COSINE-100 dark matter searches},'' {\em Astropart. Phys.},
  vol.~130, p.~102581, arXiv:2005.13784 [physics.ins-det].

\bibitem{Choi:2020qcj}
J.~Choi, B.~Park, C.~Ha, K.~Kim, S.~Kim, Y.~Kim, Y.~Ko, H.~Lee, S.~Lee, and
  S.~Olsen, ``{Improving the light collection using a new NaI(Tl)crystal
  encapsulation},'' {\em Nucl. Instrum. Meth. A}, vol.~981, p.~164556,
  arXiv:2006.02573 [physics.ins-det].

\bibitem{Park:2020fsq}
COSINE Collaboration, B.~Park {\em et~al.}, ``{Development of ultra-pure
  NaI(Tl) detectors for the COSINE-200 experiment},'' {\em Eur. Phys. J. C},
  vol.~80, no.~9, p.~814, arXiv:2004.06287 [physics.ins-det].

\bibitem{Adhikari:2018fpo}
COSINE-100 Collaboration, G.~Adhikari {\em et~al.}, ``{The COSINE-100 Data
  Acquisition System},'' {\em JINST}, vol.~13, no.~09, p.~P09006,
  arXiv:1806.09788 [physics.ins-det].

\bibitem{PhysRevC.75.044610}
B.~Mosconi, P.~Ricci, E.~Truhl\'{\i}k, and P.~Vogel, ``Model dependence of the
  neutrino-deuteron disintegration cross sections at low energies,'' {\em Phys.
  Rev. C}, vol.~75, p.~044610, Apr 2007.

\bibitem{Giampa:2021wte}
SBC Collaboration, P.~Giampa, ``{The Scintillating Bubble Chamber (SBC)
  Experiment for Dark Matter and Reactor CEvNS},'' {\em PoS}, vol.~ICHEP2020,
  p.~632, 2021.

\bibitem{SBC:2021yal}
SBC, CE\ensuremath{\nu}NS Theory Group at IF-UNAM Collaboration, L.~J. Flores
  {\em et~al.}, ``{Physics reach of a low threshold scintillating argon bubble
  chamber in coherent elastic neutrino-nucleus scattering reactor
  experiments},'' {\em Phys. Rev. D}, vol.~103, no.~9, p.~L091301,
  arXiv:2101.08785 [hep-ex].

\bibitem{Scholberg:2005qs}
K.~Scholberg, ``{Prospects for measuring coherent neutrino-nucleus elastic
  scattering at a stopped-pion neutrino source},'' {\em Phys. Rev. D}, vol.~73,
  p.~033005, arXiv:hep-ex/0511042 [hep-ex].

\bibitem{Katsioulas:2018pyh}
I.~Katsioulas {\em et~al.}, ``{A sparkless resistive glass correction electrode
  for the spherical proportional counter},'' {\em JINST}, vol.~13, no.~11,
  p.~P11006, arXiv:1809.03270 [physics.ins-det].

\bibitem{Giomataris:2020rna}
I.~Giomataris {\em et~al.}, ``{A resistive ACHINOS multi-anode structure with
  DLC coating for spherical proportional counters},'' {\em JINST}, vol.~15,
  no.~11, p.~11, arXiv:2003.01068 [physics.ins-det].

\bibitem{Giganon:2017isb}
A.~Giganon {\em et~al.}, ``{A multiball read-out for the spherical proportional
  counter},'' {\em JINST}, vol.~12, no.~12, p.~P12031, arXiv:1707.09254
  [physics.ins-det].

\bibitem{Bougamont:2010mj}
E.~Bougamont {\em et~al.}, ``{Ultra low energy results and their impact to dark
  matter and low energy neutrino physics},'' arXiv:1010.4132 [physics.ins-det].

\bibitem{giove}
G.~Heusser and other, ``{GIOVE - A new detector setup for high sensitivity
  germanium spectroscopy at shallow depth},'' {\em Eur. Phys. J. C}, vol.~75,
  no.~11, p.~531, arXiv:1507.03319 [astro-ph.IM].

\bibitem{conus}
C.~Buck {\em et~al.}, ``{A novel experiment for coherent elastic neutrino
  nucleus scattering: CONUS},'' {\em J. Phys. Conf. Ser.}, vol.~1342, no.~1,
  p.~012094, 2020.

\bibitem{https://doi.org/10.48550/arxiv.1803.09183}
D.~Akimov {\em et~al.}, ``Coherent 2018 at the spallation neutron source,''
  2018.

\bibitem{Akimov:2019ogx}
RED-100 Collaboration, D.~Y. Akimov {\em et~al.}, ``{First ground-level
  laboratory test of the two-phase xenon emission detector RED-100},'' {\em
  JINST}, vol.~15, no.~02, p.~P02020, arXiv:1910.06190 [physics.ins-det].

\bibitem{Lenardo:2019fcn}
B.~Lenardo {\em et~al.}, ``{Measurement of the ionization yield from nuclear
  recoils in liquid xenon between 0.3 - 6 keV with single-ionization-electron
  sensitivity},'' arXiv:1908.00518 [physics.ins-det].

\bibitem{https://doi.org/10.48550/arxiv.2203.07361}
M.~Abdullah {\em et~al.}, ``Coherent elastic neutrino-nucleus scattering:
  Terrestrial and astrophysical applications,'' 2022.

\bibitem{Cabrera:1984rr}
B.~Cabrera, L.~M. Krauss, and F.~Wilczek, ``{Bolometric Detection of
  Neutrinos},'' {\em Phys. Rev. Lett.}, vol.~55, p.~25, 1985.

\bibitem{Drukier:1986tm}
A.~K. Drukier, K.~Freese, and D.~N. Spergel, ``{Detecting Cold Dark Matter
  Candidates},'' {\em Phys. Rev. D}, vol.~33, pp.~3495--3508, 1986.

\bibitem{Monroe:2007xp}
J.~Monroe and P.~Fisher, ``{Neutrino Backgrounds to Dark Matter Searches},''
  {\em Phys. Rev. D}, vol.~76, p.~033007, arXiv:0706.3019 [astro-ph].

\bibitem{Vergados:2008jp}
J.~D. Vergados and H.~Ejiri, ``{Can Solar Neutrinos be a Serious Background in
  Direct Dark Matter Searches?},'' {\em Nucl. Phys. B}, vol.~804, pp.~144--159,
  arXiv:0805.2583 [hep-ph].

\bibitem{Strigari:2009bq}
L.~E. Strigari, ``{Neutrino Coherent Scattering Rates at Direct Dark Matter
  Detectors},'' {\em New J. Phys.}, vol.~11, p.~105011, arXiv:0903.3630
  [astro-ph.CO].

\bibitem{Gutlein:2010tq}
A.~Gutlein {\em et~al.}, ``{Solar and atmospheric neutrinos: Background sources
  for the direct dark matter search},'' {\em Astropart. Phys.}, vol.~34,
  pp.~90--96, arXiv:1003.5530 [hep-ph].

\bibitem{Gelmini:2018ogy}
G.~B. Gelmini, V.~Takhistov, and S.~J. Witte, ``{Casting a Wide Signal Net with
  Future Direct Dark Matter Detection Experiments},'' {\em JCAP}, vol.~1807,
  no.~07, p.~009, arXiv:1804.01638 [hep-ph].

\bibitem{OHare:2021utq}
C.~A.~J. O'Hare, ``{New Definition of the Neutrino Floor for Direct Dark Matter
  Searches},'' {\em Phys. Rev. Lett.}, vol.~127, no.~25, p.~251802,
  arXiv:2109.03116 [hep-ph].

\bibitem{Dent:2016wor}
J.~B. Dent, B.~Dutta, J.~L. Newstead, and L.~E. Strigari, ``{Dark matter, light
  mediators, and the neutrino floor},'' {\em Phys. Rev. D}, vol.~95, no.~5,
  p.~051701, arXiv:1607.01468 [hep-ph].

\bibitem{Davis:2014ama}
J.~H. Davis, ``{Dark Matter vs. Neutrinos: The effect of astrophysical
  uncertainties and timing information on the neutrino floor},'' {\em JCAP},
  vol.~1503, p.~012, arXiv:1412.1475 [hep-ph].

\bibitem{Ruppin:2014bra}
F.~Ruppin, J.~Billard, E.~Figueroa-Feliciano, and L.~Strigari,
  ``{Complementarity of dark matter detectors in light of the neutrino
  background},'' {\em Phys. Rev. D}, vol.~90, no.~8, p.~083510, arXiv:1408.3581
  [hep-ph].

\bibitem{Gaspert:2021gyj}
A.~Gaspert, P.~Giampa, and D.~E. Morrissey, ``{Neutrino backgrounds in future
  liquid noble element dark matter direct detection experiments},'' {\em Phys.
  Rev. D}, vol.~105, no.~3, p.~035020, arXiv:2108.03248 [hep-ph].

\bibitem{OHare:2020lva}
C.~A.~J. O'Hare, ``{Can we overcome the neutrino floor at high masses?},'' {\em
  Phys. Rev. D}, vol.~102, no.~6, p.~063024, arXiv:2002.07499 [astro-ph.CO].

\bibitem{OHare:2015utx}
C.~A.~J. O'Hare, A.~M. Green, J.~Billard, E.~Figueroa-Feliciano, and L.~E.
  Strigari, ``{Readout strategies for directional dark matter detection beyond
  the neutrino background},'' {\em Phys. Rev. D}, vol.~92, no.~6, p.~063518,
  arXiv:1505.08061 [astro-ph.CO].

\bibitem{Grothaus:2014hja}
P.~Grothaus, M.~Fairbairn, and J.~Monroe, ``{Directional Dark Matter Detection
  Beyond the Neutrino Bound},'' {\em Phys. Rev. D}, vol.~90, no.~5, p.~055018,
  arXiv:1406.5047 [hep-ph].

\bibitem{Mayet:2016zxu}
F.~Mayet {\em et~al.}, ``{A review of the discovery reach of directional Dark
  Matter detection},'' {\em Phys. Rept.}, vol.~627, pp.~1--49, arXiv:1602.03781
  [astro-ph.CO].

\bibitem{OHare:2017rag}
C.~A.~J. O'Hare, B.~J. Kavanagh, and A.~M. Green, ``{Time-integrated
  directional detection of dark matter},'' {\em Phys. Rev. D}, vol.~96, no.~8,
  p.~083011, arXiv:1708.02959 [astro-ph.CO].

\bibitem{Franarin:2016ppr}
T.~Franarin and M.~Fairbairn, ``{Reducing the solar neutrino background in dark
  matter searches using polarized helium-3},'' {\em Phys. Rev. D}, vol.~94,
  no.~5, p.~053004, arXiv:1605.08727 [hep-ph].

\bibitem{Vahsen:2021gnb}
S.~E. Vahsen, C.~A.~J. O'Hare, and D.~Loomba, ``{Directional recoil
  detection},'' {\em Ann. Rev. Nucl. Part. Sci.}, vol.~71, pp.~189--224,
  arXiv:2102.04596 [physics.ins-det].

\bibitem{Sassi:2021umf}
S.~Sassi, A.~Dinmohammadi, M.~Heikinheimo, N.~Mirabolfathi, K.~Nordlund,
  H.~Safari, and K.~Tuominen, ``{Solar neutrinos and dark matter detection with
  diurnal modulation},'' arXiv:2103.08511 [hep-ph].

\bibitem{akerib2020projected}
LUX-ZEPLIN Collaboration, D.~S. Akerib {\em et~al.}, ``{Projected WIMP
  sensitivity of the LUX-ZEPLIN dark matter experiment},'' {\em Phys. Rev. D},
  vol.~101, no.~5, p.~052002, arXiv:1802.06039 [astro-ph.IM].

\bibitem{Baudis:2013qla}
L.~Baudis, A.~Ferella, A.~Kish, A.~Manalaysay, T.~Marrodan~Undagoitia, and
  M.~Schumann, ``{Neutrino physics with multi-ton scale liquid xenon
  detectors},'' {\em JCAP}, vol.~01, p.~044, arXiv:1309.7024 [physics.ins-det].

\bibitem{Aalbers:2016jon}
DARWIN Collaboration, J.~Aalbers {\em et~al.}, ``{DARWIN: towards the ultimate
  dark matter detector},'' {\em JCAP}, vol.~1611, p.~017, arXiv:1606.07001
  [astro-ph.IM].

\bibitem{Newstead:2020fie}
J.~L. Newstead, R.~F. Lang, and L.~E. Strigari, ``{Atmospheric neutrinos in
  next-generation xenon and argon dark matter experiments},'' {\em Phys. Rev.
  D}, vol.~104, no.~11, p.~115022, arXiv:2002.08566 [astro-ph.CO].

\bibitem{Lang:2016zhv}
R.~F. Lang, C.~McCabe, S.~Reichard, M.~Selvi, and I.~Tamborra, ``{Supernova
  neutrino physics with xenon dark matter detectors: A timely perspective},''
  {\em Phys. Rev. D}, vol.~94, no.~10, p.~103009, arXiv:1606.09243
  [astro-ph.HE].

\bibitem{Raj:2019sci}
N.~Raj, ``{Neutrinos from Type Ia and failed core-collapse supernovae at dark
  matter detectors},'' {\em Phys. Rev. Lett.}, vol.~124, no.~14, p.~141802,
  arXiv:1907.05533 [hep-ph].

\bibitem{Suliga:2021hek}
A.~M. Suliga, J.~F. Beacom, and I.~Tamborra, ``{Towards Probing the Diffuse
  Supernova Neutrino Background in All Flavors},'' arXiv:2112.09168
  [astro-ph.HE].

\bibitem{supercdms_sensitivity_2017}
R.~{Agnese}, A.~J. {Anderson}, T.~{Aramaki}, I.~{Arnquist}, W.~{Baker},
  D.~{Barker}, R.~{Basu Thakur}, D.~A. {Bauer}, A.~{Borgland}, M.~A. {Bowles},
  P.~L. {Brink}, R.~{Bunker}, B.~{Cabrera}, D.~O. {Caldwell}, R.~{Calkins},
  C.~{Cartaro}, D.~G. {Cerde{\~n}o}, H.~{Chagani}, Y.~{Chen}, J.~{Cooley},
  B.~{Cornell}, P.~{Cushman}, M.~{Daal}, P.~C.~F. {Di Stefano}, T.~{Doughty},
  L.~{Esteban}, S.~{Fallows}, E.~{Figueroa-Feliciano}, M.~{Fritts},
  G.~{Gerbier}, M.~{Ghaith}, G.~L. {Godfrey}, S.~R. {Golwala}, J.~{Hall}, H.~R.
  {Harris}, T.~{Hofer}, D.~{Holmgren}, Z.~{Hong}, E.~{Hoppe}, L.~{Hsu}, M.~E.
  {Huber}, V.~{Iyer}, D.~{Jardin}, A.~{Jastram}, M.~H. {Kelsey}, A.~{Kennedy},
  A.~{Kubik}, N.~A. {Kurinsky}, A.~{Leder}, B.~{Loer}, E.~{Lopez Asamar},
  P.~{Lukens}, R.~{Mahapatra}, V.~{Mandic}, N.~{Mast}, N.~{Mirabolfathi}, R.~A.
  {Moffatt}, J.~D. {Morales Mendoza}, J.~L. {Orrell}, S.~M. {Oser}, K.~{Page},
  W.~A. {Page}, R.~{Partridge}, M.~{Pepin}, A.~{Phipps}, S.~{Poudel},
  M.~{Pyle}, H.~{Qiu}, W.~{Rau}, P.~{Redl}, A.~{Reisetter}, A.~{Roberts}, A.~E.
  {Robinson}, H.~E. {Rogers}, T.~{Saab}, B.~{Sadoulet}, J.~{Sander},
  K.~{Schneck}, R.~W. {Schnee}, B.~{Serfass}, D.~{Speller}, M.~{Stein},
  J.~{Street}, H.~A. {Tanaka}, D.~{Toback}, R.~{Underwood}, A.~N. {Villano},
  B.~{von Krosigk}, B.~{Welliver}, J.~S. {Wilson}, D.~H. {Wright}, S.~{Yellin},
  J.~J. {Yen}, B.~A. {Young}, X.~{Zhang}, and X.~{Zhao}, ``{Projected
  sensitivity of the SuperCDMS SNOLAB experiment},'' {\em Phys. Rev. D},
  vol.~95, no.~8, pp.~082002/1--17, 2017.

\bibitem{supercdms_snowmass_2022}
{SuperCDMS Collaboration}, M.~{Al-Bakry}, I.~{Alkhatib}, D.~{Praia do Amaral},
  T.~{Aralis}, T.~{Aramaki}, I.~{Arnquist}, I.~{Ataee Langroudy},
  E.~{Azadbakht}, S.~{Banik}, C.~{Bathurst}, D.~{Bauer}, L.~{Bezerra},
  R.~{Bhattacharyya}, P.~{Brink}, R.~{Bunker}, B.~{Cabrera}, R.~{Calkins},
  R.~{Cameron}, C.~{Cartaro}, D.~{Cerdeno}, Y.-Y. {Chang}, M.~{Chaudhuri},
  R.~{Chen}, N.~{Chott}, J.~{Cooley}, H.~{Coombes}, J.~{Corbett}, P.~{Cushman},
  F.~{De Brienne}, S.~{Dharani}, M.~L. {di Vacri}, M.~{Diamond}, E.~{Fascione},
  E.~{Figueroa}, C.~{Fink}, K.~{Fouts}, M.~{Fritts}, G.~{Gerbier},
  R.~{Germond}, M.~{Ghaith}, S.~{Golwala}, J.~{Hall}, N.~{Hassan}, B.~{Hines},
  M.~{Hollister}, Z.~{Hong}, E.~{Hoppe}, L.~{Hsu}, M.~{Huber}, V.~{Iyer},
  D.~{Jardin}, A.~{Jastram}, V.~{Kashyap}, M.~{Kelsey}, A.~{Kubik},
  N.~{Kurinsky}, R.~{Lawrence}, M.~{Lee}, A.~{Li}, J.~{Liu}, Y.~{Liu},
  B.~{Loer}, P.~{Lukens}, D.~{MacFarlane}, R.~{Mahapatra}, V.~{Mandic},
  N.~{Mast}, A.~{Mayer}, H.~M.~z. {Theenhausen}, {\'E}.~{Michaud},
  E.~{Michielin}, N.~{Mirabolfathi}, B.~{Mohanty}, S.~{Nagorny}, J.~{Nelson},
  H.~{Neog}, V.~{Novati}, J.~{Orrell}, M.~{Osborne}, S.~{Oser}, W.~{Page},
  R.~{Partridge}, D.~S. {Pedreros}, R.~{Podviianiuk}, F.~{Ponce}, S.~{Poudel},
  A.~{Pradeep}, M.~{Pyle}, W.~{Rau}, E.~{Reid}, T.~{Ren}, T.~{Reynolds},
  A.~{Roberts}, A.~{Robinson}, T.~{Saab}, B.~{Sadoulet}, I.~{Saikia},
  J.~{Sander}, A.~{Sattari}, B.~{Schmidt}, R.~{Schnee}, S.~{Scorza},
  B.~{Serfass}, S.~{Sharma Poudel}, D.~{Sincavage}, C.~{Stanford}, J.~{Street},
  H.~{Sun}, F.~{Thasrawala}, D.~{Toback}, R.~{Underwood}, S.~{Verma},
  A.~{Villano}, B.~{von Krosigk}, S.~{Watkins}, O.~{Wen}, Z.~{Williams},
  M.~{Wilson}, J.~{Winchell}, C.-p. {Wu}, K.~{Wykoff}, S.~{Yellin}, B.~{Young},
  T.~C. {Yu}, B.~{Zatschler}, S.~{Zatschler}, A.~{Zaytsev}, E.~{Zhang},
  L.~{Zheng}, and S.~{Zuber}, ``{A Strategy for Low-Mass Dark Matter Searches
  with Cryogenic Detectors in the SuperCDMS SNOLAB Facility}.'' to be submitted
  to arXiv, 2022.
\newblock to be submitted to the Proceedings of the US Community Study on the
  Future of Particle Physics (Snowmass 2021.

\bibitem{RES-NOVA:2021gqp}
RES-NOVA Collaboration, L.~Pattavina {\em et~al.}, ``{RES-NOVA sensitivity to
  core-collapse and failed core-collapse supernova neutrinos},'' {\em JCAP},
  vol.~10, p.~064, arXiv:2103.08672 [astro-ph.IM].

\bibitem{PhysRevD.61.101301}
D.~P. Snowden-Ifft, C.~J. Martoff, and J.~M. Burwell, ``{Low pressure negative
  ion drift chamber for dark matter search},'' {\em Phys. Rev. D}, vol.~61,
  p.~101301, 2000.

\bibitem{uhe_wp}
M.~Ackermann {\em et~al.}, ``{High-Energy and Ultra-High-Energy Neutrinos},''
  arXiv:2203.08096 [hep-ph].

\bibitem{ANTARES:2017bia}
ANTARES, IceCube, Pierre Auger, LIGO Scientific, Virgo Collaboration, A.~Albert
  {\em et~al.}, ``{Search for High-energy Neutrinos from Binary Neutron Star
  Merger GW170817 with ANTARES, IceCube, and the Pierre Auger Observatory},''
  {\em Astrophys. J. Lett.}, vol.~850, no.~2, p.~L35, arXiv:1710.05839
  [astro-ph.HE].

\bibitem{IceCube-Gen2:2020qha}
IceCube-Gen2 Collaboration, M.~G. Aartsen {\em et~al.}, ``{IceCube-Gen2: the
  window to the extreme Universe},'' {\em J. Phys. G}, vol.~48, no.~6,
  p.~060501, arXiv:2008.04323 [astro-ph.HE].

\bibitem{IceCube-Gen2:2021rkf}
{S. Hallmann, B. Clark, C. Glaser and D. Smith for the IceCube-Gen2
  Collaboration}, ``{Sensitivity studies for the IceCube-Gen2 radio array},''
  {\em PoS}, vol.~ICRC2021, p.~1183, arXiv:2107.08910 [astro-ph.HE].

\bibitem{IceCubePointSource}
``{All-sky point-source IceCube data: years 2010-2012}.'' \url{
  https://icecube.wisc.edu/data-releases/2018/10/all-sky-point-source-icecube-data-years-2010-2012/}.
\newblock Accessed: 2021-02-16.

\bibitem{IceCube:2016tpw}
IceCube Collaboration, M.~G. Aartsen {\em et~al.}, ``{All-sky Search for
  Time-integrated Neutrino Emission from Astrophysical Sources with 7 yr of
  IceCube Data},'' {\em Astrophys. J.}, vol.~835, no.~2, p.~151,
  arXiv:1609.04981 [astro-ph.HE].

\bibitem{PUEO:2020bnn}
PUEO Collaboration, Q.~Abarr {\em et~al.}, ``{The Payload for Ultrahigh Energy
  Observations (PUEO): a white paper},'' {\em JINST}, vol.~16, no.~08,
  p.~P08035, arXiv:2010.02892 [astro-ph.IM].

\bibitem{Venters:2019xwi}
T.~M. Venters, M.~H. Reno, J.~F. Krizmanic, L.~A. Anchordoqui, C.~Gu\'epin, and
  A.~V. Olinto, ``{POEMMA's Target of Opportunity Sensitivity to Cosmic
  Neutrino Transient Sources},'' {\em Phys. Rev. D}, vol.~102, p.~123013,
  arXiv:1906.07209 [astro-ph.HE].

\bibitem{GRAND:2018iaj}
GRAND Collaboration, J.~\'Alvarez-Mu\~niz {\em et~al.}, ``{The Giant Radio
  Array for Neutrino Detection (GRAND): Science and Design},'' {\em Sci. China
  Phys. Mech. Astron.}, vol.~63, no.~1, p.~219501, arXiv:1810.09994
  [astro-ph.HE].

\bibitem{Kimura:2017kan}
S.~S. Kimura, K.~Murase, P.~M\'esz\'aros, and K.~Kiuchi, ``{High-Energy
  Neutrino Emission from Short Gamma-Ray Bursts: Prospects for Coincident
  Detection with Gravitational Waves},'' {\em Astrophys. J. Lett.}, vol.~848,
  no.~1, p.~L4, arXiv:1708.07075 [astro-ph.HE].

\bibitem{vanVliet:2019nse}
A.~van Vliet, R.~Alves~Batista, and J.~R. Hörandel, ``{Determining the
  fraction of cosmic-ray protons at ultrahigh energies with cosmogenic
  neutrinos},'' {\em Phys. Rev. D}, vol.~100, no.~2, p.~021302,
  arXiv:1901.01899 [astro-ph.HE].

\bibitem{Muzio:2021zud}
M.~S. Muzio, G.~R. Farrar, and M.~Unger, ``{Probing the environments
  surrounding ultrahigh energy cosmic ray accelerators and their implications
  for astrophysical neutrinos},'' {\em Phys. Rev. D}, vol.~105, no.~2,
  p.~023022, arXiv:2108.05512 [astro-ph.HE].

\bibitem{Fang:2017zjf}
K.~Fang and K.~Murase, ``{Linking High-Energy Cosmic Particles by Black Hole
  Jets Embedded in Large-Scale Structures},'' {\em Nature Phys.}, vol.~14,
  no.~4, p.~396, arXiv:1704.00015 [astro-ph.HE].

\bibitem{Biehl:2017hnb}
D.~Biehl, D.~Boncioli, C.~Lunardini, and W.~Winter, ``{Tidally disrupted stars
  as a possible origin of both cosmic rays and neutrinos at the highest
  energies},'' {\em Sci. Rep.}, vol.~8, no.~1, p.~10828, arXiv:1711.03555
  [astro-ph.HE].

\bibitem{IceCube:2021uhz}
IceCube Collaboration, R.~Abbasi {\em et~al.}, ``{Improved Characterization of
  the Astrophysical Muon-Neutrino Flux with 9.5 Years of IceCube Data},''
  arXiv:2111.10299 [astro-ph.HE].

\bibitem{IceCube:2020acn}
IceCube Collaboration, M.~G. Aartsen {\em et~al.}, ``{Characteristics of the
  diffuse astrophysical electron and tau neutrino flux with six years of
  IceCube high energy cascade data},'' {\em Phys. Rev. Lett.}, vol.~125,
  no.~12, p.~121104, arXiv:2001.09520 [astro-ph.HE].

\bibitem{PierreAuger:2019ens}
Pierre Auger Collaboration, A.~Aab {\em et~al.}, ``{Probing the origin of
  ultra-high-energy cosmic rays with neutrinos in the EeV energy range using
  the Pierre Auger Observatory},'' {\em JCAP}, vol.~10, p.~022,
  arXiv:1906.07422 [astro-ph.HE].

\bibitem{ARA}
ARA Collaboration, P.~Allison {\em et~al.}, ``{Constraints on the diffuse flux
  of ultrahigh energy neutrinos from four years of Askaryan Radio Array data in
  two stations},'' {\em Phys. Rev. D}, vol.~102, no.~4, p.~043021,
  arXiv:1912.00987 [astro-ph.HE].

\bibitem{Anker:2019rzo}
A.~Anker {\em et~al.}, ``{A search for cosmogenic neutrinos with the ARIANNA
  test bed using 4.5 years of data},'' {\em JCAP}, vol.~03, p.~053,
  arXiv:1909.00840 [astro-ph.IM].

\bibitem{Gorham:2019guw}
ANITA Collaboration, P.~Gorham {\em et~al.}, ``{Constraints on the
  ultrahigh-energy cosmic neutrino flux from the fourth flight of ANITA},''
  {\em Phys. Rev. D}, vol.~99, no.~12, p.~122001, arXiv:1902.04005
  [astro-ph.HE].

\bibitem{IceCube:2016zyt}
IceCube Collaboration, M.~G. Aartsen {\em et~al.}, ``{The IceCube Neutrino
  Observatory: Instrumentation and Online Systems},'' {\em JINST}, vol.~12,
  no.~03, p.~P03012, arXiv:1612.05093 [astro-ph.IM].

\bibitem{Wissel:2020sec}
S.~Wissel {\em et~al.}, ``{Prospects for high-elevation radio detection of
  \ensuremath{>}100 PeV tau neutrinos},'' {\em JCAP}, vol.~11, p.~065,
  arXiv:2004.12718 [astro-ph.IM].

\bibitem{Romero-Wolf:2020pzh}
A.~Romero-Wolf {\em et~al.}, ``{An Andean Deep-Valley Detector for High-Energy
  Tau Neutrinos},'' in {\em {Latin American Strategy Forum for Research
  Infrastructure}}, 2 2020, 2002.06475.

\bibitem{Otte:2019aaf}
A.~N. Otte, A.~M. Brown, M.~Doro, A.~Falcone, J.~Holder, E.~Judd, P.~Kaaret,
  M.~Mariotti, K.~Murase, and I.~Taboada, ``{Trinity: An Air-Shower Imaging
  Instrument to detect Ultrahigh Energy Neutrinos},'' arXiv:1907.08727
  [astro-ph.IM].

\bibitem{Prohira:2019glh}
S.~Prohira {\em et~al.}, ``{Observation of Radar Echoes From High-Energy
  Particle Cascades},'' {\em Phys. Rev. Lett.}, vol.~124, no.~9, p.~091101,
  arXiv:1910.12830 [astro-ph.HE].

\bibitem{POEMMA:2020ykm}
POEMMA Collaboration, A.~V. Olinto {\em et~al.}, ``{The POEMMA (Probe of
  Extreme Multi-Messenger Astrophysics) observatory},'' {\em JCAP}, vol.~06,
  p.~007, arXiv:2012.07945 [astro-ph.IM].

\bibitem{RNO-G:2021hfx}
RNO-G Collaboration, J.~A. Aguilar {\em et~al.}, ``{The Radio Neutrino
  Observatory Greenland (RNO-G)},'' {\em PoS}, vol.~ICRC2021, p.~001, 2021.

\bibitem{Hooper:2007jr}
D.~Hooper, ``{Detecting MeV Gauge Bosons with High-Energy Neutrino
  Telescopes},'' {\em Phys.\ Rev.\ D}, vol.~75, p.~123001, 2007.

\bibitem{Ioka:2014kca}
K.~Ioka and K.~Murase, ``{IceCube PeV--EeV neutrinos and secret interactions of
  neutrinos},'' {\em PTEP}, vol.~2014, no.~6, p.~061E01, arXiv:1404.2279
  [astro-ph.HE].

\bibitem{Ng:2014pca}
K.~C.~Y. Ng and J.~F. Beacom, ``{Cosmic neutrino cascades from secret neutrino
  interactions},'' {\em Phys. Rev. D}, vol.~90, no.~6, p.~065035,
  arXiv:1404.2288 [astro-ph.HE].
\newblock [Erratum: Phys.Rev.D 90, 089904 (2014)].

\bibitem{Dhuria:2017ihq}
M.~Dhuria and V.~Rentala, ``{PeV scale Supersymmetry breaking and the IceCube
  neutrino flux},'' {\em JHEP}, vol.~09, p.~004, arXiv:1712.07138 [hep-ph].

\bibitem{Bustamante:2020mep}
M.~Bustamante, C.~Rosenstrøm, S.~Shalgar, and I.~Tamborra, ``{Bounds on secret
  neutrino interactions from high-energy astrophysical neutrinos},'' {\em
  Phys.\ Rev.\ D}, vol.~101, no.~12, p.~123024, arXiv:2001.04994 [astro-ph.HE].

\bibitem{Creque-Sarbinowski:2020qhz}
C.~Creque-Sarbinowski, J.~Hyde, and M.~Kamionkowski, ``{Resonant neutrino
  self-interactions},'' {\em Phys. Rev. D}, vol.~103, no.~2, p.~023527,
  arXiv:2005.05332 [hep-ph].

\bibitem{Esteban:2021tub}
I.~Esteban, S.~Pandey, V.~Brdar, and J.~F. Beacom, ``{Probing secret
  interactions of astrophysical neutrinos in the high-statistics era},'' {\em
  Phys. Rev. D}, vol.~104, no.~12, p.~123014, arXiv:2107.13568 [hep-ph].

\bibitem{IceCube:2015gsk}
IceCube Collaboration, M.~G. Aartsen {\em et~al.}, ``{A combined
  maximum-likelihood analysis of the high-energy astrophysical neutrino flux
  measured with IceCube},'' {\em Astrophys. J.}, vol.~809, no.~1, p.~98,
  arXiv:1507.03991 [astro-ph.HE].

\bibitem{IceCube:2020wum}
IceCube Collaboration, R.~Abbasi {\em et~al.}, ``{The IceCube high-energy
  starting event sample: Description and flux characterization with 7.5 years
  of data},'' {\em Phys. Rev. D}, vol.~104, p.~022002, arXiv:2011.03545
  [astro-ph.HE].

\bibitem{Huang:2019etr}
Y.~Huang, H.~Li, and B.-Q. Ma, ``{Consistent Lorentz violation features from
  near-TeV IceCube neutrinos},'' {\em Phys. Rev. D}, vol.~99, no.~12,
  p.~123018, arXiv:1906.07329 [hep-ph].

\bibitem{Jacob:2006gn}
U.~Jacob and T.~Piran, ``{Neutrinos from gamma-ray bursts as a tool to explore
  quantum-gravity-induced Lorentz violation},'' {\em Nature Phys.}, vol.~3,
  pp.~87--90, 2007.

\bibitem{Addazi:2021xuf}
A.~Addazi {\em et~al.}, ``{Quantum gravity phenomenology at the dawn of the
  multi-messenger era -- A review},'' arXiv:2111.05659 [hep-ph].

\bibitem{Li:2016kra}
S.~W. Li, M.~Bustamante, and J.~F. Beacom, ``{Echo Technique to Distinguish
  Flavors of Astrophysical Neutrinos},'' {\em Phys. Rev. Lett.}, vol.~122,
  no.~15, p.~151101, arXiv:1606.06290 [astro-ph.HE].

\bibitem{Murase:2016gly}
K.~Murase and E.~Waxman, ``{Constraining High-Energy Cosmic Neutrino Sources:
  Implications and Prospects},'' {\em Phys. Rev. D}, vol.~94, no.~10,
  p.~103006, arXiv:1607.01601 [astro-ph.HE].

\bibitem{Ahlers:2014ioa}
M.~Ahlers and F.~Halzen, ``{Pinpointing Extragalactic Neutrino Sources in Light
  of Recent IceCube Observations},'' {\em Phys.Rev.}, vol.~D90, p.~043005,
  arXiv:1406.2160 [astro-ph.HE].

\bibitem{Fang:2016hop}
K.~Fang, K.~Kotera, M.~C. Miller, K.~Murase, and F.~Oikonomou, ``{Identifying
  Ultrahigh-Energy Cosmic-Ray Accelerators with Future Ultrahigh-Energy
  Neutrino Detectors},'' {\em JCAP}, vol.~12, p.~017, arXiv:1609.08027
  [astro-ph.HE].

\bibitem{Bartos:2016wud}
I.~Bartos, M.~Ahrens, C.~Finley, and S.~Marka, ``{Prospects of Establishing the
  Origin of Cosmic Neutrinos using Source Catalogs},'' {\em Phys. Rev. D},
  vol.~96, no.~2, p.~023003, arXiv:1611.03861 [astro-ph.HE].

\bibitem{Bartos:2021tok}
I.~Bartos, D.~Veske, M.~Kowalski, Z.~Marka, and S.~Marka, ``{The IceCube Pie
  Chart: Relative Source Contributions to the Cosmic Neutrino Flux},'' {\em
  Astrophys. J.}, vol.~921, no.~1, p.~45, arXiv:2105.03792 [astro-ph.HE].

\bibitem{IceCube:2021jhz}
IceCube Collaboration, R.~Abbasi {\em et~al.}, ``{Measuring total neutrino
  cross section with IceCube at intermediate energies ($\sim$100 GeV to a few
  TeV)},'' {\em PoS}, vol.~ICRC2021, p.~1132, arXiv:2107.09764 [astro-ph.HE].

\bibitem{Bustamante:2017xuy}
M.~Bustamante and A.~Connolly, ``{Extracting the Energy-Dependent
  Neutrino-Nucleon Cross Section above 10 TeV Using IceCube Showers},'' {\em
  Phys.\ Rev.\ Lett.}, vol.~122, no.~4, p.~041101, arXiv:1711.11043
  [astro-ph.HE].

\bibitem{IceCube:2017roe}
IceCube Collaboration, M.~G. Aartsen {\em et~al.}, ``{Measurement of the
  multi-TeV neutrino cross section with IceCube using Earth absorption},'' {\em
  Nature}, vol.~551, pp.~596--600, arXiv:1711.08119 [hep-ex].

\bibitem{Denton:2020jft}
P.~B. Denton and Y.~Kini, ``{Ultra-High-Energy Tau Neutrino Cross Sections with
  GRAND and POEMMA},'' {\em Phys. Rev. D}, vol.~102, p.~123019,
  arXiv:2007.10334 [astro-ph.HE].

\bibitem{Connolly:2011vc}
A.~Connolly, R.~S. Thorne, and D.~Waters, ``{Calculation of High Energy
  Neutrino-Nucleon Cross Sections and Uncertainties Using the MSTW Parton
  Distribution Functions and Implications for Future Experiments},'' {\em Phys.
  Rev. D}, vol.~83, p.~113009, arXiv:1102.0691 [hep-ph].

\bibitem{IceCube:2018pgc}
IceCube Collaboration, M.~G. Aartsen {\em et~al.}, ``{Measurements using the
  inelasticity distribution of multi-TeV neutrino interactions in IceCube},''
  {\em Phys. Rev. D}, vol.~99, no.~3, p.~032004, arXiv:1808.07629 [hep-ex].

\bibitem{Safa:2019ege}
I.~Safa, A.~Pizzuto, C.~A. Argüelles, F.~Halzen, R.~Hussain, A.~Kheirandish,
  and J.~Vandenbroucke, ``{Observing EeV neutrinos through Earth: GZK and the
  anomalous ANITA events},'' {\em JCAP}, vol.~01, p.~012, arXiv:1909.10487
  [hep-ph].

\bibitem{KM3Net:2016zxf}
KM3Net Collaboration, S.~Adrian-Martinez {\em et~al.}, ``{Letter of intent for
  KM3NeT 2.0},'' {\em J. Phys. G}, vol.~43, no.~8, p.~084001, arXiv:1601.07459
  [astro-ph.IM].

\bibitem{Avrorin:2019dli}
Baikal-GVD Collaboration, A.~Avrorin {\em et~al.}, ``{Neutrino Telescope in
  Lake Baikal: Present and Future},'' {\em PoS}, vol.~ICRC2019, p.~1011,
  arXiv:1908.05427 [astro-ph.HE].

\bibitem{P-ONE:2020ljt}
P-ONE Collaboration, M.~Agostini {\em et~al.}, ``{The Pacific Ocean Neutrino
  Experiment},'' {\em Nature Astron.}, vol.~4, no.~10, pp.~913--915,
  arXiv:2005.09493 [astro-ph.HE].

\bibitem{Askaryan:1961pfb}
G.~A. Askar'yan, ``{Excess negative charge of an electron-photon shower and its
  coherent radio emission},'' {\em Zh. Eksp. Teor. Fiz.}, vol.~41,
  pp.~616--618, 1961.

\bibitem{Anchordoqui:2019omw}
L.~A. Anchordoqui {\em et~al.}, ``{Performance and science reach of the Probe
  of Extreme Multimessenger Astrophysics for ultrahigh-energy particles},''
  {\em Phys. Rev. D}, vol.~101, no.~2, p.~023012, arXiv:1907.03694
  [astro-ph.HE].

\bibitem{Garcia-Fernandez:2020dhb}
D.~Garc\'\i{}a-Fern\'andez, A.~Nelles, and C.~Glaser, ``{Signatures of
  secondary leptons in radio-neutrino detectors in ice},'' {\em Phys. Rev. D},
  vol.~102, no.~8, p.~083011, arXiv:2003.13442 [astro-ph.HE].

\bibitem{Lai:2013kja}
K.-C. Lai, C.-C. Chen, and P.~Chen, ``{The Strategy of Discrimination between
  Flavors for Detection of Cosmogenic Neutrinos},'' {\em Nucl. Phys. B Proc.
  Suppl.}, vol.~246-247, p.~95, arXiv:1303.1949 [hep-ph].

\bibitem{Gerhardt:2010bj}
L.~Gerhardt and S.~R. Klein, ``{Electron and Photon Interactions in the Regime
  of Strong LPM Suppression},'' {\em Phys. Rev. D}, vol.~82, p.~074017,
  arXiv:1007.0039 [hep-ph].

\bibitem{Allison:2014kha}
ARA Collaboration, P.~Allison {\em et~al.}, ``{First Constraints on the
  Ultra-High Energy Neutrino Flux from a Prototype Station of the Askaryan
  Radio Array},'' {\em Astropart. Phys.}, vol.~70, p.~62, arXiv:1404.5285
  [astro-ph.HE].

\bibitem{ARIANNA:2019scz}
ARIANNA Collaboration, A.~Anker {\em et~al.}, ``{Targeting ultra-high energy
  neutrinos with the ARIANNA experiment},'' {\em Adv. Space Res.}, vol.~64,
  pp.~2595--2609, arXiv:1903.01609 [astro-ph.IM].

\bibitem{Anker:2020lre}
A.~Anker {\em et~al.}, ``{White Paper: ARIANNA-200 high energy neutrino
  telescope},'' arXiv:2004.09841 [astro-ph.IM].

\bibitem{Aguilar:2019jay}
J.~Aguilar {\em et~al.}, ``{The Next-Generation Radio Neutrino Observatory --
  Multi-Messenger Neutrino Astrophysics at Extreme Energies},''
  arXiv:1907.12526 [astro-ph.HE].

\bibitem{deVries:2013qwa}
K.~D. de~Vries, K.~Hanson, and T.~Meures, ``{On the feasibility of RADAR
  detection of high-energy neutrino-induced showers in ice},'' {\em Astropart.
  Phys.}, vol.~60, p.~25, arXiv:1312.4331 [astro-ph.HE].

\bibitem{Prohira:2017nyr}
S.~Prohira and D.~Besson, ``{Particle-level model for radar based detection of
  high-energy neutrino cascades},'' {\em Nucl. Instrum. Meth. A}, vol.~922,
  p.~161, arXiv:1710.02883 [physics.ins-det].

\bibitem{Fargion:2000iz}
D.~Fargion, ``{Discovering Ultra High Energy Neutrinos by Horizontal and Upward
  $\tau$ Air-Showers: Evidences in Terrestrial Gamma Flashes?},'' {\em
  Astrophys. J.}, vol.~570, p.~909, 2002.

\bibitem{Abbasi:2019fmh}
Telescope Array Collaboration, R.~U. Abbasi {\em et~al.}, ``{Search for
  Ultra-High-Energy Neutrinos with the Telescope Array Surface Detector},''
  {\em J. Exp. Theor. Phys.}, vol.~131, no.~2, pp.~255--264, arXiv:1905.03738
  [astro-ph.HE].

\bibitem{Vargas:2016hcp}
H.~León~Vargas, A.~Sandoval, E.~Belmont, and R.~Alfaro, ``{Capability of the
  HAWC Gamma-Ray Observatory for the Indirect Detection of Ultrahigh-Energy
  Neutrinos},'' {\em Adv.\ Astron.}, vol.~2017, p.~1932413, arXiv:1610.04820
  [astro-ph.IM].

\bibitem{Nam:2016cib}
J.~Nam {\em et~al.}, ``{Design and implementation of the TAROGE experiment},''
  {\em Int. J. Mod. Phys. D}, vol.~25, no.~13, p.~1645013, 2016.

\bibitem{Nam:2020hng}
J.~Nam {\em et~al.}, ``{High-elevation synoptic radio array for detection of
  upward moving air-showers, deployed in the Antarctic mountains},'' {\em PoS},
  vol.~ICRC2019, p.~967, 2020.

\bibitem{Deaconu:2019rdx}
C.~Deaconu {\em et~al.}, ``{Searches for Ultra-High Energy Neutrinos with
  ANITA},'' {\em PoS}, vol.~ICRC2019, p.~867, arXiv:1908.00923 [astro-ph.HE].

\bibitem{Aramo:2004pr}
C.~Aramo, A.~Insolia, A.~Leonardi, G.~Miele, L.~Perrone, O.~Pisanti, and
  D.~Semikoz, ``{Earth-skimming UHE Tau neutrinos at the fluorescence detector
  of Pierre Auger observatory},'' {\em Astropart.\ Phys.}, vol.~23, p.~65,
  2005.

\bibitem{Gora:2014lya}
D.~G\'ora, E.~Bernardini, and A.~Kappes, ``{Searching for tau neutrinos with
  Cherenkov telescopes},'' {\em Astropart. Phys.}, vol.~61, p.~12,
  arXiv:1402.4243 [astro-ph.IM].

\bibitem{Gora:2016mmy}
D.~Góra and E.~Bernardini, ``{Detection of tau neutrinos by Imaging Air
  Cherenkov Telescopes},'' {\em Astropart. Phys.}, vol.~82, p.~77,
  arXiv:1606.01676 [astro-ph.IM].

\bibitem{MAGIC:2018gza}
MAGIC Collaboration, M.~L. Ahnen {\em et~al.}, ``{Limits on the flux of tau
  neutrinos from 1 PeV to 3 EeV with the MAGIC telescopes},'' {\em Astropart.
  Phys.}, vol.~102, pp.~77--88, arXiv:1805.02750 [astro-ph.IM].

\bibitem{Fiorillo:2020xst}
D.~F. Fiorillo, G.~Miele, and O.~Pisanti, ``{Tau Neutrinos with Cherenkov
  Telescope Array},'' arXiv:2007.13423 [hep-ph].

\bibitem{Weekes:1989tc}
T.~C. Weekes {\em et~al.}, ``{Observation of TeV gamma rays from the Crab
  nebula using the atmospheric Cerenkov imaging technique},'' {\em Astrophys.
  J.}, vol.~342, pp.~379--395, 1989.

\bibitem{Otte:2018uxj}
A.~N. Otte, ``{Studies of an air-shower imaging system for the detection of
  ultrahigh-energy neutrinos},'' {\em Phys. Rev. D}, vol.~99, no.~8, p.~083012,
  arXiv:1811.09287 [astro-ph.IM].

\bibitem{Sasaki:2014mwa}
M.~Sasaki and G.~W.-S. Hou, ``{Neutrino Telescope Array Letter of Intent: A
  Large Array of High Resolution Imaging Atmospheric Cherenkov and Fluorescence
  Detectors for Survey of Air-showers from Cosmic Tau Neutrinos in the PeV-EeV
  Energy Range},'' arXiv:1408.6244 [astro-ph.IM].

\bibitem{Adams:2017fjh}
J.~H. Adams {\em et~al.}, ``{White paper on {EUSO-SPB2}},'' arXiv:1703.04513
  [astro-ph.HE].

\bibitem{Aad:2008zzm}
ATLAS Collaboration, G.~Aad {\em et~al.}, ``{The ATLAS Experiment at the CERN
  Large Hadron Collider},'' {\em JINST}, vol.~3, p.~S08003, 2008.

\bibitem{Anchordoqui:2021ghd}
L.~A. Anchordoqui {\em et~al.}, ``{The Forward Physics Facility: Sites,
  Experiments, and Physics Potential},'' arXiv:2109.10905 [hep-ph].

\bibitem{Feng:2022inv}
J.~L. Feng {\em et~al.}, ``{The Forward Physics Facility at the High-Luminosity
  LHC},'' arXiv:2203.05090 [hep-ex].

\bibitem{FASER:2018bac}
FASER Collaboration, A.~Ariga {\em et~al.}, ``{Technical Proposal for FASER:
  ForwArd Search ExpeRiment at the LHC},'' arXiv:1812.09139 [physics.ins-det].

\bibitem{snomassfaserdoc}
H.~Abreu {\em et~al.}, ``Neutrino/dark particle detectors for the hl-lhc
  forward beam; snowmass 2021 loi,'' 2021.

\bibitem{fpfworkshops}
``Forward physics facility workshops.'' https://indico.cern.ch/category/13966/.

\bibitem{FASER:2021mtu}
FASER Collaboration, H.~Abreu {\em et~al.}, ``{First neutrino interaction
  candidates at the LHC},'' {\em Phys. Rev. D}, vol.~104, no.~9, p.~L091101,
  arXiv:2105.06197 [hep-ex].

\bibitem{DUNE:2018tke}
DUNE Collaboration, B.~Abi {\em et~al.}, ``{The DUNE Far Detector Interim
  Design Report Volume 1: Physics, Technology and Strategies},''
  arXiv:1807.10334 [physics.ins-det].

\bibitem{Mann:1973pr}
W.~A. Mann {\em et~al.}, ``{Study of the reaction $\nu n\rightarrow \mu^-
  p$},'' {\em Phys. Rev. Lett.}, vol.~31, pp.~844--847, 1973.

\bibitem{Barish:1977qk}
S.~J. Barish {\em et~al.}, ``{Study of Neutrino Interactions in Hydrogen and
  Deuterium. 1. Description of the Experiment and Study of the Reaction $\nu d
  \rightarrow \mu^- p p_s$},'' {\em Phys. Rev.}, vol.~D16, p.~3103, 1977.

\bibitem{Miller:1982qi}
K.~L. Miller {\em et~al.}, ``{Study of the reaction $\nu_\mu d\rightarrow\mu^-
  p p_s$},'' {\em Phys. Rev.}, vol.~D26, pp.~537--542, 1982.

\bibitem{Baker:1981su}
N.~J. Baker, A.~M. Cnops, P.~L. Connolly, S.~A. Kahn, H.~G. Kirk, M.~J.
  Murtagh, R.~B. Palmer, N.~P. Samios, and M.~Tanaka, ``{Quasielastic Neutrino
  Scattering: A Measurement of the Weak Nucleon Axial Vector Form-Factor},''
  {\em Phys. Rev.}, vol.~D23, pp.~2499--2505, 1981.

\bibitem{Kitagaki:1983px}
T.~Kitagaki {\em et~al.}, ``{High-Energy Quasielastic $\nu_\mu
  n\rightarrow\mu^- p$ Scattering in Deuterium},'' {\em Phys. Rev.}, vol.~D28,
  pp.~436--442, 1983.

\bibitem{Wilkinson:2014yfa}
C.~Wilkinson, P.~Rodrigues, S.~Cartwright, L.~Thompson, and K.~McFarland,
  ``{Reanalysis of bubble chamber measurements of muon-neutrino induced single
  pion production},'' {\em Phys. Rev. D}, vol.~90, no.~11, p.~112017,
  arXiv:1411.4482 [hep-ex].

\bibitem{Lu:2015hea}
X.~G. Lu, D.~Coplowe, R.~Shah, G.~Barr, D.~Wark, and A.~Weber,
  ``{Reconstruction of Energy Spectra of Neutrino Beams Independent of Nuclear
  Effects},'' {\em Phys. Rev. D}, vol.~92, no.~5, p.~051302, arXiv:1507.00967
  [hep-ex].

\bibitem{Hamacher-Baumann:2020ogq}
P.~Hamacher-Baumann, X.~Lu, and J.~Mart{\'i}n-Albo, ``{Neutrino-hydrogen
  interactions with a high-pressure time projection chamber},'' {\em Phys. Rev.
  D}, vol.~102, no.~3, p.~033005, arXiv:2005.05252 [physics.ins-det].

\bibitem{SAKAI198289}
Y.~Sakai, H.~B{\"{o}}ttcher, and W.~Schmidt, ``Excess electrons in liquid
  hydrogen, liquid neon, and liquid helium,'' {\em Journal of Electrostatics},
  vol.~12, pp.~89 -- 96, 1982.

\bibitem{HARRISON1971418}
H.~Harrison and B.~Springett, ``Electron mobility variation in dense hydrogen
  gas,'' {\em Chemical Physics Letters}, vol.~10, no.~4, pp.~418 -- 421, 1971.

\bibitem{DUNE:2021tad}
DUNE Collaboration, A.~Abed~Abud {\em et~al.}, ``{Deep Underground Neutrino
  Experiment (DUNE) Near Detector Conceptual Design Report},'' {\em
  Instruments}, vol.~5, no.~4, p.~31, arXiv:2103.13910 [physics.ins-det].

\bibitem{Bradner:1960iok}
H.~Bradner, ``Bubble chambers,'' {\em Annual Reviews of Nuclear Science},
  vol.~10, pp.~109--160,  [hep-ex/nuc-ex].

\bibitem{Dykes:1981dvb}
M.~D. \emph{et al.}, ``Holographic photography of bubble chamber tracks: A
  feasibility test,'' {\em Nuclear Instruments and Methods}, vol.~179,
  pp.~487--493, 1981.

\bibitem{Crabb:1997cy}
D.~G. Crabb and W.~Meyer, ``{Solid polarized targets for nuclear and particle
  physics experiments},'' {\em Ann. Rev. Nucl. Part. Sci.}, vol.~47,
  pp.~67--109, 1997.

\bibitem{Dael:1992xh}
A.~Dael, D.~Cacaut, H.~Desportes, R.~Duthil, B.~Gallet, F.~Kircher, C.~Lesmond,
  Y.~Pabot, and J.~Thinel, ``{A Superconducting 2.5-T high accuracy solenoid
  and a large 0.5-T dipole magnet for the SMC target},'' {\em IEEE Trans.
  Magnetics}, vol.~28, no.~1, pp.~560--563, 1992.

\bibitem{Berryhill:2019gan}
A.~Berryhill and J.~Ritter, ``{A Dual 5T Superconducting Magnet System for the
  Brookhaven National Lab Electron Beam Ion Source},'' {\em IEEE Trans. Appl.
  Supercond.}, vol.~29, no.~5, p.~4100104, 2019.

\bibitem{BUNYATOVA200422}
E.~Bunyatova, ``Free radicals and polarized targets,'' {\em Nuclear Instruments
  and Methods in Physics Research Section A: Accelerators, Spectrometers,
  Detectors and Associated Equipment}, vol.~526, no.~1, pp.~22--27, 2004.
\newblock Proceedings of the ninth International Workshop on Polarized Solid
  Targets and Techniques.

\bibitem{vandenBrandt:2002ab}
B.~van~den Brandt, P.~Hautle, J.~A. Konter, and E.~I. Bunyatova, ``{Progress in
  scintillating polarized targets for spin physics},'' in {\em {2nd
  International Symposium on the Gerasimov-Drell-Hearn Sum Rule and the Spin
  Structure of the Nucleon (GDH 2002)}}, pp.~183--187, 7 2002.

\bibitem{vandenBrandt:2002rs}
B.~van~den Brandt, E.~I. Bunyatova, P.~Hautle, J.~A. Konter, S.~Mango, and
  I.~B. Nemchonok, ``{An 'active' target for spin physics: Polarizing nuclei in
  plastic scintillators},'' {\em Czech. J. Phys.}, vol.~52, pp.~C689--C694,
  2002.

\end{thebibliography}




\end{document}